%% file: diss.tex
\documentclass[12pt]{book}

\usepackage{a4wide}
\usepackage{fancyhdr}
\usepackage{graphicx}
\usepackage[bookmarks]{hyperref}
\usepackage{amssymb,amsmath,color,comment,float}

\def\be{\begin{equation}}
\def\ee{\end{equation}}
\def\ba#1\ea{\begin{align}#1\end{align}}
\newcommand{\vs}{\nonumber\\}
\newcommand{\refeq}[1]{eq.~(\ref{eq:#1})}
\newcommand{\refeqs}[2]{eqs.~(\ref{eq:#1})--(\ref{eq:#2})}
\newcommand{\refEq}[1]{Eq.~(\ref{eq:#1})}
\newcommand{\refEqs}[2]{Eqs.~(\ref{eq:#1})--(\ref{eq:#2})}
\newcommand{\reffig}[1]{figure~\ref{fig:#1}}
\newcommand{\refFig}[1]{Figure~\ref{fig:#1}}
\newcommand{\refsec}[1]{section~\ref{sec:#1}}
\newcommand{\refapp}[1]{appendix~\ref{app:#1}}
\newcommand{\refchp}[1]{chapter~\ref{chp:#1}}
\newcommand{\reftab}[1]{table~\ref{tab:#1}}
\newcommand{\refTab}[1]{Table~\ref{tab:#1}}

\renewcommand{\chaptermark}[1]%
         {\markboth{\thechapter.\ #1}{}}
\renewcommand{\sectionmark}[1]%
         {\markright{\thesection\ #1}}
\newcommand{\LMUTitle}[9]{
  \thispagestyle{empty}
  \vspace*{\stretch{1}}
  {\parindent0cm
   \rule{\linewidth}{.7ex}}
  \begin{flushright}

    \vspace*{\stretch{1}}
    \sffamily\bfseries\Huge
    #1\\
    \vspace*{\stretch{1}}
    \sffamily\bfseries\large
    #2
    \vspace*{\stretch{1}}
  \end{flushright}
  \rule{\linewidth}{.7ex}
  \vspace*{\stretch{5}}
  \begin{center}
    \includegraphics[width=2in]{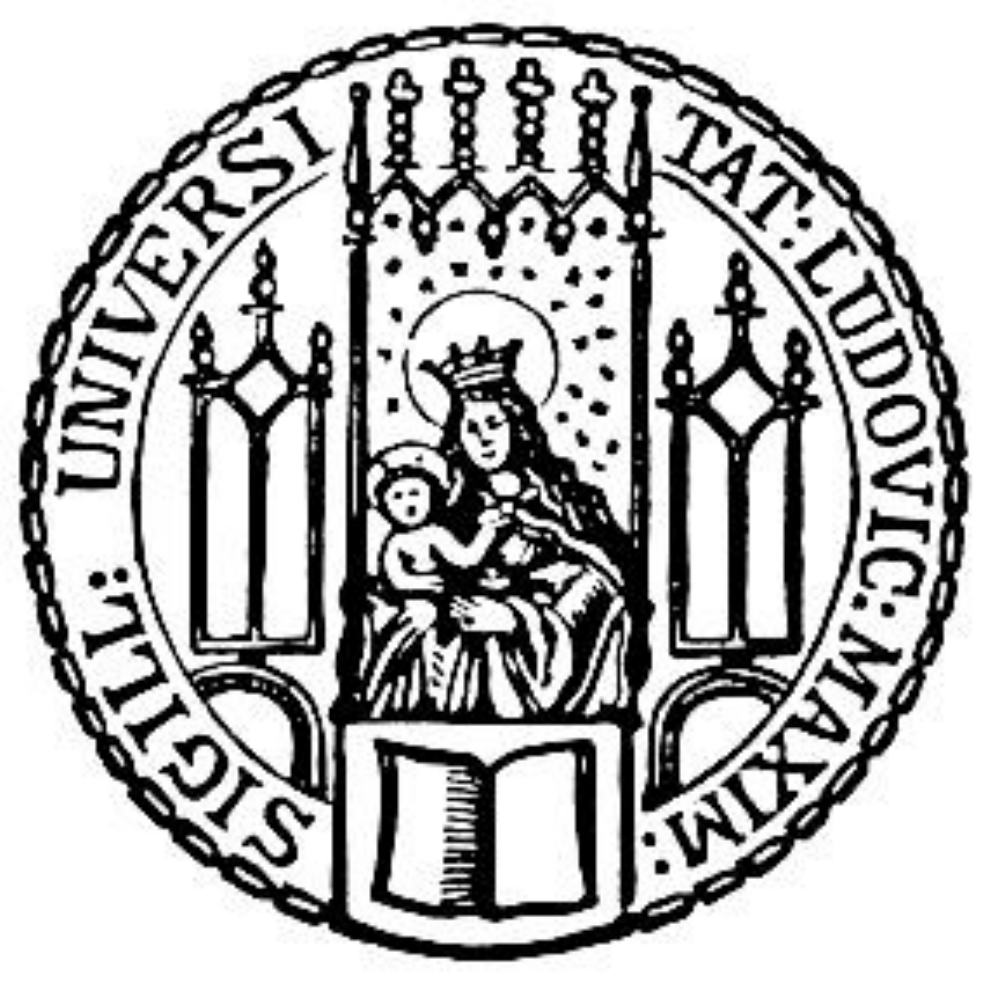}
  \end{center}
  \vspace*{\stretch{1}}
  \begin{center}\sffamily\LARGE{#5}\end{center}
  \newpage
  \thispagestyle{empty}

  \cleardoublepage
  \thispagestyle{empty}

  \vspace*{\stretch{1}}
  {\parindent0cm
  \rule{\linewidth}{.7ex}}
  \begin{flushright}
    \vspace*{\stretch{1}}
    \sffamily\bfseries\Huge
    #1\\
    \vspace*{\stretch{1}}
    \sffamily\bfseries\large
    #2
    \vspace*{\stretch{1}}
  \end{flushright}
  \rule{\linewidth}{.7ex}

  \vspace*{\stretch{3}}
  \begin{center}
    \Large Dissertation\\
    \Large an der #4\\
    \Large der Ludwig--Maximilians--Universit\"at\\
    \Large M\"unchen\\
    \vspace*{\stretch{1}}
    \Large vorgelegt von\\
    \Large #2\\
    \Large aus #3\\
    \vspace*{\stretch{2}}
    \Large M\"unchen, den #6
  \end{center}

  \newpage
  \thispagestyle{empty}

  \vspace*{\stretch{1}}

  \begin{flushleft}
    \large Erstgutachter:  #7 \\[1mm]
    \large Zweitgutachter: #8 \\[1mm]
    \large Tag der m\"undlichen Pr\"ufung: #9\\
  \end{flushleft}

  \cleardoublepage
}

\begin{document}

  \frontmatter

  \LMUTitle
      {Position-dependent power spectrum: a new
       observable in the large-scale structure}
      {Chi-Ting Chiang}                        
      {Taipei, Taiwan}                         
      {Fakult\"at f\"ur Physik}                
      {M\"unchen 2015}                         
      {Abgabedatum}                            
      {Prof. Dr. Eiichiro Komatsu}             
      {Prof. Dr. Jochen Weller}                
      {22 June 2015}                           

	\tableofcontents
	\markboth{CONTENTS}{CONTENTS}

	\listoffigures
	\markboth{LIST OF FIGURES}{LIST OF FIGURES}

	\listoftables
	\markboth{LIST OF TABLES}{LIST OF TABLES}
	\cleardoublepage

	\cleardoublepage
	\phantomsection
	\addcontentsline{toc}{chapter}{Zusammenfassung}
	\markboth{Zusammenfassung}{Zusammenfassung}
	\include{zusammenfassung_german}

	\cleardoublepage
	\phantomsection
	\addcontentsline{toc}{chapter}{Abstract}
	\markboth{ABSTRACT}{ABSTRACT}
	\include{zusammenfassung}

	\mainmatter\setcounter{page}{1}
	\include{kap_01}
	\include{kap_02}
	\include{kap_03}
	\include{kap_04}
	\include{kap_05}
	\include{kap_06}
	\include{kap_07}

	\begin{appendix}
	\include{anh_01}
	\include{anh_02}
	\include{anh_03}
	\include{anh_04}
	\end{appendix}

	\backmatter
	\cleardoublepage
	\phantomsection
	\addcontentsline{toc}{chapter}{Bibliography}
	\bibliographystyle{jkthesis}
	\bibliography{literatur}
	\markboth{}{}

	\cleardoublepage
	\phantomsection
	\addcontentsline{toc}{chapter}{Acknowledgment}
	\markboth{Acknowledgment}{Acknowledgment}
	\include{danksagung}

\end{document}

%% file: zusammenfassung_german.tex
\chapter*{Zusammenfassung}
In dieser Dissertation f\"uhre ich eine neue Observable der grossr\"aumigen
Struktur des Universums ein, das ortsabh\"angige Leistungsspektrum. Diese
Gr\"o{\ss}e bietet ein Mass f\"ur den ``gequetschten'' Limes der
Dreipunktfunktion (Bispektrum), das heisst, eine Wellenzahl ist wesentlich
kleiner als die beiden \"ubrigen.  Physikalisch beschreibt dieser Limes
der Dreipunktfunktion die Modulation des Leistungsspektrums auf kleinen
Skalen durch grossr\"aumige Moden.  Diese Modulation wird sowohl durch
die Schwerkraft bewirkt als auch (m\"oglicherweise) durch die kosmologische
Inflation im fr\"uhen Universum.

F\"ur die Messung teilen wir das Gesamtvolumen der Himmelsdurchmusterung,
oder kosmologischen Simulation, in Teilvolumina ein.  In jedem Teilvolumen
messen wir die \"Uberdichte relativ zur mittleren Dichte der Materie (oder
Anzahldichte der Galaxien) und das lokale Leistungsspektrum.  Anschliessend
messen wir die Korrelation zwischen \"Uberdichte und Leistungsspektrum.
Ich zeige, dass diese Korrelation einem Integral \"uber die Dreipunktfunktion
entspricht.  Wenn die Skala, an der das Leistungsspektrum ausgewertet wird
(inverse Wellenzahl, um genau zu sein), viel kleiner als die Gr\"o{\ss}e des
Teilvolumens ist, dann ist das Integral \"uber die Dreipunktfunktion vom
gequetschten Limes dominiert.

Um physikalisch zu verstehen, wie eine grossr\"aumige Dichtefluktuation das
lokale Leistungsspektrum beeinflusst, wenden wir das Bild vom ``unabh\"angigen Universum''
(``separate universe'') an. Im Kontext der allgemeinen Relativit\"atstheorie kann eine langwellige
Dichtefluktuation exakt durch eine Friedmann-Robertson-Walker-(FRW-)Raumzeit
beschrieben werden, deren Parameter sich von der ``wahren'' FRW-Raumzeit
unterscheiden und eindeutig von der Dichtefluktuation bestimmt werden. Die
Modulation des lokalen Leistungsspektrums kann dann durch die Strukturbildung
innerhalb der modifizierten FRW-Raumzeit beschrieben werden. Insbesondere
zeige ich, dass die Dreipunktfunktion im gequetschten Limes durch diesen
Ansatz einfacher und besser beschrieben wird als durch die herk\"ommliche
Herangehensweise mittels St\"orungstheorie.

Diese neue Observable ist nicht nur einfach zu interpretieren (sie stellt die
Antwort des lokalen Leistungsspektrums auf eine gro{\ss}skalige Dichtest\"orung
dar), sie erm\"oglicht zudem die komplexe Berechnung der vollen Dreipunktsfunktion
zu umgehen, weil das Leistungsspektrum genauso wie die mittlere Dichte wesentlich
leichter als die Dreipunktsfunktion zu bestimmen sind.

Anschlie{\ss}end wende ich die gleiche Methodik auf die Daten der
Himmelsdurchmustering SDSS-III Baryon Oscillation Spectroscopic Survey (BOSS)
an, insbesondere den Data Release 10 CMASS Galaxienkatalog.  Wie ich zeige,
stimmt das in den wirklichen Daten gemessene ortsabh\"angige Leistungsspektrum
mit den sogenannten ``mock'' (also simulierten)
Galaxienkatalogen \"uberein, die auf dem PTHalo-Algorithmus basieren und die
r\"aumlichde Verteilung der wirklichen Galaxien im statistischen Sinne m\"oglichst
genau beschreiben wollen.
Genauer gesagt, liegen die Daten innerhalb der Streuung, die das
ortsabh\"angige Leistungsspektrum zwischen den verschiedenen Realisierungen
von ``mock'' Katalogen aufweist.  Diese Streuung betr\"agt ca. 10\% des
Mittelewerts.  In Kombination mit dem (anisotropen) globalen Leistungsspektrum
der Galaxien sowie dem Signal im schwachen Gravitationslinseneffekt,
benutze ich diese 10\%-Messung des ortsabh\"angigen Leistungsspektrums,
um den quadratischen Bias-Parameter der von BOSS gemessenen Galaxien
zu bestimmen, mit dem Ergebnis $b_2 = 0.41 \pm 0.41$ (68\% Vertrauensintervall).

Schlie{\ss}lich verallgemeinern wir die Analyse der Antwort des lokalen
Leistungsspektrums auf eine H\"aufung von $m$ gro{\ss}r\"aumigen Wellenl\"angenmoden, wobei $m \le 3$.
In Analogie zum vorherigen Fall, kann die resultierende Modulation des
Leistungsspektrums mit der $m+2$-Punktskorrelationsfunktion im Limes
gequetschter Konfigurationen (so dass immer zwei Wellenl\"angen wesentlich
l\"anger sind als die anderen), gemittled \"uber die auftretenden Winkel,
in Verbindung gebracht werden. Mit Hilfe von Simulationen ``unabh\"angiger Universen'',
dass hei{\ss}t $N$-body-Simulationen in Anwesenheit von Dichtest\"orungen unendlicher
L\"ange, vergleichen wir unsere semianalytischen Modelle, die auf dem Bild der
unabh\"angigen Universen basieren, mit den vollst\"andig nichtlinearen Simulationen
bei bisher unerreichter Genauigkeit. Zudem testen wir die Annahme der gew\"ohnlichen
St\"orungstheorie, dass die nichtlineare N-Punktskorrelationsfunktion vollst\"andig
durch das lineare Leistungsspektrum bestimmt ist. Wir finden bereits Abweichungen
von 10\% bei Wellenzahlen von $k \simeq 0.2 - 0.5~h~{\rm Mpc}^{-1}$ f\"ur die Drei-
bis F\"unf-Punktskorrelationsfunktion bei Rotverschiebung $z = 0$. Dieses Ergebnis
deutet darauf hin, dass die gew\"ohnlichye St\"orungstheorie nicht ausreicht um die
Dynamik kollissionsloser Teilchen f\"ur Wellenzahlen gr\"o{\ss}er als diese korrekt
vorherzusagen, selbst wenn alle h\"oheren Ordnungen in die Berechnung mit einbezogen werden.

%% file: zusammenfassung.tex
\chapter*{Abstract}

We present a new observable, position-dependent power spectrum, to measure
the large-scale structure bispectrum in the so-called squeezed configuration,
where one wavenumber, say $k_3$, is much smaller than the other two, i.e.
$k_3\ll k_1\approx k_2$. The squeezed-limit bispectrum measures how the
small-scale power spectrum is modulated by a long-wavelength scalar density
fluctuation, and this modulation is due to gravitational evolution and also
possibly due to the inflationary physics.

We divide a survey volume into smaller subvolumes. We compute the local power
spectrum and the mean overdensity in each smaller subvolume, and then measure
the correlation between these two quantities. We show that this correlation
measures the integral of the bispectrum, which is dominated by the squeezed
configurations if the wavenumber of the local power spectrum is much larger
than the corresponding wavenumber of the size of the subvolumes. This integrated
bispectrum measures how small-scale power spectrum responds to a long-wavelength
mode.

To understand theoretically how the small-scale power spectrum is affected by
a long-wavelength overdensity gravitationally, we use the ``separate universe
picture.'' A long-wavelength overdensity compared to the scale of interest can
be absorbed into the change of the background cosmology, and then the small-scale
structure formation evolves in this modified cosmology. We show that this approach
models nonlinearity in the bispectrum better than the traditional approach based
on the perturbation theory.

Not only this new observable is straightforward to interpret (the response of
the small-scale power spectrum to a long-wavelength overdensity), but it also
sidesteps the complexity of the full bispectrum estimation because both power
spectrum and mean overdensity are significantly easier to estimate than the
full bispectrum.

We report on the first measurement of the bispectrum with the position-dependent
correlation function from the SDSS-III Baryon Oscillation Spectroscopic Survey
(BOSS) Data Release 10 CMASS sample. We detect the amplitude of the bispectrum
of the BOSS CMASS galaxies at $7.4\sigma$, and constrain their nonlinear bias
to be $b_2=0.41\pm0.41$ (68\% C.L.) combining our bispectrum measurement with
the anisotropic clustering and the weak lensing signal.

We finally generalize the study to the response of the small-scale power spectrum
to $m$ long-wavelength overdensities for $m\le3$. Similarly, this response can be
connected to the angle-average $(m+2)$-point function in the squeezed configurations
where two wavenumbers are much larger than the other ones. Using separate universe
simulations, i.e. $N$-body simulations performed in the presence of an infinitely
long-wavelength overdensity, we compare our semi-analytical models based on the
separate universe approach to the fully nonlinear simulations to unprecedented
accuracy. We also test the standard perturbation theory hypothesis that the nonlinear
$n$-point function is completely predicted by the linear power spectrum at the same
time. We find discrepancies of 10\% at $k\simeq0.2-0.5~h~{\rm Mpc}^{-1}$ for three-
to five-point functions at $z=0$. This result suggests that the standard perturbation
theory fails to describe the correct dynamics of collisionless particles beyond these
wavenumbers, even if it is calculated to all orders in perturbations.


%% file: kap_01.tex
{

\renewcommand{\v}[1]{\mathbf{#1}}
\newcommand{\vx}{\v{x}}
\newcommand{\vr}{\v{r}}
\newcommand{\vk}{\v{k}}
\newcommand{\vq}{\v{q}}

\newcommand{\hMpc}{~h^{-1}~{\rm Mpc}}
\newcommand{\ihMpc}{~h~{\rm Mpc}^{-1}}

\chapter{Introduction}
\section{Why study the position-dependent power spectrum of the large-scale structure?}
The standard cosmological paradigm has been well developed and tested by the
observations of the cosmic microwave background (CMB) and the large-scale
structure. The inhomogeneities seen in the universe originate from quantum
fluctuations in the early universe, and these quantum fluctuations were
stretched to macroscopic scales larger than the horizon during the cosmic
inflation \cite{guth:1981,sato:1981,albrecht/steinhardt:1982,linde:1982},
which is the early phase with exponential growth of the scale factor. After
the cosmic inflation, the hot Big-Bang universe expanded and cooled down,
and the macroscopic inhomogeneities entered into horizon and seeded all
the structures we observe today.

With the success of connecting the quantum fluctuations in the early universe
to the structures we see today, the big questions yet remain: What is the
physics behind inflation? Also, what is nature of dark energy, which causes
the accelerated expansion in the late-time universe (see \cite{frieman/turner/huterer:2008}
for a review)? As the standard cosmological paradigm passes almost all the
tests from the current observables, especially the two-point statistics of
CMB and galaxy surveys, it is necessary to go to higher order statistics to
obtain more information and critically test the current model. In particular,
the mode coupling between a long-wavelength scalar density fluctuation and
the small-scale structure formation receives much attention in the past few
years. This coupling is due to the nonlinear gravitational evolution (see
\cite{bernardeau/etal:2002} for a review), and possibly the inflationary
physics. Therefore, this provides a wonderful opportunity to test our
understanding of gravity, as well as to probe the properties of inflation.

Traditionally, the $n$-point function with $n>2$ is used to characterize the
mode coupling. Specifically, if one is interested in the coupling between one
long-wavelength mode and two short-wavelength modes, we measure the three-point
correlation function or its Fourier counterpart, the bispectrum, in the so-called
``squeezed configurations,'' in which one wavenumber, say $k_3$, is much smaller
than the other two, i.e. $k_3\ll k_1\approx k_2$.

In the simplest model for the primordial non-Gaussianity (see \cite{chen:2010}
for a review on the general primordial non-Gaussianities from various inflation
models), the primordial scalar potential is given by
\be
 \Phi(\vr)=\phi(\vr)+f_{\rm NL}\left[\phi^2(\vr)-\langle\phi^2(\vr)\rangle\right] ~,
\ee
where $\phi(\vr)$ is a Gaussian field and $f_{\rm NL}$ is a constant characterizing
the amplitude of the non-Gaussianity, which encodes the properties of inflation.
Note that $\langle\phi^2(\vr)\rangle$ assures $\langle\Phi(\vr)\rangle=0$. This
simple model is known as the local-type primordial non-Gaussianity because
$\Phi(\vr)$ depends locally on $\phi(\vr)$. The bispectrum of this local model is
\be
 B_{\Phi}(\vk_1,\vk_2,\vk_3)=2f_{\rm NL}\left[P_{\Phi}(k_1)P_{\Phi}(k_2)+{\rm 2~cyclic}\right] ~,
\ee
where $P_{\Phi}(k)\propto k^{n_s-4}$ is the power spectrum of the primordial scalar
potential and $n_s\simeq0.96$ is its spectral index \cite{komatsu/etal:2011}. We can
rewrite $B_{\Phi}$ of this local model by fixing one wavenumber, say $k_1$, as
\ba
 B_{\Phi}(\vk_1,\vk_2,\vk_3)\:&\propto k_1^{2(n_s-4)}
 \left[\left(\frac{k_2}{k_1}\right)^{n_s-4}+\left(\frac{k_3}{k_1}\right)^{n_s-4}
 +\left(\frac{k_2}{k_1}\right)^{n_s-4}\left(\frac{k_3}{k_1}\right)^{n_s-4}\right] \vs
 \:&\propto k_1^{2(n_s-4)}
 \left[\left(\frac{k_2}{k_1}\right)^{n_s-4}+\left(\frac{|\vk_1+\vk_2|}{k_1}\right)^{n_s-4}
 +\left(\frac{k_2}{k_1}\right)^{n_s-4}\left(\frac{|\vk_1+\vk_2|}{k_1}\right)^{n_s-4}\right] ~,
\ea
where $\vk_3=-\vk_1-\vk_2$ because of the assumption of homogeneity. $B_{\Phi}$
apparently peaks at $k_1\approx k_2\gg k_3\approx0$, so the local-type primordial
non-Gaussianity is the most prominent in the squeezed-limit bispectrum.

Constraining the physics of inflation using the squeezed-limit bispectrum of
CMB is a solved problem \cite{komatsu:2010}. With the Planck satellite, the
current constraint on the local-type primordial non-Gaussianity is $f_{\rm NL}=2.5\pm5.7$
(68\% C.L.) using the temperature data alone and $f_{\rm NL}=0.8\pm5.0$ using
the temperature and polarization data \cite{planck/nonG:2015}. These are close
to the best limits obtainable from CMB. To improve upon them, we must go beyond
CMB to the large-scale structure, where observations are done in three-dimensional
space (unlike CMB embedded on a two-dimensional sphere). Thus, in principle, the
large-scale structure contains more information to improve the constraint on the
physics of inflation.

Measuring the three-point function from the large-scale structure (e.g. distribution
of galaxies), however, is considerably more challenging compared to CMB. From the
measurement side, the three-point function measurements are computationally expensive.
In configuration space, the measurements rely on finding particle triplets with the
naive algorithm scaling as $N_{\rm par}^3$ where $N_{\rm par}$ is the number of
particles. Current galaxy redshift surveys contain roughly a million galaxies, and
we need 50 times as many random samples as the galaxies for characterizing the survey
window function accurately. Similarly, in Fourier space, the bispectrum measurements
require counting all possible triangle configurations formed by different Fourier
modes, which is also computationally expensive. From the modeling side, galaxy surveys
have more complicated survey selection function, which can bias the estimation (see
e.g. \cite{chiang/etal:2013}). Additionally, the nonlinear gravitational evolution
of matter density field and the complexity of galaxy formation make it challenging
to extract the primordial signal. The above difficulties explain why only few measurements
of the three-point function of the large-scale structure have been reported in the
literature \cite{scoccimarro/etal:2001,feldman/etal:2001,verde/etal:2002,kayo/etal:2004,
nishimichi/etal:2007,mcbride/etal:2011a,mcbride/etal:2011b,marin/etal:2013,gilmarin/etal:2014b,guo/etal:2015}.

Since our main interest is to measure the three-point function of the squeezed
configurations, there is a simpler way to sidestep all the above complexities of
the three-point function estimation. As we stated, in the presence of the mode
coupling between long- and short-wavelength modes, a long-wavelength density
fluctuation modifies the small-scale structure formation, and so the observables
become \emph{position-dependent}. \refFig{ch1_ks_kl} sketches the short-wavelength
modes with (red) and without (blue) correlation with the long-wavelength mode. As
a consequence, for example, the $n$-point statistics and the halo mass function
would depend on the local long-wavelength overdensity, or equivalently the position
in space. Measurements of spatially-varying observables capture the effect of mode
coupling, and can be used to test our understanding of gravity and the physics of
inflation. A similar idea of measuring the shift of the peak position of the baryonic
acoustic oscillation in different environments has been studied in \cite{roukema/etal:2015}.

\begin{figure}[t!]
\centering\includegraphics[width=0.8\textwidth]{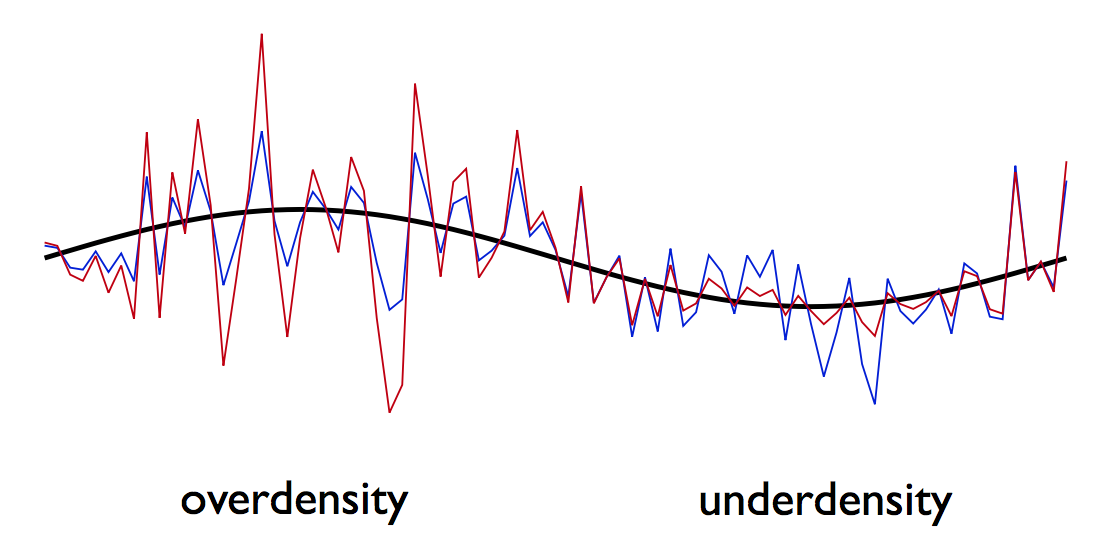}
\caption[Sketch of the correlation between long and short wavelength modes]
{In the absence of the correlation between long and short wavelength modes, the
blue fluctuation have the same statistical property in the overdense and underdense
regions. On the other hand, in the presence of the positive correlation between
long and short wavelength modes, the red fluctuation has larger (smaller)
variance in the overdense (undersense) region.}
\label{fig:ch1_ks_kl}
\end{figure}

In this dissertation, we focus on the position-dependent two-point statistics
(see \cite{cole/kaiser:1989,mo/white:1996} for the mass function). Consider a
galaxy redshift survey or simulation. Instead of measuring the power spectrum
within the entire volume, we divide the volume into many subvolumes, within
which we measure the power spectrum. These power spectra of subvolumes vary
spatially, and the variation is correlated with the mean overdensities of the
subvolumes with respect to the entire volume. This correlation measures an
integral of the bispectrum (or the three-point function in configuration space),
which represents the \emph{response} of the small-scale clustering of galaxies
(as measured by the position-dependent power spectrum) to the long-wavelength
density perturbation (as measured by the mean overdensity of the subvolumes).

Not only is this new observable, \emph{position-dependent power spectrum}, of
the large-scale structure conceptually straightforward to interpret, but it is
also simpler to measure than the full bispectrum, as the machineries for the two-point
statistics estimation are well developed (see \cite{feldman/kaiser/peacock:1994}
for power spectrum and \cite{landy/szalay:1993} for two-point function) and the
measurement of the overdensity is simple. In particular, the computational
requirement is largely alleviated because we explore a subset of the three-point
function corresponding to the squeezed configurations. More precisely, in Fourier
space we only need to measure the power spectrum, and in configuration space the
algorithm of measuring the two-point function by finding particle pairs scales
as $N_s(N_{\rm par}/N_s)^2=N_{\rm par}^2/N_s$ for the entire volume with $N_s$
being the number of subvolumes. In addition, for a fixed size of the subvolume,
the measurement depends on only one wavenumber or one separation, so estimating
the covariance matrix is easier than that of the full bispectrum from a realistic
number of mock catalogs. The position-dependent power spectrum can thus be regarded
as a useful compression of information of the squeezed-limit bispectrum.

As this new observable uses basically the existing and routinely applied
machineries to measure the two-point statistics, one can easily gain extra
information of the three-point function, which is sensitive to the nonlinear
bias of the observed tracers, from the current spectroscopic galaxy surveys.
Especially, since the position-dependent power spectrum picks up the signal
of the squeezed-limit bispectrum, it is sensitive to the primordial non-Gaussianity
of the local type.

This dissertation is organized as follows. In the rest of this chapter,
we review the status of the observations of the large-scale structure,
and the current theoretical understanding.

In \refchp{ch2_posdep_pk_xi}, we introduce the main topic of this dissertation:
position-dependent power spectrum and correlation function. We show how the
correlation between the position-dependent two-point statistics and the
long-wavelength overdensity is related to the three-point statistics. We also
make theoretical template for this correlation using various bispectrum models.

In \refchp{ch3_sepuni}, we introduce the ``separate universe approach,'' in which
a long-wavelength overdensity is absorbed into the background, and the small-scale
structure formation evolves in the corresponding modified cosmology. This is the
basis for modeling the response of the small-scale structure formation to the
long-wavelength overdensity. We consider the fiducial cosmology to be flat
$\Lambda$CDM, and show that the overdensity acts as the curvature in the
separate universe. 

In \refchp{ch4_posdeppk}, we measure the position-dependent power spectrum
from cosmological $N$-body simulations. We compare various theoretical
approaches to modeling the measurements from simulations, particularly the
separate universe approach when the scales of the position-dependent power
spectrum are much smaller than that of the long-wavelength overdensity. We
also study the dependences of the position-dependent power spectrum on the
cosmological parameters, as well as using the Fisher matrix to predict the
expected constraints on biases and local-type primordial non-Gaussianity
for current and future galaxy surveys.

In \refchp{ch5_posdepxi}, we report on the first measurement of the three-point
function with the position-dependent correlation function from the SDSS-III
Baryon Oscillation Spectroscopic Survey Data Release 10 (BOSS DR10) CMASS sample.
We detect the amplitude of the three-point function of the BOSS CMASS galaxies
at $7.4\sigma$. Combining the constraints from position-dependent correlation
function, global two-point function, and the weak lensing signal, we determine
the quadratic (nonlinear) bias of BOSS CMASS galaxies.

In \refchp{ch6_npt_sq_sep}, we generalize the study to the response of the
small-scale power spectrum in the presence of $m$ long-wavelength modes for
$m\le3$. This response can be linked to the angular-averaged squeezed limit
of $(m+2)$-point functions. We shall also introduce the separate universe
simulations, in which $N$-body simulations are performed in the presence
of a long-wavelength overdensity by modifying the cosmological parameters.
The separate universe simulations allow unprecedented measurements for the
squeezed-limit $n$-point function. Finally, we test the standard perturbation
theory hypothesis that the nonlinear $n$-point function is completely predicted
by the linear power spectrum at the same time. We find discrepancies of 10\% at
$k\simeq0.2-0.5~h~{\rm Mpc}^{-1}$ for five- to three-point functions at $z=0$.
This suggests the breakdown of the standard perturbation theory, and quantifies
the scales that the corrections of the effective fluid become important for
the higher order statistics.

In \refchp{ch7_summary}, we summarize this dissertation, and present the outlook.

\section{Observations and measurements of the large-scale structure}
In the 1960s, before the invention of the automatic plate measuring machine and
the densitometer, the galaxy catalogs such as Zwicky \cite{zwicky/etal:1961} and
Lick \cite{shane/wirtanen:1967} relied on visual inspection of poorly calibrated
photographic plates. These surveys consisted of different neighboring photographic
plates, so the uniformity of the calibration, which might cause large-scale gradients
in the observed area, was a serious issue. Because of the lack of the redshifts
(depths) of galaxies, only the angular clustering studies were possible. In addition,
the sizes of the surveys were much smaller than the ones today, thus only the clustering
on small scales, where the nonlinear effect is strong, can be studied. Nevertheless,
in the 1970s Peebles and his collaborators did the first systematic study on galaxy
clustering using the catalogs at that time. The series of studies, starting with
\cite{peebles:1973}, considered galaxies as the tracers of the large-scale structure
for the first time, which was a ground-breaking idea. These measurements confirmed
the power-law behavior of the angular two-point function, and the interpretation was
done in the framework of Einstein-de Sitter universe, i.e. matter-dominated flat
Friedmann-Lema{\^i}tre-Robertson-Walker universe.

In the 1980s, the invention of the automatic scanning machines as well as CCDs
revolutionized the large-scale structure surveys, and resulted in a generation
of wide-field surveys with better calibration and a three-dimensional view of
the universe. Photographic plates became obsolete for the large-scale structure
studies, and nowadays photometric surveys use large CCD cameras with millions of
pixels. The galaxy redshift surveys, which generally require target selections
with photometric detection and then spectroscopic follow-up, thus open a novel
avenue to study the universe. It was shown that the redshift-space two-point
correlation function in the CfA survey \cite{huchra/etal:1983} agreed well with
the previous studies on angular clustering, if the redshift direction is integrated
over \cite{davis/peebles:1983}. In the 1990s, the number of galaxies in surveys
was $\sim10^3-10^4$, but with these data it was already shown that the large-scale
power spectrum was inconsistent with the CDM model \cite{efstathiou/etal:1990,
saunders/etal:1991,vogeley/etal:1992}, in agreement with the study done in the
angular clustering \cite{maddox/etal:1990}.

\begin{figure}[t!]
\centering
\includegraphics[width=0.495\textwidth]{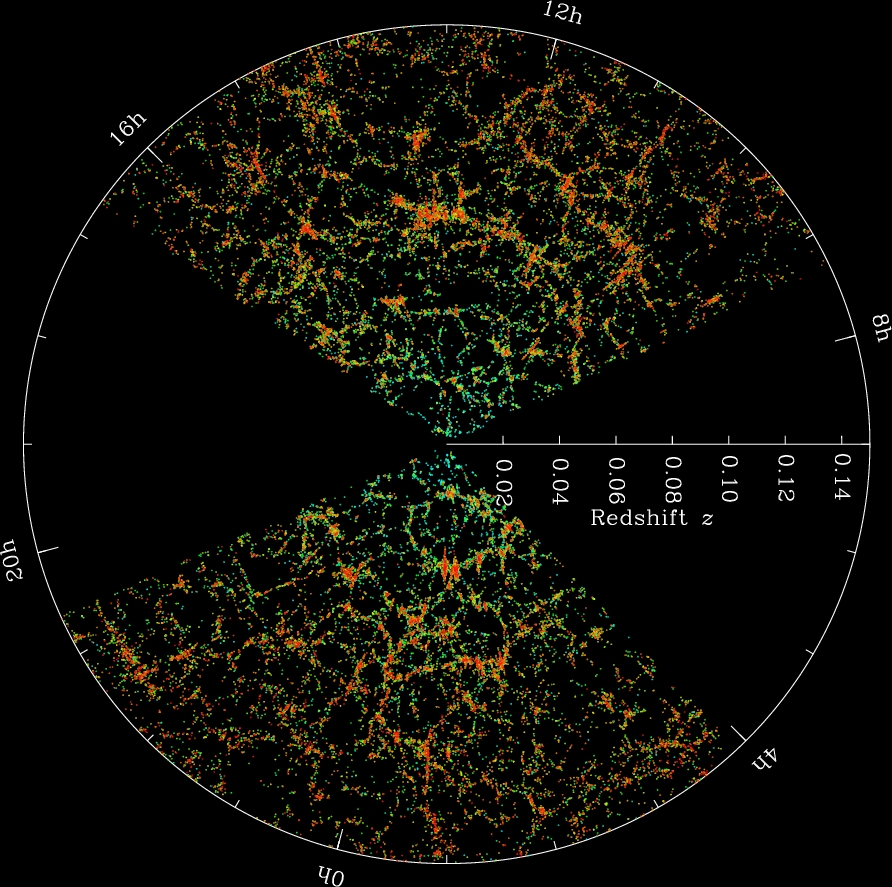}
\includegraphics[width=0.495\textwidth]{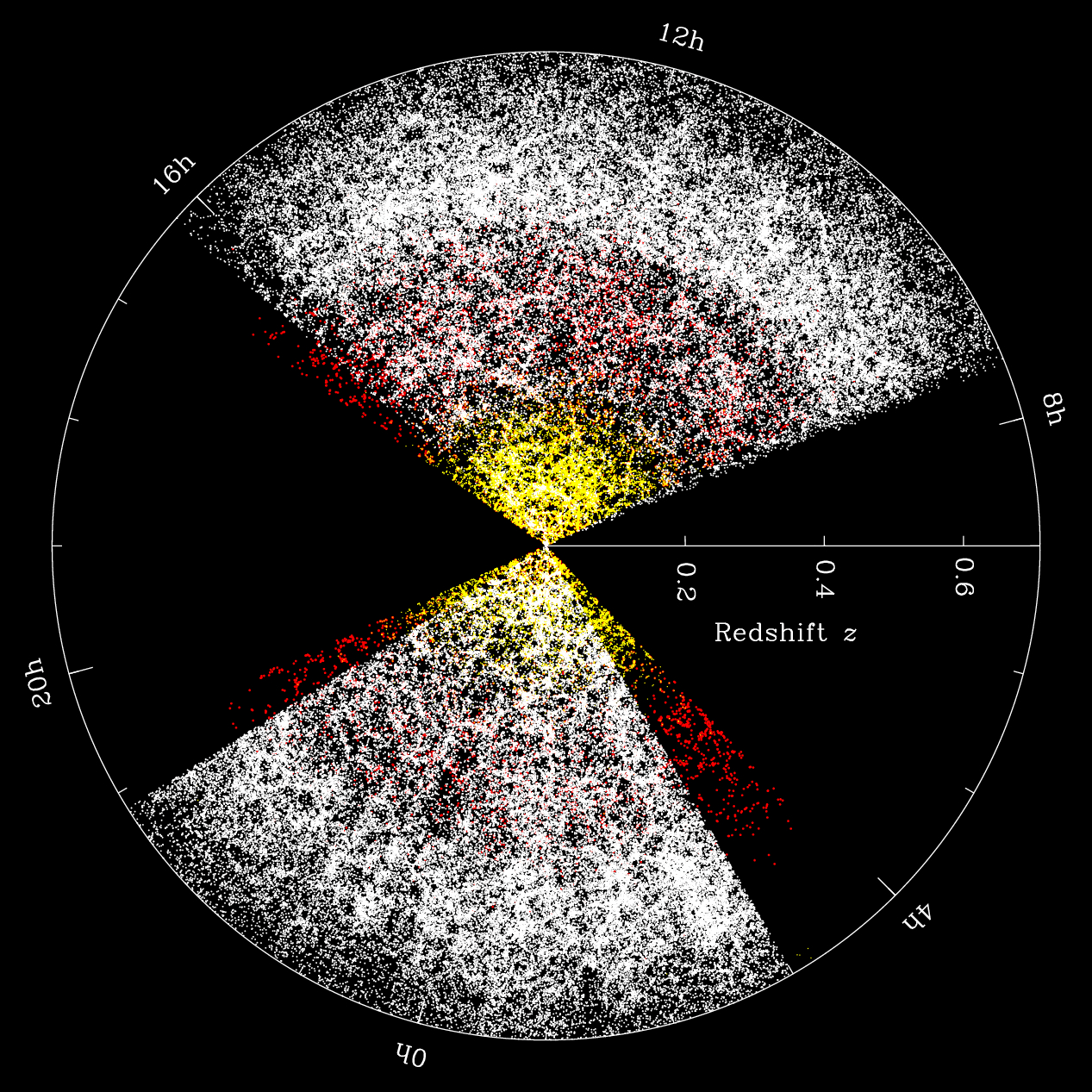}
\caption[SDSS galaxy maps projected on the RA-redshift plane]
{(Credit: Michael Blanton and SDSS collaboration) SDSS galaxy maps projected on the
RA-redshift plane. (Left) The SDSS main galaxy sample out to $z\sim0.15$. (Right)
Yellow, red, and white dots are the SDSS main galaxy sample, luminous red galaxies,
and CMASS sample, respectively, out to $z\sim0.7$.}
\label{fig:ch1_sdss}
\end{figure}

Another quantum leap of the sizes of the galaxy surveys happened in the 2000s,
when the technology of the massive multi-fiber or multi-slit spectroscopy became
feasible. Surveys such as Two-degree-Field Galaxy Redshift Survey (2dFGRS)
\cite{colless/etal:2001} and Sloan Digital Sky Survey (SDSS) \cite{york/etal:2000}
targeted at obtaining spectra of $\sim10^5-10^6$ galaxies. The left panel of
\reffig{ch1_sdss} shows the SDSS main galaxy sample out to $z\sim0.15$, which
corresponds to roughly $440\hMpc$. It is clear even visually that the distribution
of galaxies follows filamentary structures, with voids in between the filaments.
These data contain precious information of the properties of the universe. For
example, in 2005, the baryonic acoustic oscillations (BAO) in the two-point
statistics were detected for the first time by 2dFGRS \cite{cole/etal:2005} in
the power spectrum and SDSS \cite{eisenstein/etal:2005} in the two-point correlation
function, as shown in \reffig{ch1_bao}. The detection is phenomenal given the fact
that the BAOs are of order a few percent features on the smooth functions. Galaxy
surveys in this era started probing the weakly nonlinear regime, where the theoretical
understanding is better, so we can extract cosmological information.

\begin{figure}[t!]
\centering
\includegraphics[width=0.47\textwidth,trim=0cm 2cm 0cm 0cm]{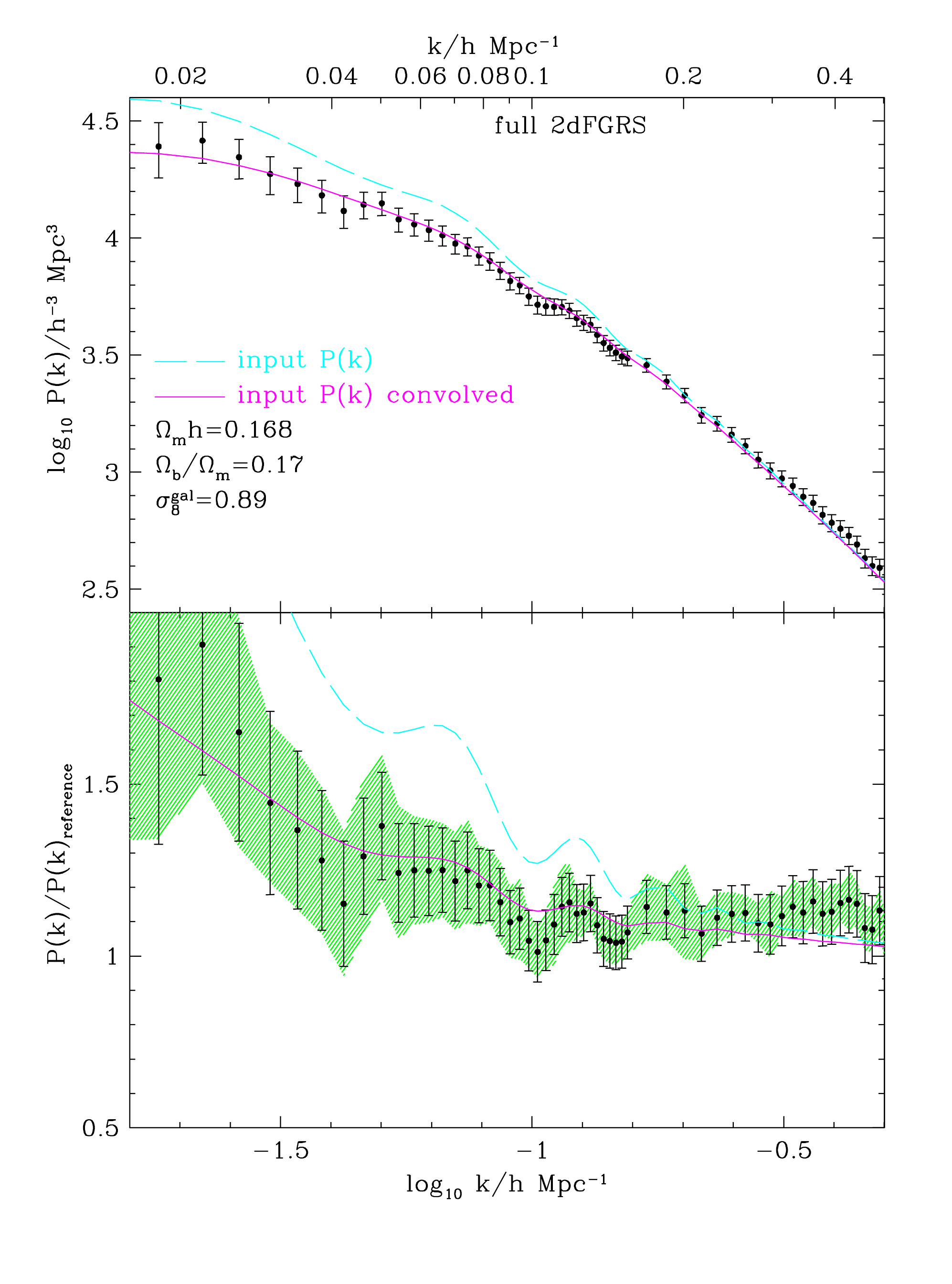}
\includegraphics[width=0.52\textwidth]{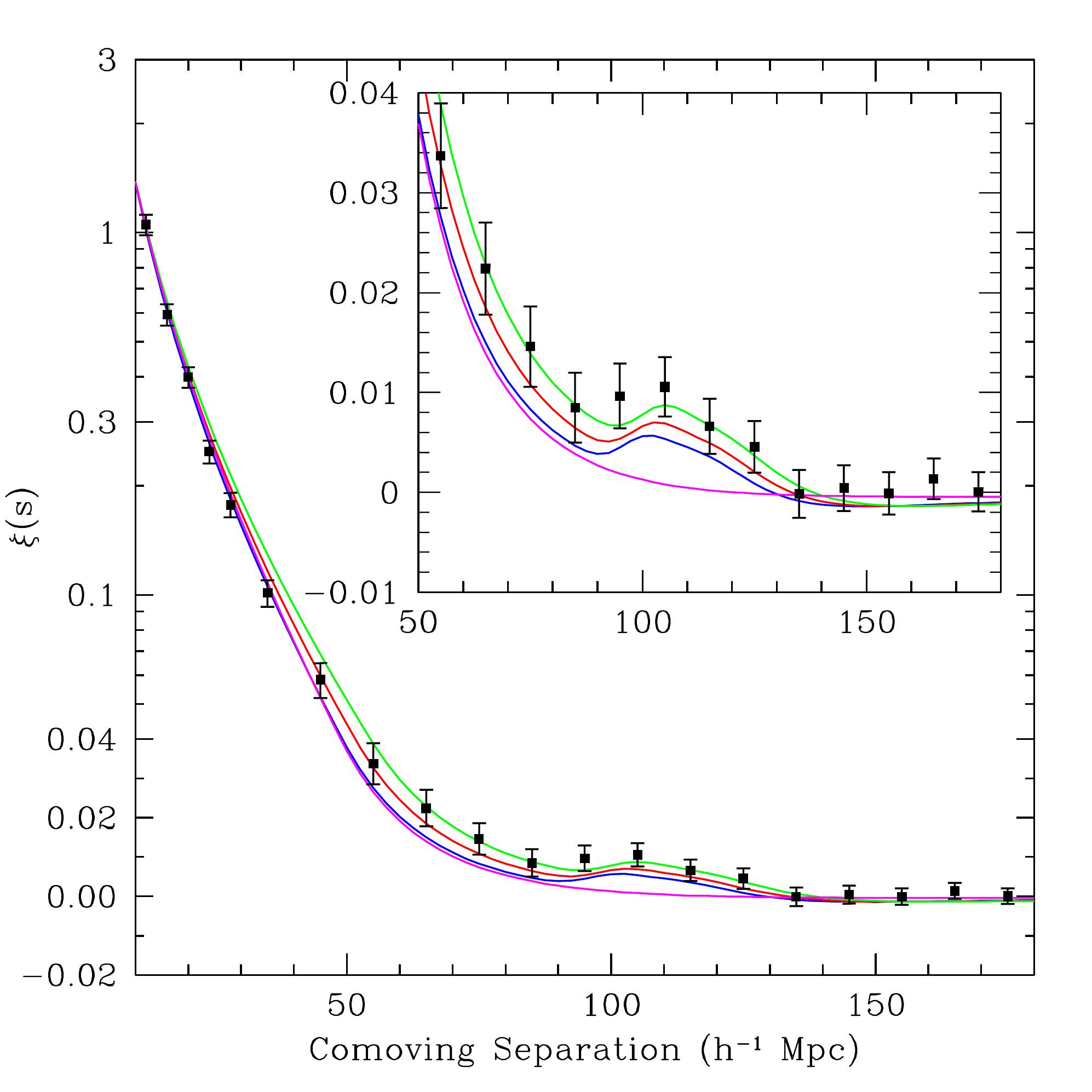}
\caption[The first detection of BAOs in 2dFGRS and SDSS]
{(Credit: 2dFGRS and SDSS collaborations) The first detection of BAOs in 2dFGRS
(left) and SDSS (right). The analysis of 2dFGRS was done in Fourier space, and
the BAOs are the wiggles in the power spectrum at $0.05\ihMpc\lesssim k\lesssim0.2\ihMpc$;
the analysis of SDSS was done in configuration space, and the BAOs are the bump
in the two-point correlation function at $r\sim100\hMpc$.}
\label{fig:ch1_bao}
\end{figure}

With the success of the first BAO detection, more galaxy redshift surveys, such
as WiggleZ \cite{blake/etal:2011a,blake/etal:2011b} and SDSS-III Baryon Oscillation
Spectroscopic Survey (BOSS) \cite{anderson/etal:2014a,anderson/etal:2014}, followed
and extended to higher redshifts. The right panel of \reffig{ch1_sdss} shows the
galaxy map of SDSS out to $z\sim0.7$, which corresponds to roughly $1800\hMpc$.
The BAO feature can be used as a standard ruler to measure the angular diameter
distance and Hubble expansion rate. This is particularly useful for studying the
time-dependence of dark energy, which began to dominate the universe at $z\sim0.4$
where the galaxy clustering is measured. The BAOs in the galaxy clustering thus
becomes a powerful probe of dark energy.

Thus far, most of the studies have focused on the two-point statistics. However,
there is much more information in the higher-order statistics. Especially, ongoing
galaxy surveys such as Dark Energy Survey \cite{des:2005} and the extended BOSS,
as well as upcoming galaxy surveys such as Hobby Eberly Telescope Dark Energy
eXperiment (HETDEX) \cite{hill/etal:2008} and Subaru Prime Focus Spectrograph
\cite{takada/etal:2014} will measure the galaxies at even higher redshift.
For instance, HETDEX will use Lyman-alpha emitters as tracers to probe the
matter distribution at $1.9\lesssim z\lesssim3.5$. At such high redshift, the
gravitational evolution is relatively weak and can still be predicted analytically,
hence this is an ideal regime to critically test our understanding of gravity,
as well as the physics of inflation via the primordial non-Gaussianity.

Currently, most of the constraint on the local-type primordial non-Gaussianity from
the large-scale structure is through the scale-dependent bias \cite{dalal/etal:2008,
matarrese/verde:2008,slosar/etal:2008}. That is, dark matter halos (or galaxies) are
biased tracers of the underlying matter distribution forming at density peaks, so
the formation of halos would be modulated by the additional correlation between the
long and short wavelength modes due to the primordial non-Gaussianity. As a result,
the halo bias contains a $k^{-2}$ scale-dependent correction, and this correction
is prominent on large scale. Measurements of the large-scale galaxy power spectrum,
which is proportional to bias squared, can thus be used to constrain the primordial
non-Gaussianity of the local type.

This distinct feature appears in the galaxy power spectrum at very large scales,
hence it is crucial to have a huge survey volume to beat down the cosmic variance.
Moreover, if the galaxies are highly biased, then the signal-to-noise ratio would
also increase. Thus, many studies have used quasars at $0.5\lesssim z\lesssim3.5$
from SDSS to constrain the primordial non-Gaussianity \cite{ho/etal:2013,agarwal/ho/shandera:2014,
leistedt/peiris/roth:2014}. Similar methods, such as combining the abundances and
clustering of the galaxy clusters \cite{mana/etal:2013} as well as the correlation
between CMB lensing and large-scale structure \cite{giannantonio/etal:2014,
giannantonio/percival:2014}, have also been proposed to study the primordial
non-Gaussianity from large-scale structure. It is predicted in \cite{leistedt/peiris/roth:2014}
that for Large Synoptic Survey Telescope \cite{lsst:2012} the constraint on
$f_{\rm NL}$, parametrization of the local-type primordial non-Gaussianity,
using the scale-dependent bias can reach $\sigma(f_{\rm NL})\sim5$ (95\% C.L.).

The error bar on $f_{\rm NL}$ from the scale-dependent bias is limited by the number
of Fourier modes on large scales. On the other hand, for the bispectrum analysis, we
are looking for triangles formed by different Fourier modes, so the bispectrum contains
more information and will have a tighter constraint on $f_{\rm NL}$. The difficulty for
using the large-scale structure bispectrum to constrain $f_{\rm NL}$ is that gravity
produces non-zero squeezed-limit bispectrum even without primordial non-Gaussianity,
and the signal from gravity dominates for the current limit on $f_{\rm NL}$. This is
why recent measurements of the large-scale structure bispectrum have focused on
constraining the growth and galaxy biases \cite{marin/etal:2013,gilmarin/etal:2014b}.

The difficulty in modeling nonlinear effects can be alleviated if the observations are
done in the high-redshift universe, where the gravitational evolution on quasi-linear
scales can still be described by the perturbation theory approach. While theoretically
we are reaching the stage for studying the higher-order statistics, e.g. the three-point
correlation function or the bispectrum, if the data are obtained at high redshift, in
practice the measurements and analyses are still computational challenging. It is thus
extremely useful to find a way to compress the information, such that studying the
three-point correlation function of the galaxy clustering is feasible.

Another complexity of the bispectrum measurement from galaxy surveys is the window
function effect. Namely, galaxy surveys almost always have non-ideal survey geometry,
e.g. masking around the close and bright objects or the irregular boundaries, as
well as the spatial changes in the extinction, transparency, and seeing. These
effects would bias the measurement, and extracting the true bispectrum signal
becomes difficult. While the observational systematics would enters into the
estimation of both power spectrum and bispectrum, the technique of deconvolving
the window function effects, e.g. \cite{sato/huetsi/yamamoto:2011,sato/etal:2013},
has been relatively well developed for the two-point statistics.

The subject of this dissertation is to find a method to more easily extract the
bispectrum in the squeezed configurations, where one wavenumber, say $k_3$, is
much smaller than the other two, i.e. $k_3\ll k_1\approx k_2$. The squeezed-limit
bispectrum measures the correlation between one long-wavelength mode ($k_3$) and
two short-wavelength modes ($k_1$, $k_2$), which is particularly sensitive to the
local-type primordial non-Gaussianity. Specifically, we divide a survey into subvolumes,
and measure the correlation between the position-dependent two-point statistics
and the long-wavelength overdensity. This correlation measures an integral on the
bispectrum, and is dominated by the squeezed-limit signal if the wavenumber of the
position-dependent two-point statistics is much larger than the wavenumber corresponding
to the size of the subvolumes. Therefore, without employing the three-point function
estimator, we can extract the squeezed-limit bispectrum by the position-dependent
two-point statistics technique. Furthermore, nonlinearity of the correlation between
the position-dependent two-point statistics and a long-wavelength mode can be well
modeled by the separate universe approach, in which the long-wavelength overdensity
is absorbed into the background cosmology; the window function effect can also be
well taken care of because this technique measures essentially the two-point function
and the mean overdensity, for which the procedures for removing the window function
effects are relatively well developed. With the above advantages, the position-dependent
two-point statistics is thus a novel and promising method to study the squeezed-limit
bispectrum of the large-scale structure.

Galaxy redshift surveys has entered a completely new era at which the sizes (e.g.
survey volume and number of observed galaxies) are huge, and redshifts are high.
As the signal-to-noise ratio of the higher-order statistics, especially for the
primordial non-Gaussianity, will be much higher in the upcoming galaxy surveys
than the previous ones, we should do our best to extract the precious signal for
improving our understanding of the universe. The new observable, position-dependent
power spectrum, proposed in the dissertation would help us achieve this goal.

\section{Theoretical understanding of the large-scale structure}
\subsection{Simulations}
How do we understand the gravitational evolution of the large-scale structure?
Because of the process is nonlinear, the gold standard is the cosmological
$N$-body simulations of collisionless particles.

Using $N$-body simulations to solve for gravitational dynamics has a long history
(see e.g. \cite{dehnen/read:2011} for a review). The first computer calculation
was done back in 1963 by \cite{aareth:1963} with $N=16$. Later, $N$ has roughly
doubled every two years following Moore's law, and nowadays state-of-the-art
simulations for collisionless particles have $N=10^9-5\times10^{10}$, e.g.
\cite{springel/etal:2005, teyssier/etal:2009,stadel/etal:2009,klypin/etal:2014}.
The $N$-body codes, such as GADGET-2 \cite{springel:2005}, solve dynamics of dark
matter particles, and dark matter particles are grouped into dark matter halos
by algorithms such as friends-of-friends or spherical overdensity. We can thus
study the properties of these halos, such as the clustering and the mass function.
\refFig{ch1_millennium} shows a slice of the Millennium simulation \cite{springel/etal:2005}.
The cosmic web of the large-scale structure is obvious, and we also find the
similarity between the simulation and the observation, i.e. \reffig{ch1_sdss}.

\begin{figure}[t!]
\centering
\includegraphics[width=0.5\textwidth]{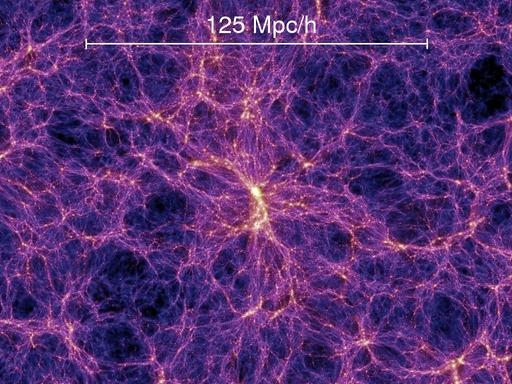}
\caption[A slice of the Millennium simulation]
{(Credit: Volker Springel) A slice of the Millennium simulation shows the cosmic
web of the large-scale structure.}
\label{fig:ch1_millennium}
\end{figure}

Direct observations, however, can only be done for luminous objects, e.g. galaxies,
so we must relate galaxies to the halos. As baryonic physics is more complicated
than gravity-only dynamics, simulations of galaxy formation can only be done with
the usage of sub-grid physics models. That is, the empirical relations of the feedback
from baryonic physic such as supernovae and active galactic nuclei are used in the
galaxy formation simulations. State-of-the-art hydrodynamic simulations for the galaxy
formation are Illustris \cite{vogelsberger/etal:2014} and EAGLE \cite{schaye/etal:2015}.
Since these simulations are extremely computationally intensive, the simulation box
size cannot be too large (e.g. $\sim100\hMpc$ for both Illustris and EAGLE). On the
other hand, galaxy redshift surveys at present day have sizes of order $1-10~h^{-3}~{\rm Gpc}^3$.

Alternatively, we can link the simulated dark matter halos to galaxies using techniques
such as semi-analytic models \cite{kauffmann/white/guiderdoni:1993}, which use the
merging histories of dark matter halos, or halo occupation distribution \cite{berlind/etal:2003,
kravtsov/etal:2004}, which uses the statistical relation between halos and galaxies.
As both methods contain free parameters in the models, these parameters can be tuned
such that multiple properties (e.g. clustering and environmental dependence) of the
``simulated'' galaxies match the observed ones (see \cite{guo/etal:2013} for the recent
comparison between semi-analytic models and the observations). These thus provide a
practically feasible way to populate galaxies in dark matter halos in the simulations.

The remaining task, especially for the clustering analysis of galaxy redshift surveys,
is to generate a large suite of simulations with halos, so they can be used to estimate
the covariance matrices of the correlation function or the power spectrum, which are
the necessary ingredient for the statistical interpretation of the cosmological information.
In particular, the present-day galaxy redshift surveys contain a huge volume, and if
high enough mass resolution for halos (which depend on the properties of the observed
galaxies) is required, $N$-body simulation are not practical. Thus, most analyses for
the two-point statistics of galaxy surveys use algorithms such that the scheme of solving
dynamics is simplified to generate ``mock'' halos. For example, COLA \cite{tassev/zaldarriaga/eisenstein:2013}
solves the long-wavelength modes analytically and short-wavelength modes by $N$-body
simulations, and PTHalos \cite{scoccimarro/sheth:2002,manera/etal:2013,manera/etal:2015}
is based on the second-order Lagrangian perturbation theory.

As the full $N$-body simulations are regarded as the standard, we can input identical
initial conditions to various codes for generating mock catalogs (halos) and compare
the performances. Currently most of the observations now have been focused on the
two-point statistics and the mass function, so these algorithms are designed to recover
these two quantities. On the other hand, for the three-point function, which is more
sensitive to nonlinear effects, more careful and systematic studies are necessary. One
recent comparison between major methodologies for generating mock catalogs shows that
the differences between $N$-body simulations and mock generating codes are much larger
for the three-point function than for the two-point function \cite{chuang/etal:2014}.
This suggests that at this moment the full $N$-body simulations are still required
to understand the three-point function, or the bispectrum in Fourier space.

The full bispectrum contains triangles formed by different Fourier modes. While the
configurations of the bispectrum can be simulated easily if three wavenumbers are similar,
the squeezed triangles are more difficult because a large volume is needed to simulate the
coupling between long-wavelength modes and small-scale structure formation. In particular,
if high enough mass resolution is required, the simulations become computationally demanding.
In this dissertation, we provide a solution to this problem. Specifically, we absorb the
long-wavelength overdensity into the modified background cosmology (which is the subject
in \refchp{ch3_sepuni}) and perform the $N$-body simulations in the separate universe (which
is the subject in \refsec{ch6_sepuni_sim}). This setting simulates how the small-scale
structure is affected by a long-wavelength mode. As the box size of the separate universe
simulations can be small ($\sim500\hMpc$), increasing the mass resolution becomes feasible.
This technique is therefore useful for understanding the nonlinear coupling between long
and short wavelength modes.

Another computational challenge to the bispectrum analysis is the number of Fourier bins.
Specifically, the bispectrum contains all kinds of triangles, so the number of bins is much
larger than that of the power spectrum. If the mock catalogs are used to estimate the covariance
matrix, many more realizations are required to characterize the bispectrum than the power
spectrum. The lack of realizations of mock catalogs would result in errors in the covariance
matrix estimation, and the parameter estimation would be affected accordingly \cite{dodelson/schneider:2013}.
Therefore, even if there is an algorithm to generate mock catalogs with accurate two- and
three-point statistics, we still need a huge amount of them for data analysis, which can
be computational challenging. The advantage of the position-dependent power spectrum is
that it depends only on one wavenumber, so the number of bins is similar to that of the
power spectrum. This means that we only need a reasonable number of realizations ($\sim1000$)
for analyzing the power spectrum and position-dependent power spectrum jointly.

\subsection{Theory}
While $N$-body simulations are the gold standard for understanding nonlinearity of the
large-scale structure, it is impractical to run various simulations with different cosmological
parameters or models. It is therefore equally important to develop analytical models so they
are easier to compute. We can then use them in the cosmological inferences, e.g. the
Markov chain Monte Carlo methods.

The most commonly used technique is the perturbation theory approach, in which the fluctuations
are assumed to be small so they can be solved recursively (see \cite{bernardeau/etal:2002}
for a review). For example, in the standard perturbation theory (SPT), the density fluctuations
and peculiar velocity fields are assumed to be small. Using this assumption, we can expand
the continuity, Euler, and Poisson equations at different orders, and solve the coupled
differential equations order by order (see \refapp{trz_bi} and \cite{jeong/komatsu:2006}
for a brief overview on SPT). The SPT power spectrum at the first order contains the product
of two first order fluctuations ($P_{11}$); the next-to-leading order SPT power spectrum
contains the products of first and third order fluctuations ($P_{13}$) as well as two second
order fluctuations ($P_{22}$). A similar approach can be done in Lagrangian space, in which
the displacement field mapping the initial (Lagrangian) position to the final (Eulerian)
position of fluid element is solved perturbatively \cite{matsubara:2008a,matsubara:2008b}.

Generally, the perturbation theory approach works well on large scales and at high redshift,
where nonlinearity is small. On small scale and at low redshift, the nonlinear effect becomes
more prominent, including higher order corrections is thus necessary. However, the nonlinear
effect can be so large that even including more corrections does not help. The renormalized
perturbation theory (RPT) was introduced to alleviate the problem \cite{crocce/scoccimarro:2006,
bernardeau/crocce/scoccimarro:2012}. Specifically, RPT categorizes the corrections into two
kinds: the mode-coupling effects and the renormalization of the propagator (of the gravitational
dynamics). Thus, in RPT the corrections for nonlinearity become better defined, and so the
agreement with the nonlinear power spectrum extends to smaller scales compared to SPT. Another
approach is the effective field theory (EFT) \cite{carrasco/hertzberg/senatore:2012}: on
large-scale the matter fluid is characterized by parameters such as sound speed and viscosity,
and these parameters are determined by the small-scale physics that is described by the Boltzmann
equation. In practice, these parameters are measured from $N$-body simulations with a chosen
smoothing radius. As for RPT, EFT also gives better agreement with the nonlinear power spectrum
on smaller scales compared to SPT.

A different kind of approach to compute the clustering properties of the large-scale structure
is to use some phenomenological models. The well known phenomenological model is the halo model
(see \cite{cooray/sheth:2002} for a review), where all matter is assumed to be contained inside
halos, which are characterized by the density profile (e.g. NFW profile \cite{navarro/frenk/white:1997})
and the mass function (e.g. Sheth-Tormen mass function \cite{sheth/tormen:1999}). The matter
$n$-point functions is then the sum of one-halo term (all $n$ positions are in one halo), two-halo
term ($(n-1)$ positions are in one halo and the other one is in a different halo), to $n$-halo
term ($n$ positions are in $n$ different halos). The small-scale nonlinear matter power spectrum
should thereby be described by the halo properties, i.e. the one-halo term. Some recent work
attempted to extend the halo model to better describe the nonlinear matter power spectrum. For
example, in \cite{mohammed/seljak:2014}, the Zeldovich approximation \cite{zeldovich:1970} is
added in the two-halo term, and a polynomial function ($A_0+A_2k^2+A_4k^4+\cdots$) is added to
model the baryonic effects; in \cite{seljak/vlah:2015}, the two-halo term with the Zeldovich
approximation is connected to the SPT one-loop power spectrum ($P_{11}+P_{13}+P_{22}$), but a
more more complicated function is used for the one-halo term.

One can also construct fitting functions based on results of $N$-body simulations. The most
famous fitting functions of the nonlinear matter power spectrum are the halofit prescription
\cite{smith/etal:2003} and the Coyote emulator \cite{heitmann/etal:2014}. More specifically,
the Coyote emulator was constructed with a suite of $N$-body simulations with a chosen range
of cosmological parameters (e.g. $\Omega_b$, $\Omega_m$). Then, the power spectrum is computed 
based on the interpolation of the input cosmological parameters. Thus, the apparent limitation
of these simulation-calibrated fitting formulae is that they are only reliable within limited
range of cosmological parameters and restricted cosmological models.

Similar to that of simulations, most of the theory work has been focused on precise description
of nonlinearity of the matter power spectrum, while relatively few work has been devoted to
investigate nonlinearity of the bispectrum. Therefore, the matter bispectrum is normally computed
at the SPT tree-level, i.e. the product of two first order fluctuations and a second order
fluctuation. Some studies \cite{scoccimarro/couchman:2001,gilmarin/etal:2012,gilmarin/etal:2014a}
considered nonlinearity of the bispectrum by replacing the SPT kernel with fitting formulae
containing some parameters, which are then obtained by fitting to $N$-body simulations. These
models, however, lack the theoretical foundation, and it is unclear how the fitting parameters
would depend on the cosmological parameters.

In this dissertation, we provide a semi-analytical model for describing the bispectrum in the
squeezed configurations. Specifically, we show that the real-space angle-average squeezed-limit
bispectrum is the response of the power spectrum to an isotropically infinitely long-wavelength
overdensity. Due to the presence of this overdensity, the background cosmology is modified, and
the small-scale power spectrum evolves as if matter is in the separate universe. In \refchp{ch4_posdeppk},
we show it is straightforward to combine the separate universe approach and the power spectrum
computed from perturbation theory approach, phenomenological models, or simulation-calibrated
fitting formulae. More importantly, the results of the separate universe approach agree better
with the $N$-body simulation measurements in the squeezed limit than that of the real-space
bispectrum fitting formula. In \refchp{ch6_npt_sq_sep}, we generalize the separate universe
approach to the response of the small-scale power spectrum to $m$ infinitely long-wavelength
overdensities for $m\le3$. As expected, this response is related to the squeezed-limit $(m+2)$-point
function with a specific configuration shown in \reffig{ch6_squ_conf}. The separate universe
approach is thus extremely useful for modeling the squeezed-limit $n$-point functions, and
its analytical form can be added to the fitting formulae.

As future galaxy surveys contain data with unprecedented amount and quality which can
be used to test our understanding of gravity and the physics of inflation, accurate
theoretical model, especially for the bispectrum, is required to achieve the goal.
While the full bispectrum contains various triangles formed by different Fourier modes,
in this dissertation we present the theoretical model specifically for the squeezed
triangles, so more work needs to be done for the other configurations.

}

%% file: kap_02.tex
{

\renewcommand{\v}[1]{\mathbf{#1}}
\newcommand{\vx}{\v{x}}
\newcommand{\vr}{\v{r}}
\newcommand{\vk}{\v{k}}
\newcommand{\vq}{\v{q}}

\newcommand{\Om}{\Omega_m}
\newcommand{\rhob}{\bar\rho}
\def\rhocr{\rho_{\rm cr,0}}
\renewcommand{\d}{\delta}
\newcommand{\hMpc}{~h^{-1}~{\rm Mpc}}
\newcommand{\ihMpc}{~h~{\rm Mpc}^{-1}}
\newcommand{\hd}{\hat{\delta}}
\newcommand{\hdb}{\hat{\bar\delta}}
\newcommand{\hP}{\hat{P}}
\newcommand{\hib}{\hat{iB}}
\newcommand{\cO}{\mathcal O}
\newcommand{\iz}{i\zeta}

\newcommand{\Dr}{\Delta r}
\newcommand{\Dk}{\Delta k}
\newcommand{\vl}{\v{l}}
\newcommand{\vm}{\v{m}}
\newcommand{\vn}{\v{n}}
\newcommand{\vj}{\v{j}}
\newcommand{\vL}{\v{L}}
\newcommand{\vJ}{\v{J}}
\newcommand{\vS}{\v{S}}
\newcommand{\vg}{\v{g}}
\newcommand{\tpc}{(2\pi)^3}

\def\approxprop{%
  \def\p{%
    \setbox0=\vbox{\hbox{$\propto$}}%
    \ht0=0.6ex \box0 }%
  \def\s{%
    \vbox{\hbox{$\sim$}}%
  }%
  \mathrel{\raisebox{0.7ex}{%
      \mbox{$\underset{\s}{\p}$}%
    }}%
}

\chapter{Position-dependent two-point statistics}
\label{chp:ch2_posdep_pk_xi}

\section{In Fourier space}
\label{sec:ch2_kspace}
\subsection{Position-dependent power spectrum}
\label{sec:ch2_posdeppk}
Consider a density fluctuation field, $\delta(\vr)$, in a survey (or simulation)
of volume $V_r$. The mean overdensity of this volume vanishes by construction,
i.e.
\be
 \bar\delta=\frac1{V_r}\int_{V_r}d^3r~\delta(\vr)=0 ~.
\ee
The global power spectrum of this volume can be estimated as
\be
 \hP(\vk)=\frac1{V_r}|\delta(\vk)|^2 ~,
\ee
where $\delta(\vk)$ is the Fourier transform of $\delta(\vr)$.

We now identify a subvolume $V_L$ centered at $\vr_L$. The mean overdensity
of this subvolume is
\be
 \bar\delta(\vr_L)=\frac1{V_L}\int_{V_L}d^3r~\delta(\vr)
 =\frac1{V_L}\int d^3r~\delta(\vr)W(\vr-\vr_L) ~,
\label{eq:ch2_bd}
\ee
where $W(\vr)$ is the window function. For simplicity and a straightforward
application to the $N$-body simulation box, throughout this dissertation
we use a cubic window function given by
\be
W(\vr)=W_L(\vr)=\prod_{i=1}^3\:\theta(r_i), \quad
\theta(r_i) = \left\{
\begin{array}{cc}
1, & |r_i|\le L/2, \\
0,  & \mbox{otherwise}~.
\end{array}\right.
\label{eq:ch2_WL}
\ee
where $L$ is the side length of $V_L$. The results are not sensitive to
the exact choice of the window function, provided that the scale of
interest is much smaller than $L$. While $\bar\delta=0$, $\bar\delta(\vr_L)$
is non-zero in general. In other words, if $\bar\delta(\vr_L)$ is positive
(negative), then this subvolume is overdense (underdense) with respect to
the mean density in $V_r$.

Similar to the definition of the global power spectrum in $V_r$, we define
the position-dependent power spectrum in $V_L$ as
\be
 \hP(\vk,\vr_L)\equiv \frac1{V_L}|\delta(\vk,\vr_L)|^2 ~,
\label{eq:ch2_posdep_pk}
\ee
where
\be
 \delta(\vk,\vr_L)\equiv \int_{V_L}d^3r~\delta(\vr)e^{-i\vr\cdot\vk}
\label{eq:ch2_local_ft}
\ee
is the \emph{local} Fourier transform of the density fluctuation field.
The integral ranges over the subvolume centered at $\vr_L$. With
this quantity, the mean density perturbation in the subvolume centered
at $\vr_L$ is given by
\be
 \bar\delta(\vr_L)=\frac1{V_L}\delta(\vk=0,\vr_L) ~.
\ee
One can use the window function $W_L$ to extend the integration boundaries
to infinity as
\be
 \delta(\vk,\vr_L)=\int d^3r~\delta(\vr)W_L(\vr-\vr_L)e^{-i\vr\cdot\vk}
 =\int\frac{d^3q}{(2\pi)^3}~\delta(\vk-\vq)W_L(\vq)e^{-i\vr_L\cdot\vq} ~,
\label{eq:ch2_localdeltak}
\ee
where $W_L(\vq)=L^3\prod_{i=1}^3{\rm sinc}(q_iL/2)$ is the Fourier transform of
the window function and ${\rm sinc}(x)=\sin(x)/x$. Therefore, the position-dependent
power spectrum of the subvolume $V_L$ centered at $\vr_L$ is
\be
 \hP(\vk,\vr_L)=\frac{1}{V_L}\int\frac{d^3q_1}{(2\pi)^3}\int\frac{d^3q_2}{(2\pi)^3}~
 \delta(\vk-\vq_1)\delta(-\vk-\vq_2)W_L(\vq_1)W_L(\vq_2)e^{-i\vr_L\cdot(\vq_1+\vq_2)} ~.
\label{eq:ch2_localpk}
\ee

\subsection{Integrated bispectrum}
\label{sec:ch2_ib}
The correlation between $\hP(\vk,\vr_L)$ and $\bar\delta(\vr_L)$ is given by
\ba
 \langle\hP(\vk,\vr_L)\bar\delta(\vr_L)\rangle\:&=
 \frac1{V_L^2}\int\frac{d^3q_1}{(2\pi)^3}\int\frac{d^3q_2}{(2\pi)^3}
 \int\frac{d^3q_3}{(2\pi)^3}~\langle\delta(\vk-\vq_1)
 \delta(-\vk-\vq_2)\delta(-\vq_3)\rangle \vs
 \:&\hspace{3.5cm}\times W_L(\vq_1)W_L(\vq_2)W_L(\vq_3)e^{-i\vr_L\cdot(\vq_1+\vq_2+\vq_3)} ~,
\label{eq:ch2_corr_localpk_deltabar}
\ea
where $\langle\ \rangle$ denotes the ensemble average over many universes.
In the case of a simulation or an actual survey, the average is taken instead
over all the subvolumes in the simulation or the survey volume. Through the
definition of the bispectrum,
$\langle\delta(\vq_1)\delta(\vq_2)\delta(\vq_3)\rangle=B(\vq_1,\vq_2,\vq_3)(2\pi)^3\delta_D(\vq_1+\vq_2+\vq_3)$
where $\delta_D$ is the Dirac delta function, \refeq{ch2_corr_localpk_deltabar}
can be rewritten as
\ba
 \langle\hP(\vk,\vr_L)\bar\delta(\vr_L)\rangle
 \:&=\frac{1}{V_L^2}\int\frac{d^3q_1}{(2\pi)^3}\int\frac{d^3q_3}{(2\pi)^3}~
 B(\vk-\vq_1,-\vk+\vq_1+\vq_3,-\vq_3) \vs
 \:&\hspace{4.3cm}\times W_L(\vq_1)W_L(-\vq_1-\vq_3)W_L(\vq_3) \vs
 \:&\equiv iB_L(\vk) ~.
\label{eq:ch2_ib}
\ea
As anticipated, the correlation of the position-dependent power spectrum and
the local mean density perturbation is given by an integral of the bispectrum,
and we will therefore refer to this quantity as the \emph{integrated bispectrum},
$iB_L({\bf k})$.

As expected from homogeneity, the integrated bispectrum is independent of the
location ($\vr_L$) of the subvolumes. Moreover, for an isotropic window function
and bispectrum, the result is also independent of the direction of $\vk$. The
cubic window function \refeq{ch2_WL} is of course not entirely spherically
symmetric,\footnote{We choose the cubic subvolumes merely for simplicity. In
general one can use any shapes. For example, one may prefer to divide the subvolumes
into spheres, which naturally lead to an isotropic integrated bispectrum $iB_L(k)$.}
and there is a residual dependence on $\hat k$ in \refeq{ch2_ib}. In the following,
we will focus on the angle average of \refeq{ch2_ib},
\ba
 iB_L(k)\:&\equiv\int\frac{d^2\hat k}{4\pi}~iB(\vk)
 =\left\langle\left(\int\frac{d^2\hat k}{4\pi}~\hP(\vk,\vr_L)\right)\bar\delta(\vr_L)\right\rangle \vs
 \:&=\frac1{V_L^2}\int\frac{d^2\hat k}{4\pi}\int\frac{d^3q_1}{(2\pi)^3}
 \int\frac{d^3q_3}{(2\pi)^3}~B(\vk-\vq_1,-\vk+\vq_1+\vq_3,-\vq_3) \vs
 \:&\hspace{5.7cm}\times  W_L(\vq_1)W_L(-\vq_1-\vq_3)W_L(\vq_3) ~. 
\label{eq:ch2_ib_angle_average}
\ea
The integrated bispectrum contains integrals of three sinc functions, ${\rm sinc}(x)$,
which are damped oscillating functions and peak at $|x|\lesssim \pi$. Most of the
contribution to the integrated bispectrum thus comes from values of $q_1$ and $q_3$
at approximately $1/L$. If the wavenumber $\vk$ we are interested in is much larger
than $1/L$ (e.g., $L=300~h^{-1}~{\rm Mpc}$ and $k\gtrsim0.3~h~{\rm Mpc}^{-1}$),
then the dominant contribution to the integrated bispectrum comes from the bispectrum
in squeezed configurations, i.e., $B(\vk-\vq_1,-\vk+\vq_1+\vq_3,-\vq_3) \to
B(\vk,-\vk,-\vq_3)$ with $q_1\ll k$ and $q_3\ll k$.

\subsection{Linear response function}
\label{sec:ch2_resp}
Consider the following general separable bispectrum,
\be
 B(\vk_1, \vk_2, \vk_3) = f(\vk_1,\vk_2) P(k_1) P(k_2) + 2~{\rm cyclic},
\label{eq:ch2_bis_gen}
\ee
where $f(\vk_1,\vk_2) = f(k_1, k_2, \hat k_1\cdot\hat k_2)$ is a dimensionless
symmetric function of two $\vk$ vectors and the angle between them. If $f$ is
non-singular as one of the $k$ vectors goes to zero, we can write, to lowest
order in $q_1/k$ and $q_3/k$,
\ba
 B(\vk-\vq_1,-\vk+\vq_1+\vq_3,-\vq_3) =\:& f(\vk-\vq_1,-\vq_3) P(|\vk-\vq_1|) P(q_3) \vs
 & + f(-\vk+\vq_1+\vq_3,-\vq_3) P(|-\vk+\vq_1+\vq_3|) P(q_3) \vs
 & + f(\vk-\vq_1,-\vk+\vq_1+\vq_3) P(|\vk-\vq_1|)P(|-\vk+\vq_1+\vq_3|) \vs
 =\:& 2 f(\vk, 0) P(k) P(q_3) + f(\vk,-\vk) [P(k)]^2 + \cO\left(\frac{q_{1,3}}k\right) ~.
\ea
For matter, momentum conservation requires that $f(\vk,-\vk)=0$ \cite{peebles:1974}, as
can explicitly be verified for the $F_2$ kernel of perturbation theory. We then obtain
\be
 \int\frac{d^2\hat k}{4\pi}~B(\vk-\vq_1,-\vk+\vq_1+\vq_3,-\vq_3)
 = \check{f}(k) P(k) P(q_3) + \cO\left(\frac{q_{1,3}}{k}\right)^2~,
\label{eq:ch2_bis_sq}
\ee
where $\check{f}(k) \equiv2f(0,k)$. Note that the terms linear in $q_{1,3}$
cancel after angular average. Since the window function in real space satisfies
$W_L^2(\vr) = W_L(\vr)$, we have
\be
 \int\frac{d^3q_1}{(2\pi)^3}~W_L(\vq_1)W_L(-\vq_1-\vq_3)=W_L(\vq_3) ~.
\ee
Performing the $\vq_1$ integral in \refeq{ch2_ib_angle_average} then yields
\be
 iB_L(k) \stackrel{k L\to\infty}{=}\frac{1}{V_L^2}\int\frac{d^3q_3}{(2\pi)^3}~
 W_L^2(\vq_3) P(q_3) \check{f}(k)P(k)=\sigma_L^2\check{f}(k)P(k) ~,
\label{eq:ch2_ib_sq}
\ee
where $\sigma_L^2$ is the variance of the density field on the subvolume scale,
\be
\sigma_L^2\equiv\frac{1}{V_L^2}\int\frac{d^3q_3}{(2\pi)^3}~W_L^2(\vq_3) P(q_3) ~.
\label{eq:ch2_sigmaL}
\ee
\refEq{ch2_ib_sq} shows that the integrated bispectrum measures how the small-scale
power spectrum, $P(k)$, responds to a large-scale density fluctuation with variance
$\sigma_L^2$, with a response function given by $\check{f}(k)$.

An intuitive way to arrive at the same expression is to write the response of the
small-scale power spectrum to a large-scale density fluctuation as
\be
 \hP(\vk,\vr_L)=\left.P(\vk)\right|_{\bar\delta=0}+
 \left.\frac{dP(\vk)}{d\bar\delta}\right|_{\bar\delta=0}\bar\delta(\vr_L)+\dots ~,
\label{eq:ch2_Pkexp}
\ee
where we have neglected gradients and higher derivatives of $\bar\delta(\vr_L)$.
We then obtain, to leading order,
\be
 iB_L(k)=\sigma_L^2\left.\frac{d\ln P(k)}{d\bar\delta}\right|_{\bar\delta=0}P(k).
\label{eq:ch2_CZ}
\ee
Comparing this result with \refeq{ch2_ib_sq}, we find that $\check{f}(k)$
indeed corresponds to the \emph{linear response} of the small-scale power to
the large-scale density fluctuation, $d\ln P(k)/\bar\delta$. Inspired by
\refeq{ch2_CZ}, we define another quantity, the \emph{normalized integrated
bispectrum},
\be
 \frac{iB_L(k)}{\sigma_L^2\hP(k)} ~.
\ee
This quantity is equal to $\check{f}(k)$ and the linear response function
in the limit of $kL\to \infty$.

For the standard perturbation theory kernel
\be
 f(\vk_1,\vk_2)=F_2(\vk_1,\vk_2)=\frac57+\frac12\frac{\vk_1\cdot\vk_2}{k_1k_2}
 \left(\frac{k_1}{k_2}+\frac{k_2}{k_1}\right)
 +\frac27\left(\frac{\vk_1\cdot\vk_2}{k_1k_2}\right)^2 ~,
\ee
in the squeezed limit the integrated bispectrum becomes (see \refapp{trz_kernel}
for the detailed derivation)
\be
 iB_L(k)\stackrel{k L\to\infty}{=}
 \left[\frac{47}{21}-\frac13\frac{d\ln P_l(k)}{d\ln k}\right]P_l(k)\sigma_L^2 ~,
\label{eq:ch2_ib_sq_spt}
\ee
and the response function is
\be
 \check{f}(k)=\frac{47}{21}-\frac13\frac{d\ln P_l(k)}{d\ln k} ~.
\label{eq:ch2_resp_spt}
\ee
We shall discuss more details in \refsec{ch4_bi_spt}.

\subsection{Integrated bispectrum of various bispectrum models}
\label{sec:ch2_ib_models}
To evaluate the integrated bispectrum, we insert the bispectrum models into
\refeq{ch2_ib_angle_average} and perform the eight-dimensional integral.
Because of the high dimensionality of the integral, we use the Monte Carlo
integration routine in GNU Scientific Library to numerically evaluate $iB_L(k)$.
Let us consider the simplest model of galaxy bispectrum with local-type
primordial non-Gaussianity
\be
 B_g(\vk_1,\vk_2,\vk_3)=b_1^3B_{\rm SPT}(\vk_1,\vk_2,\vk_3)
 +b_1^2b_2B_{b_2}(\vk_1,\vk_2,\vk_3)+b_1^3f_{\rm NL}B_{f_{\rm NL}}(\vk_1,\vk_2,\vk_3) ~,
\label{eq:ch2_bi_png}
\ee
where $b_1$ is the linear bias, $b_2$ is the quadratic nonlinear bias, and
$f_{\rm NL}$ is the parametrization for the local-type primordial non-Gaussianity.
Note that the scale-dependent bias due to the local type non-Gaussianity
\cite{dalal/etal:2008,matarrese/verde:2008,slosar/etal:2008} is neglected in
\refeq{ch2_bi_png}, and the latest bispectrum model with primordial non-Gaussianity
can be found in \cite{baldauf/seljak/senatore:2011,tasinato/etal:2014}.

The first two terms of \refeq{ch2_bi_png} are due to the nonlinear gravitational
evolution. Specifically, the standard perturbation theory (SPT) with local bias
model predicts (see \refapp{trz_kernel} for detailed derivation)
\ba
 B_{\rm SPT}(\vk_1,\vk_2,\vk_3)=\:&2F_2(\vk_1,\vk_2)P_l(k_1,a)P_l(k_2,a)+{\rm 2~cyclic} \vs
 B_{b_2}(\vk_1,\vk_2,\vk_3)=\:&P_l(k_1,a)P_l(k_2,a)+{\rm 2~cyclic} ~,
\label{eq:ch2_bi_hl}
\ea
where $P_l(k)$ is the linear bispectrum. For the bispectrum of local-type primordial
non-Gaussianity, we consider the local ansatz for the primordial scalar potential as
\cite{komatsu/spergel:2001}
\be
 \Phi(\vr)=\phi(\vr)+f_{\rm NL}[\phi^2(\vr)-\langle\phi^2(\vr)\rangle] ~,
\label{eq:ch2_fnl_model}
\ee
where $\phi(\vr)$ is a Gaussian field, and $f_{\rm NL}$ is a constant characterizing
the amplitude of the primordial non-Gaussianity. As the density fluctuations are
linked to the scalar potential through the Poisson equation
\be
 \delta(\vk,a)=M(k,a)\Phi(\vk) ~,~~
 M(k,a)=\frac{2}{3}\frac{D(a)}{H_0^2\Om}k^2T(k) ~,
\label{eq:ch2_poisson}
\ee
with $D(a)$ and $T(k)$ being the linear growth factor and the transfer function
respectively, in the leading order the primordial non-Gaussianity appears in
matter bispectrum as
\be
 f_{\rm NL}B_{f_{\rm NL}}(\vk_1,\vk_2,\vk_3)=2f_{\rm NL}
 M(k_1,a)M(k_2,a)M(k_3,a)[P_{\Phi}(k_1)P_{\Phi}(k_2)+{\rm 2~cyclic}] ~,
\label{eq:ch2_bi_fnl}
\ee
where $P_{\Phi}(k)$ is the power spectrum of the scalar potential.

\begin{figure}[t!]
\centering
\includegraphics[width=1\textwidth]{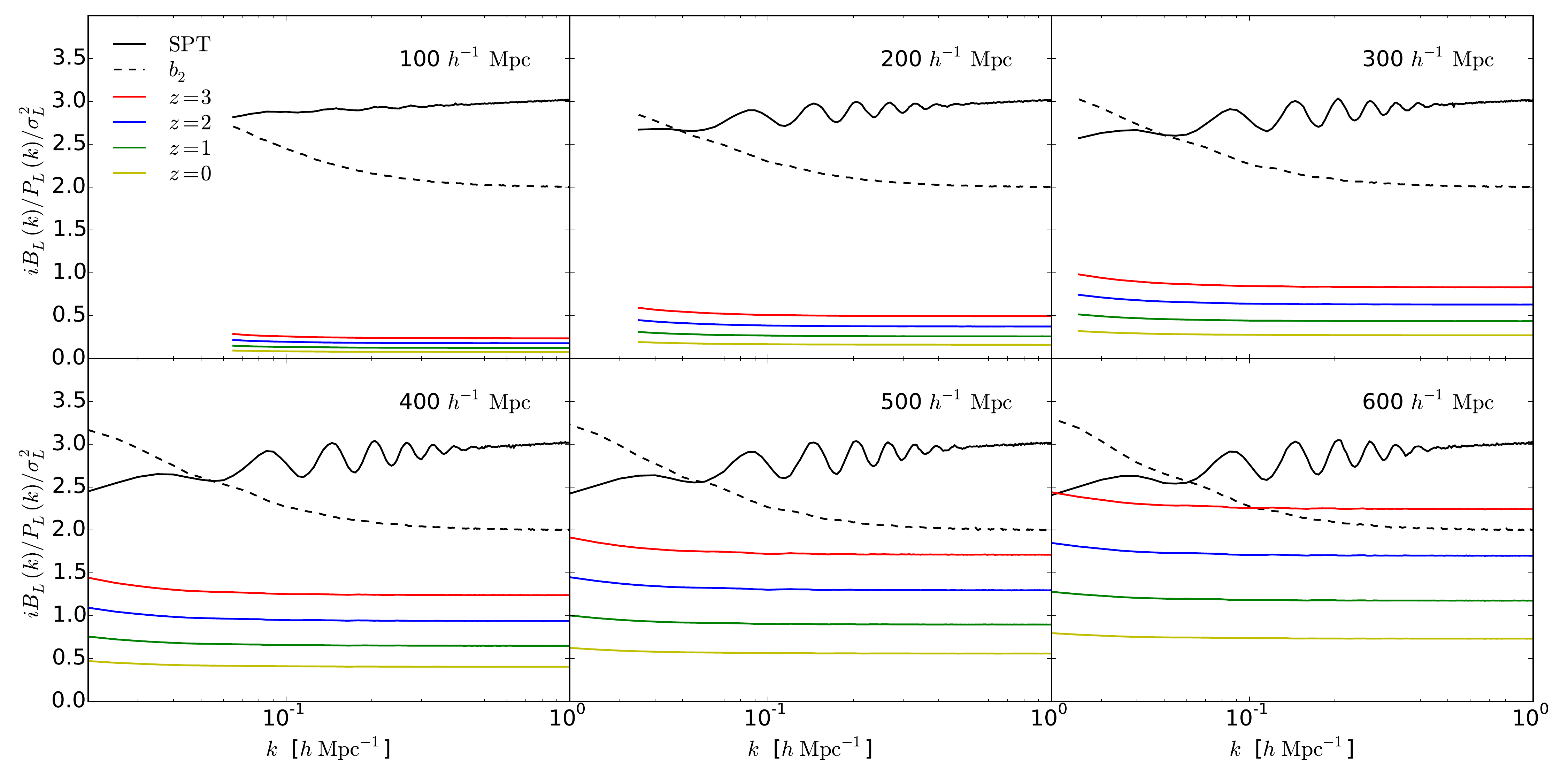}
\caption[Normalized integrated bispectrum of the simplest $f_{\rm NL}$ model]
{Normalized integrated bispectrum in different sizes of subvolumes. The colored
lines show non-Gaussian components at different redshifts assuming $f_{\rm NL}=50$,
while the solid and dashed lines show linear and nonlinear bias components assuming
$b_1=b_2=1$. The cut-off at low-$k$ corresponds to the fundamental frequencies
of the subvolumes, $2\pi/L$.}
\label{fig:ch2_ib_norm_png}
\end{figure}

\refFig{ch2_ib_norm_png} shows the normalized integrated bispectrum, for which
we shall denote as $ib_L$ in this section, in different sizes of subvolumes. For
the parameters we assume $b_1=b_2=1$ and $f_{\rm NL}=50$. We find that contributions
from late-time evolution (black solid and dashed lines for $ib_{L,{\rm SPT}}(k)$ and
$ib_{L,b_2}(k)$, respectively) are redshift-independent, but the ones from primordial
non-Gaussianity (colored lines at different redshifts) increase with increasing redshift.
This is due to the redshift-dependence in $M(k,a)$. Namely, while $ib_{L,{\rm SPT}}(k)$
and $ib_{L,b_2}(k)$ are independent of $D(a)$, $ib_{L,f_{\rm NL}}(k)$ is proportional
to $1/D(a)$. This means that it is more promising to hunt for primordial non-Gaussianity
in high-redshift galaxy surveys. We also find that for a given subvolume size $ib_{L,f_{\rm NL}}(k)$
is fairly scale-independent, as $ib_{L,{\rm SPT}}(k)$ and $ib_{L,b_2}(k)$. This is somewhat
surprising because when $k$ (the scale of the position-dependent power spectrum) is large
we reach the squeezed limit, and this should be the ideal region to search for primordial
non-Gaussianity. However, it turns out that what really determines the amplitude of $ib_{L,f_{\rm NL}}(k)$
is the subvolumes size, as we can see from different panels in \reffig{ch2_ib_norm_png}.
One can understand this by considering the long-wavelength mode, $k_l$, and the short-wavelength
modes, $k_s$ ($k_s\gg k_s$). In the squeezed limit,
\ba
 B_{f_{\rm NL}}(k_l,k_s,k_s)\:&\propto M(k_l)M^2(k_s)
 [2P_{\Phi}(k_l)P_{\Phi}(k_s)+P_{\Phi}^2(k_s)] \vs
 \:&\approxprop M(k_l)M^2(k_s)P_{\Phi}(k_l)P_{\Phi}(k_s) ~,
\label{eq:ch2_bfnl_kdep}
\ea
as $P_{\Phi}(k)\propto k^{n_s-4}$ with $n_s=0.95$. Also $M(k)\propto k^2$ and $k^0$ on
large and small scales respectively, hence
\be
 ib_{L,f_{\rm NL}}(k)\approxprop\frac{k_l^{n_s-2}k_s^{ns-4}}{\sigma_{L,l}^2M^2(k_s)P_{\Phi}(k_s)}
 \approxprop\frac{k_l^{n_s-2}}{\sigma_{L,l}^2} ~,
\label{eq:ch2_ibfnl_kdep}
\ee
which is $k_s$-independent and the amplitude of $ib_{L,f_{\rm NL}}(k)$ is solely
determined by $k_l$, the subvolume size. This means that to hunt for primordial
non-Gaussianity, it is necessary to use different sizes of subvolumes to break
the degeneracy between $ib_{L,f_{\rm NL}}(k)$ and the late-time contributions.

\section{In configuration space}
\label{sec:ch2_rspace}
\subsection{Position-dependent correlation function}
\label{sec:ch2_posdepxi}
We now turn to the position-dependent two-point statistics in configuration space,
i.e. position-dependent correlation function. Consider a density fluctuation field,
$\delta(\vr)$, in a survey (or simulation) volume $V_r$.
The global two-point function is defined as
\be
 \xi(r)=\langle\delta(\vx)\delta(\vx+\vr)\rangle ~,
\ee
where we assume that $\delta(\vr)$ is statistically homogeneous and isotropic,
so $\xi(r)$ depends only on the separation $r$. As the ensemble average cannot
be measured directly, we estimate the global two-point function as
\be
 \hat{\xi}(r)=\frac{1}{V_r} \int \frac{d^2\hat{\vr}}{4\pi}
 \int_{\vx,\vx+\vr\in V_r}d^3x~\d(\vr+\vx)\d(\vx) ~.
\label{eq:ch2_hat_xi_g}
\ee
The ensemble average of \refeq{ch2_hat_xi_g} is not equal to $\xi(r)$. Specifically,
\ba
 \langle\hat\xi(r)\rangle=\:&\frac{1}{V_r}\int\frac{d^2\hat{r}}{4\pi}
 \int_{\vx,\vx+\vr\in V_r}d^3x~\langle\d(\vr+\vx)\d(\vx)\rangle
 =\xi(r)\frac{1}{V_r}\int\frac{d^2\hat{r}}{4\pi}\int_{\vx,\vx+\vr\in V_r} d^3x ~.
\label{eq:ch2_ensavg_hat_xi_g}
\ea
The second integral in \refeq{ch2_ensavg_hat_xi_g} is $V_r$ only if $\vr=0$, and
the fact that it departs from $V_r$ is due to the finite boundary of $V_r$.
We shall quantify this boundary effect later in \refeq{ch2_ensavg_hat_xi}.

We now define the position-dependent correlation function of a cubic subvolume
$V_L$ centered at $\vr_L$ to be
\ba
 \hat\xi(\vr,\vr_L)=\:&\frac{1}{V_L}\int_{\vx,\vr+\vx \in V_L}d^3x~\delta(\vr+\vx)\delta(\vx) \vs
 =\:&\frac{1}{V_L}\int d^3x~\delta(\vr+\vx)\delta(\vx)W_L(\vr+\vx-\vr_L)W_L(\vx-\vr_L) ~,
\label{eq:ch2_posdep_xi}
\ea
where $W_L(\vr)$ is the window function given in \refeq{ch2_WL}. In this dissertation,
we consider only the angle-averaged position-dependent correlation function
(i.e. monopole) defined by
\be
 \hat\xi(r,\vr_L)=\int\frac{d^2\hat{r}}{4\pi}~\hat\xi(\vr,\vr_L)
 =\frac{1}{V_L}\int\frac{d^2\hat{r}}{4\pi}\int d^3x~
 \delta(\vr+\vx)\delta(\vx)W_L(\vr+\vx-\vr_L)W_L(\vx-\vr_L) ~.
\label{eq:ch2_hat_xi}
\ee
Again, while the overdensity in the entire volume $\bar\delta=\int_{V_r}d^3r~\delta(\vr)$
is zero by construction, the overdensity in the subvolume
$\bar\delta(\vr_L)=\int_{V_L}d^3r~\delta(\vr)=\int_{V_r}d^3r~\delta(\vr)W_L(\vr-\vr_L)$
is in general non-zero.

Similarly to that of the global two-point function, the ensemble average of
\refeq{ch2_hat_xi} is not equal to $\xi(r)$. Specifically,
\ba
 \langle\hat\xi(r,\vr_L)\rangle=\:&\frac{1}{V_L}\int\frac{d^2\hat{r}}{4\pi}\int d^3x~
 \langle\d(\vr+\vx)\d(\vx)\rangle W_L(\vr+\vx-\vr_L)W_L(\vx-\vr_L) \vs
 =\:&\xi(r)\frac{1}{V_L}\int\frac{d^2\hat{r}}{4\pi}\int d^3x'~
 W_L(\vr+\vx')W_L(\vx')\equiv\xi(r)f_{L,\rm bndry}(r) ~,
\label{eq:ch2_ensavg_hat_xi}
\ea
where $f_{L,\rm bndry}(r)$ is the boundary effect due to the finite size of
the subvolume. While $f_{L,\rm bndry}(r)=1$ for $r=0$, the boundary effect
becomes larger for larger separations. The boundary effect can be computed
by the five-dimensional integral in \refeq{ch2_ensavg_hat_xi}. Alternatively,
it can be evaluated by the ratio of the number of the random particle pairs
of a given separation in a finite volume to the expected random particle pairs
in the shell with the same separation in an infinite volume. We have evaluated
$f_{L,\rm bndry}(r)$ in both ways, and the results are in an excellent agreement.

As the usual two-point function estimators based on pair counting (such as
Landy-Szalay estimator which will be discussed in \refsec{ch5_sub_quan}) or
grid counting (which will be discussed in \refapp{mock_halo}) do not contain
the boundary effect, when we compare the measurements to the model which is
calculated based on \refeq{ch2_hat_xi}, we shall divide the model by
$f_{L,\rm bndry}(r)$ to correct for the boundary effect.

\subsection{Integrated three-point function}
\label{sec:ch2_iz}
The correlation between $\hat\xi(r,\vr_L)$ and $\bar\delta(\vr_L)$ is given by
\ba
 \langle\hat\xi(r,\vr_L)\bar\delta(\vr_L)\rangle=\:&
 \frac1{V_L^2}\int\frac{d^2\hat{r}}{4\pi}\int d^3x_1\int d^3x_2
 ~\langle\delta(\vr+\vx_1)\delta(\vx_1)\delta(\vx_2)\rangle \vs
 &\hspace{4cm}\times W_L(\vr+\vx_1-\vr_L)W_L(\vx_1-\vr_L)W_L(\vx_2-\vr_L) \vs
 =\:&\frac{1}{V_L^2}\int\frac{d^2\hat{r}}{4\pi}\int d^3x_1\int d^3x_2
 ~\zeta(\vr+\vx_1+\vr_L,\vx_1+\vr_L,\vx_2+\vr_L) \vs
 &\hspace{4cm}\times W_L(\vr+\vx_1)W_L(\vx_1)W_L(\vx_2) ~,
\label{eq:ch2_iz}
\ea
where $\zeta(\vr_1,\vr_2,\vr_3)\equiv\langle\delta(\vr_1)\delta(\vr_2)\delta(\vr_3)\rangle$
is the three-point correlation function. Because of the assumption of homogeneity
and isotropy, the three-point function depends only on the separations $|\vr_i-\vr_j|$
for $i\neq j$, and so $\langle\hat\xi(r,\vr_L)\bar\delta(\vr_L)\rangle$ is independent
of $\vr_L$. Furthermore, as the right-hand-side of \refeq{ch2_iz} is an integral
of the three-point function, we will refer to this quantity as the ``integrated
three-point function,'' $\iz_L(r)\equiv\langle\hat\xi(r,\vr_L)\bar\delta(\vr_L)\rangle$.

$\iz_L(r)$ can be computed if $\zeta(\vr_1,\vr_2,\vr_3)$ is known. For example,
SPT with the local bias model at the tree level in real space gives
\be
 \zeta(r)=b_1^3\zeta_{\rm SPT}(r)+b_1^2b_2\zeta_{b_2}(r) ~,
\ee
where $\zeta_{\rm SPT}$ and $\zeta_{b_2}$ are given below. Here, $b_1$ and $b_2$
are the linear and quadratic (nonlinear) bias parameters, respectively. Because
of the high dimensionality of the integral, we use the Monte Carlo integration
routine in the GNU Scientific Library to numerically evaluate the eight-dimensional
integral for $\iz_L(r)$.

The first term, $\zeta_{\rm SPT}$, is given by \cite{jing/boerner:1997,barriga/gaztanaga:2002}
\ba
 \zeta_{\rm SPT}(\vr_1,\vr_2,\vr_3)=\:&\frac{10}{7}\xi_l(r_{12})\xi_l(r_{23})
 +\mu_{12,23}[\xi_l'(r_{12})\phi_l'(r_{23})+\xi_l'(r_{23})\phi_l'(r_{12})] \vs
 \:&+\frac{4}{7}\Bigg\lbrace-3\frac{\phi_l'(r_{12})\phi_l'(r_{23})}{r_{12}r_{13}}
 -\frac{\xi_l(r_{12})\phi_l'(r_{23})}{r_{23}}-\frac{\xi_l(r_{23})\phi_l'(r_{12})}{r_{12}} \vs
 &~~~~~~~~ +\mu_{12,23}^2\left[\xi_l(r_{12})+\frac{3\phi_l'(r_{12})}{r_{12}}\right]
 \left[\xi_l(r_{23})+\frac{3\phi_l'(r_{23})}{r_{23}}\right]\Bigg\rbrace \vs
 \:&+2~{\rm cyclic} ~,
\label{eq:ch2_zeta_spt}
\ea
where $r_{12}=|\vr_1-\vr_2|$, $\mu_{12,23}$ is the cosine between $\vr_{12}$
and $\vr_{23}$, $'$ refers to the spatial derivative, and
\be
 \xi_l(r)\equiv\int\frac{dk}{2\pi^2}~k^2P_l(k){\rm sinc}(kr)\\, ~~~~~~
 \phi_l(r)\equiv\int\frac{dk}{2\pi^2}~P_l(k){\rm sinc}(kr)\\,
\label{eq:ch2_xil_phil}
\ee
with $P_l(k)$ being the linear matter power spectrum. The subscript $l$ denotes the
quantities in the linear regime. The second term, $\zeta_{b_2}$, is the nonlinear
local bias three-point function. The nonlinear bias three-point function is then
\be
 \zeta_{b_2}(\vr_1,\vr_2,\vr_3)=\xi_l(r_{12})\xi_l(r_{23})+2~{\rm cyclic} ~.
\label{eq:ch2_zeta_b2}
\ee
Note that $\zeta_{\rm SPT}$ and $\zeta_{b_2}$ are simply Fourier transform of
$B_{\rm SPT}$ and $B_{b_2}$ respectively, as shown in \refeq{ch2_bi_hl}.

\begin{figure}[t!]
\centering
\includegraphics[width=1\textwidth]{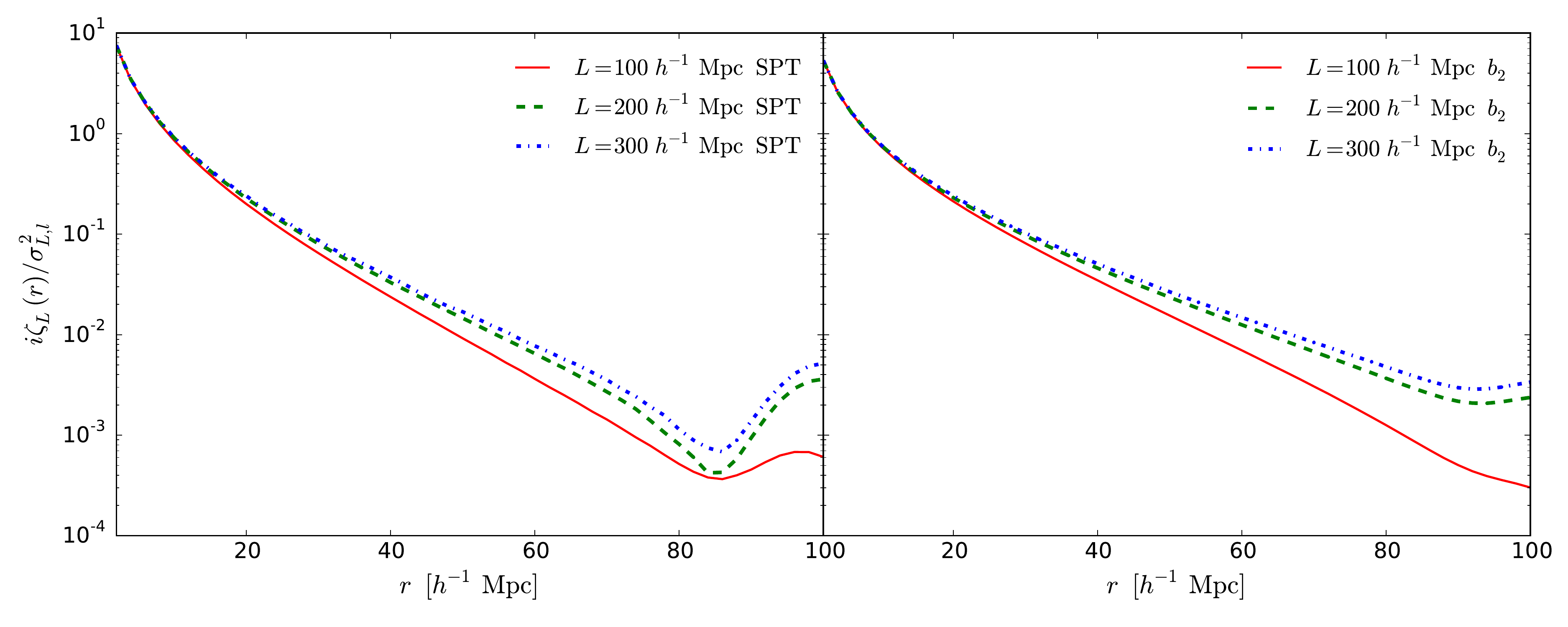}
\caption[Normalized $\iz_{L,{\rm SPT}}$ and $\iz_{L,b_2}$]
{Normalized $\iz_{L,{\rm SPT}}$ (left) and $\iz_{L,b_2}$ (right) for $L=100\hMpc$
(red solid), $200\hMpc$ (green dashed), and 300$\hMpc$ (blue dotted) at $z=0$.}
\label{fig:ch2_iz_norm}
\end{figure}

\refFig{ch2_iz_norm} shows the scale-dependencies of $\iz_{L,{\rm SPT}}$ and
$\iz_{L,b_2}$ at $z=0$ with $P_l(k)$ computed by CLASS \cite{lesgourgues:2011}.
We normalize $\iz_L(r)$ by $\sigma_{L,l}^2$, where
\be
 \sigma_{L,l}^2\equiv\langle\bar\delta_l(\vr_L)^2\rangle
 =\frac{1}{V_L^2}\int\frac{d^3k}{(2\pi)^3}~P_l(k)|W_L(\vk)|^2 ~,
\label{eq:ch2_sigmalL2}
\ee
is the variance of the linear density field in the subvolume $V_L$. The choice of
this normalization is similar to that of the integrated bispectrum as discussed in
\refsec{ch2_ib}, and we shall discuss more details in \refsec{ch2_iz_sq}. We find
that the scale-dependencies of $\iz_{L,{\rm SPT}}(r)$ and $\iz_{L,b_2}(r)$ are similar
especially on small scales.
This is because the scale-dependence of the bispectrum in the squeezed limit is
(see appendix of \refapp{trz_sq})
\be
 B_{\rm SPT}\to\left[\frac{68}{21}-\frac13\frac{d\ln k^3P_l(k)}{d\ln k}\right]P_l(k)P_l(q) ~,~~
 B_{b_2}\to2P_l(k)P_l(q) ~,
\ee
where $k$ and $q$ are the short- and long-wavelength modes, respectively. For
a power-law power spectrum without features, the squeezed-limit $B_{\rm SPT}$
and $B_{b_2}$ have exactly the same scale dependence and cannot be distinguished.
This results in a significant residual degeneracy between
$b_1$ and $b_2$, and will be discussed in \refchp{ch5_posdepxi} where we measure the
integrated three-point function of real data and fit to the models. When $r$ is small,
$\iz_L(r)/\sigma_{L,l}^2$ becomes independent of the subvolume size. We derive this
feature when we discuss the squeezed limit, where $r\ll L$, in \refsec{ch2_iz_sq}.

\subsection{Connection to the integrated bispectrum}
\label{sec:ch2_iz_to_ib}
Fourier transforming the density fields, the integrated three-point function
can be written as
\ba
 \iz_L(\vr) \:&=\frac{1}{V_L^2}\int\frac{d^3q_1}{(2\pi)^3}\cdots\int\frac{d^3q_6}{(2\pi)^3}~(2\pi)^9
 \delta_D(\vq_1+\vq_2+\vq_3)\delta_D(\vq_1+\vq_2+\vq_4+\vq_5)\delta_D(\vq_3+\vq_6) \vs
 \:&~~~~~~~~~~~~~~~~~~\times B(\vq_1,\vq_2,\vq_3)W_L(\vq_4)W_L(\vq_5)W_L(\vq_6)
 e^{i[\vr\cdot(\vq_1+\vq_4)-\vr_L\cdot(\vq_4+\vq_5+\vq_6)]} \vs
 \:&=\int\frac{d^3k}{(2\pi)^3}~iB_L(\vk)e^{i\vr\cdot\vk} ~,
\label{eq:ch2_iz_ft_d}
\ea
and it is simply the Fourier transform of the integrated bispectrum. Similarly,
the angle-averaged integrated three-point function is related to the angle-averaged
integrated bispectrum (\refeq{ch2_ib_angle_average}) as
\be
 \iz_L(r)=\int\frac{k^2dk}{2\pi^2}~iB_L(k){\rm sinc}(kr) ~.
\label{eq:ch2_iz_ib_ang_avg}
\ee

In \refchp{ch5_posdepxi}, we measure the integrated three-point function of real data,
and thus we need the model for $\iz_L(r)$ in redshift space. Unlike the three-point
function in real space (\refeq{ch2_zeta_spt} and \refeq{ch2_zeta_b2}), we do not have
the analytical expression for redshift-space three-point function in configuration
space. Since the integrated three-point function is the Fourier transform of the
integrated bispectrum, we compute the redshift-space integrated three-point function
by first evaluating the redshift-space angle-averaged integrated bispectrum with
\refeq{ch2_ib_angle_average}\footnote{The explicit expression of the SPT redshift-space
bispectrum is in \refapp{trz_kernel}.}, and then performing the one-dimensional
integral \refeq{ch2_iz_ib_ang_avg}. This operation thus requires a nine-dimensional
integral.

\begin{figure}[t!]
\centering
\includegraphics[width=1\textwidth]{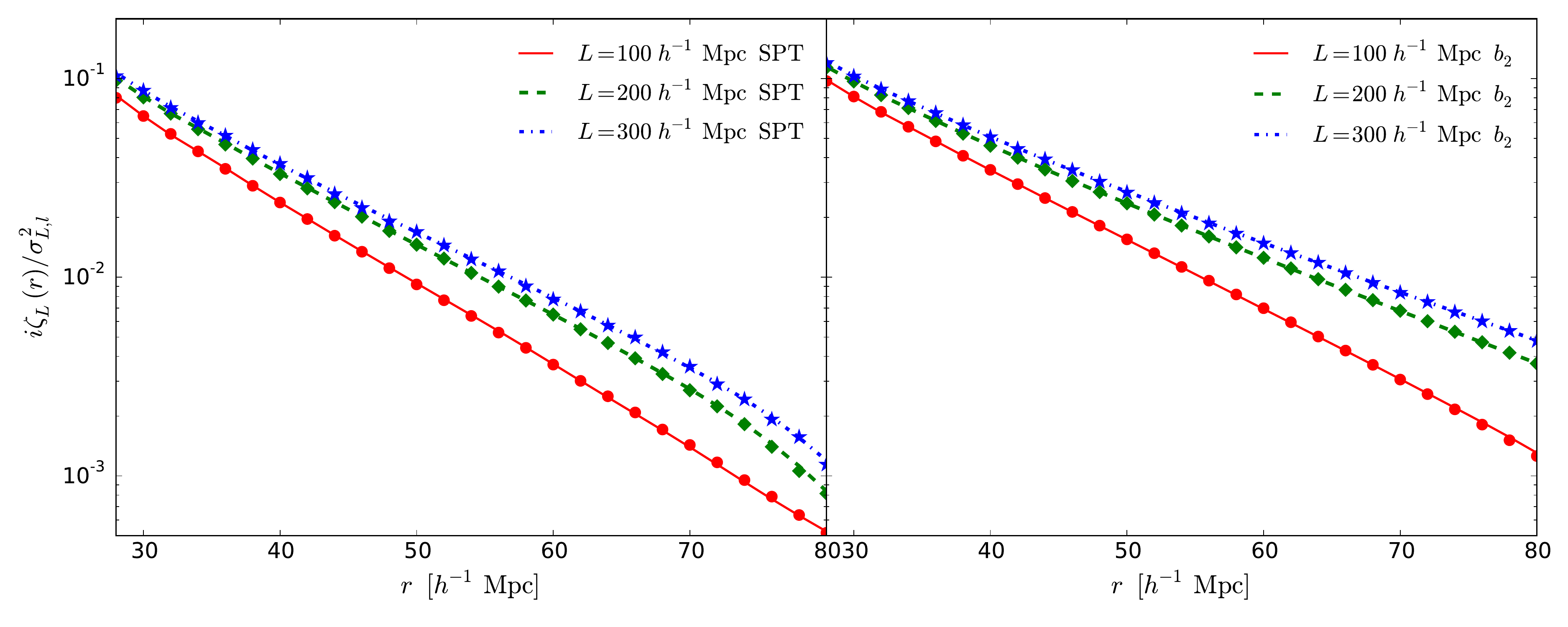}
\caption[Normalized $\iz_{L,{\rm SPT}}$ and $\iz_{L,b_2}$ evaluated from Fourier space]
{Normalized $\iz_{L,{\rm SPT}}$ (left) and $\iz_{L,b_2}$ (right) for $L=100\hMpc$
(red solid), $200\hMpc$ (green dashed), and 300$\hMpc$ (blue dotted) at $z=0$
evaluated from Fourier space of a nine-dimensional integral (\refeq{ch2_ib_angle_average}
and \refeq{ch2_iz_ib_ang_avg}). The same quantities evaluated from configuration
space of an eight-dimensional integral (\refeq{ch2_iz}) are shown with symbols.}
\label{fig:ch2_iz_norm_kspace}
\end{figure}

To check the precision of numerical integration, we compare the results from
the eight-dimensional integral in \refeq{ch2_iz} with the nine-dimensional integral
\refeq{ch2_ib_angle_average} and \refeq{ch2_iz_ib_ang_avg} for both $\iz_{L,{\rm SPT}}$
and $\iz_{L,b_2}$. As the latter gives a noisy result, we apply a Savitzky-Golay
filter (with window size 9 and polynomial order 4) six times, and the results
are shown in \reffig{ch2_iz_norm_kspace}. We find that, on the scales of interest
($30\hMpc\le r\le78\hMpc$, which we will justify in \refsec{ch5_mock_r}), both
results are in agreement to  within 2\%. As the current uncertainty on the measured
integrated correlation function presented in this dissertation is of order 10\%
(see \refsec{ch5_data} for more details), the numerical integration yields sufficiently
precise results.

\subsection{Squeezed limit}
\label{sec:ch2_iz_sq}
In the squeezed limit, where the separation of the position-dependent correlation
function is much smaller than the size of the subvolume ($r\ll L$), the integrated
three-point function has a straightforward physical interpretation, as for the
integrated bispectrum. In this case, the mean density in the subvolume acts effectively
as a constant ``background'' density (see \refchp{ch3_sepuni} for more details).
Consider the position-dependent correlation function, $\hat\xi(\vr,\vr_L)$, is
measured in a subvolume with overdensity $\bar\delta(\vr_L)$. If the overdensity
is small, we may Taylor expand $\hat\xi(\vr,\vr_L)$ in orders of $\bar\delta$ as
\be
 \hat\xi(\vr,\vr_L) = \left.\xi(\vr)\right|_{\bar\delta=0}+
 \left.\frac{d\xi(\vr)}{d\bar\delta}\right|_{\bar\delta=0}\bar\delta+\cO(\bar\delta^2) ~.
\label{eq:ch2_xi_sepuni}
\ee
The integrated three-point function in the squeezed limit is then, at leading order
in the variance $\langle\bar\delta^2\rangle$, given by
\be
 \iz_L(\vr)=\langle\hat\xi(\vr,\vr_L)\bar\delta(\vr_L)\rangle
 =\left.\frac{d\xi(\vr)}{d\bar\delta}\right|_{\bar\delta=0}
 \langle\bar\delta^2\rangle+\cO(\bar\delta^3) ~.
\label{eq:ch2_iz_sq}
\ee
As $\langle\bar\delta^2\rangle=\sigma_L^2$\footnote{If $\bar\delta=\bar\delta_l$ then
$\sigma_L^2=\sigma_{L,l}^2$. But $\bar\delta$ can in principle be nonlinear or the mean
overdensity of the biased tracers, so here we denote the variance to be $\sigma_L^2$.},
$\iz_L(\vr)$ normalized by $\sigma_L^2$ is $d\xi(\vr)/d\bar\delta$ at leading order,
which is the linear response of the correlation function to the overdensity. Note that
in \refeq{ch2_iz_sq} there is no dependence on the subvolume size apart from $\sigma_L^2$,
as shown also by the asymptotic behavior of the solid lines in \reffig{ch2_iz_norm} for
$r\to0$.

As $\iz_L(r)$ is the Fourier transform of $iB_L(k)$, the response of the correlation
function, $d\xi(r)/d\bar\delta$, is also the Fourier transform of the response of
the power spectrum, $dP(k)/d\bar\delta$. For example, we can calculate the response
of the linear matter correlation function, $d\xi_l(r)/d\bar\delta$, by Fourier transforming
the response of the linear power spectrum, which is given in \refeqs{ch2_ib_sq_spt}{ch2_resp_spt}.
In \reffig{ch2_iz_norm_sq}, we compare the normalized $\iz_{L,{\rm SPT}}(r)$ with
$d\xi_l(r)/d\bar\delta$. Due to the large dynamic range of the correlation function,
we divide all the predictions by $\xi(r)$. As expected, the smaller the subvolume
size, the smaller the $r$ for $\iz_{L,{\rm SPT}}(r)$ to be close to
$[1/\xi_l(r)][d\xi_l(r)/d\bar\delta]$, i.e., reaching the squeezed limit. Specifically,
for $100\hMpc$, $200\hMpc$, and $300\hMpc$ subvolumes, the squeezed limit is reached
to 10\% level at $r\sim10\hMpc$, $18\hMpc$, and $25\hMpc$, respectively.

\begin{figure}[t!]
\centering
\includegraphics[width=0.8\textwidth]{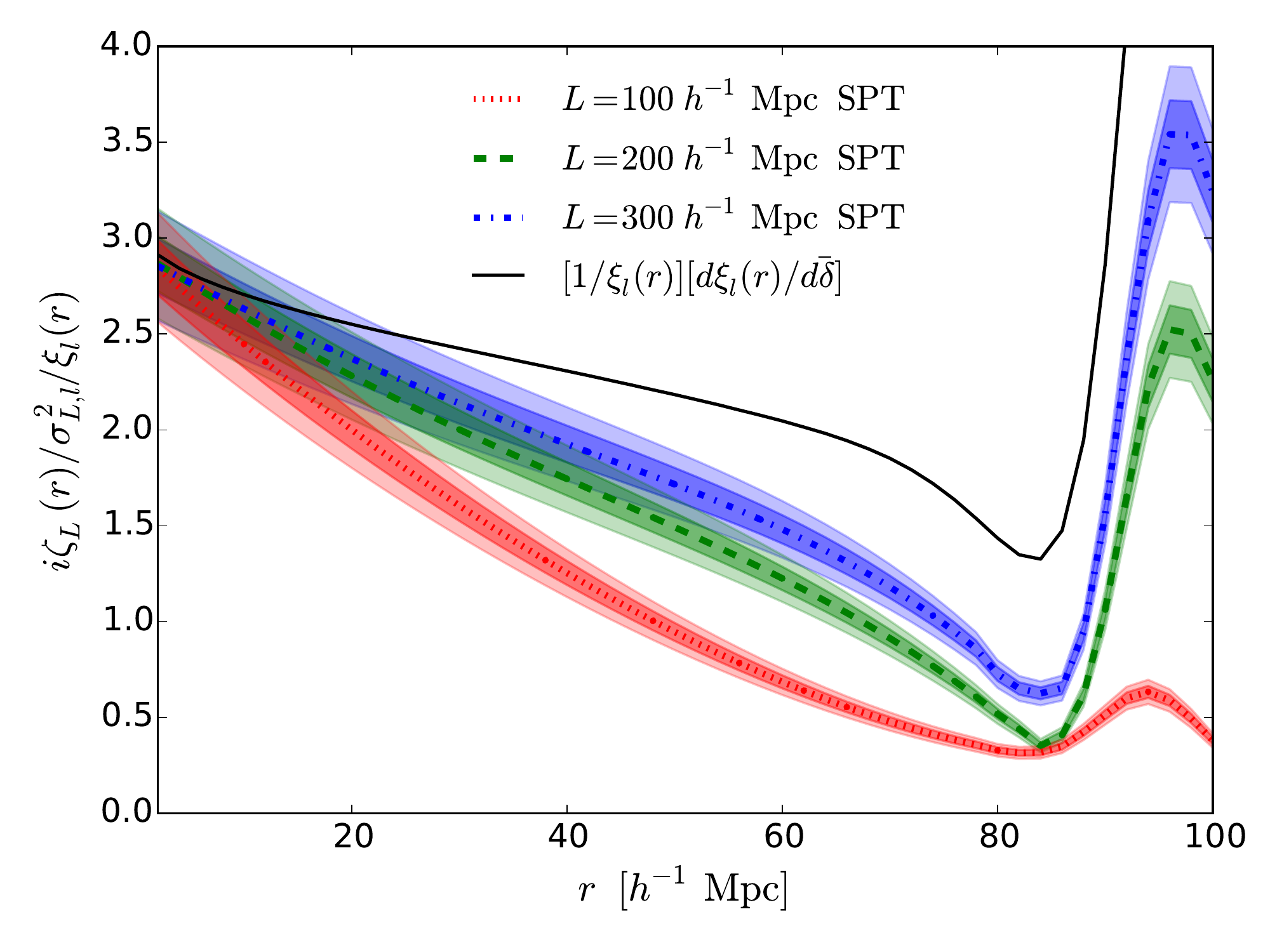}
\caption[Linear response of the correlation function and normalized $\iz_{L,{\rm SPT}}(r)$]
{The linear response function $[1/\xi_l(r)][d\xi_l(r)/d\bar\delta]$ (black solid)
and the normalized $\iz_{L,{\rm SPT}}(r)$ for $L=100\hMpc$ (red dotted), $200\hMpc$
(green dashed), and 300$\hMpc$ (blue dot-dashed). The light and dark bands
correspond to $\pm5\%$ and $\pm10\%$ of the predictions, respectively.}
\label{fig:ch2_iz_norm_sq}
\end{figure}

\subsection{Shot noise}
\label{sec:ch2_shot}
If the density field is traced by discrete particles, $\d_d(\vr)$, then the three-point
function contains a shot noise contribution given by
\ba
 \langle\d_d(\vr_1)\d_d(\vr_2)\d_d(\vr_3)\rangle\:&=\langle\d(\vr_1)\d(\vr_2)\d(\vr_3)\rangle \vs
 \:&+\left[\frac{\langle\d(\vr_1)\d(\vr_2)\rangle}{\bar n(\vr_3)}\d_D(\vr_1-\vr_3)+2~{\rm cyclic}\right]
 +\frac{\delta_D(\vr_1-\vr_2)\delta_D(\vr_1-\vr_3)}{\bar n(\vr_2)\bar n(\vr_3)} ~,
\label{eq:ch2_shot}
\ea
where $\bar n(r)$ is the mean number density of the discrete particles. The shot noise
can be safely neglected for the three-point function because it only contributes when
$\vr_1=\vr_2$, $\vr_1=\vr_3$, or $\vr_2=\vr_3$. On the other hand, the shot noise of
the integrated three-point function can be computed by inserting \refeq{ch2_shot} into
\refeq{ch2_iz}, which yields
\ba
 \iz_{\rm shot}(r)=\xi(r)\frac1{V_L^2}\int\frac{d^2\hat r}{4\pi}\int d^3x~
 \left[\frac1{\bar n(\vx+\vr+\vr_L)}+\frac1{\bar n(\vx+\vr_L)}\right]W_L(\vx+\vr)W_L(\vx) ~,
\ea
where we have assumed $r>0$. If we further assume that the mean number
density is constant, then the shot noise of the integrated three-point function can be
simplified as
\be
 \iz_{\rm shot}(r)=2\xi(r)\frac1{V_L\bar n}f_{L,\rm bndry}(r) ~.
\ee
For the measurements of PTHalos mock catalogs and the BOSS DR10 CMASS sample shown
in \refchp{ch5_posdepxi}, the shot noise is subdominant (less than 7\% of the total
signal on the scales of interest).

}

%% file: kap_03.tex
{

\newcommand{\Om}{\Omega_m}
\newcommand{\tO}{\tilde{\Omega}}
\newcommand{\tH}{\tilde{H}}
\renewcommand{\th}{\tilde{h}}
\newcommand{\tK}{\tilde{K}}
\newcommand{\tD}{\tilde{D}}
\newcommand{\tf}{\tilde{f}}
\newcommand{\ta}{\tilde{a}}
\newcommand{\td}{\tilde{\delta}}
\newcommand{\drho}{\delta_{\rho}}
\newcommand{\da}{\delta_a}
\newcommand{\dH}{\delta_H}
\newcommand{\trho}{\tilde{\rho}}
\newcommand{\rhob}{\bar{\rho}}
\newcommand{\trhob}{\tilde{\bar\rho}}
\newcommand{\tp}{\tilde{p}}
\newcommand{\dotd}{\dot{\delta}}
\newcommand{\ddotd}{\ddot{\delta}}
\newcommand{\tou}{t_{\rm out}}

\chapter{Separate universe picture}
\label{chp:ch3_sepuni}
Mode coupling plays a fundamental role in cosmology. Even if the initial density
fluctuations generated by inflation are perfectly Gaussian, the subsequent nonlinear
gravitational evolution couples long and short wavelength modes as well as generates
non-zero bispectrum. A precise understanding of how a long-wavelength density affects
the small-scale structure formation gravitationally is necessary, especially for
extracting the signal of bispectrum due to primordial non-Gaussianity.

A useful way to describe the behavior of the small-scale structure formation in an
overdense (underdense) environment is the separate universe picture \cite{barrow/saich:1993,
mcdonald:2003, sirko:2005,baldauf/etal:2011,sherwin/zaldarriaga:2012,li/hu/takada:2014,
li/hu/takada:2014b,dai/pajer/schmidt:2015a,dai/pajer/schmidt:2015b}. Imagine a local
observer who can only access to the comoving distance of the short-wavelength modes
$\sim1/k_S$. If there exists a long-wavelength mode such that $1/k_L\gg1/k_S$, then
the local small-scale physics would be interpreted with a Friedmann-Lema{\^i}tre-Robertson-Walker
(FLRW) background \cite{lemaitre:1933}. That is, in the separate universe picture,
the overdensity is absorbed into the modified background cosmology, and the small-scale
structure formation evolves in this modified cosmology.

The separate universe picture can be proven from the general relativistic approach:
one locally constructs a frame, Conformal Fermi Coordinates, such that it is valid
across the scale of a coarse-grained universe $\Lambda^{-1}$ ($k_L<\Lambda<k_S$) at
all times \cite{pajer/schmidt/zaldarriaga:2013,dai/pajer/schmidt:2015a,dai/pajer/schmidt:2015b}.
In Conformal Fermi Coordinates, the small-scale structure around an observer is
interpreted as evolving in FLRW universe modified by the long-wavelength overdensity.
The separate universe picture is restricted to scales larger than the sound horizons
of all fluid components, where all fluid components are comoving.

In the presence of $\bar\delta$, the background cosmology changes, and the position-dependent
power spectrum is affected accordingly. Since the response of the position-dependent
power spectrum to the long-wavelength overdensity, $[dP(k)/d\bar\delta]_{\bar\delta=0}$,
can be related to the bispectrum in the squeezed configurations, the separate universe
picture is useful for modeling the squeezed-limit bispectrum generated by nonlinear
gravitational evolution. Not only this is conceptually straightforward to interpret,
but it also captures more nonlinear effects than the direct bispectrum modeling,
as we will show in \refchp{ch4_posdeppk}. 

In this chapter, which serves as the basis for the later separate universe approach
modeling, we derive the mapping between the overdense universe in a spatially flat
background cosmology (but with cosmological constant) to the modified cosmology in
\refsec{ch3_mapping}. In \refsec{ch3_sep_uni_eds}, we show that if the background
cosmology is Einstein-de Sitter, i.e. matter dominated, then the changes in the
scale factor and the linear growth factor can be solved analytically.

\section{Mapping the overdense universe to the modified cosmology}
\label{sec:ch3_mapping}
Consider the universe with mean density $\rhob(t)$ and a region with overdensity
$\drho(t)$, then the mean density $\trhob(t)$ in this region is
\be
 \trhob(t)=\rhob(t)[1+\drho(t)] ~.
\label{eq:ch3_trhob}
\ee
In this section, we shall derive the mapping of the cosmological parameters
between the fiducial (overdense) and modified cosmologies as a function of
the linearly extrapolated (Lagrangian) present-day overdensity
\be
 \delta_{L0}=\drho(t_i)\frac{D(t_0)}{D(t_i)} ~,
\label{eq:ch3_dL0}
\ee
where $D$ is the linear growth function of the fiducial cosmology, $t_0$
is the present time, and $t_i$ is some initial time at which $\drho(t_i)$
is still small. However, we shall not assume $\delta_{L0}$ to be small.

The mean overdensity of the overdense region can be expressed in terms of
the standard cosmological parameters, i.e. $\rhob(t_0)=\Om\frac{3H_0^2}{8\pi G}$
and $H_0=100~h~{\rm km}~{\rm s}^{-1}~{\rm Mpc}^{-1}$, as
\be
 \frac{\Om h^2}{a^3(t)}[1+\drho(t)]=\frac{\tO_m\th^2}{\ta^3(t)} ~
\label{eq:ch3_Omh2_1}
\ee
where the parameters in the modified cosmology are denoted with tilde.
For the fiducial cosmology, we adopt the standard convention for the scale
factor such that $a(t_0)=1$. In contrast, for the modified cosmology, it is
convenient to choose $\ta(t\to0)=a$ and $\drho(a\to0)=0$, which then leads to
\be
 \Om h^2=\tO_m\th^2 ~.
\label{eq:ch3_Omh2_2}
\ee
We also introduce
\be
 \ta(t)=a(t)[1+\da(t)] ~,
\label{eq:ch3_ta_1}
\ee
so that
\be
 1+\drho(t)=[1+\da(t)]^{-3} ~.
\label{eq:ch3_da}
\ee
Using \refeq{ch3_ta_1}, the first and second time derivatives, represented with
dots, of the scale factor are
\ba
 \frac{\dot{\ta}}{\ta}\:&=\tH=\frac{\dot{a}[1+\da]+a\dotd_a}{a[1+\da]}=H+\frac{\dotd_a}{1+\da} ~, \vs
 \frac{\ddot{\ta}}{\ta}\:&=\frac{\ddot{\ta}[1+\da]+2\dot{\ta}\dotd_a+a\ddotd_a}{a[1+\da]}
 =\frac{\ddot{a}}{a}+\frac{\ddotd_a+2H\dotd_a}{1+\da} ~.
\label{eq:ch3_da_dot}
\ea

The two Friedmann equations in the flat fiducial cosmology are
\ba
 \left(\frac{\dot{a}}{a}\right)^2\:&=H^2(t)=\frac{8\pi G}{3}[\rhob(t)+\rho_X(t)] ~, \vs
 \frac{\ddot{a}}{a}\:&=-\frac{4\pi G}{3}[\rhob(t)+\rho_X(t)+3p_x(t)] ~,
\label{eq:ch3_fr1}
\ea
where $\rho_X$ and $p_x$ are the energy density and pressure for dark energy,
respectively. For the modified cosmology, the Friedmann equations hold, but
with non-zero curvature $\tK$ and modified densities and scale factor as
\ba
 \left(\frac{\dot{\ta}}{\ta}\right)^2\:&=\tH^2(t)=\frac{8\pi G}{3}[\trhob(t)+\trho_X(t)]-\frac{\tK}{\ta^2(t)} ~, \vs
 \frac{\ddot{\ta}}{\ta}\:&=-\frac{4\pi G}{3}[\trhob(t)+\trho_X(t)+3\tp_x(t)] ~.
\label{eq:ch3_fr2}
\ea
Before we derive the relation for the curvature $\tK$, let us first
discuss the dark energy component.

If dark energy is not a cosmological constant, then there are also perturbations
in dark energy fluid, i.e. $\delta\rho_X=\trho_X-\rho_X$ and $\delta p_X=\tp_X-p_X$.
In order for the separate universe approach to work, matter and dark energy have
to be comoving and follow geodesics of the FLRW metric. Since this requires negligible
pressure gradients, the separate universe approach is only applicable to density
perturbations with wavelength $2\pi/k$ that are much larger than the dark energy
sound horizon, $k\ll H_0/|c_s|$, where the sound speed is defined as $c_s^2=\delta p_X/\delta\rho_X$ \cite{creminelli/etal:2010}.
This means that the region with the overdensity that we consider here has to be
much larger than the dark energy sound horizon. For simplicity, in the following
we shall assume that dark energy is just a cosmological constant $\Lambda$, thereby
$\trho_{\Lambda}=\rho_{\Lambda}$ and $\tp_{\Lambda}=p_{\Lambda}=-\rho_{\Lambda}$.

In order to be a valid Friedmann model, the curvature has to be conserved, i.e.
$\dot{\tK}=0$. To show this, we express the curvature using the first Friedmann
equation and take the time derivative:
\ba
 \dot{\tK}\:&=\frac{16\pi G}{3}\ta\dot{\ta}[\trhob+\trho_{\Lambda}]
 +\frac{8\pi G}{3}\ta^2\dot{\trhob}-2\dot{\ta}\ddot{\ta}
 =\frac{16\pi G}{3}\ta\dot{\ta}[\trhob+\trho_{\Lambda}]
 -\frac{24\pi G}{3}\ta\dot{\ta}\trhob-2\dot{\ta}\ddot{\ta} \nonumber\\
 \:&=\frac{8\pi G}{3}\ta\dot{\ta}[-\trhob+2\trho_{\Lambda}]-2\dot{\ta}\ddot{\ta}
 =2\ta\dot{\ta}\frac{\ddot{\ta}}{\ta}-2\dot{\ta}\ddot{\ta}=0 ~,
\label{eq:ch3_tK_dot}
\ea
where we use $\dot{\trho}_{\Lambda}=0$ and the continuity equation
$\dot{\trhob}=-3\trhob\tH$.

To solve the curvature of the modified cosmology in terms of the fiducial
cosmological parameters and $\delta_{L0}$, we express $\tK$ by the difference
of the first Friedmann equation between the modified and fiducial cosmologies as
\ba
 \frac{\tK}{a^2}\:&=\frac{8\pi G}{3}\rhob\drho(1+\da)^2
 -2H\dotd_a(1+\da)-\dotd_a^2 \nonumber\\
 \:&=\frac{8\pi G}{3}\rhob\left[(1+\da)^{-1}-(1+\da)^2\right]
 -2H\dotd_a(1+\da)-\dotd_a^2 ~,
\label{eq:ch3_tK_1}
\ea
where we use \refeq{ch3_da} and \refeq{ch3_da_dot} to represent $\drho$ and $\tH$,
respectively. Since the curvature is conserved (\refeq{ch3_tK_dot}), \refeq{ch3_tK_1}
can be evaluated at an early time $t_i$ at which the perturbations $\da(t_i)$
and $\drho(t_i)$ are infinitesimal as well as the universe is in matter domination.
In this regime, we have
\be
 H^2=H_0^2\Om a^{-3} ~,~~~ \da=-\drho/3 ~,~~~ \dotd_a=H\da ~,
\label{eq:ch3_md}
\ee
with which we can derive
\be
 \frac{\tK}{a^2(t_i)}=\frac{8\pi G}{3}\rhob(t_i)[-3\da(t_i)]-2H^2(t_i)\da(t_i)
 =\frac{5\Om H_0^2\drho(t_i)}{3a^3(t_i)} ~.
\label{eq:ch3_tK_2}
\ee
This then to express $\tK$ in terms of the fiducial cosmological parameters and
$\delta_{L0}$ as
\be
 \frac{\tK}{H_0^2}=\frac53\frac{\Om}{a(t_i)}\drho(t_i)
 =\frac53\frac{\Om}{D(t_0)}\delta_{L0} ~,
\label{eq:ch3_tK_3}
\ee
where we use the fact that in the matter dominated regime $D(t_i)=a(t_i)$.

We now derive the cosmological parameters, $\tO_m$, $\tO_{\Lambda}$, and $\tO_K$
of the modified cosmology. Note that by convention these parameters are defined
through the first Friedmann equation at $\tilde{t}_0$ where $\ta(\tilde{t}_0)=1$
and $\tH(\tilde{t}_0)=\tH_0$, therefore
\be
 \tO_K=-\frac{\tK}{\tH_0^2} ~;~~~ \tO_m=\frac{8\pi G}{3\tH_0^2}\trhob(\tilde{t}_0)
 ~;~~~ \tO_{\Lambda}=\frac{8\pi G}{3\tH_0^2}{\rho_{\Lambda}} ~.
\label{eq:ch3_Omega_1}
\ee
The cosmological parameters in the modified cosmology can be related to the ones
in the fiducial cosmology through $\tH_0=H_0(1+\dH)$, which become
\be
 \tO_m=\Om(1+\dH)^{-2} ~,~~~ \tO_{\Lambda}=\Omega_{\Lambda}(1+\dH)^{-2}
 ~,~~~ \tO_K=1-\tO_m-\tO_{\Lambda}=1-(1+\dH)^{-2} ~,
\label{eq:ch3_Omega_2}
\ee
where we use $\Om+\Omega_{\Lambda}=1$ because the fiducial cosmology is flat.

Alternatively, $\dH$ can be expressed in terms of $\tK$ as
\be
 \dH=\left(1-\frac{\tK}{H_0^2}\right)^{1/2}-1 ~.
\label{eq:ch3_dH}
\ee
There is no solution for $\dH$ if $\tK/H_0^2>1$, or equivalently
$\delta_{L0}>(\frac53\frac{\Om}{D(t_0)})^{-1}$. This is because for such
a large positive curvature, the universe reaches turn around before $\ta=1$,
at which the modified cosmological parameters are defined. In other words,
this is not a physical problem, but merely a parameterization issue under
the standard convention.

Finally, we shall derive the equation for $\da(t)$, so that the observables
of different cosmologies can be compared at the same physical time $t$.
Inserting the second Friedmann equation of the fiducial and modified cosmologies
into \refeq{ch3_da_dot} yields an ordinary differential equation for the
perturbation to the scale factor
\be
 \ddotd_a+2H\dotd_a+\frac{4\pi G}{3}\rhob\left[(1+\da)^{-2}-(1+\da)\right]=0 ~,
\label{eq:ch3_da_diff}
\ee
or equivalently
\be
 \ddotd_{\rho}+2H\dotd_{\rho}-\frac{4}{3}\frac{\dotd_{\rho}^2}{1+\drho}
 -4\pi G\rho\drho(1+\drho)=0 ~.
\label{eq:ch3_drho_diff}
\ee
When linearizing ($\drho\ll1$) \refeq{ch3_drho_diff}, one obtains the equation
for the linear growth factor. More generally, \refeq{ch3_drho_diff} is exactly
the equation for the interior density of a spherical top-hat perturbation in
a $\Lambda$CDM universe \cite{schmidt/etal:2009}. For a certain $\tou$, one
can numerically calculate $\da(\tou)$ through \refeq{ch3_da_diff} to get
$\ta(\tou)=a(\tou)[1+\da(\tou)]$. Alternatively, one can numerically evaluate
$\ta(\tou)$ by
\be
 \tou=\int_0^{a(\tou)}~\frac{da}{aH(a)}
 =\int_0^{\ta(\tou)}~\frac{d\ta}{\ta\tH(\ta)}
 =\int_0^{a(\tou)[1+\da(\tou)]}~\frac{d\ta}{\ta\tH(\ta)} ~.
\label{eq:ch3_da_int}
\ee

\section{The modified cosmology in Einstein-de Sitter background}
\label{sec:ch3_sep_uni_eds}
In the Einstein-de Sitter (EdS) universe we have
\be
 \Om=1 ~;~ \rho_{\Lambda}=0 ~;~ H(a)=H_0a^{-3/2} ~;~
 a(t)=\left(\frac{3}{2}H_0t\right)^{2/3} ~;~ D(t)=a(t) ~.
\label{eq:ch3_eds}
\ee
In this section, we shall show that if the background cosmology is EdS,
then the scale factor $\da(t)$ and the Eulerian overdensity $\drho(t)$,
as well as the linear growth factor $\tD(t)$ and the logarithmic growth
rate $\tf(t)$ in the modified cosmology (overdense region) can be expressed
in series of $\delta_{L0}$.

\subsection{Scale factor and Eulerian overdensity}
\label{sec:ch3_scale_factor}
In order to solve $\da(t)$, we consider the spherical collapse model for
a overdense region \cite{gunn/gott:1972,peebles:1974,padmanabhan:1993/struform}.
Since the scale of this region is much smaller than the horizon size, we
can use the Newtonian dynamics and write the equation of motion of a shell
of particles at radius $\tilde{r}$ as
\be
 \ddot{\tilde{r}}=-\frac{G\tilde{M}}{\tilde{r}^2} ~,~~~
 \tilde{M}=\frac{4\pi}{3}\trhob(t_i)\tilde{r}^3(t_i)={\rm constant} ~,
\label{eq:ch3_sph_col_eom}
\ee
where $t_i$ is some initial time. Note that in \refeq{ch3_sph_col_eom} we
neglect the shell crossing, i.e. if $\tilde{r}_1(t_i)>\tilde{r}_2(t_i)$
then $\tilde{r}_1(t)>\tilde{r}_2(t)$ for all $t$.

\refEq{ch3_sph_col_eom} is known as the ``cycloid'' and the parametric
solution is
\be
 \ta(\theta)=\frac{\tilde{r}(\theta)}{\tilde{x}}=A(1-\cos\theta) ~,~~~
 t(\theta)=B(\theta-\sin\theta) ~,
\label{eq:ch3_sph_col_1}
\ee
where $\tilde{x}$ is some comoving distance for the normalization.
Using the Leibniz rule, one finds that
\be
 \dot{\tilde{r}}=\frac{d\tilde{r}}{d\theta}\frac{d\theta}{dt}
 =\frac{A\tilde{x}}{B}\frac{\sin\theta}{(1-\cos\theta)} ~,~~~~
 \ddot{\tilde{r}}=-\frac{A\tilde{x}}{B^2}\frac1{(1-\cos\theta)^2} ~,
\label{eq:ch3_sph_col_2}
\ee
and the equation of motion thus requires
\be
 \frac{A^3\tilde{x}^3}{B^2}=G\tilde{M}
 =\frac{4\pi G}{3}\trhob(t_i)\ta^3(t_i)\tilde{x}^3
 \quad\Leftrightarrow\quad
 \frac{A^3}{B^2}=\frac{\tH_0^2\tO_m}{2} ~.
\label{eq:ch3_sph_col_3}
\ee
Note that since this region is overdense, $\tO_m-1=-\tO_K>0$ and it
is positively curved. For an underdense (negatively curved) region,
the similar parameterization works with cos and sin in \refeq{ch3_sph_col_2}
replaced by cosh and sinh.

To determine the constants $A$ and $B$, we can write the first
Friedmann equation using the parametric solution as
\ba
 \:&\tH^2=\left(\frac{\dot\ta}{\ta}\right)^2
 =\frac1{B^2}\frac{(1-\cos^2\theta)}{(1-\cos\theta)^4}
 =\tH_0^2\left\lbrace\tO_m[A(1-\cos\theta)]^{-3}
 +(1-\tO_m)[A(1-\cos\theta)]^{-2}\right\rbrace \vs
 \:&\frac{(1-\cos^2\theta)}{B^2}=\frac{\tH_0^2}{A^2}\left\lbrace
 \left[\frac{\tO_m}{A}-(\tO_m-1)\right]
 -\cos\theta\left[\frac{\tO_m}{A}-2(\tO_m-1)\right]
 -\cos^2\theta(\tO_m-1)\right\rbrace ~.
\label{eq:ch3_sph_col_4}
\ea
Since \refeq{ch3_sph_col_4} is valid for all $\theta$, the coefficient
multiplied by $\cos\theta$ must be zero, which then allows us to solve
\be
 A=\frac12\frac{\tO_m}{(\tO_m-1)} ~,~~~
 B=\frac12\frac{\tO_m}{\tH_0(\tO_m-1)^{3/2}} ~.
\label{eq:ch3_sph_col_5}
\ee
One finds that \refeq{ch3_sph_col_5} indeed yields $A^3/B^2=\tH_0^2\tO_m/2$.

For simplicity, we define
\be
 \epsilon=\frac{\tO_m-1}{\tO_m}=1-(1+\dH)^2=\frac{\tK}{H_0^2}=\frac53\delta_{L0} ~,
\label{eq:ch3_epsilon}
\ee
so that
\be
 \ta(\theta)=\frac12\epsilon^{-1}(1-\cos\theta) ~;~~~
 t(\theta)=\frac34t_0\epsilon^{-3/2}(\theta-\sin\theta) ~.
\ee
The goal is to obtain
\be
 \ta(t)=a(t)[1+\da(t)]=a(t)\left[1+\sum_{n=1}^{\infty}e_n[a(t)\delta_{L0}]^n\right] ~,~~~
 \ta(t_0)=1+\sum_{n=1}^{\infty}e_n\delta_{L0}^n ~,
\label{eq:ch3_ta_2}
\ee
we thus need to solve $\theta_0$ such that
\be
 t(\theta_0)=t_0=\frac34t_0\epsilon^{-3/2}(\theta_0-\sin\theta_0)
 \quad\Leftrightarrow\quad
 \theta_0-\sin\theta_0=\frac34\epsilon^{-3/2} ~.
\label{eq:ch3_theta0_1}
\ee

We perform a series expansion,
\be
 \theta_0-\sin\theta_0=\frac16\theta_0^3-\frac1{120}\theta_0^5+\cdots
 =\sum_{n=1}^{\infty}b_n\theta_0^{2n+1} ~,
\label{eq:ch3_theta0_2}
\ee
and solve $\theta_0$ order by order. That is, for the $n^{\rm th}$ order
solution, we solve
\be
 \frac43\epsilon^{3/2}=\sum_{k=1}^nb_k\left[\theta_0^{(n-k+1)}\right]^{2k+1} ~,
\label{eq:ch3_theta0_3}
\ee
and $\theta_0=\lim_{n\to\infty}\theta_0^{(n)}$. Note that in order to trust
the final solution at order $\delta_{L0}^m$, the series solution needs to be
expanded to $n=m+1$. In the following we choose $m=5$, which yields
\be
 \theta_0=2\epsilon^{1/2}\left[1+\frac1{15}\epsilon+\frac2{175}\epsilon^2
 +\frac4{1575}\epsilon^3+\frac{43}{67375}\epsilon^4+\cdots\right] ~.
\label{eq:ch3_theta0_4}
\ee

Finally, we insert \refeq{ch3_theta0_4} into $\ta(\theta)$, expand in $\epsilon$,
replace $\epsilon$ with $\frac53\delta_{L0}$, and match with \refeq{ch3_ta_2}
for the coefficients $e_n$. For the first five coefficients, we get
\be
 e_1=-\frac13 ~;~ e_2=-\frac1{21} ~;~ e_3=-\frac{23}{1701} ~;~
 e_4=-\frac{1894}{392931} ~;~ e_5=-\frac{3293}{1702701} ~.
\label{eq:ch3_en}
\ee
Once $\da(t)$ is known, we can use \refeq{ch3_da} to solve the Eulerian
overdensity $\drho(t)=\sum_{n=1}^{\infty}f_n[\delta_{L0}a(t)]^n$, and
the first five coefficients are
\be
 f_1=1 ~;~ f_2=\frac{17}{21} ~;~ f_3=\frac{341}{567} ~;~
 f_4=\frac{55805}{130997} ~;~ f_5=\frac{213662}{729729} ~.
\label{eq:ch3_fn}
\ee

\begin{figure}[t!]
\centering
\includegraphics[width=1\textwidth]{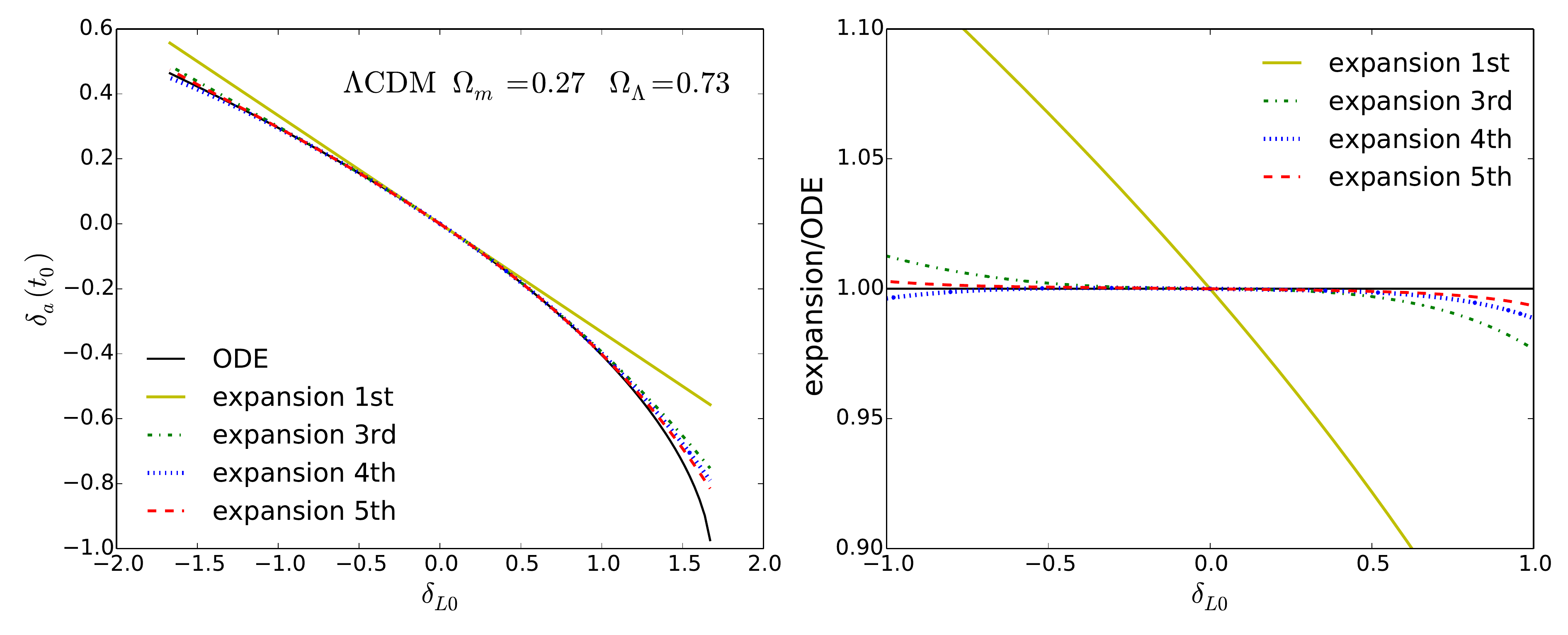}
\caption[Perturbation in the scale factor in the modified cosmology with $\Lambda$CDM background]
{(Left) Perturbation in the scale factor in the modified cosmology, $\da(t_0)$,
as a function of $\delta_{L0}$. The present-day scale factor is $a(t_0)=1$ in
the background $\Lambda$CDM universe, with $\Om=0.27$ and $\Omega_{\Lambda}=0.73$.
The black solid line shows the numerical solution to the ordinary differential
equation, \refeq{ch3_da_diff}; the yellow solid, green dot-dashed, blue dotted,
and red dashed lines show the series solutions in EdS, \refeq{ch3_da_drho_eds},
at the first, third, fourth, and fifth order, respectively. (Right) Same as the
left panel, but for the ratios of the series solutions expansion to the solution
of the ordinary differential equation.}
\label{fig:ch3_dL0_da_lcdm}
\end{figure}

While \refeqs{ch3_en}{ch3_fn} are strictly valid only if the background
is EdS, we find that they are also accurate in $\Lambda$CDM background
if $a(t)$ is replaced with $D(t)/D(t_0)$, where $D(t)$ is the growth
factor in the fiducial cosmology. In other words,
\be
 \da(t)=\sum_{n=1}^{\infty}e_n\left[\delta_{L0}\frac{D(t)}{D(t_0)}\right]^n ~;~~~
 \drho(t)=\sum_{n=1}^{\infty}f_n\left[\delta_{L0}\frac{D(t)}{D(t_0)}\right]^n ~.
\label{eq:ch3_da_drho_eds}
\ee
\refFig{ch3_dL0_da_lcdm} quantifies the performance of the EdS expansion with
$a(t)$ being replaced by $D(t)/D(t_0)$ for $\da(t)$, i.e. \refeq{ch3_da_drho_eds}.
For a large range of $\delta_{L0}$, the results of EdS expansion agree very well
with the numerical solution of the differential equation \refeq{ch3_da_diff}.
Specifically, at the third (fifth) order expansion, the fractional difference
is at the sub-percent level for $|\delta_{L0}|\sim1$.

\subsection{Small-scale growth}
\label{sec:ch3_linear_growth}
In this subsection, we shall iteratively solve the linear growth factor $\tD(t)$ and
logarithmic growth rate $\tf(t)$ in series of $\delta_{L0}$, as \refeqs{ch3_en}{ch3_fn},
in the modified cosmology with the EdS background. We consider a small-scale (short-wavelength)
density perturbation in the overdense region such that $\td_s=\trho/\trhob-1$. Note
that $\td_s$ should not be confused with the long-wavelength density perturbation
$\drho$ of this entirely overdense region with respect to the EdS background. Moreover,
$\td_s$ is defined with respect to the background density of the overdense universe
$\trhob$ instead of the EdS background $\rhob$.

The small-scale growth equation for $\td_s$ in the modified cosmology is given by
\be
 \ddot{\td}_s+2\tH\dot{\td}_s-4\pi G\trhob\td_s=0 ~.
\label{eq:ch3_growth_1}
\ee
Using
\ba
 \tH^2\:&=\tH_0^2\left[\tO_m\ta^{-3}+(1-\tO_m)\ta^{-2}\right]
 =H_0^2\left(\ta^{-3}-\frac53\delta_{L0}\ta^{-2}\right) ~, \vs
 4\pi G\trhob\:&=\frac32\tO_m\tH_0^2\ta^{-3}=\frac32H_0^2\ta^{-3} ~,
\label{eq:ch3_tH_trhob}
\ea
the growth equation can be rewritten as
\be
 \ddot{\td}_s+2H_0\left(\ta^{-3}-\frac53\delta_{L0}\ta^{-2}\right)^{1/2}\dot{\td}_s
 -\frac32H_0^2\ta^{-3}\td_s=0 ~.
\label{eq:ch3_growth_2}
\ee
Note that although we map the overdense universe to a positively curved
universe, the curvature contribution to the Poisson equation is neglected.
If $\tK/H_0^2\sim1$, the correction in the Poisson equation becomes relevant
for the small-scale modes $\td_s$ that are around the horizon size. Since
we are studying the subhorizon evolution of the small-scale modes, and
moreover we are mostly interested in the overdensity such that
$\tK/H_0^2\sim\delta_{L0}\ll1$, the correction is entirely negligible.
Thus, the curvature contributes to the growth only through the expansion
rate $\tH$.

Replacing the time coordinate $t$ with $y=\ln a(t)$ where $a(t)$ is the
scale factor in the EdS background, we rewrite the growth equation as
\be
 \frac{d^2}{dy^2}\td_s+\left[2(1+\da)^{-3/2}\left(1-\frac53\delta_{L0}[1+\da]\right)^{1/2}-\frac32\right]
 \frac{d}{dy}\td_s-\frac32(1+\da)^{-3}\td_s=0 ~,
\label{eq:ch3_growth_3}
\ee
Thus far, all the derivations are exact. To see that \refeq{ch3_growth_3} makes
sense, we consider the zeroth order approximation, i.e. $\delta_{L0}\to0$. In
this regime, $\da\to0$ and the growth equation becomes
\be
 \frac{d^2}{dy^2}\td^{(0)}_s+\frac12\frac{d}{dy}\td^{(0)}_s-\frac32\td^{(0)}_s=0 ~,
\label{eq:ch3_growth_4}
\ee
where the subscript $(0)$ denotes that it is the zeroth order solution. There
are two solutions to $\td^{(0)}_s$, the growing mode proportional to $a$ and
the decaying mode proportional to $a^{-3/2}$. As expected, because $\delta_{L0}\to0$,
the result is identical to the growth in the background EdS cosmology. In the
following, we shall drop the decaying mode following the standard practice.
Furthermore, we shall normalize $\td^{(0)}_s$ to $a(t)$ at early times, and
replace it with $\tD(t)$ to denote the small-scale growth factor. This means
$\tD^{(0)}(t)=a(t)$.

To solve higher order solutions, we insert the expansion of $\da$ in terms of
$a(t)\delta_{L0}=e^y\delta_{L0}$ (\refeq{ch3_ta_2} and \refeq{ch3_en}) into
the growth equation and obtain
\be
 \frac{d^2}{dy^2}\tD(y)+\left[\sum_{m=0}^{\infty}c_m\delta^m_{L0}e^{my}\right]
 \frac{d}{dy}\tD(y)-\left[\sum_{m=0}^{\infty}d_m\delta^m_{L0}e^{my}\right]\tD(y)=0 ~,
\label{eq:ch3_growth_5}
\ee
with coefficients $c_m$ and $d_m$ given by
\ba
 2(1+\da)^{-3/2}\left(1-\frac53\delta_{L0}[1+\da]\right)^{1/2}-\frac32
 \:&=\sum_{m=0}^{\infty}c_m[a(t)\delta_{L0}]^m \vs
 \frac32(1+\da)^{-3}\:&=\sum_{m=0}^{\infty}d_m[a(t)\delta_{L0}]^m ~.
\label{eq:ch3_cm_dm}
\ea
Correspondingly, we write the pure growing-mode solution in series of $\delta_{L0}$ as
\be
 \tD=\sum_{n=0}^{\infty}g_n\delta^n_{L0}e^{(n+1)y}
 =D(t)\sum_{n=0}^{\infty}g_n[a(t)\delta_{L0}]^n
\label{eq:ch3_tD_gn_1}
\ee
with coefficients $g_n$. Given our normalization, i.e. $\tD^{(0)}=a(t)$,
$g_0=1$. Thus,
\be
 \frac{d}{dy}\tD(y)=\sum_{n=0}^{\infty}(n+1)g_n\delta^n_{L0}e^{(n+1)y} ~;~~~
 \frac{d^2}{dy^2}\tD(t)=\sum_{n=0}^{\infty}(n+1)^2g_n\delta^n_{L0}e^{(n+1)y} ~.
\label{eq:ch3_dtD_dy}
\ee

Supposed that we have solutions of $\tD(y)$ to the $(n-1)^{\rm th}$ order,
then the solution at the $n^{\rm th}$ order has to satisfy
\be
 (n+1)^2g_n\delta^n_{L0}e^{(n+1)y}+\sum_{m=0}^ng_{n-m}\delta^{n-m}_{L0}e^{(n-m+1)y}
 \left[(n-m+1)c_m-d_m\right]e^{my}\delta^m_{L0}=0 ~.
\label{eq:ch3_tD_sol}
\ee
The term $\delta^n_{L0}e^{(n+1)y}$ factors out, and we obtain a simple
algebraic relation for $g_n$ in terms of $c_m$ and $d_m$ for $0\le m\le n$,
and $g_m$ for $0\le m \le n-1$ as
\be
 (n+1)^2g_n+\sum_{m=0}^ng_{n-m}[(n-m+1)c_m-d_m]=0 ~.
\label{eq:ch3_gn_cm_dm}
\ee
Using $e_n$ in \refeq{ch3_en} to get $c_m$ and $d_m$ through \refeq{ch3_cm_dm},
we obtain the first five $g_n$ to be
\be
 g_1=\frac{13}{21} ~;~ g_2=\frac{71}{189} ~;~ g_3=\frac{29609}{130977} ~;~
 g_4=\frac{691858}{5108103} ~;~ g_5=\frac{8682241}{107270163} ~.
\label{eq:ch3_gn}
\ee

Similar to the previous subsection, the expansion of $\tD$ in terms of $\delta_{L0}$
with the coefficients (\refeq{ch3_gn}) is strictly valid in the EdS background
cosmology. In order to generalize from EdS to other cosmologies, we replace $a(t)$
with $D(t)/D(t_0)$, so that
\be
 \tD(t)=D(t)\left\lbrace1+\sum_{n=1}^{\infty}g_n\left[\delta_{L0}\frac{D(t)}{D(t_0)}\right]^n\right\rbrace ~.
\label{eq:ch3_tD_gn_2}
\ee
\refFig{ch3_dL0_tD_lcdm} quantifies the performance of the expansion with $a(t)$
being replaced by $D(t)/D(t_0)$ for $\tD(t)$, i.e. \refeq{ch3_tD_gn_2}. The
agreement is not as good as for $\da(t)$, nevertheless the fifth order expansion
gives 5\% fractional differences at $|\delta_{L0}|\sim1$. Note also that while
for positive $\delta_{L0}$ the EdS expansion is always smaller than (but converging
to) the numerical solution in $\Lambda$CDM, for negative $\delta_{L0}$ the fourth
order expansion has a different trend compared to the other orders. This is because
at the $n^{\rm th}$ order $\delta_{L0}^n$ dominates when $|\delta_{L0}|>1$, and
at the negative $\delta_{L0}$ end the result would thus depend on the parity of
the expansion order. It is clear though the higher the expansion order, the
better the agreement.

\begin{figure}[t!]
\centering
\includegraphics[width=1\textwidth]{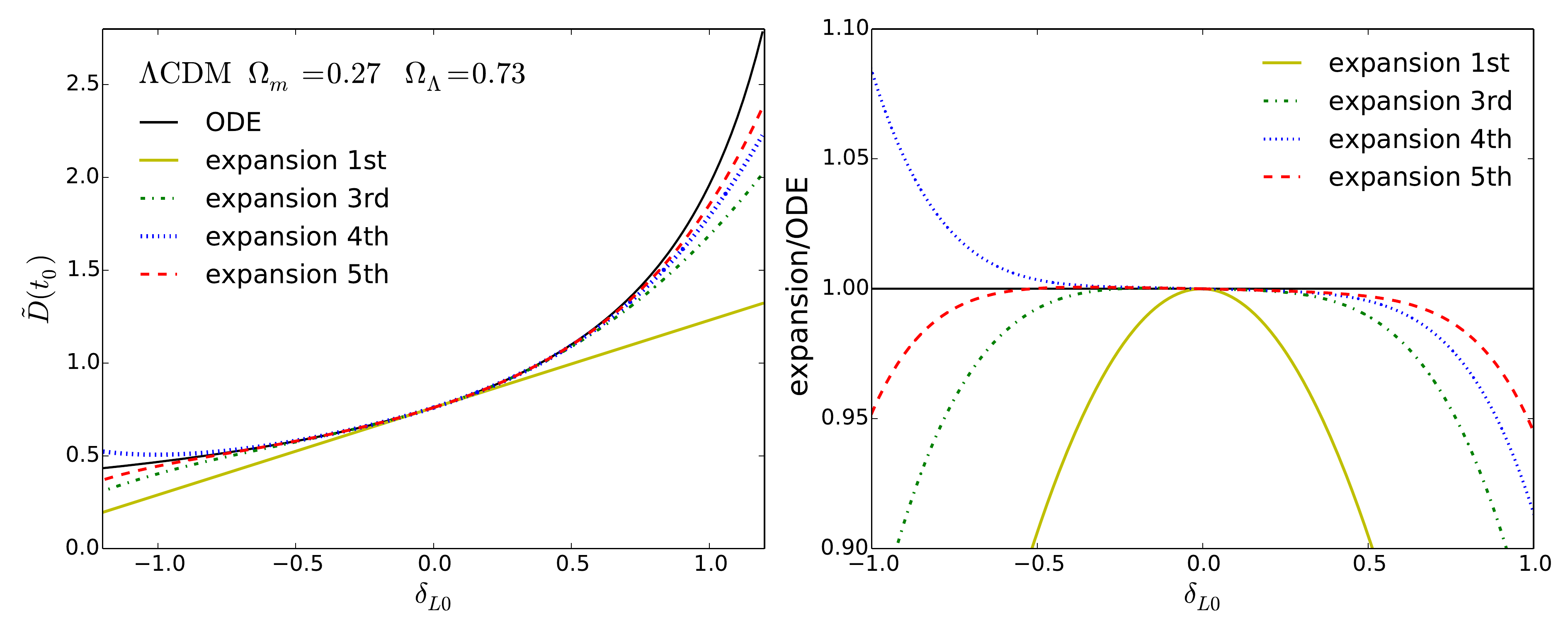}
\caption[Linear growth factor in the modified cosmology with $\Lambda$CDM background]
{(Left) The linear growth factor in the modified cosmology, $\tD(t_0)$ with $a(t_0)=1$,
in the background $\Lambda$CDM universe with $\Om=0.27$ and $\Omega_{\Lambda}=0.73$.
The black solid line shows the numerical solution to \refeq{ch3_growth_2}; the yellow
solid, green dot-dashed, blue dotted, and red dashed lines show the series solutions,
\refeq{ch3_tD_gn_2}, at the first, third, fourth, and fifth order, respectively. (Right)
Same as the left panel, but for the ratios of the series solutions to the solution of the
differential equation of the small-scale growth.}
\label{fig:ch3_dL0_tD_lcdm}
\end{figure}

With the expansions of $\da(t)$ (\refeq{ch3_da_drho_eds}) and $\tD(t)$ (\refeq{ch3_tD_gn_2}),
we can finally derive the series expansion for the logarithmic growth rate,
\be
 f(t)=\frac{d\ln D(t)}{d\ln a(t)}=\frac{\dot{D}(t)}{D(t)}\frac{a(t)}{\dot{a}(t)} ~.
\label{eq:ch3_f_def}
\ee
In the modified cosmology, we have (defining $e_0=1$ as for $g_0$)
\ba
 \tf(t)\:&=\frac{\dot{\tD}}{\tD}\frac{\ta}{\dot{\ta}}
 =\frac{\dot D\sum_{n=0}^\infty(n+1)g_n\delta_L^n(t)}{D\sum_{n=0}^\infty g_n\delta_L^n(t)}
 \frac{a\sum_{n=0}^\infty e_n\delta_L^n(t)}{\dot a\sum_{n=0}^\infty e_n\delta_L^n(t)
 +a\frac{\dot D}{D}\sum_{n=0}^\infty ne_n\delta_L^n(t)} \vs
 \:&=f\frac{\sum_{n=0}^\infty(n+1)g_n\delta_L^n(t)}{\sum_{n=0}^\infty g_n\delta_L^n(t)}
 \frac{\sum_{n=0}^\infty e_n\delta_L^n(t)}{\sum_{n=0}^\infty e_n\delta_L^n(t)+f\sum_{n=0}^\infty ne_n\delta_L^n(t)} ~,
\label{eq:ch3_tf}
\ea
where $\delta_L(t)=\delta_{L0}\frac{D(t)}{D(t_0)}$. Note that in the EdS background
$f=1$, and so \refeq{ch3_tf} can be simplified as
\be
 \tf(t)=\frac{\sum_{n=0}^\infty(n+1)g_n\delta_L^n(t)}{\sum_{n=0}^\infty g_n\delta_L^n(t)}
 \frac{\sum_{n=0}^\infty e_n\delta_L^n(t)}{\sum_{n=0}^\infty(n+1)e_n\delta_L^n(t)} ~.
\label{eq:ch3_tf_eds}
\ee
\refFig{ch3_dL0_tf_lcdm} shows the performance of the expansion of $\tf$ in the $\Lambda$CDM
background. It is not as good as for $\tD$, but for $|\delta_{L0}|\sim0.8$ the fifth order
expansion gives approximately 5\% fractional difference with respect to the numerical solution
of the differential equation. One also notes the parity-feature at the negative $\delta_{L0}$,
as for $\tD$.

\begin{figure}[t!]
\centering
\includegraphics[width=1\textwidth]{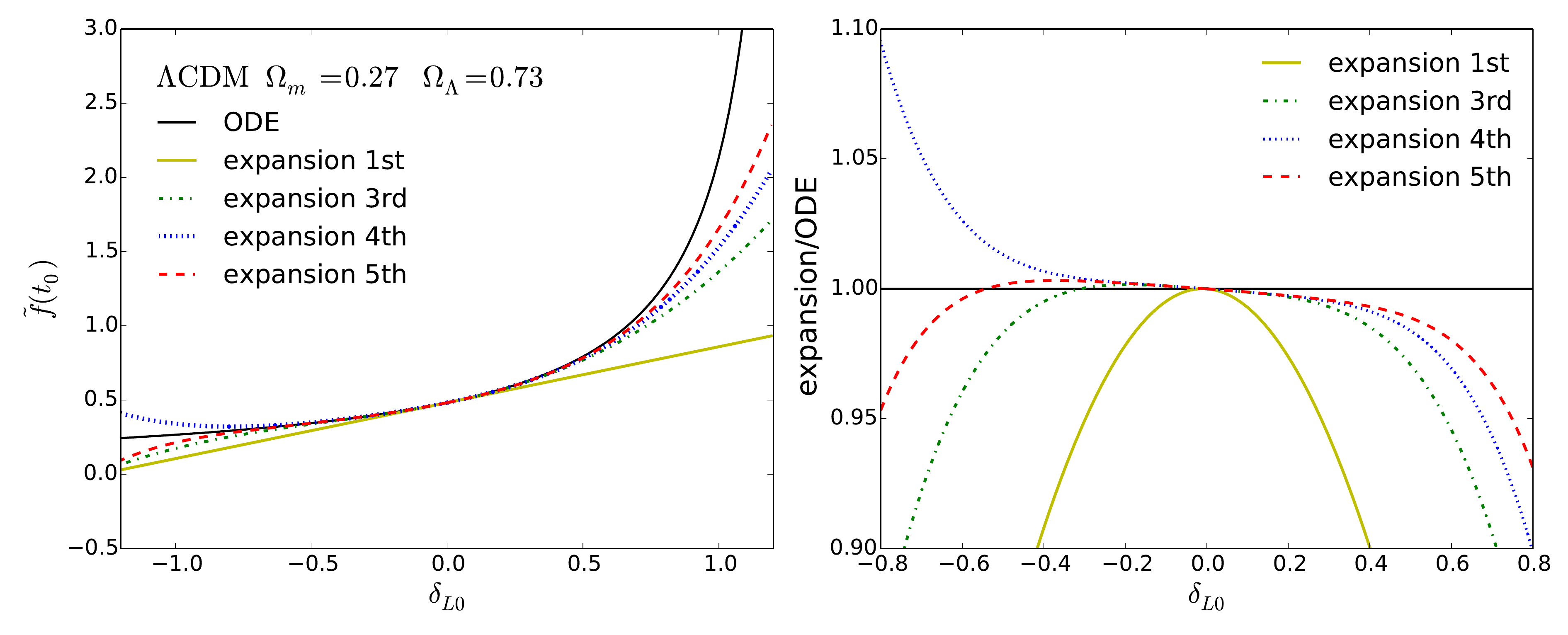}
\caption[Logarithmic growth rate in the modified cosmology with $\Lambda$CDM background]
{(Left) The logarithmic growth rate in the modified cosmology, $\tf(t_0)$
with $a(t_0)=1$, in the background $\Lambda$CDM universe  with $\Om=0.27$
and $\Omega_{\Lambda}=0.73$. The black solid line shows the numerical
solution to \refeq{ch3_growth_2}; the yellow solid, green dot-dashed,
blue dotted, and red dashed lines show the series solutions, \refeq{ch3_tf},
at the first, third, fourth, and fifth order, respectively. (Right)
Same as the left panel, but for the ratios of the series solutions to
the solution of the differential equation of the small-scale growth.}
\label{fig:ch3_dL0_tf_lcdm}
\end{figure}

}

%% file: kap_04.tex
{

\renewcommand{\v}[1]{\mathbf{#1}}
\newcommand{\vx}{\v{x}}
\newcommand{\vr}{\v{r}}
\newcommand{\vk}{\v{k}}
\newcommand{\vq}{\v{q}}

\newcommand{\Om}{\Omega_m}
\newcommand{\rhob}{\bar\rho}
\def\rhocr{\rho_{\rm cr,0}}
\renewcommand{\d}{\delta}
\newcommand{\hMpc}{~h^{-1}~{\rm Mpc}}
\newcommand{\ihMpc}{~h~{\rm Mpc}^{-1}}
\newcommand{\hd}{\hat{\delta}}
\newcommand{\hdb}{\hat{\bar\delta}}
\newcommand{\hP}{\hat{P}}
\newcommand{\hib}{\hat{iB}}
\newcommand{\bd}{\bar\delta}
\newcommand{\cO}{\mathcal O}
\newcommand{\Dk}{\Delta k}

\chapter{Measurement of position-dependent power spectrum}
\label{chp:ch4_posdeppk}
As introduced in \refsec{ch2_kspace}, the correlation between
the position-dependent power spectrum,
\be
 \hP(\vk,\vr_L)=\frac{1}{V_L}\int\frac{d^3q_1}{(2\pi)^3}\int\frac{d^3q_2}{(2\pi)^3}~
 \delta(\vk-\vq_1)\delta(-\vk-\vq_2)W_L(\vq_1)W_L(\vq_2)e^{-i\vr_L\cdot(\vq_1+\vq_2)} ~,
\label{eq:ch4_pos_dep_pk}
\ee
and the mean overdensity,
\be
 \bd(\vr_L)=\frac{1}{V_L}\int\frac{d^3q}{(2\pi)^3}~\delta(-\vq)W_L({\vq})e^{-i\vr_L\cdot\vq} ~,
\label{eq:ch4_bd}
\ee
is the integrated bispectrum,
\ba
 iB_L(k)\:&=\langle\hP(k,\vr_L)\bd(\vr_L)\rangle \vs
 \:&=\frac{1}{V_L^2}\int\frac{d^2\hat k}{4\pi}\int\frac{d^3q_1}{(2\pi)^3}
 \int\frac{d^3q_3}{(2\pi)^3}~B(\vk-\vq_1,-\vk+\vq_1+\vq_3,-\vq_3) \vs
 \:&\hspace{5.4cm}\times W_L(\vq_1)W_L(-\vq_1-\vq_3)W_L(\vq_3) ~,
\label{eq:ch4_ib_ang_avg}
\ea
with $V_L$ being the size of the subvolume.

In the squeezed limit where the scale of the position-dependent power spectrum is
much smaller than the subvolume size, i.e. $k\gg1/L$, the integrated bispectrum
can be simplified as
\be
 iB_L(k)\stackrel{k L \to \infty}{=}\frac{1}{V_L^2}\int\frac{d^3q}{(2\pi)^3}~
 W_L^2(\vq)P(q)\check{f}(k)P(k)=\sigma_L^2\check{f}(k)P(k) ~,
\label{eq:ch4_ib_sq}
\ee
where $\check{f}(k)=2f(0,k)$ with $f(\vk_1,\vk_2)$ being a dimensionless symmetric
function for the separable bispectrum, and $\sigma_L^2$ is the variance of the density
fluctuation in $V_L$,
\be
 \sigma_L^2=\frac{1}{V_L^2}\int\frac{d^3q}{(2\pi)^3}~P(q)W_L^2(\vq) ~.
\label{eq:ch4_sigmaL2}
\ee
An intuitive way to arrive at \refeq{ch4_ib_sq} is to consider the expansion of
the position-dependent power spectrum in the presence of a long-wavelength density
fluctuation $\bd$ as
\be
 \hP(k,\vr_L) = \left.P(k)\right|_{\bd=0} +
 \left.\frac{dP(k)}{d\bd}\right|_{\bd=0}\bd+\dots ~,
\label{eq:ch4_Pkexp}
\ee
and the leading-order correlation between $P(k,\vr_L)$ and $\bd$ is
\be
 iB_L(k)=\sigma_L^2\left.\frac{d\ln P(k)}{d\bar{\delta}}\right|_{\bar{\delta}=0}P(k) ~.
\label{eq:ch4_ib_sq_sepuni}
\ee
Inspired by \refeq{ch4_ib_sq} and \refeq{ch4_ib_sq_sepuni}, we define the normalized
integrated bispectrum to be
\be
 \frac{iB_L(k)}{\hP(k)\sigma_L^2} ~,
\label{eq:ch4_nib}
\ee
and it is equal to the linear response function, $\check{f}(k)$ or $d\ln P(k)/d\bd$,
in the limit of $kL\to\infty$.

In this chapter, we measure the position-dependent power spectrum and the integrated
bispectrum from $N$-body simulations in \refsec{ch4_nbody}, and compare with the
theoretical modeling of the measurements in \refsec{ch4_bi_mod} and \refsec{ch4_sep_uni}.
At the end of this chapter, we shall discuss the dependence of the integrated bispectrum
on the cosmological parameters in \refsec{ch4_cosmodep}, and the expected constraint on
the primordial non-Gaussianity using the Fisher matrix calculation in \refsec{ch4_fisher}.
We conclude in \refsec{ch4_conclusion}.

\section{$N$-body simulations and the estimators}
\label{sec:ch4_nbody}
We now present measurements of the position-dependent power spectrum from 160
collisionless $N$-body simulations of a $2400\hMpc$ box with $768^3$ particles
(which corresponds to $2.29\times10^{12}M_{\odot}$). The same simulations are
used in \cite{deputter/etal:2012}, and we refer to section~3 of \cite{deputter/etal:2012}
for more details. In short,the initial conditions are set up using different
realizations of Gaussian random fields with the linear power spectrum computed
by CAMB \cite{lewis/challinor/lasenby:2000,lewis/bridle:2002}. We adopt a flat
$\Lambda$CDM cosmology, and the cosmological parameters are $\Om=0.27$, $\Omega_bh^2=0.023$,
$h=0.7$, $n_s=0.95$, and $\sigma_8=0.7913$. The particles are displaced from
the initial grid points using the second-order Lagrangian perturbation theory
\cite{crocce/pueblas/scoccimarro:2006} at the initial redshift $z_i=19$. The
simulations are carried out using the Tree-PM code Gadget-2 \cite{springel:2005},
taking only the gravitational force into account.

To construct the density fluctuation field on grid points, we first distribute
all the particles in the $2400\hMpc$ box onto a $1000^3$ grid by the cloud-in-cell
(CIC) density assignment scheme. Then the density fluctuation field at the grid
point $\vr_g$ is
\be
 \hd(\vr_g)=\frac{N(\vr_g)}{\bar N}-1 ~,
\label{eq:ch4_hd}
\ee
where hat denotes the estimated quantities, $N(\vr_g)$ is the fractional number
of particles after the CIC assignment at $\vr_g$, and $\bar N=768^3/1000^3$ is
the mean number of particles in each grid cell.

We then divide the $2400\hMpc$ box in each dimension by $N_{\rm cut}=4$,
8, and 20, so that there are 64, 512, and 8000 subvolumes with a side
length of 600, 300, and $120~h^{-1}~{\rm Mpc}$, respectively. The mean
density perturbation in a subvolume centered at $\vr_L$ is
\be
 \hdb(\vr_L)=\frac1{N_{\rm grid}^3}\sum_{\vr_g\in V_L}\hd(\vr_g) ~,
\label{eq:ch4_hdb}
\ee
where $(N_{\rm grid})^3=(1000/N_{\rm cut})^3$ is the number of grid points
within the subvolume. To compute the position-dependent power spectrum, we
use \texttt{FFTW}\footnote{Fast Fourier Transformation library: \url{www.fftw.org}}
to Fourier transform $\hd(\vr_g)$ in each subvolume with the grid size
$(N_{\rm grid})^3$. While the fundamental frequency of the subvolume,
$k_F=2\pi/L$, decreases with the subvolume size $L$, the Nyquist frequency
of the FFT grid, $k_{Ny}=k_FN_{\rm grid}/2\approx1.3\ihMpc$,
is the same in all cases.

The position-dependent power spectrum is then computed as
\be
 \hP(k,\vr_L)=\frac1{V_LN_{\rm mode}}
 \sum_{k-\Delta k/2\le|\vk_i|\le k+\Delta k/2}|\hd(\vk_i,\vr_L)|^2 ~,
\label{eq:ch4_hP}
\ee
where $N_{\rm mode}$ is the number of Fourier modes in the bin
$\left[k-\Delta k/2, k+\Delta k/2\right]$, and we set $\Delta k\approx0.01\ihMpc$
in all cases. We choose this $\Delta k$ for all $N_{\rm cut}$ to sample well the
baryon acoustic oscillations (BAO) and thereby are able to show how the window
function of the different subvolumes damps the BAO (see \reffig{ch4_nbody_meas}).
We follow the procedures in \cite{jing:2005} to correct for the smoothing due to
the CIC density assignment and also for the aliasing effect in the power spectrum.
Note, however, that this correction is only important for wavenumbers near the
Nyquist frequency $1.31~\ihMpc$, and we are interested in scales $k\lesssim0.4\ihMpc$.

\begin{figure}[t!]
\centering
\includegraphics[width=1\textwidth]{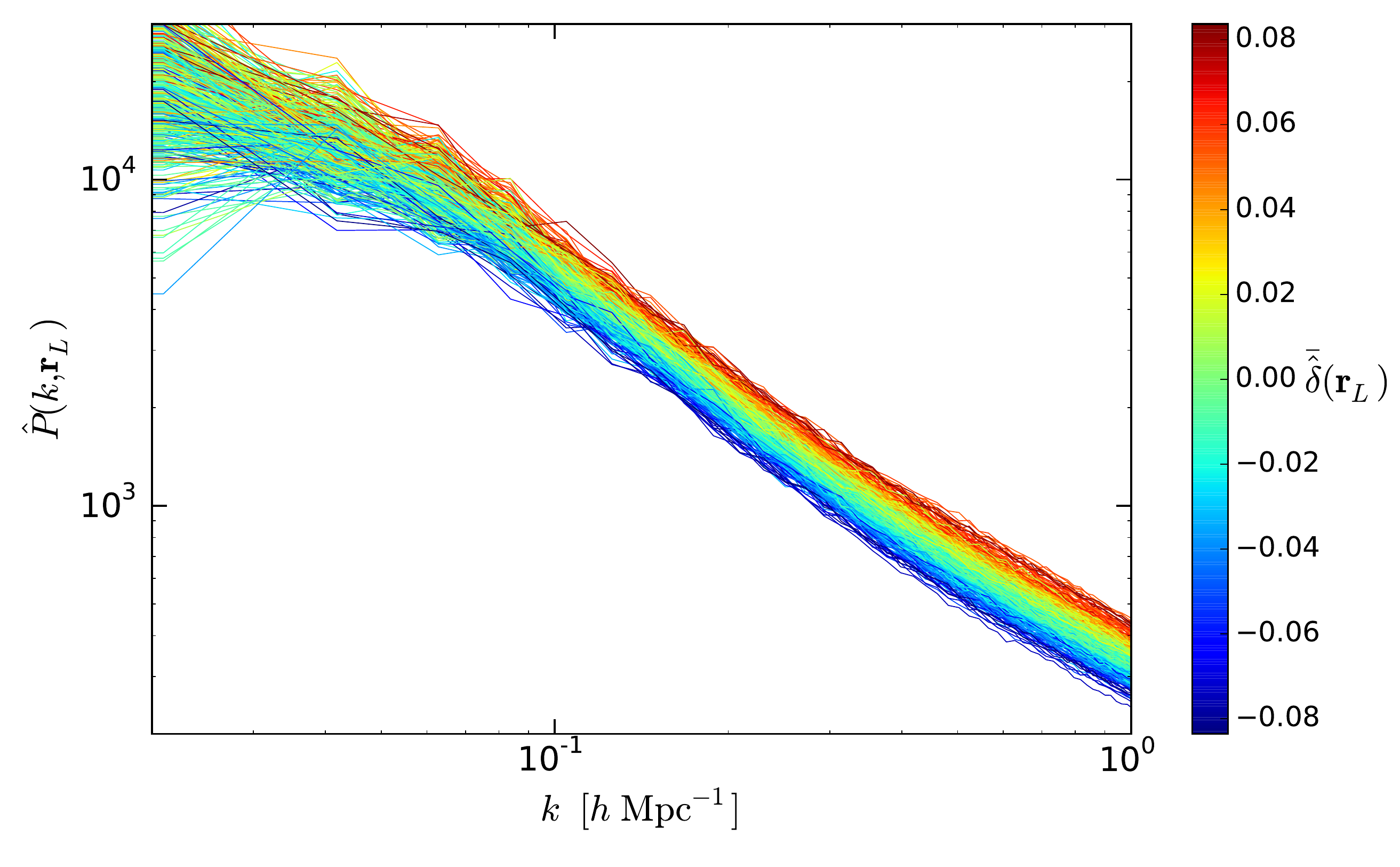}
\caption[Position-dependent power spectra measured from 512
subvolumes with $L=300\hMpc$ in one realization]
{Position-dependent power spectra measured from 512 subvolumes with $L=300\hMpc$
in one realization at $z=0$. The color represents $\hdb(\vr_L)$ of each subvolume.}
\label{fig:ch4_pk_dm_color}
\end{figure}

\refFig{ch4_pk_dm_color} shows the position-dependent power spectrum measured
from 512 subvolumes with $L=300\hMpc$ in one realization at $z=0$. The color
represents $\hdb(\vr_L)$ of each subvolume. The positive correlation between
the subvolume power spectra and $\hdb(\vr_L)$ is obvious. The response of the
position-dependent power spectrum to the long-wavelength density fluctuation
is clearly measurable at high significance in the simulations.

We measure the integrated bispectrum through
\be
 \hib_L(k)=\frac1{N_{\rm cut}^3}\sum_{i=1}^{N_{\rm cut}^3}\hP(k,\vr_{L,i})\hdb(\vr_{L,i}),
\label{eq:ch4_hib}
\ee
where $\hP(k,\vr_{L,i})$ and $\hdb(\vr_{L,i})$ are measured in the $i^{\rm th}$
subvolume. Further, motivated by \refeq{ch4_ib_sq}, we normalize the integrated
bispectrum by the mean power spectrum in the subvolumes,
\be
 \bar\hP_L(k)=\frac{1}{N_{\rm cut}^3}\sum_{i=1}^{N_{\rm cut}^3}\hP(k,\vr_{L,i}) ~,
\label{eq:ch4_hPb}
\ee
and the variance of the mean density fluctuation in the subvolumes,
\be
 \hat\sigma_L^2=\frac{1}{N_{\rm cut}^3}\sum_{i=1}^{N_{\rm cut}^3}\hdb^2(\vr_{L,i}) ~.
\label{eq:ch4_hsigmaL2}
\ee
Note that by construction
\be
 \bar{\hdb}_L=\frac{1}{N_{\rm cut}^3}\sum_{i=1}^{N_{\rm cut}^3}\hdb(\vr_{L,i})=0 ~.
\label{eq:ch4_hdbb}
\ee
This quantity
\be
 \frac{\hib_L(k)}{\bar\hP_L(k)\hat\sigma_L^2} ~,
\label{eq:ch4_hibn}
\ee
is the estimator of the normalized integrated bispectrum (\refeq{ch4_nib}), and is
equal to the linear response function, $d\ln P(k)/d\bd$, given in \refeq{ch4_ib_sq_sepuni}
in the limit of $kL\to \infty$.

\begin{figure}[t!]
\centering
\includegraphics[width=1\textwidth]{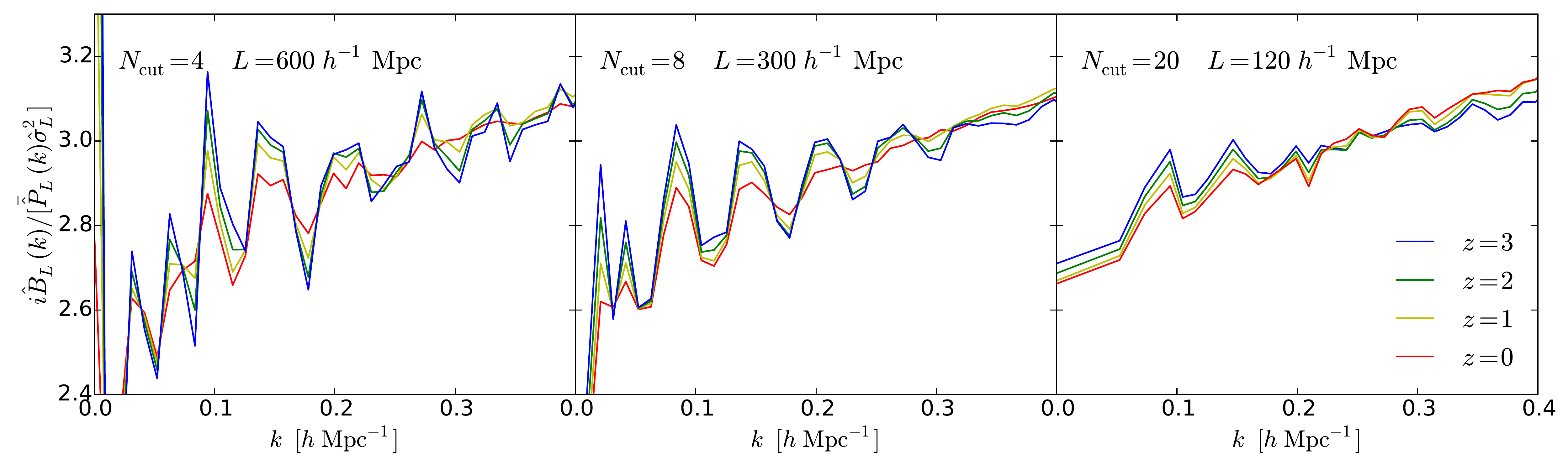}
\caption[Normalized integrated bispectrum,averaged over 160 collisionless
$N$-body simulations with Gaussian initial conditions.]
{Normalized integrated bispectrum averaged over 160 collisionless $N$-body
simulations with Gaussian initial conditions. From left to right are $N_{\rm cut}=4$
($L=600\hMpc$), 8 ($300\hMpc$), and 20 ($120\hMpc$); the blue, green, yellow,
and red lines are $z=3,$ 2, 1, and 0, respectively. For clarity, we do not
show the error bars.}
\label{fig:ch4_nbody_meas}
\end{figure}

\refFig{ch4_nbody_meas} shows the normalized integrated bispectrum, averaged over
160 collisionless $N$-body simulations at different redshifts. For clarity, no error
bars are shown in this figure. We have compared the results with a higher-resolution
simulation with $1536^3$ particles and starting at higher redshift ($z_i=49$ compared
to $z_i=19$ for our 160 simulations). For the scales and redshifts shown in
\reffig{ch4_nbody_meas}, the differences are less than 1\%. However, we expect an up
to 5\% uncertainty in the integrated bispectrum at $z=3$ (less at lower $z$) due to
transients which affect the bispectrum more strongly than the power spectrum
\cite{crocce/pueblas/scoccimarro:2006,mccullagh/jeong}, as well as other systematics
such as mass resolution.

Since the initial conditions are Gaussian, the bispectrum is generated entirely by
nonlinear gravitational evolution. We thus measure the effect of a long-wavelength
density perturbation on the evolution of small-scale structures. The wiggles visible
in each panel of \reffig{ch4_nbody_meas} are due to the BAOs. The BAOs in the right
panel are strongly damped because the box size ($120\hMpc$) approaches the BAO scale,
and the window function smears the BAO feature \cite{chiang/etal:2013}. Further, BAO
amplitudes are larger at higher redshifts as they are less damped by nonlinear evolution
\cite{eisenstein/seo/white:2007}. The broad-band shape of the normalized integrated
bispectrum evolves on small scales due to nonlinear evolution, leading to an effective
steepening of its slope. We now turn to the theoretical modeling of the results shown
in \reffig{ch4_nbody_meas}.

\section{Bispectrum modeling}
\label{sec:ch4_bi_mod}
We use two different approaches to model the integrated bispectrum. In the first
approach, we model the bispectrum and compute the integral to obtain the integrated
bispectrum. In the second approach, we model the response of the small-scale power
spectrum to a long wavelength perturbation directly using the ``separate universe''
picture. For clarity, we will show the comparison between model prediction and
simulations only for the $L=300\hMpc$ subvolumes ($N_{\rm cut}=8$).  The agreement
with simulations is independent of subvolume size as long as the subvolume size is
large enough for $\bar{\delta}$ to be in the linear regime, and the window function
is taken into account.

We first compute the integrated bispectrum by using a model for the bispectrum
in \refeq{ch4_ib_ang_avg} and perform the eight-dimensional integral. Because
of the high dimensionality, we use the Monte Carlo integration routine in GNU
Scientific Library to evaluate the angular-averaged integrated bispectrum. In
the following, we consider two different models for the matter bispectrum.

\subsection{Standard perturbation theory}
\label{sec:ch4_bi_spt}
The standard perturbation theory (SPT) \cite{bernardeau/etal:2002} gives
the tree-level matter bispectrum as
\be
 B_{\rm SPT}(\vk_1,\vk_2,\vk_3)=2[P_l(k_1)P_l(k_2)F_2(\vk_1,\vk_2)+2 \:\rm cyclic],
\label{eq:ch4_bi_spt}
\ee
where $P_l(k)$ is the linear matter power spectrum, and
\be
 F_2(\vk_1,\vk_2)=\frac57+\frac12\frac{\vk_1\cdot\vk_2}{k_1k_2}
 \left(\frac{k_1}{k_2}+\frac{k_2}{k_1}\right)+\frac27
 \left(\frac{\vk_1\cdot\vk_2}{k_1k_2}\right)^2 ~.
\label{eq:ch4_F2}
\ee
In order to normalize the integrated bispectrum, we need an expression for the
mean subvolume power spectrum $\bar\hP_{L}(k)$. For this we use the linear power
spectrum convolved with the window function,
\be
 P_{l,L}(k)=\frac{1}{V_L}\int\frac{d^3q}{(2\pi)^3}~P_l(|\vk-\vq|)|W_L(\vq)|^2 ~,
\label{eq:ch4_pk_modeling}
\ee
while the variance of the mean density fluctuation in the subvolumes is given by
\refeq{ch4_sigmaL2}. Both quantities are calculated through Monte Carlo integration.

\begin{figure}[t!]
\centering
\includegraphics[width=1\textwidth]{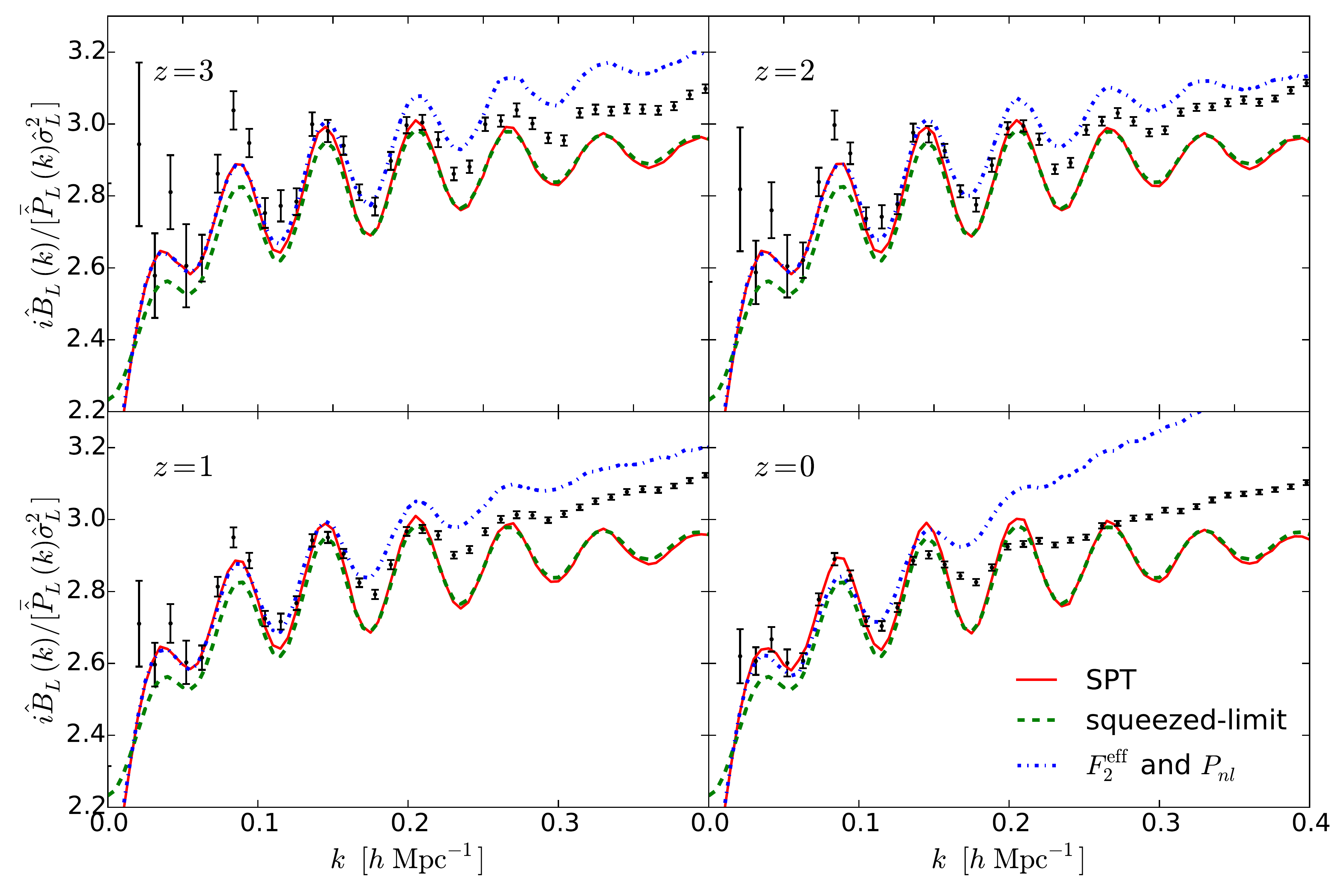}
\caption[Bispectrum modeling for the measured integrated bispectrum in $300\hMpc$ subvolumes]
{The SPT and the $F_2^{\rm eff}(\vk_1,\vk_2)$ predictions for the normalized
integrated bispectrum at different redshifts. The red solid and blue dot-dashed
lines are computed by the direct integration of the eight-dimensional integral
(\refeq{ch4_ib_ang_avg}) with the standard $F_2(\vk_1,\vk_2)$ kernel and the
linear power spectrum, and $F_2^{\rm eff}(\vk_1,\vk_2)$ and the nonlinear power
spectrum, respectively. The green dashed lines show the squeezed-limit approximation
(\refeq{ch4_int_bi_approx}) to the SPT results. The $N$-body simulation results
are shown by the black data points with the error bars showing the standard
deviation on the mean measured from 160 simulations.}
\label{fig:ch4_spt_f2eff_modeling}
\end{figure}

We compare the normalized integrated bispectrum measured from the simulations
with the SPT prediction in \reffig{ch4_spt_f2eff_modeling} (red lines). The
SPT prediction is independent of redshift. This is because the linear power
spectra at various redshifts are only different by the wavenumber-independent
linear growth factor, $D^2(z)$. Therefore, the linear growth factor cancels
out in the normalized integrated bispectrum. The SPT predictions agree with
the simulations relatively well at $z\ge1$ and $k\lesssim0.2\ihMpc$, whereas
they fail at lower redshifts as well as on smaller scales, where nonlinearities
become too strong to be described by SPT. Especially, the BAO amplitudes at
$k\gtrsim0.2\ihMpc$ are affected: while the SPT predictions are redshift-independent,
the simulations show smaller BAO amplitudes at lower redshifts.

The eight-dimensional integral in \refeq{ch4_ib_ang_avg} simplifies greatly if we
focus on the squeezed-limit bispectrum. In \refapp{trz_bi}, we show (note that
$B_{\rm SPT}\equiv B_{\rm SQ1,1}$)
\ba
 \:&~~~\int\frac{d^2\Omega_{\hat\vk}}{4\pi}~B_{\rm SPT}(\vk-\vq_1,-\vk+\vq_1+\vq_3,-\vq_3) \vs
 \:&=\left[\frac{68}{21}-\frac{1}{3}\frac{d\ln k^3 P_l(k)}{d\ln k}\right]P_l(k)P_l(q_3) 
 + \cO\left[\left(\frac{q_{1,3}}{k}\right)^2\right] ~,
\label{eq:ch4_bi_approx}
\ea
for  $k\gg q_1,q_3$. We can then apply \refeq{ch4_ib_sq} and perform all the
integrals analytically in the limit of $kL\to \infty$ to obtain
\ba
 iB_{L,\rm SPT}(k)\:&=\frac{1}{V_L^2}\int\frac{d^2\Omega_{\hat\vk}}{4\pi}
 \int\frac{d^3q_1}{(2\pi)^3}\int\frac{d^3q_3}{(2\pi)^3}~
 B_{\rm SPT}(\vk-\vq_1,-\vk+\vq_1+\vq_3,-\vq_3) \vs
 \:&\hspace{5.8cm}\times W_L(\vq_1)W_L(-\vq_1-\vq_3)W_L(\vq_3) \vs
 \:&\hspace{-0.3cm}\stackrel{k L \to \infty}{=}
 \left[\frac{68}{21}-\frac{1}{3}\frac{d\ln k^3 P_l(k)}{d\ln k}\right]P_l(k)\sigma_L^2 ~.
\label{eq:ch4_int_bi_approx}
\ea
Comparing this result with \refeq{ch4_ib_sq_sepuni}, we find that the linear response
of the power spectrum in SPT is given by
\be
\left.\frac{d\ln P_l(k)}{d\bar{\delta}}\right|_{\rm SPT}
=\frac{68}{21}-\frac{1}{3}\frac{d\ln k^3 P_l(k)}{d\ln k}.
\label{eq:ch4_sptresponse}
\ee

The green dashed lines in \reffig{ch4_spt_f2eff_modeling} show the squeezed-limit
approximation given in \refeq{ch4_int_bi_approx}. While they are different from
the full integration (red solid lines) at $k\lesssim0.2\ihMpc$, for which the
squeezed-limit approximation fails and the direct integration is required, they
agree well, with the fractional difference being less than 1.5\% (1\% for $L=600\hMpc$),
at $k\gtrsim0.2\ihMpc$, corresponding to a value of $1/(kL)\lesssim0.02$. Thus,
the squeezed-limit is reached already with good precision for $kL\gtrsim50$.

\refEq{ch4_int_bi_approx} does not contain any window function effect apart from that in
the variance $\sigma_L^2$. While this is a good approximation for the slowly-varying part
of the integrated bispectrum, it does not capture the smearing of the BAO features due to
the window function. We incorporate this effect by replacing $d\ln P_l(k)/d\ln k$ with
appropriately convolved forms, ${\rm conv}[d P_l(k)/d\ln k]\:/\:{\rm conv}[P_{l}(k)]$,
in \refeq{ch4_int_bi_approx}. This form is motivated by the separate universe approach
discussed in \refsec{ch4_sep_uni}, and provides an accurate result as shown in
\reffig{ch4_spt_f2eff_modeling}.

\subsection{Bispectrum fitting formula}
\label{sec:ch4_bi_fit}
The SPT predictions fail on smaller scales as well as at lower redshifts where
nonlinearity becomes too strong to be described by SPT. An empirical fitting
formula for nonlinear evolution of the matter bispectrum was proposed in
\cite{scoccimarro/couchman:2001} and further improved in \cite{gilmarin/etal:2012}.
In short, the form is the same as the tree-level matter bispectrum, but
$F_2(\vk_1,\vk_2)$ is replaced by an effective kernel, $F_2^{\rm eff}(\vk_1,\vk_2)$,
which contains nine fitting parameters, $\{a_1, \cdots  a_9\}$, to account for
nonlinearity (see eqs. 2.6 and 2.12 in \cite{gilmarin/etal:2012} for details).
Therefore, we use $F_2^{\rm eff}(\vk_1,\vk_2)$ and compute the integrated bispectrum
by performing the eight-dimensional integral numerically with Monte Carlo integration.
We use the same values of the best-fit parameters provided in table 2 of \cite{gilmarin/etal:2012},
which were calibrated by fitting to simulation results between $z=0$ and $z=1.5$.
In contrast to the SPT formalism that uses the linear power spectrum in \refeq{ch4_bi_spt},
the fitting formula uses the nonlinear power spectrum, for which we use the mean power
spectrum measured from the 160 simulation boxes. For the normalization of the integrated
bispectrum, we convolve the nonlinear power spectrum with the subvolume window function
as in \refeq{ch4_pk_modeling}. Note that the $F_2^{\rm eff}$ fitting formula is not
specifically designed for the squeezed configuration, but instead was calibrated to
a wide range of triangle configurations of the matter bispectrum.

The blue dot-dashed lines in \reffig{ch4_spt_f2eff_modeling} show the normalized
integrated bispectrum computed with $F_2^{\rm eff}$, which clearly depends on redshift.
At $z\gtrsim1$, the $F_2^{\rm eff}$ modeling and the simulations are in good agreement
at $k\lesssim0.2\ihMpc$. At $k>0.2\ihMpc$, although the $F_2^{\rm eff}$ modeling predicts
larger broad-band power of the normalized integrated bispectrum, the BAO amplitudes still
agree well with the simulations. This is most obvious for the two BAO peaks at $0.25\ihMpc\le k\le0.35\ihMpc$.
On the other hand, at $z=0$, the $F_2^{\rm eff}$ modeling predicts much larger normalized
integrated bispectrum on small scales than measured in the simulations, so that the
fitting formula does not perform much better than tree-level perturbation theory at $z=0$.

\section{Separate universe approach}
\label{sec:ch4_sep_uni}
In the second approach, we compute the effects of a long-wavelength density fluctuation
on the small-scale power spectrum by treating each over- and under dense region as a
separate universe with a different background density. This approach thus neglects
the finite size of the subvolumes and is valid for wavenumbers which satisfy $kL\gg1$
(specifically, $kL\gtrsim 50$ for percent-level accuracy).

The power spectrum in a separate universe with an infinite-wavelength density
perturbation, $\bd$, with respect to the global flat $\Lambda$CDM cosmology can
be expanded as in \refeq{ch4_Pkexp}. Through \refeqs{ch4_ib_sq_sepuni}{ch4_nib},
the normalized integrated bispectrum is equal to the linear response of the
nonlinear matter power spectrum at wavenumber $k$ to $\bd$:
\be
 \frac{iB_L(k)}{P(k)\sigma_L^2}=\frac{d\ln P(k)}{d\bd} ~.
\label{eq:ch4_ib_resp}
\ee
This is not exactly true if the subvolumes for which $iB_L(k)$ is measured are
not spherical. For example, since the cubic window function is anisotropic, the
integrated bispectrum might pick up contributions from the tidal field. However,
we have verified that the anisotropy of the cubic window function has a negligible
effect, by computing the dipole and quadrupole of the integrated bispectrum
through \refeq{ch4_ib_ang_avg}. The ratios to the monopole are less than $10^{-5}$
on the scales of interest.

A universe with an infinite-wavelength density perturbation with respect to a flat
fiducial cosmology is equivalent to a universe with non-zero curvature. This alters
the scale factor, Hubble rate, and linear growth as shown in \refchp{ch3_sepuni},
and thus affects the power spectrum. Say this long-wavelength overdensity is
\be
 \bd(t)=\frac{\tilde{\bar\rho}(t)}{\bar\rho(t)}-1=\frac{D(t)}{D(t_0)}\bd(t_0) ~,
\label{eq:ch4_bd}
\ee
where $\bar\rho(t)$ is the background matter density in the fiducial cosmology,
$D(t)$ is the linear growth factor in the same cosmology, $\tilde{\rhob}$ is
the background matter density in a slightly curved universe, $t_0$ is a reference
time such that $a(t_0)=1$, and $\bd_0$ is the density perturbation at $t_0$.
Note that as in \refchp{ch3_sepuni}, we denote the quantities in the modified
(curved) cosmology with a tilde.

In \refeq{ch4_bd} and in the following of this section, we assume that $\bd(t)$
is small and evolves linearly. This is justified because here we consider the
subvolume to be $300\hMpc$, and so $\sigma_L^2(z=0)\sim9\times10^{-4}$. One
can also see this in \reffig{ch4_pk_dm_color}: $|\hdb(\vr_L)|\lesssim0.08$
in 512 subvolumes with $L=300\hMpc$ of one realization at $z=0$. Therefore,
we shall consider the effect on the power spectrum only to the linear order
in $\bd(t)$, and drop $\bd^n(t)$ for $n\ge2$.

With this assumption, the scale factor in the modified cosmology is given by
\be
 \tilde a(t) = a(t) \left[1 - \frac13 \bar\d(t) \right] ~.
\label{eq:ch4_ta}
\ee
Since the physical coordinates are the same in two cosmologies, this implies
that the comoving coordinates of the two cosmologies are related by
\be
 \tilde\vx=\frac{a(t)}{\tilde a(t)}\vx=\left[1+\frac13\bd(t)\right]\vx ~.
\label{eq:ch4_tx}
\ee
What we want is to compare the observables between the fiducial and modified
cosmologies, so the quantities computed in the comoving coordinates of the
modified cosmology have to be mapped to that with respect to the comoving
coordinates of the fiducial cosmology. Specifically, in order to match to
the comoving coordinates of the fiducial cosmology, we transform the comoving
coordinates in the modified cosmology as
\be
 \check\vx=\left[1-\frac13\bd(t)\right]\tilde\vx=c\tilde\vx
\label{eq:ch4_tx_2}
\ee
where c is a constant at a given time. This assures $\check\vx=\vx$.

Let us now consider how the transform in the coordinates affects the two-point
statistics. Since the correlation function is a dimensionless scalar quantity,
in the new coordinates $\check\vx$ with the transformation of $\check x_i=c_ix_i$
(for $i=0$, 1, and 2) the correlation function $\check\xi(\check\vx)$ must be
\be
 \check\xi(\check x_0,\check x_1,\check x_2)=\xi(x_0,x_1,x_2) ~.
\label{eq:ch4_hxi}
\ee
The power spectrum is then transformed as
\ba
 \check P(\check k_0,\check k_1,\check k_2)\:&=\int d^3\check x~
 \check\xi(\check x_0,\check x_1,\check x_2)
 e^{-i(\check x_0\check k_0+\check x_1\check k_1+\check x_2\check k_2)} \vs
 \:&=c_0c_1c_2\int d^3x~\xi(x_0,x_1,x_2)
 e^{-i(c_0x_0\check k_0+c_1x_1\check k_1+c_2x_2\check k_2)} \vs
 \:&=c_0c_1c_2\int d^3x~\xi(x_0,x_1,x_2)e^{-i(x_0k_0+x_1k_1+x_2k_2)} \vs
 \:&=c_0c_1c_2P(k_0,k_1,k_2)=
 c_0c_1c_2P(c_0\check k_0,c_1\check k_1c_2\check k_2) ~,
\label{eq:ch4_hpk_k_1}
\ea
where we define $\check k_i=c_i^{-1}k_i$. This makes sense because $k\sim x^{-1}$
and so $\check k_i\sim\check{x}_i^{-1}=c_i^{-1}x_i^{-1}\sim c_i^{-1}k_i$ (see also
appendix~A of \cite{pajer/schmidt/zaldarriaga:2013}).

Inserting $c$ through \refeq{ch4_tx_2}, we have the change in power spectrum up
to the linear order of $\bd(t)$ as
\ba
 \tilde P(k,t)\:&\rightarrow\left[1-\frac13\bd(t)\right]^3
 P\left(k\left[1-\frac13\bd(t)\right],t\right) \vs
 \:&=\left[1-\bd(t)\right]P(k,t)\left[1-\frac13\frac{d\ln P(k,t)}{d\ln k}\bd(t)\right] \vs
 \:&=P(k,t)\left[1-\frac13\frac{d\ln k^3P(k,t)}{d\ln k}\bd(t)\right] ~.
\label{eq:ch4_tpk_k_2}
\ea
\refEq{ch4_tpk_k_2} is also known as the ``dilation'' effect in
\cite{li/hu/takada:2014,li/hu/takada:2014b}, which is the consequence that
the presence of the long-wavelength overdensity perturbation slows down
the local expansion rate.

Another effect arises due to the change in the ``reference density''. That is,
the background density in two cosmologies are related by $\tilde{\bar\rho}(t)=\rho(t)\left[1+\bd(t)\right]$,
and since power spectrum is proportional to density squared, in the overdense
universe the power spectrum at the linear order of $\bd(t)$ becomes
\be
 \tilde P(\tilde k,t)\rightarrow\left[1+\bd(t)\right]^2\tilde P(\tilde k,t)
 =\left[1+2\bd(t)\right]\tilde P(\tilde k,t) ~.
\label{eq:ch4_tpk_rho}
\ee
Combining the effects of dilation and the reference density, and using the scale
factor instead of time, the power spectrum in the presence of $\bd$ is given by
\ba
 P(k,a |\bd)\:&=\left[1+2\bd(t)\right]\tilde P\left(k,\tilde a\right)
 \left[1-\frac13\frac{d\ln k^3P(k,t)}{d\ln k}\bd(t)\right] \vs
 \:&=\tilde P\left(k,a\left[1-\frac13 \bd(a)\right]\right)
 \left[1 +\left(2-\frac13\frac{d\ln k^3 P(k,a)}{d\ln k}\right)\bd(a)\right] ~.
\label{eq:ch4_Pkd0}
\ea
Note that this expression is only valid to linear order in $\bd$.

Both $P(k)$ and $\bd$ are measured in a finite volume, described by the window
function $W_L$. In order to take this into account, \refeq{ch4_Pkd0} is convolved
by the window function. Note that we take the convolution {\it after} applying
the derivative $d\ln k^3P(k)/d\ln k$, rather than taking the derivative of the
convolved power spectrum. This is because the window function is fixed in terms
of observed coordinates  (in the fiducial cosmology), i.e., it is not subject to
the rescaling of \refeq{ch4_tx}. Taking the slope of the convolved power spectrum
would correspond to a window function defined in the ``local'' curved cosmology.

\subsection{Linear power spectrum}
\label{sec:ch4_spt_res}
For the linear power spectrum, $P_l$, we have
\be
 \tilde P_l\left(k, a\left[1-\frac13 \bar\d(a)\right]\right)
 =\left(\frac{\tilde D\left(a\left[1-\frac13  \bar\d(a)\right]\right)}{D(a)}\right)^2 P_l(k, a)\,.
\label{eq:ch4_tP}
\ee
As shown in \refsec{ch3_linear_growth} (see also appendix~D of
\cite{baldauf/etal:2011}), the linear growth factor is changed following
\be
 \tilde D\left(a\left[1-\frac13 \bar\d(a)\right]\right)
 = D(a) \left[1 + \frac{13}{21} \bar\d(a) \right]\,,
\label{eq:ch4_Dtilde}
\ee
where $D(a)$ is the growth factor in the fiducial cosmology. The prefactor
$13/21$ is only strictly valid for an Einstein-de Sitter cosmology; however,
the cosmology dependence is very mild. The fractional difference of $d\ln D(a)/d\bar\d$
between $\Lambda$CDM cosmology and Einstein-de Sitter universe at $z=0$ is
at the 0.1\% level.

Putting everything together, \refeq{ch4_Pkd0} yields for the linear
response function of the linear power spectrum
\be
 \frac{d \ln P_l(k, a)}{d \bar\d(a)} = 
 \frac{68}{21} - \frac13 \frac{d\ln k^3 P_l(k,a)}{d\ln k}\,.
\label{eq:ch4_Pkd0L}
\ee
This result (which again is only exact for Einstein-de Sitter) matches
the expression derived from the $F_2$ kernel given in \refeq{ch4_sptresponse}.

\subsection{SPT 1-loop power spectrum}
\label{sec:ch4_1loop_res}
Expanding matter density fluctuations to third order, one obtains the so-called
``SPT 1-loop power spectrum'' given by $P(k,a)=P_l(k,a)+P_{22}(k,a)+2P_{13}(k,a)$,
where \cite{bernardeau/etal:2002}
\ba
 P_{22}(k,a) \:&= 
 2 \int \frac{d^3 q}{(2\pi)^3}~P_{l}(q,a) P_{l}(|\vk-\vq|,a)
            \left[F_2(\vq,\vk-\vq)\right]^2, \\
 2P_{13}(k,a) \:&=
 \frac{2\pi k^2}{252} P_{l}(k,a) \int_{0}^{\infty}\frac{dq}{(2\pi)^3}~P_{l}(q,a) \vs
 \:&\times\Biggl[100\frac{q^2}{k^2} -158 + 12\frac{k^2}{q^2} -42 \frac{q^4}{k^4}
 +\frac{3}{k^5q^3}(q^2-k^2)^3(2k^2+7q^2)\ln\left( \frac{k+q}{|k-q|} \right)\Biggl] \nonumber ~.
\label{eq:ch4_P22_P13}
\ea
Both $P_{22}$ and $P_{13}$ are proportional to $D^4(a)$. Modifying the growth
factor as described in \refsec{ch4_spt_res}, we obtain the linear response
function of the SPT 1-loop power spectrum as
\be
 \frac{d \ln P(k, a)}{d \bar\d(a)} =
 \frac{68}{21} - \frac13 \frac{d\ln k^3 P(k,a)}{d\ln k}
 +\frac{26}{21}\frac{P_{22}(k,a)+2P_{13}(k,a)}{P(k,a)} ~.
\label{eq:ch4_Pkd01loop}
\ee
Note that this can easily be generalized to $n$ loops in perturbation theory by
using that $d\ln P_{(n-{\rm loop})}(k,a)/d\ln D(a) = 2n+2$. We include the window
function effect by computing  ${\rm conv}[dP(k)/d\bar{\delta}]/{\rm conv}[P(k)]$.

\refFig{ch4_sep_uni_pert} compares the linear theory and the SPT 1-loop predictions
with the $N$-body simulation results. The SPT 1-loop prediction captures the damping
of BAOs due to nonlinear evolution, and agrees well with the simulation results at
$z=1$, 2, and 3. This is expected from the excellent performance of the 1-loop matter
power spectrum at high redshifts as demonstrated by \cite{jeong/komatsu:2006}. The
agreement degrades rapidly at $z=0$, also as expected. Note that comparing $z=2$
and 3, the 1-loop prediction seems to agree better with the measurements at $z=2$.
However, as mentioned in \refsec{ch4_nbody}, transients and other systematics might
have an impact of up to 5\% on the measurements at $z=3$, which is larger than the
difference shown in the top left panel of \reffig{ch4_sep_uni_pert}.

\begin{figure}[t!]
\centering
\includegraphics[width=1\textwidth]{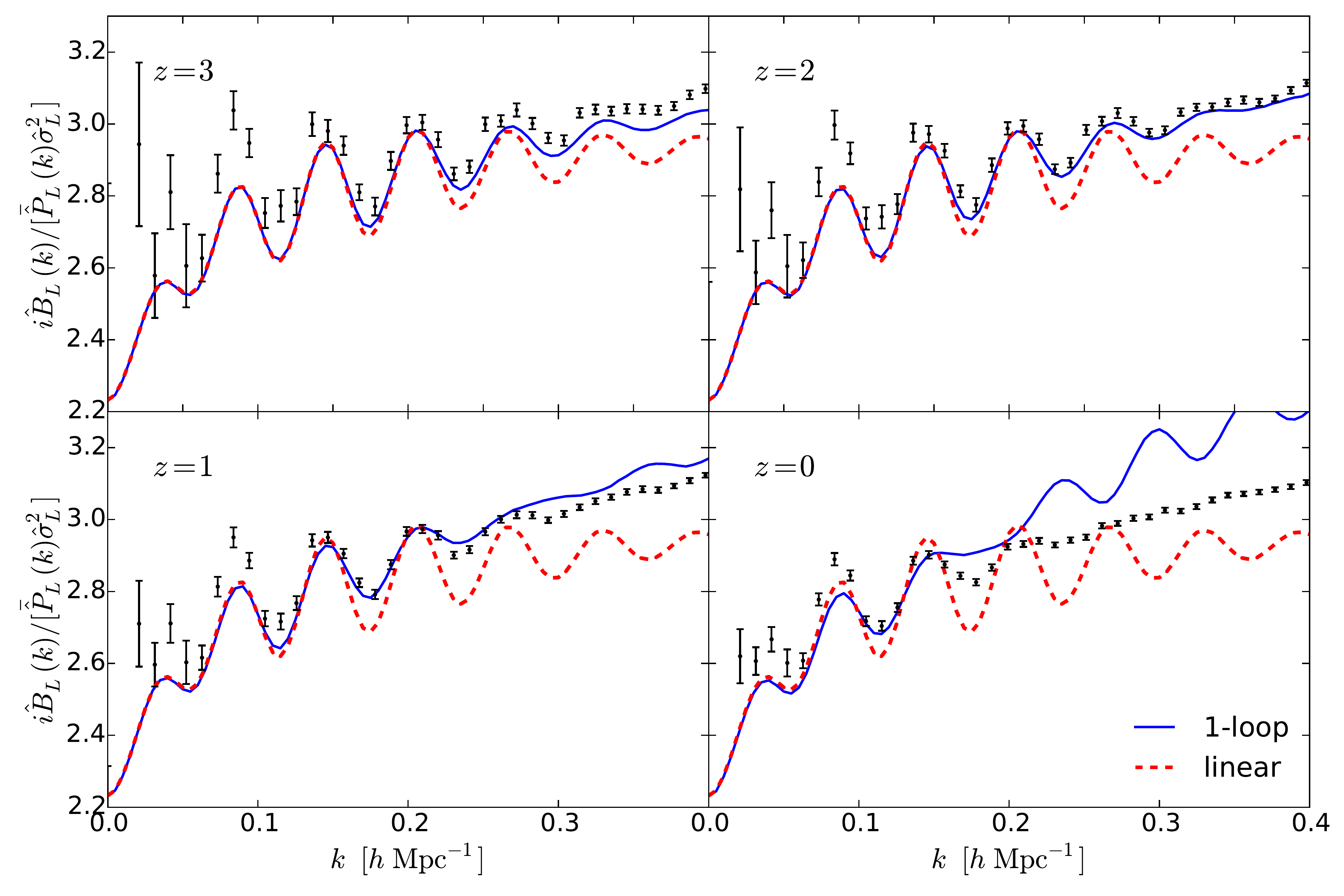}
\caption[Separate universe approach of the linear and 1-loop response
functions for the measured integrated bispectrum in $300\hMpc$ subvolumes]
{Normalized integrated bispectrum from the $N$-body simulations (points
with error bars) and the linear response functions, $d\ln P(k,a)/d\bar\d(a)$,
computed from the separate universe approach combined with perturbation
theory. The red dashed lines show the linear theory results (\refeq{ch4_Pkd0L}),
while the blue solid lines show the SPT 1-loop results (\refeq{ch4_Pkd01loop}).
The agreement between the 1-loop predictions and the simulation results
is very good at $z\ge 1$. Note that the difference between the normalized
integrated bispectrum and the linear response function at $k\lesssim 0.2\ihMpc$
is due to the squeezed limit not being reached yet (see the text below
\refeq{ch4_sptresponse}).}
\label{fig:ch4_sep_uni_pert}
\end{figure}

\subsection{halofit and Coyote emulator}
\label{sec:ch4_halofit_coyote_res}
We now apply the separate universe approach to simulation-calibrated fitting
formulae for the nonlinear matter power spectrum, specifically the halofit
prescription \cite{smith/etal:2003} and the Coyote emulator \cite{heitmann/etal:2014}.
These prescriptions yield $\tilde P(k, a)$ for a given set of cosmological
parameters, so that \refeq{ch4_Pkd0} can be immediately applied. However,
the Coyote emulator does not provide predictions for curved cosmologies,
and we hence adopt a simpler approach here.

In case of the linear power spectrum, the effect of the modified cosmology
enters only through the modified growth factor given in \refeq{ch4_Dtilde}.
Correspondingly, we can approximate the effect on the nonlinear power
spectrum by a change in the value of the power spectrum normalization
$\sigma_8$ at redshift zero,
\be
 \sigma_8 \to \left[1 + \frac{13}{21} \bar\d_0 \right] \sigma_8 ~,
\label{eq:ch4_sig8_resc}
\ee
where we have used the Einstein-de Sitter prediction. Therefore, the nonlinear
power spectrum response becomes
\be
 \frac{d\ln P_{nl}(k,a)}{d\bar\d(a)}= \frac{13}{21}\frac{d\ln P_{nl}(k,a)}{d\ln \sigma_8}
 +2 - \frac{1}{3}\frac{d\ln k^3P_{nl}(k,a)}{d\ln k} ~.
\label{eq:fig4_Pkd0NL}
\ee

\begin{figure}[t!]
\centering
\includegraphics[width=1\textwidth]{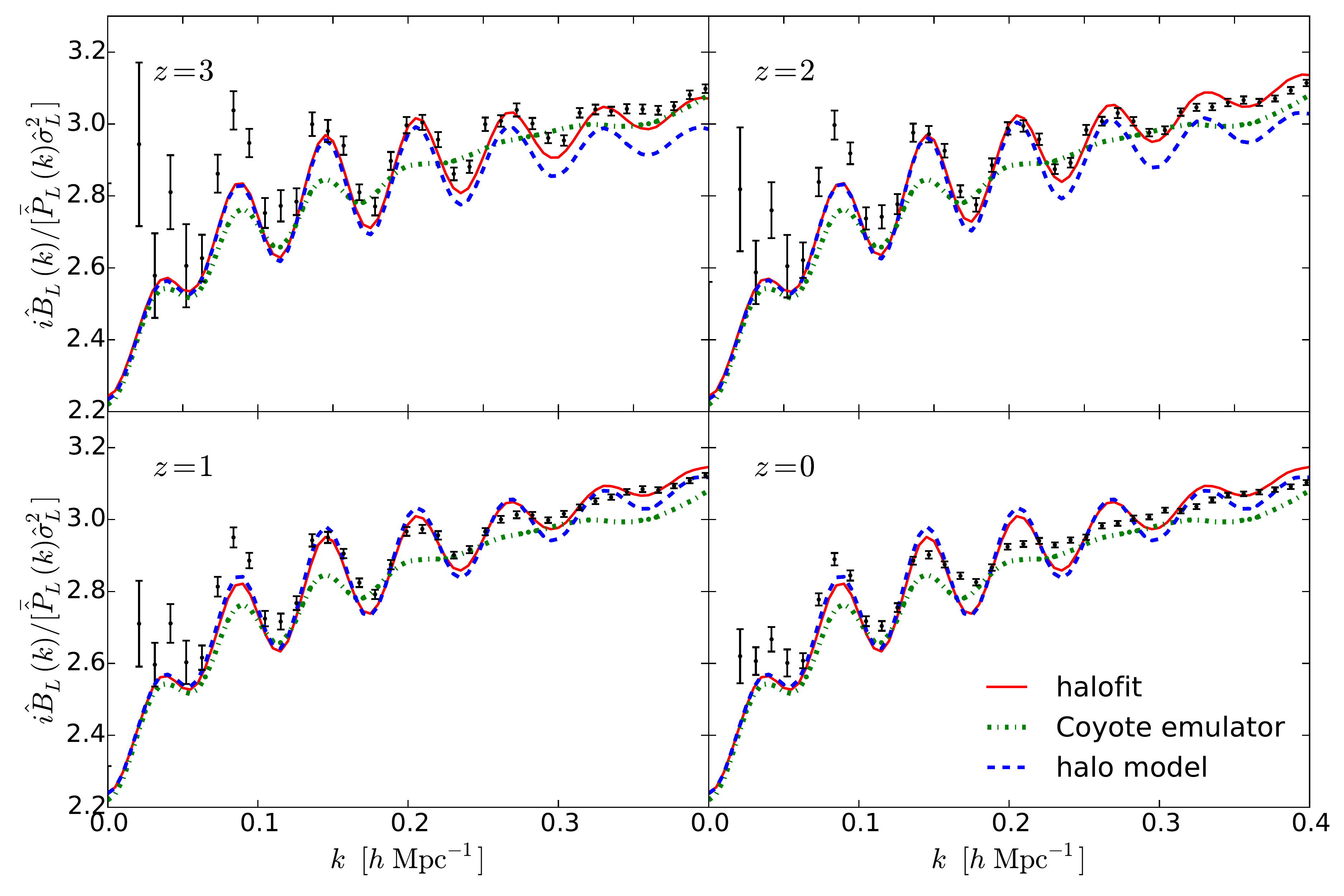}
\caption[Separate universe approach of the halofit, the Coyote emulator,
and the halo model response functions for the measured integrated bispectrum
in $300\hMpc$ subvolumes]
{Same as \reffig{ch4_sep_uni_pert}, but for the linear response functions
computed from halofit (red solid), the Coyote emulator (green dat-dashed),
and the halo model (blue dashed).}
\label{fig:ch4_sep_uni_nonpert}
\end{figure}

The results of applying \refeq{fig4_Pkd0NL} to halofit (red solid) and the
Coyote emulator (green dot-dashed) are shown in \reffig{ch4_sep_uni_nonpert}.
In terms of broad-band power, the halofit prediction provides a good match.
However, the predicted BAO amplitude are larger than the measurement, especially
at low redshift at $k\gtrsim0.3\ihMpc$. Also, while the BAO phases of halofit
follow the SPT prediction, there are some differences with respect to the measurement
of the $N$-body simulations due to the nonlinear evolution. The Coyote emulator
performs to better than $\sim 2$\% over the entire range of scales and redshifts.
It slightly underpredicts the small-scale power at $k>0.3\ihMpc$ for $z\geq1$.
For redshifts $z\geq2$ and on the scales considered, the 1-loop predictions
are of comparable accuracy to the Coyote emulator, while the latter provides
a better fit at lower redshifts. Finally, note also our previous caveat
regarding transients at the end of \refsec{ch4_1loop_res}.

\subsection{Halo model}
\label{sec:ch4_halomodel_res}
In the halo model (see \cite{cooray/sheth:2002} for a review), all matter is
assumed to be contained within halos with a certain distribution of mass given
by the mass function, and a certain density profile. Along with the clustering
properties of the halos, these quantities then determine the statistics of the
matter density field on all scales including the nonlinear regime. $N$-point
functions can be conveniently decomposed into one- through $N$-halo pieces.
In the following, we will follow the most common halo model approach and assume
a linear local bias of the halos.

Adopting the notation of \cite{takada/hu:2013}, the halo model power spectrum,
$P^{\rm HM}(k)$, is given by
\ba
 P^{\rm HM}(k) =P^{\rm 2h}(k)+P^{\rm 1h}(k) ~,~~
 P^{\rm 2h}(k) =\left[I^1_1(k)\right]^2 P_l(k) ~,~~
 P^{\rm 1h}(k) =I^0_2(k,k) ~,
\label{eq:ch4_PkHM}
\ea
where
\be
 I^n_m(k_1,\cdots ,k_m)\equiv \int d\ln M~n(\ln M)\left(\frac{M}{\rhob}\right)^mb_n(M)u(M|k_1)\cdots u(M|k_m) ~,
\label{eq:ch4_Inmdef}
\ee
and $n(\ln M)$ is the mass function (comoving number density per interval in log
mass), $M$ is the halo mass, $b_n(M)$ is the $n$-th order local bias parameter,
and $u(M|k)$ is the dimensionless Fourier transform of the halo density profile,
for which we use the NFW profile \cite{navarro/frenk/white:1997}. We normalize
$u$ so that $u(M|k\to 0)=1$. The notation given in \refeq{ch4_Inmdef} assumes
$b_0\equiv1$. $u(M|k)$ depends on $M$ through the scale radius $r_s$, which in
turn is given through the mass-concentration relation. All functions of $M$ in
\refeq{ch4_Inmdef} are also functions of $z$ although we have not shown this for
clarity. In the following, we adopt the Sheth-Tormen mass function \cite{sheth/tormen:1999}
with the corresponding peak-background split bias, and the mass-concentration
relation of \cite{bullock/etal:2001}. The exact choice of the latter has negligible
impact on the mildly nonlinear scales, but does not affect the conclusion.

We now derive how the power spectrum given in \refeq{ch4_PkHM} responds to an
infinitely long-wavelength density perturbation $\bar\d$, as was done for the
halofit and Coyote emulator approaches. For this, we consider the one-halo and
two-halo terms separately. The key physical assumption we make is that halo
profiles in {\it physical} coordinates are unchanged by the long-wavelength
density perturbation. That is, halos at a given mass $M$ in the presence of
$\bar\d$ have the same scale radius $r_s$ and scale density $\rho(r_s)$ as
in the fiducial cosmology. This assumption, which is related to the stable
clustering hypothesis, can be tested independently with simulations, but we
shall leave it for future work. Given this assumption, the density perturbation
$\bar\d$ then mainly affects the linear power spectrum, which determines the
halo-halo clustering (two-halo term), and the abundance of halos at a given mass.

We begin with the two-halo term. The response of the linear power spectrum is
given by \refeq{ch4_Pkd0L}. The expression for the two-halo term in \refeq{ch4_PkHM}
is simply the convolution (in configuration space) of the halo correlation
function in the linear bias model with the halo density profiles. By assumption,
the density profiles do not change, hence $I^1_1$ only changes through the bias
$b_1(M)$ and the mass function $n(\ln M)$. The bias $b_N(M)$ quantifies the $N$-th
order response of the mass function $n(\ln M)$ to an infinite-wavelength density
perturbation  \cite{mo/white:1996,schmidt/jeong/desjacques:2013}:
\be
 b_N(M)=\frac{1}{n(\ln M)} \frac{\partial^N n(\ln M)}{\partial \bar\d^N}\Big|_{\bar\delta=0} ~.
\label{eq:ch4_bNdef}
\ee
We then have
\ba
 \frac{\partial n(\ln M)}{\partial\bar\d}\Big|_{\bar\delta=0} =\:& b_1(M) n(\ln M) ~, \vs
 \frac{\partial b_1(M)}{\partial\bar\d}\Big|_{\bar\delta=0} =\:& - [b_1(M)]^2 + b_2(M) ~.
\label{eq:ch4_dn_db1_dbd}
\ea
Thus,
\ba
 \frac{\partial}{\partial\bar\d} I^1_1(k) =\:& \int d\ln M \:n(\ln M) \left(\frac{M}{\rhob}\right)
 \left\lbrace[b_1(M)]^2 - [b_1(M)]^2 + b_2(M) \right\rbrace u(M|k) \vs
 =\:& \int d\ln M \:n(\ln M) \left(\frac{M}{\rhob}\right) b_2(M) u(M|k) \vs
 =\:& I^2_1(k) ~.
\label{eq:ch4_dI11}
\ea
In the large-scale limit, $k\to 0$, this vanishes by way of the halo model consistency relation
\be
 \int d\ln M\: n(\ln M)\left(\frac{M}{\rhob}\right) b_N(M) =
 \left\{
 \begin{array}{ll}
 1, & N = 1\,, \\
 0, & N \geq 1\,.
 \end{array}\right.
\label{eq:ch4_hm_cons}
\ee
For finite $k$ however, \refeq{ch4_dI11} does not vanish. Thus, the linear response
function of the two-halo term becomes
\be
 \frac{d P^{\rm 2h}(k)}{d\bar\d}\Big|_{\bar\delta=0} = \left[ \frac{68}{21}
 - \frac13 \frac{d\ln k^3 P_l(k)}{d\ln k} \right] P^{\rm 2h}(k)
 + 2 I^2_1(k) I^1_1(k) P_l(k) ~.
\label{eq:ch4_Pk2hd0}
\ee

Note that we recover the tree-level result given in \refeq{ch4_Pkd0L} in the large-scale
limit. Strictly speaking, this expression is not consistent, since the term $I^2_1$
implies a non-zero $b_2$ while in \refeq{ch4_PkHM} we have assumed a pure linear bias.
Of course, if we allowed for $b_2$ in \refeq{ch4_PkHM}, we would obtain a contribution
from $b_3$ in \refeq{ch4_Pk2hd0}, and so on. This reflects the fact that the halo model
itself cannot be made entirely self-consistent. Note that in \refeq{ch4_Pk2hd0} the
slope is taken from the {\it linear}, not two-halo power spectrum. This is a consequence
of our assumption that halo profiles do not change due to $\bar\d$; in other words,
having $d\ln k^3 P^{\rm 2h}/d\ln k$ would imply that the profiles do change (in the
sense that they are fixed in comoving, rather than physical coordinates).

We now turn to the one-halo term. Given our assumption about density profiles,
this term is much simpler. The only effect is the change in the mass function,
which through \refeq{ch4_bNdef} (for $N=1$) yields
\be
 \frac{\partial}{\partial\bar\d} I^0_2(k, k) = I^1_2(k,k) ~.
\label{eq:ch4_dI02}
\ee
We thus obtain
\ba
 \frac{d P^{\rm 1h}(k)}{d\bar\d}\Big|_{\bar\delta=0} =\:& I^1_2(k,k)\,.
\label{eq:ch4_P1hd0}
\ea
Putting everything together, we obtain
\be
 \frac{d\ln P^{\rm HM}(k)}{d\bar\d}\Big|_{\bar\delta=0} = \left[P^{\rm HM}(k)\right]^{-1}
 \left[ \left(\frac{68}{21} - \frac13 \frac{d\ln k^3 P_l(k)}{d\ln k}\right) P^{\rm 2h}(k)
  + 2 I^2_1(k) I^1_1(k) P_l(k) + I^1_2(k,k) \right]\,.
\label{eq:ch4_Pkd0HM}
\ee
The prediction of \refeq{ch4_Pkd0HM} is shown as the blue dashed lines in \reffig{ch4_sep_uni_nonpert}.
The amplitude and broad-band shape agree with the simulations well. The main discrepancy
in the halo model prediction is the insufficient damping of the BAO wiggles.

An alternative approach to derive the halo model prediction for $iB_L(k)$ is to use
higher $N$-point functions \cite{kehagias/perrier/riotto:2014,li/hu/takada:2014},
which are decomposed into one-, $\dots, N-$halo terms. We now compare \refeq{ch4_Pkd0HM}
with the results of \cite{li/hu/takada:2014}, which were derived from the halo model
four-point function in the collapsed limit. Note that the squeezed limit is assumed
in both approaches. Their eq.~(27) is
\be
 \frac{d\ln P^{\rm HM}(k)}{d\bar\d}\Big|_{\bar\delta=0} = \left[P^{\rm HM}(k)\right]^{-1}
 \left[ \left(\frac{68}{21} - \frac13 \frac{d\ln k^3 P^{\rm 2h}(k)}{d\ln k}\right)P^{\rm 2h}(k) + I^1_2(k,k) \right]\,.
\label{eq:ch4_LHT}
\ee
There are two differences to \refeq{ch4_Pkd0HM}: the term $\propto I^2_1$ is absent,
and the slope is taken from from $P^{\rm 2h}$ rather than $P_l$. The $I^2_1$ term is
absent in \refeq{ch4_LHT} as by assumption $b_2$ was taken to be zero in the four-point
function of \cite{li/hu/takada:2014}; as discussed above, its inclusion is somewhat
ambiguous given the lack of self-consistency of the halo model approach. The different
power spectrum slopes are due to the different sources of this term in the two derivations.
In our case, the assumption of unchanged halo profiles dictates the form of \refeq{ch4_Pkd0HM}.
In the derivation of \refeq{ch4_LHT}, the slope originates from the integral over the
$F_2$ kernel in the three-halo term, which proceeds as described in \refapp{trz_bi} but
involves $P^{\rm 2h}$ instead of $P_l$. Note however that the numerical difference between
\refeq{ch4_LHT} and \refeq{ch4_Pkd0HM} is only at the percent level.

\section{Dependence on cosmological parameters}
\label{sec:ch4_cosmodep}
Both the matter power spectrum and (integrated) bispectrum depend on the cosmological
parameters such as $\Om,\,\sigma_8, n_s$. However, the normalized integrated bispectrum
is much less sensitive to cosmology as the leading cosmology dependence is taken out
by the normalizing denominator.

\refEq{ch4_Pkd01loop} is useful for understanding the dependence of the response function
of the power spectrum (and thus the normalized integrated bispectrum) on cosmological
parameters. The second term depends on the local spectral index of the matter power spectrum,
$d\ln k^3P(k)/d\ln k$, which depends on the initial power spectrum tilt, $n_s$, and the matter
and radiation densities which change the redshift of matter-radiation equality as well as the
BAO scale. It also depends on the shape of BAO wiggles, and increasing the amplitude of the
matter power spectrum ($\sigma_8$) leads to a stronger damping of the BAO feature. Increasing
$\sigma_8$ further increases the last term, which is proportional to $\sigma_8^2$.
\begin{figure}[t!]
\centering
\includegraphics[width=1\textwidth]{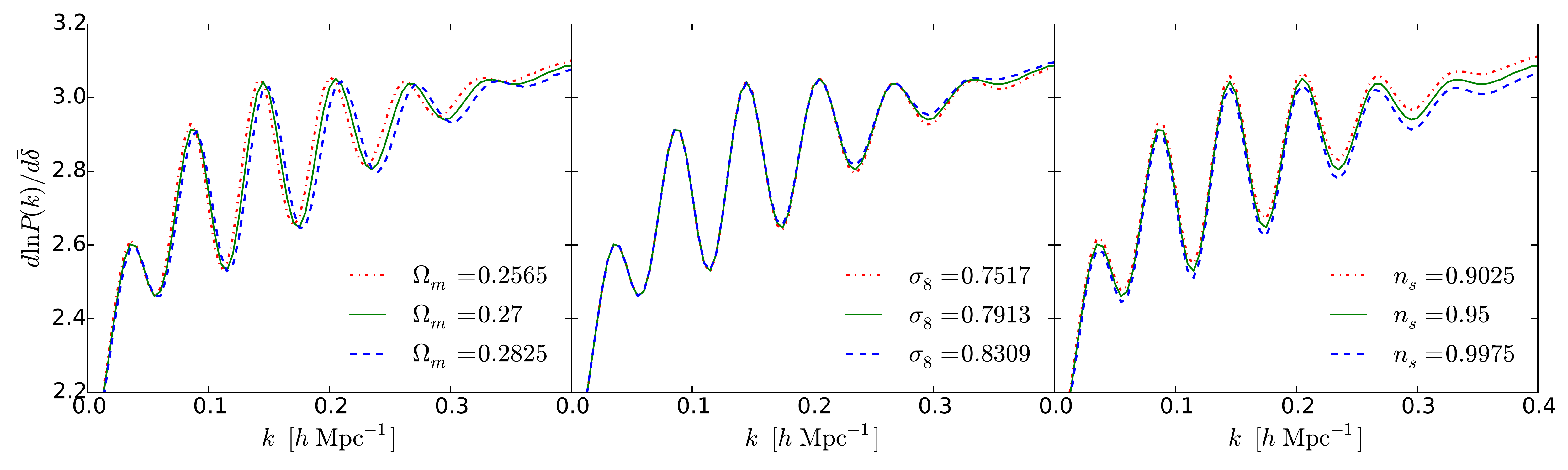}
\caption[Cosmological dependences of the integrated bispectrum]
{The linear response functions computed from the SPT 1-loop power spectrum with various
cosmological parameters at $z=2$. The fiducial cosmology ($\Omega_m=0.27$, $\sigma_8=0.7913$,
and $n_s=0.95$) is shown in green solid lines. The red dot-dashed (blue dashed) lines represent
the cosmologies with -5\% (+5\%) of the fiducial parameters, $\Omega_m$ (left), $\sigma_8$
(middle), and $n_s$ (right).}
\label{fig:ch4_cosmo_dep}
\end{figure}

\refFig{ch4_cosmo_dep} shows the linear response functions, $d\ln P(k,a)/d\bar\d(a)$
computed from the SPT 1-loop power spectrum (\refeq{ch4_Pkd01loop}) at $z=2$ when
varying cosmological parameters by $\pm 5\%$. The effects on the response functions
are at the percent level or less, illustrating the weak cosmology dependence of this
observable. On the scales considered, the shift in the BAO scale when varying $\Om$
leads to the relatively largest effect. We expect that the sensitivity to changes
in $\sigma_8$ will be higher on smaller, more nonlinear scales.

\section{Fisher matrix calculation}
\label{sec:ch4_fisher}
Now that we understand the behavior of the integrated bispectrum, how does it
compare with the full bispectrum in terms of measuring the cosmological parameters,
particularly the primordial non-Gaussianity? In this section, we perform the Fisher
matrix calculation (see e.g. \cite{tegmark/taylor/heavens:1997} for a review) for
both the full bispectrum analysis and the integrated bispectrum technique, and
compare the performances between the two methods. We shall use the simplest primordial
non-Gaussianity model as discussed in \refsec{ch2_ib_models}, i.e.
\ba
 B_g(\vk_1,\vk_2,\vk_3)=\:&b_1^3B_{\rm SPT}(\vk_1,\vk_2,\vk_3)
 +b_1^2b_2B_{b_2}(\vk_1,\vk_2,\vk_3)+b_1^3f_{\rm NL}B_{f_{\rm NL}}(\vk_1,\vk_2,\vk_3) \vs
 iB_{L,g}(k)=\:&b_1^3iB_{L,{\rm SPT}}(k)+b_1^2b_2iB_{L,b_2}(k)
 +b_1^3f_{\rm NL}iB_{L,f_{\rm NL}}(k) ~. 
\label{eq:ch4_png}
\ea

The Fisher matrix of the reduced bispectrum $Q(k_1,k_2,k_3)\equiv B(k_1,k_2,k_3)/[P(k_1)P(k_2)+{\rm 2~cyclic}]$
is given by
\be
 F_{Q,\alpha\beta}=\sum_{k_1,k_2,k_3\le k_{\rm max}}\frac{\partial Q(k_1,k_2,k_3)}{\partial p_{\alpha}}
 \frac{\partial Q(k_1,k_2,k_3)}{\partial p_{\beta}}\frac{1}{\Delta Q^2(k_1,k_2,k_3)} ~,
\label{eq:ch4_fisher_q}
\ee
where $(k_1,k_2,k_3)$ have to form a triangle, $p_{\alpha}\in[b_1,b_2,f_{\rm NL}]$
are the parameters we want to constrain, and $\Delta Q^2(k_1,k_2,k_3)$ is the variance
of the reduced bispectrum estimator. Similarly, the Fisher matrix of the integrated
bispectrum can be written as
\be
 F_{ib_L,\alpha\beta}=\sum_{L}\sum_{k\le k_{\rm max}}\frac{\partial ib_L(k)}{\partial p_{\alpha}}
 \frac{\partial ib_L(k)}{\partial p_{\beta}}\frac{1}{\Delta ib_L^2(k)} ~.
\label{eq:ch4_fisher_ib}
\ee
Here, we ignore off-diagonal elements of the covariance matrix of $Q$ or $ib_L$.
In general, nonlinear evolution generates non-vanishing covariances. This is,
however, justified at high redshift ($z\ge2$) as we show in the right panel of
\reffig{ibv_ib_corr}.

The variance of the estimator for the integrated bispectrum is computed in
\refapp{ib_var}. The variance of the estimator for the bispectrum is computed
\cite{sefusatti/komatsu:2007}
\be
 \Delta B^2(k_1,k_2,k_3)=\frac{\pi s_{123}}{k_1k_2k_3}P(k_1)P(k_2)P(k_3) ~,
\label{eq:ch4_Delta_b2}
\ee
where $s_{123}=6$, 2, 1 for equilateral, isosceles, and general triangles,
respectively (we set $\Dk$ to be the fundamental frequency). Similar to
the integrated bispectrum, we assume that the variance of the reduced
bispectrum is dominated by the numerators, so
\be
 \Delta Q^2(k_1,k_2,k_3)\approx\frac{\Delta B^2(k_1,k_2,k_3)}{[P(k_1)P(k_2)+{\rm 2~cyclic}]^2} ~.
\label{eq:ch4_Q2_ib2_1}
\ee

Since galaxies are observed in redshift space, we model the redshift-space
distortions by the simple Kaiser factor, $P_z=K_pP_r$ and $B_z=K_bB_r$, where
the subscripts $r$ and $z$ denote the real- and redshift-space quantities, and
\be
 K_p=1+\frac{2}{3}\frac{f}{b_1}+\frac{1}{5}\left(\frac{f}{b_1}\right)^2 ~,~~
 K_b=1+\frac{2}{3}\frac{f}{b_1}+\frac{1}{9}\left(\frac{f}{b_1}\right)^2
\label{eq:ch4_rsd_kaiser}
\ee
with $f=d\ln D/d\ln a$ being the growth rate. The derivatives of $Q$ and $ib_L$
with respect to $b_1$ thus contain the contributions from $dK_p/db_1$ and
$dK_b/db_1$. The variances of the redshift-space reduced bispectrum and the normalized
integrated bispectrum with the Poisson shot noise are then given by
\ba
 \Delta Q^2(k_1,k_2,k_3)\approx\:&\frac{\pi s_{123}}{k_1k_2k_3}
 \frac{[P_z(k_1)+P_{\rm shot}][P_z(k_2)+P_{\rm shot}][P_z(k_3)+P_{\rm shot}]}
 {[P_z(k_1)P_z(k_2)+{\rm 2~cyclic}]^2}  \vs
 \Delta ib_L^2(k)\approx\:&\frac{V_L}{V_rN_{kL}}
 \frac{[\sigma_{L,z}^2+P_{\rm shot}/V_L][P_{L,z}(k)+P_{\rm shot}]^2}{\sigma_{L,z}^4P_{L,z}^2(k)} ~,
\label{eq:ch4_Q2_ib2_2}
\ea
where $P_z(k)=b_1^2K_pP_l(k)$, $\sigma_{L,z}^2=b_1^2K_p\sigma_L^2$,
and $P_{L,z}(k)=b_1^2K_pP_L(k)$.

\begin{figure}[t!]
\centering
\includegraphics[width=0.495\textwidth]{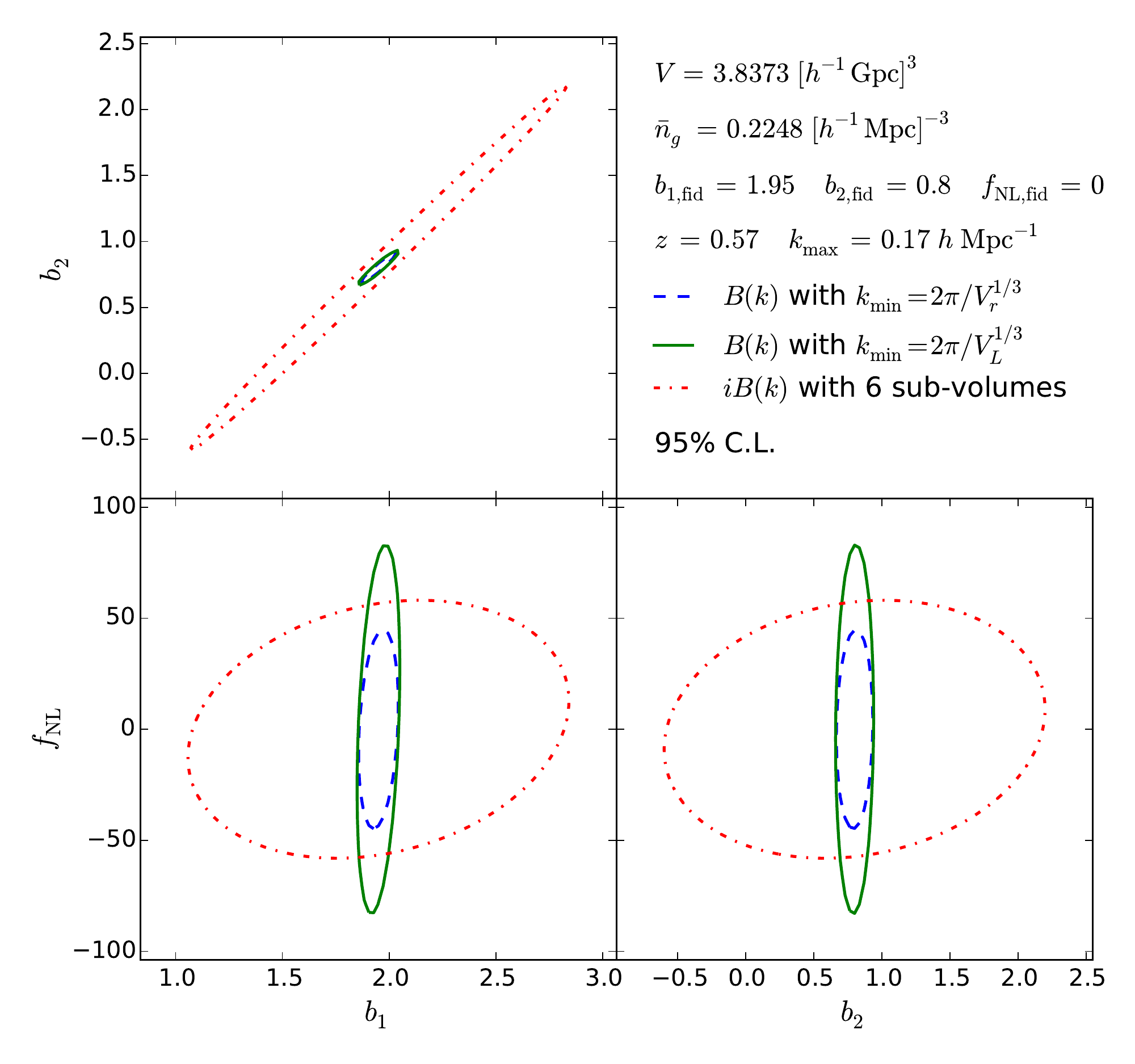}
\includegraphics[width=0.495\textwidth]{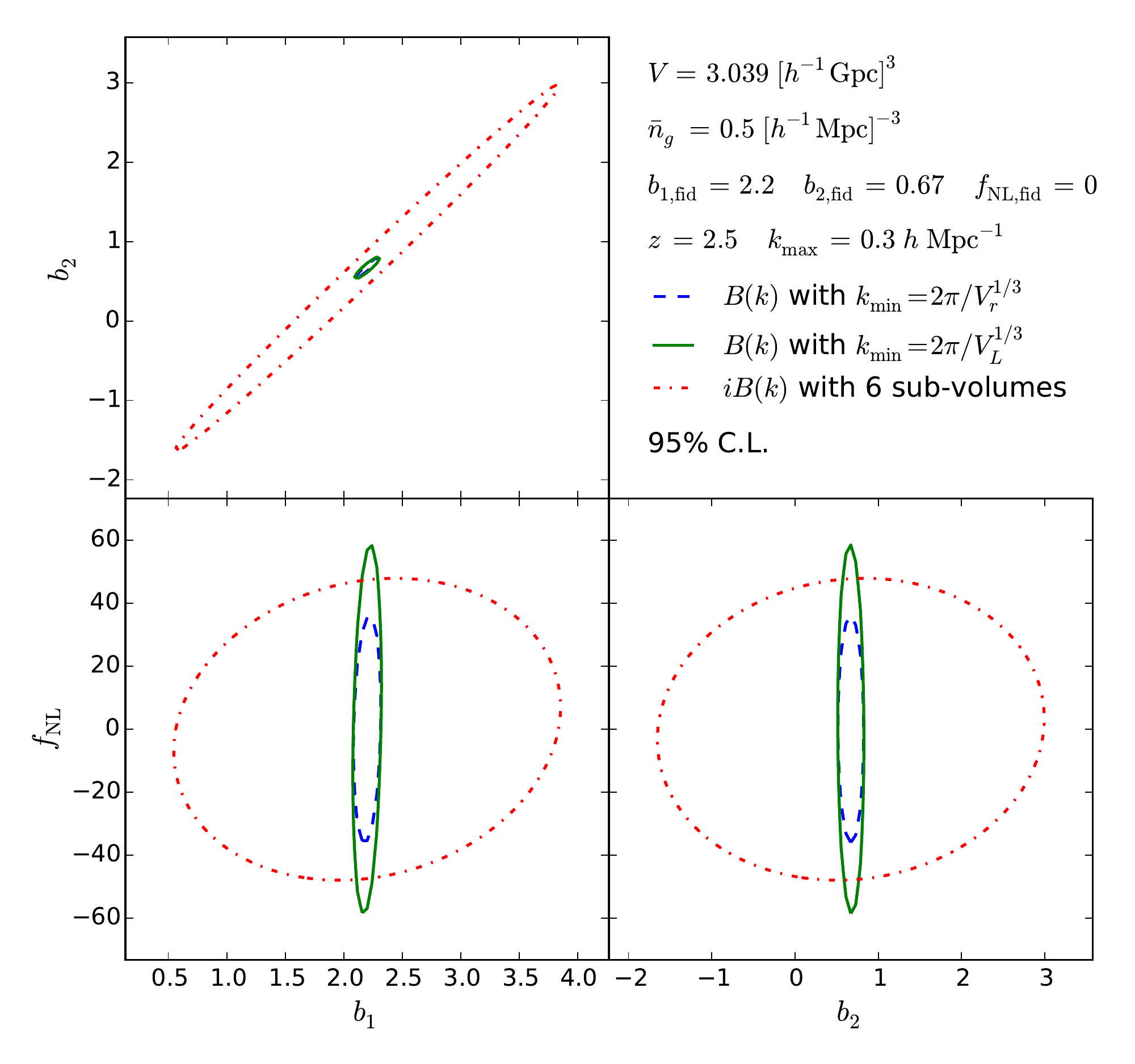}
\caption[Two-dimensional joint 95\% C.L. constraints on galaxy
bias and primordial non-Gaussianity for BOSS and HETDEX]
{Two-dimensional joint 95\% C.L. constraints on galaxy bias and primordial
non-Gaussianity for BOSS (left) and HETDEX (right). The survey parameters
are in the top right panel. The top left, bottom left, and bottom right
panels show the joint constraints on $(b_1,b_2)$, $(b_1,f_{\rm NL})$,
and $(b_2,f_{\rm NL})$ marginalized over $f_{\rm NL}$, $b_2$, and $b_1$,
respectively. The blue dashed, green solid, and red dot-dashed lines are
for full bispectrum with $k_{\rm min}=k_F=2\pi/V_r^{1/3}$, full bispectrum
with $k_{\rm min}=k_{F,L}=2\pi/L$ where $L$ is the largest subvolume size
($600\hMpc$), and integrated bispectrum for six sizes of subvolumes (100$\hMpc$
to $600\hMpc$ with an increment of $100\hMpc$), respectively.}
\label{fig:ch4_fisher_boss_hetdex}
\end{figure}

\refFig{ch4_fisher_boss_hetdex} shows the two-dimensional joint 95\% C.L.
constraints on galaxy bias and primordial non-Gaussianity for BOSS \cite{bossdr12:2015}
(left) and HETDEX \cite{hill/etal:2008} (right). The survey parameters and
the fiducial cosmological parameters are shown in the top-right of each panel.
The blue dashed line is the full bispectrum analysis with $k_{\rm min}=2\pi/V_r^{1/3}$
being the fundamental frequency of the entire survey $V_r$; the green solid
line is the full bispectrum analysis with $k_{\rm min}=2\pi/V_L^{1/3}$ being
the fundamental frequency of the largest subvolume for the integrated bispectrum
($V_L=[600\hMpc]^3$); the red dot-dashed line is the integrated bispectrum
with six sizes of subvolumes from $100\hMpc$ to $600\hMpc$ (with an increment
of $100\hMpc$) with $k_{\rm min}=2\pi/V_L^{1/3}$ being the fundamental frequency
of the corresponding subvolumes.

One finds that as long as $k_{\rm min}$ is set to be the fundamental frequency
of the largest subvolume, the integrated bispectrum technique gives the similar
constraint on $f_{\rm NL}$ compared to the full bispectrum analysis.\footnote{Note
that the numerical results are sensitive to the choices of $k_{\rm min}$ and
$k_{\rm max}$ because we count the Fourier modes in this range. For different lines,
although $k_{\rm max}$ is set to be the same, in practice we stop counting Fourier
modes if $k>k_{\rm max}$. Therefore they have different ``true'' $k_{\rm max}$, and
the contour area would be affected, especially the green solid lines seem to have
slightly worse constraints on $f_{\rm NL}$ compared to the red dot-dashed lines.
Here, however, what we are interested in is the general properties between different
methods, so we should neglect the effect due to different $k_{\rm max}$.} On the other
hand, the integrated bispectrum has poor constraints on $b_1$ and $b_2$ compared
to the full bispectrum analysis. In particular, in the top-left panel the signal
of integrated bispectrum has a strong degeneracy between $b_1$ and $b_2$. This
is somewhat expected because in \reffig{ch2_ib_norm_png} one finds that not only
the black solid and black dashed lines have similar scale dependences in a given
subvolume, but in different subvolumes they all have similar contribution, unlike
the bispectrum of the primordial non-Gaussianity. This thus makes it difficult
to break the degeneracy between $b_1$ and $b_2$ using the integrated bispectrum.

We also find that while the full bispectrum analysis and the integrated bispectrum
technique give similar constraints on $f_{\rm NL}$, the number of counted Fourier
modes differ dramatically. For example, for the BOSS parameter, the full bispectrum
analysis with $k_{\rm min}=2\pi/V_r^{1/3}$ counts 7113 configurations of triangles,
whereas the integrated bispectrum technique counts only 54 Fourier modes. Even if
$k_{\rm min}$ is set to be $2\pi/V_L^{1/3}$ for the full bispectrum analysis, there
are still 6730 configurations of triangles. This is a big advantage for the integrated
bispectrum technique because estimating the covariance matrix from mock catalogs would
require a large number of realizations, which will be difficult (but not impossible)
to obtain for the full bispectrum. This difference in the number of counted Fourier
modes also explains why the full bispectrum analysis has much better constraints on
$b_1$ and $b_2$. However, many of the triangles do not contain much more information
on $f_{\rm NL}$, so if one is interested in measuring $f_{\rm NL}$, the integrated
bispectrum technique provides an easier approach and captures most of the information.

\section{Discussion and conclusion}
\label{sec:ch4_conclusion}
In this chapter, we have demonstrated a novel method to measure the squeezed-limit
bispectrum. By measuring the correlation between the mean density fluctuation and
the position-dependent power spectrum, we obtain a measurement of an integral of
the bispectrum (integrated bispectrum) without having to actually measure three-point
correlations in the data. The integrated bispectrum is dominated by the squeezed-limit
bispectrum, which is much easier to model than the full bispectrum for all configurations.
This is evidenced by \reffig{ch4_sep_uni_pert} and \reffig{ch4_sep_uni_nonpert},
where we show model predictions accurate to a few percent using existing techniques
and without tuning any parameters.

A further, key advantage of this new observable is that both the mean density
fluctuation and the power spectrum are significantly easier to measure in actual
surveys than the bispectrum in terms of survey selection functions. In particular,
the procedures developed for power spectrum estimation can be directly applied
to the measurement of the position-dependent power spectrum. Additionally, the
position-dependent power spectrum depends on only one wavenumber (at fixed size
of the subvolume) rather than the three wavenumbers of the bispectrum. Consequently,
the covariance matrix also becomes easier to model.

We have measured the position-dependent power spectrum in 160 collisionless $N$-body
simulations with Gaussian initial conditions, and have used two different approaches,
bispectrum modeling and the separate universe approach, to model the measurements.
All of the approaches work well on large scales, $k\lesssim0.2\ihMpc$, and at high
redshift. On small scales, where nonlinearities become important, the separate
universe approach (\refsec{ch4_sep_uni}) applied through the Coyote emulator prescription
performs best at redshifts $z<2$, while the SPT 1-loop predictions perform equally well
at $z\geq 2$. Both show agreement to within a few percent up to $k=0.4\ihMpc$. Accurate
predictions for the position-dependent power spectrum on these and even smaller scales
can be obtained by applying the separate universe approach to dedicated small-box
$N$-body simulations of curved cosmologies, as described in \refsec{ch6_sepuni_sim}.

The normalized integrated bispectrum is relatively insensitive to changes in cosmological
parameters (\refsec{ch4_cosmodep}), and we do not expect that it will allow for competitive
cosmology constraints. On the other hand, this property can also be an advantage: since
this observable can be predicted accurately without requiring a precise knowledge of
the cosmology, it can serve as a useful systematics test for example in weak lensing
surveys. As an example, consider \refeq{ch4_ib_sq} applied to shear measurements. A
constant multiplicative bias $1+m$ in the shear estimation contributes a factor $(1+m)^3$
on the left hand side of the equation, and a factor $(1+m)^4$ on the right hand side.
Thus, by comparing the measured normalized integrated bispectrum with the (essentially
cosmology-independent) expectation, one can constrain the multiplicative shear bias.

The position-dependent power spectrum can also naturally be applied to the case
of spectroscopic galaxy surveys, in which case the nonlinear bias of the observed
tracers also contributes to the bispectrum and position-dependent power spectrum.
Thus, when applied to halos or galaxies, this observable can serve as an independent
probe of the bias parameters and break degeneracies between bias and growth which
are present when only considering the halo or galaxy power spectrum. We shall
exploit this to measure the nonlinear bias of the BOSS CMASS galaxies in \refchp{ch5_posdepxi}.

We finally use the Fisher matrix to show that if multiple sizes of subvolumes are
used, the position-dependent power spectrum captures most of the information of
the local-type non-Gaussianity contained in the full bispectrum analysis, but for
much less Fourier modes. This is a huge advantage for this novel technique as
the computational requirement for estimating the covariance matrix is largely
alleviated. Consequently, the position-dependent power spectrum provides an easier
approach for hunting the primordial non-Gaussianity for future galaxy surveys.

}

%% file: kap_05.tex
{

\renewcommand{\v}[1]{\mathbf{#1}}
\newcommand{\vx}{\v{x}}
\newcommand{\vr}{\v{r}}
\newcommand{\vk}{\v{k}}
\newcommand{\vq}{\v{q}}

\newcommand{\bd}{\bar\delta}
\newcommand{\iz}{i\zeta}
\newcommand{\hMpc}{~h^{-1}~{\rm Mpc}}
\newcommand{\se}{\sigma_8}
\newcommand{\sef}{\sigma_{8,\rm fid}}

\chapter{Measurement of position-dependent correlation function}
\label{chp:ch5_posdepxi}
In this chapter, we measure the position-dependent correlation function
and the integrated three-point function from real data, the SDSS-III
Baryon Oscillation Spectroscopic Survey Data Release 10 (BOSS DR10)
CMASS sample \cite{ahn/etal:2014,anderson/etal:2014}.

As introduced in \refsec{ch2_rspace}, the correlation between the position-dependent
correlation function,
\be
 \hat\xi(r,\vr_L)=\int\frac{d^2\hat{r}}{4\pi}~\hat\xi(\vr,\vr_L)
 =\frac{1}{V_L}\int\frac{d^2\hat{r}}{4\pi}\int d^3x~
 \delta(\vr+\vx)\delta(\vx)W_L(\vr+\vx-\vr_L)W_L(\vx-\vr_L) ~,
\label{eq:ch5_posdep_xi}
\ee
and the mean overdensity,
\be
 \bd(\vr_L)=\frac{1}{V_L}\int\frac{d^3q}{(2\pi)^3}~\delta(-\vq)W_L({\vq})e^{-i\vr_L\cdot\vq} ~,
\label{eq:ch5_bd}
\ee
is the integrated three-point function
\ba
 \iz_L(r)\:&=\langle\hat\xi(r,\vr_L)\bar\delta(\vr_L)\rangle \vs
 \:&=\frac{1}{V_L^2}\int\frac{d^2\hat{r}}{4\pi}\int d^3x_1\int d^3x_2
 ~\zeta(\vr+\vx_1+\vr_L,\vx_1+\vr_L,\vx_2+\vr_L) \vs
 &\hspace{5.5cm}\times W_L(\vr+\vx_1)W_L(\vx_1)W_L(\vx_2) ~,
\label{eq:ch5_iz}
\ea
where $V_L$ is the size of the subvolume. Inspired by the behavior in the
squeezed-limit where $r\ll L$, we define the normalized integrated three-point
function as $\iz_L(r)/\sigma_L^2$ with $\sigma_L^2$ being the variance of the
fluctuations in $V_L$.
Note that when comparing the model to the measurements, we shall divide the
model by $f_{L,\rm bndry}(r)$, which is given in \refeq{ch2_ensavg_hat_xi},
to correct for the boundary effect.

The integrated three-point function is simply the Fourier transform of the integrated bispectrum
\be
  \iz_L(r)=\int\frac{k^2dk}{2\pi^2}~iB_L(k){\rm sinc}(kr) ~.
\label{eq:ch5_iz_ib_ang_avg}
\ee
Therefore, if we do not have the analytical expression for the three-point
function, we can first compute the integrated bispectrum and Fourier transform
it to obtain the integrated three-point function. One example is the redshift-space
integrated three-point function, for which we first evaluate the redshift-space
integrated bispectrum with the explicit expression of the SPT redshift-space
bispectrum given in \refapp{trz_kernel}, and then apply \refeq{ch5_iz_ib_ang_avg}.
We also show that the precision of this operation (nine-dimensional integral
in total) is within 2\% on the scales of interest ($30\hMpc\le r\le78\hMpc$,
which we will justify in \refsec{ch5_mock_r}).

This chapter is organized as follows. In \refsec{ch5_mock} and \refsec{ch5_data},
we measure the position-dependent correlation function from PTHalos mock catalogs
\cite{scoccimarro/sheth:2002,manera/etal:2013,manera/etal:2015} and the BOSS DR10
CMASS sample, respectively. The cosmological interpretation of the measurement is
in \refsec{ch5_interpretation}. We conclude in \refsec{ch5_conclusion}.

\section{Measurement of PTHalos mock catalogs}
\label{sec:ch5_mock}
We first apply the position-dependent correlation function technique to the
600 PTHalos mock galaxy catalogs  of the BOSS DR10 CMASS sample in the North
Galactic Cap (NGC). From now on, we refer to the real and mock BOSS DR10
CMASS samples as the ``observations'' and ``mocks'', respectively.

We use the redshift range of $0.43<z<0.7$, and each realization of mocks
contains roughly 400,000 galaxies. We convert the positions of galaxies
in RA, DEC, and redshift to comoving distances using the cosmological
parameters of the mocks. The mocks have the same observational conditions
as the observations, and we correct the observational systematics by weighting
each galaxy differently. Specifically, we upweight a galaxy if its nearest
neighbor has a redshift failure ($w_{\rm zf}$) or a missing redshift due
to a close pair ($w_{\rm cp}$). We further apply weights to correct for
the correlation between the number density of the observed galaxies and
stellar density ($w_{\rm star}$) and seeing ($w_{\rm see}$).
We apply the same weights as done in the analyses of the BOSS collaboration,
namely FKP weighting, $w_{\rm FKP}=[1+P_w\bar n(z){\rm comp}]^{-1}$
\cite{feldman/kaiser/peacock:1994}, where $P_w=20000~h^{-3}~{\rm Mpc}^3$,
and $\bar n(z)$ and ``${\rm comp}$'' are the expected galaxy number density
and the survey completeness, respectively, provided in the catalogs.
Therefore, each galaxy is weighted by
$w_{\rm BOSS}=(w_{cp}+w_{zf}-1)w_{\rm star}w_{\rm see}w_{\rm FKP}$.

In this section, we present measurements from mocks in real space in
\refsec{ch5_mock_r} and redshift space in \refsec{ch5_mock_z}. The
application to the CMASS DR10 sample is the subject of \refsec{ch5_data}.

\subsection{Dividing the subvolumes}
\label{sec:ch5_division}
We use SDSSPix\footnote{SDSSPix: \url{http://dls.physics.ucdavis.edu/~scranton/SDSSPix}}
to pixelize the DR10 survey area. In short, at the lowest resolution (res=1)
SDSSPix divides the sphere equally into $n_x=36$ longitudinal slices across
the hemisphere (at equator each slice is 10 degrees wide), and each slice is
divided into $n_y=13$ pieces along constant latitudes with equal area. Thus,
for res=1 there are $n_x\times n_y=468$ pixels. In general the total number
of pixels is $n_x'\times n_y'=({\rm res}~n_x)\times({\rm res}~n_y)=(\rm res)^2\times468$,
and in this chapter we shall set res=1024. After the pixelization, the $i^{\rm th}$
object (a galaxy or a random sample) has the pixel number $(i_x,i_y)$.

We use two different subvolume sizes. To cut the irregular survey volume into
subvolumes with roughly the same size, we first divide the random samples at
all redshifts into 10 and 20 slices across longitudes with similar numbers of
random samples; we then divide the random samples in each slice into 5 and 10
segments across latitudes with similar numbers of random samples. \refFig{ch5_ran_div}
shows the two resolutions of our subvolumes before the redshift cuts. (Note that
this resolution is different from the resolution of SDSSPix, which we always set
to res=1024.) Each colored pattern extends over the redshift direction. Finally,
we divide the two resolutions into three ($z_{\rm cut}=0.5108$, 0.5717) and five
($z_{\rm cut}=0.48710$, 0.52235, 0.55825, 0.60435) redshift bins.

As a result, there are 150 and 1000 subvolumes for the low and high resolution
configurations, respectively. The sizes of the subvolumes are  approximately
$V_L^{1/3}=220\hMpc$ and $120\hMpc$, respectively\footnote{The shapes of the
subvolumes are not exactly cubes. For example, for the high resolution, the ratios
of square root of the area to the depth, $\sqrt{L_xL_y}/L_z$, are roughly 0.78,
1.42, 1.51, 1.28, and 0.71, from the lowest to the highest redshift bins. The
results are not sensitive to the exact shape of the subvolumes, as long as the
separation of the position-dependent correlation function that we are interested
in is sufficiently smaller than $L_x$, $L_y$, and $L_z$.}. The fractional differences
between the numbers of the random samples in  subvolumes for the low and high
resolutions are within $^{+0.68\%}_{-0.58\%}$ and $^{+1.89\%}_{-1.83\%}$,
respectively. Since the number of random samples represents the effective volume,
all subvolumes at a given resolution have similar effective volumes. We assign
galaxies into subvolumes following the division of random samples.

\begin{figure}[t!]
\centering
\includegraphics[width=0.495\textwidth]{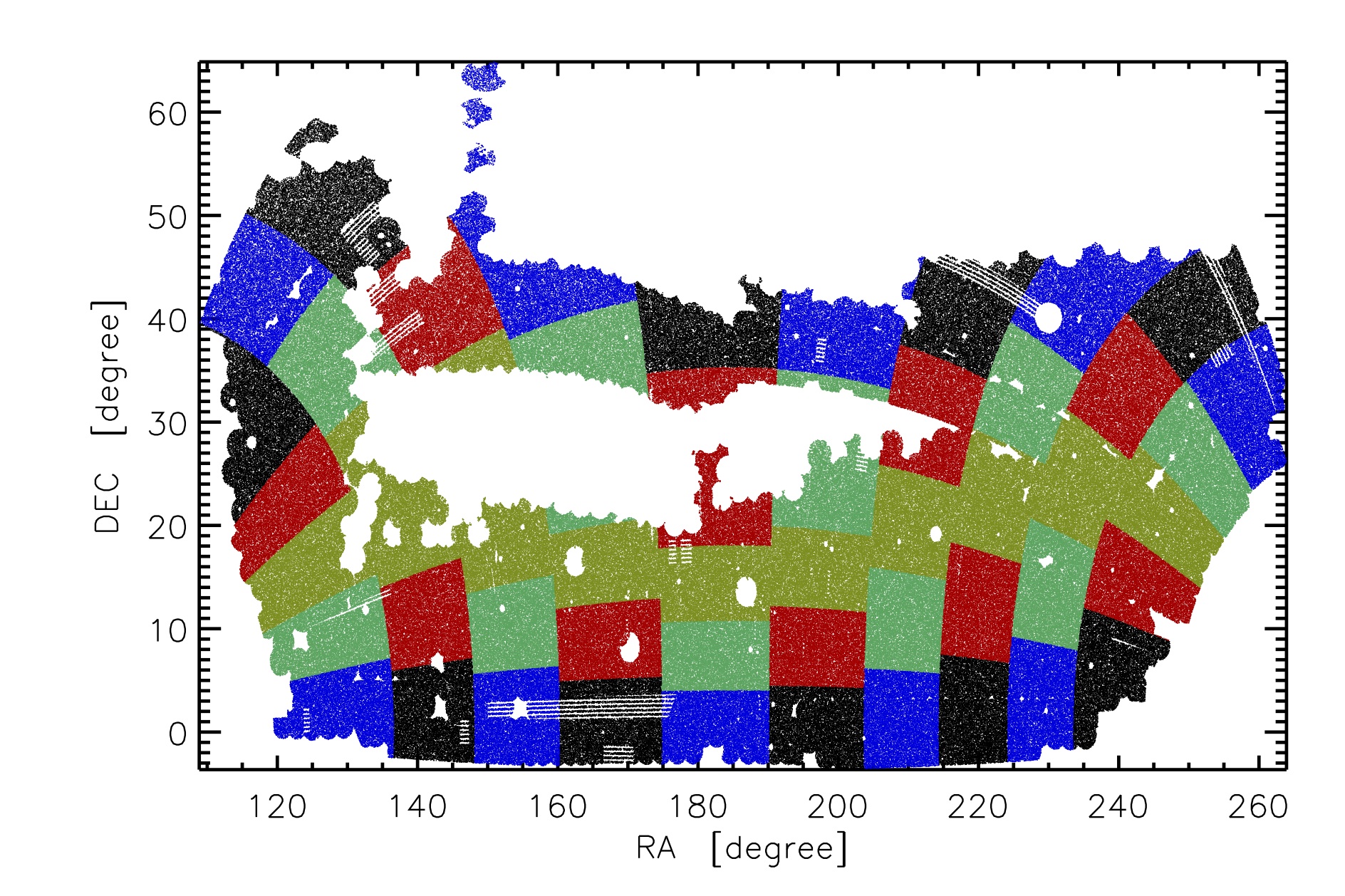}
\includegraphics[width=0.495\textwidth]{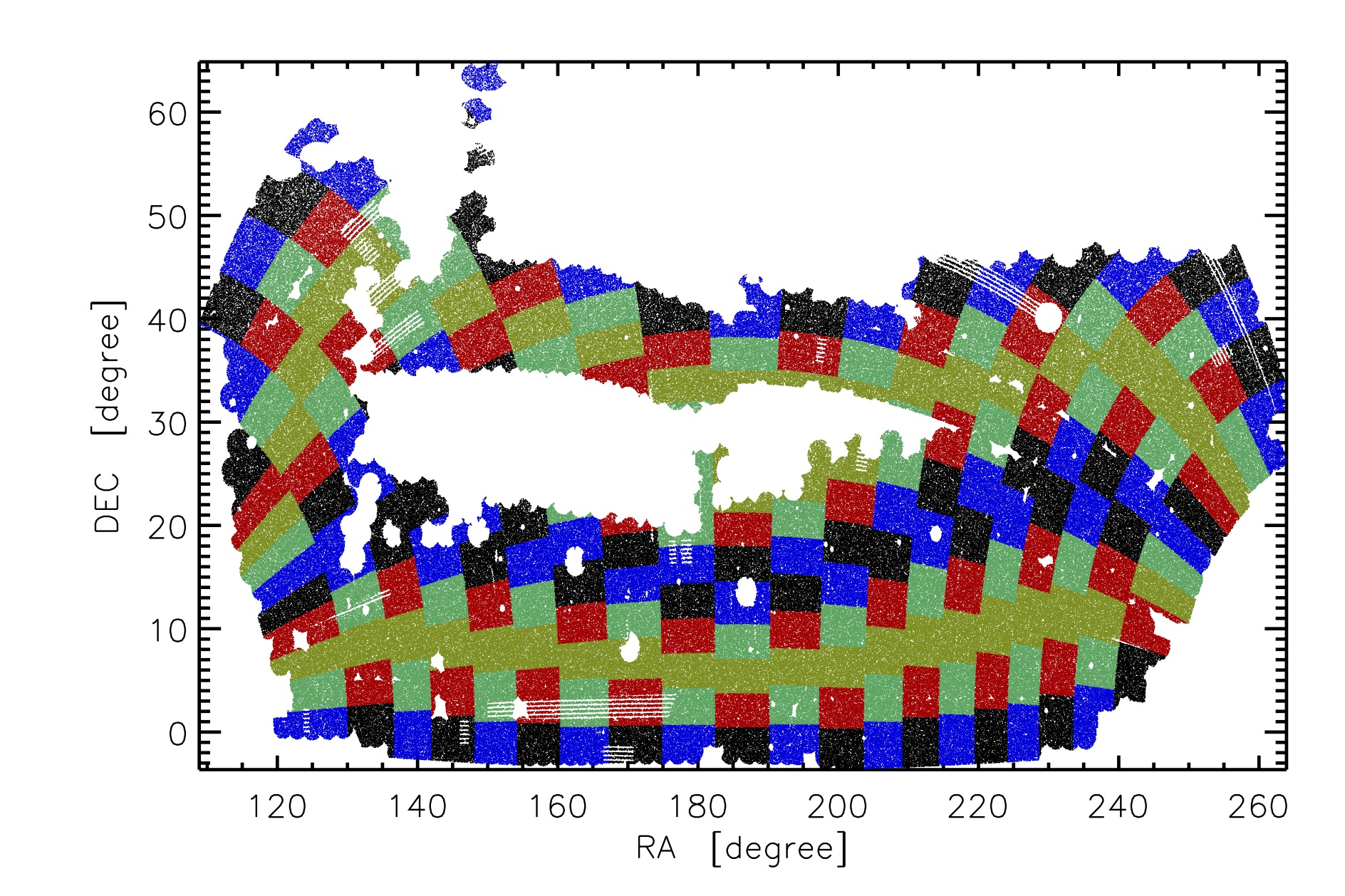}
\caption[Division of random samples into subvolumes in the RA-DEC plane]
{Division of random samples into subvolumes with two resolutions
 in the RA-DEC plane. Each colored pattern extends over the redshift direction.}
\label{fig:ch5_ran_div}
\end{figure}

\subsection{Estimators in the subvolumes}
\label{sec:ch5_sub_quan}
In the $i^{\rm th}$ subvolume, we measure the mean overdensity with respect to the
entire NGC, $\bd_i$, and the position-dependent correlation function, $\hat\xi_i(r)$.
The mean overdensity is estimated by comparing the total weighted galaxies to the
expected number density given by the random samples, i.e.,
\be
 \bd_i=\frac{1}{\alpha}\frac{w_{g,i}}{w_{r,i}}-1\,, ~~~~~
 \alpha\equiv\frac{\sum_{i=1}^{N_s}w_{g,i}}{\sum_{i=1}^{N_s}w_{r,i}}
 =\frac{w_{g,{\rm tot}}}{w_{r,{\rm tot}}} ~,
\ee
where $w_{g,i}$ and $w_{r,i}$ are the total weights ($w_{\rm BOSS}$) of galaxies
and random samples in the $i^{\rm th}$ subvolume, respectively, and $N_s$ is the
number of subvolumes.

We use the Landy-Szalay estimator \cite{landy/szalay:1993} to estimate
the position-dependent correlation function as
\be
 \hat\xi_{{\rm LS},i}(r,\mu)=\frac{DD_i(r,\mu)}{RR_i(r,\mu)}
\left( \frac{[\sum_r w_{r,i}]^2-\sum_r w_{r,i}^2}{[\sum_g w_{g,i}]^2-\sum_g w_{g,i}^2}\right)
 -\frac{DR_i(r,\mu)}{RR_i(r,\mu)}\frac{([\sum_r w_{r,i}]^2-\sum_r w_{r,i}^2)}{\sum_g w_{g,i} \sum_r w_{r,i}}+1 ~,
\label{eq:ch5_ls_xi_est}
\ee
where $DD_i(r,\mu)$, $DR_i(r,\mu)$, and $RR_i(r,\mu)$ are the weighted numbers
of galaxy-galaxy, galaxy-random, and random-random pairs within the $i^{\rm th}$
subvolume, respectively, and $\mu$ is the cosine between the line-of-sight vector
and the vector connecting galaxy pairs ($\vr_1-\vr_2$). The summations such as
$\sum_{r}w_{r,i}$ and $\sum_{g}w_{g,i}$ denote the sum over all the random samples
and galaxies within the $i^{\rm th}$ subvolume, respectively. The angular average
correlation function is then $\hat\xi_{{\rm LS},i}(r)=\int_0^1d\mu~\hat\xi_{{\rm LS},i}(r,\mu)$.

\refEq{ch5_ls_xi_est} estimates the correlation function assuming that the
density fluctuation is measured relative to the ${\it local}$ mean. However,
the position-dependent correlation function defined in \refeq{ch5_posdep_xi}
uses the density fluctuation relative to the {\it global} mean. These two
fluctuations can be related by $\delta_{\rm global}=(1+\bd)\delta_{\rm local}+\bd$
with $\bd=\bar n_{\rm local}/\bar n_{\rm global}-1$. Thus, the position-dependent
correlation function, $\hat\xi_i(r)$, is related to the Landy-Szalay estimator as
\be
 \hat\xi_i(r)=(1+\bd_i)^2\hat\xi_{{\rm LS},i}(r)+\bd_i^2 ~.
\label{eq:ch5_xi_corr}
\ee

To compute the average quantities over all subvolumes, we weight by $w_{r,i}$
in the corresponding subvolume. For example, for a given variable $g_i$ in the
$i^{\rm th}$ subvolume, the average over all subvolumes, $\bar g$, is defined by
\be
 \bar g=\frac{1}{w_{r,{\rm tot}}}\sum_{i=1}^{N_s}g_iw_{r,i} ~.
\label{eq:ch5_ens_avg}
\ee
Since the number of random samples in each subvolume represents the effective
volume, the average quantities are effective-volume weighted. \refEq{ch5_ens_avg}
assures that the mean of the individual subvolume overdensities is zero,
\be
 \bd=\frac{1}{w_{r,{\rm tot}}}\sum_{i=1}^{N_s}\bd_iw_{r,i}
 =\frac{1}{w_{r,{\rm tot}}}\sum_{i=1}^{N_s}\left[\frac{1}{\alpha}w_{g,i}-w_{r,i}\right]
 =\frac{\alpha}{\alpha}-1=0 ~.
\ee
We also confirm that $\bar{\hat{\xi}}(r)$ from \refeq{ch5_xi_corr} agrees
with the two-point function of all galaxies in the entire survey, on scales
smaller than the subvolume size.
With $\hat\xi_i(r)$ and $\bd_i$, we estimate the shot-noise-corrected
integrated three-point function in the subvolume of size $L$ as
\be
 \iz(r)=\frac{1}{w_{r,{\rm tot}}}\sum_{i=1}^{N_s}\left[\hat\xi_i(r)\bd_i
 -2\bar{\hat\xi}(r)\frac{(1+\alpha)}{\alpha}\frac{\sum_rw^2_{r,i}}{\sum_r\bar n_{r,i}{\rm comp}_{r,i}w^2_{r,i}}
 \left(\sum_r\frac{1}{\bar n_{r,i}{\rm comp}_{r,i}}\right)^{-1}\right]w_{r,i} ~,
\ee
where the second term in the parentheses is the shot noise contribution, and
$\bar n_{r,i}$ and ${\rm comp}_{r,i}$ are the expected galaxy number density
and the survey completeness, respectively, of the random samples. Similarly,
we estimate the shot-noise-corrected variance of the fluctuations in the
subvolumes of size $L$ as
\be
 \sigma_L^2=\frac{1}{w_{r,{\rm tot}}}\sum_{i=1}^{N_s}
 \left[\bd_i^2-\frac{(1+\alpha)}{\alpha}\frac{\sum_rw^2_{r,i}}
 {\sum_r\bar n_{r,i}{\rm comp}_{r,i}w^2_{r,i}}
 \left(\sum_r\frac{1}{\bar n_{r,i}{\rm comp}_{r,i}}\right)^{-1}\right]w_{r,i} ~,
\ee
where the second term in the parentheses is the shot noise contribution.
We find that the shot noise is subdominant (less than 10\%) in both $\iz(r)$
and $\sigma_L^2$.

\subsection{Measurements in real space}
\label{sec:ch5_mock_r}
\refFig{ch5_mock_r} shows the measurements of the two-point function $\xi(r)$
from the entire survey (top left) and the normalized integrated three-point
functions (bottom panels),
\be
 \frac{\iz_L(r)}{\sigma_L^2}=\left(\frac{1}{w_{r,{\rm tot}}}\sum_{i=1}^{N_s}
 \left[\hat\xi_i(r)\bd_iw_{r,i}\right]\right)\frac{1}{\sigma_L^2} ~,
\ee
for the subvolumes of two sizes ($220\hMpc$ in the bottom-left and $120\hMpc$
in the bottom-right panels). The gray lines show individual realizations,
while the dashed lines show the mean.

\begin{figure}[t!]
\centering
\includegraphics[width=0.495\textwidth]{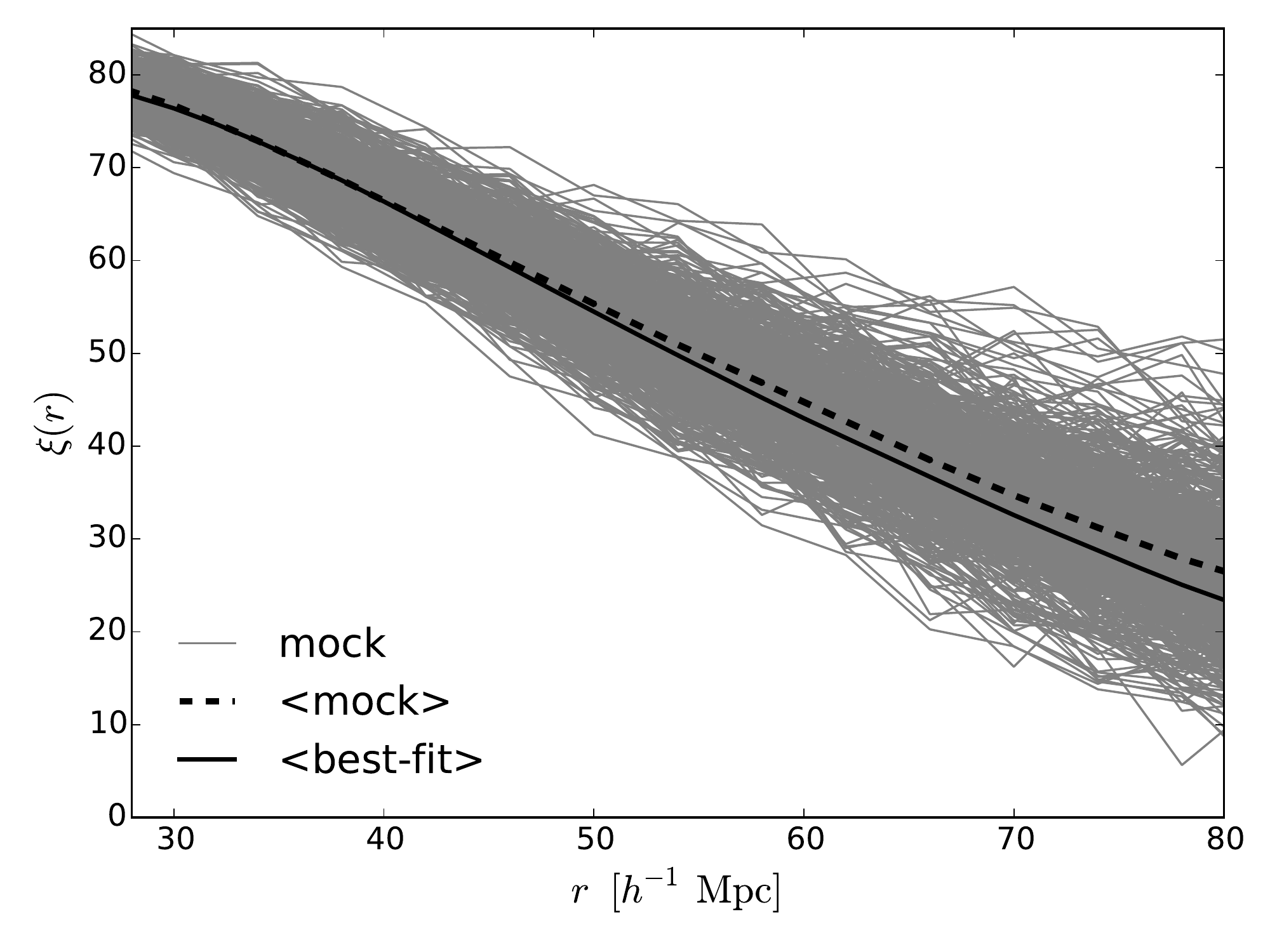}
\includegraphics[width=0.495\textwidth]{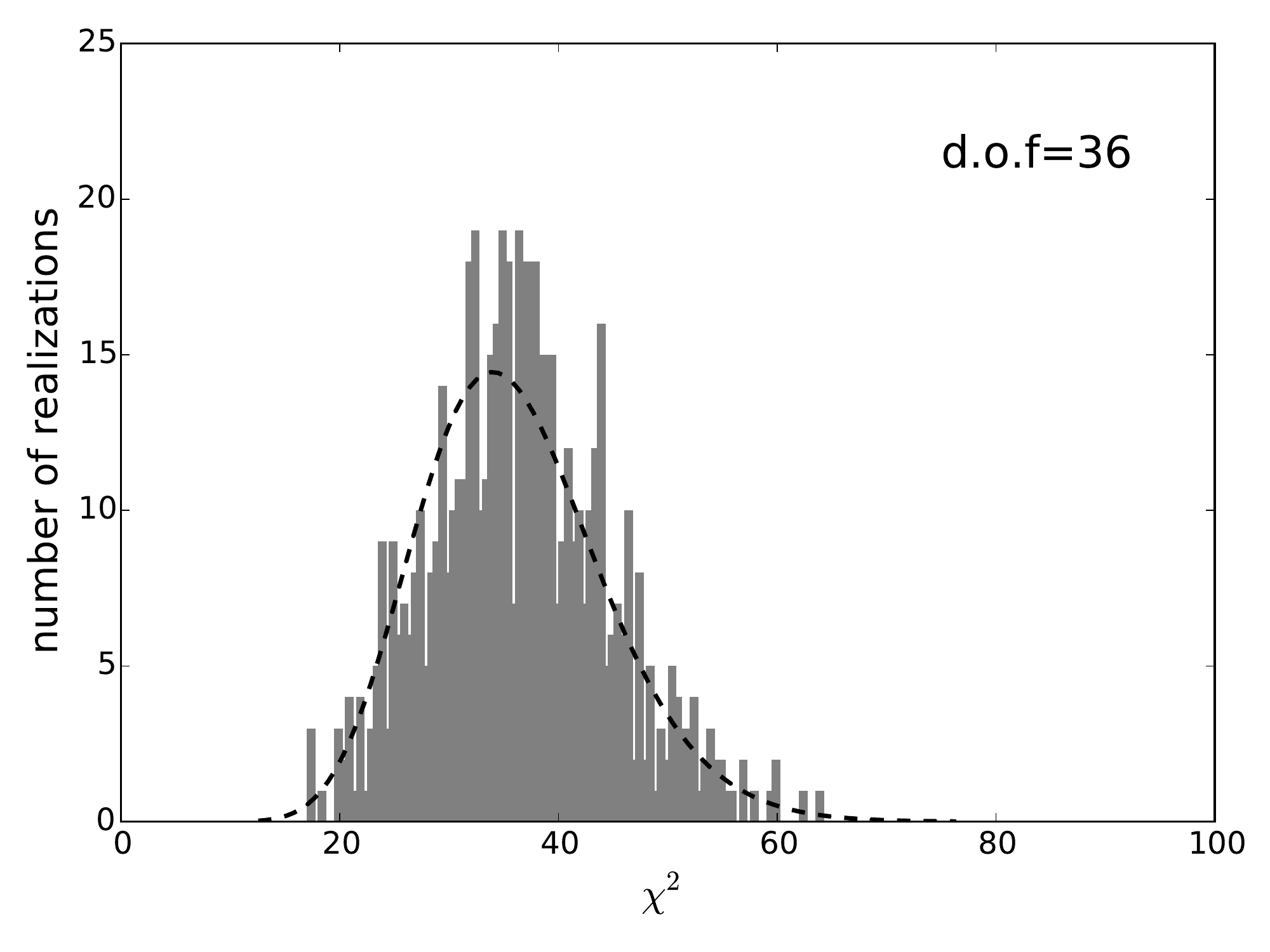}
\includegraphics[width=0.495\textwidth]{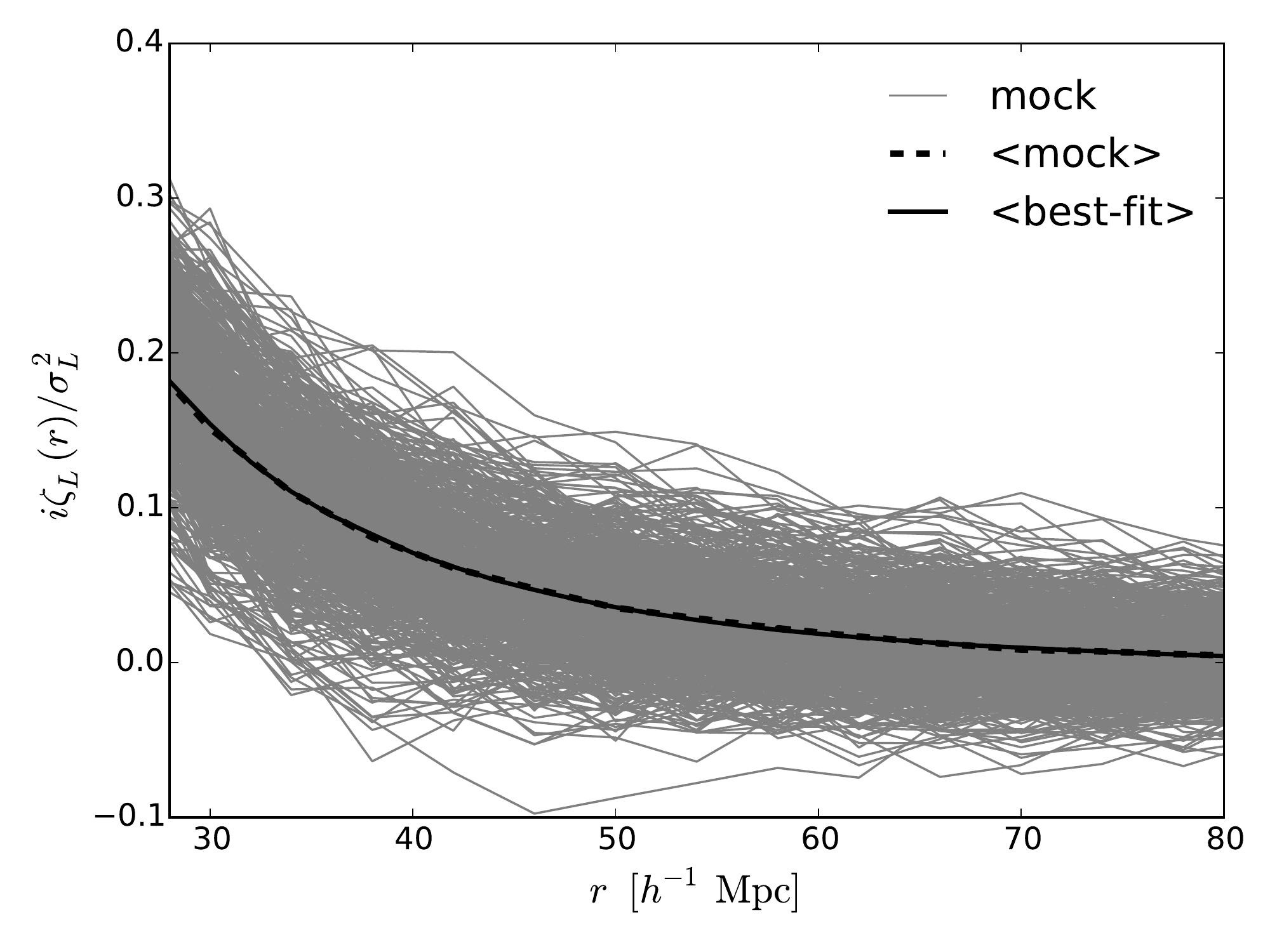}
\includegraphics[width=0.495\textwidth]{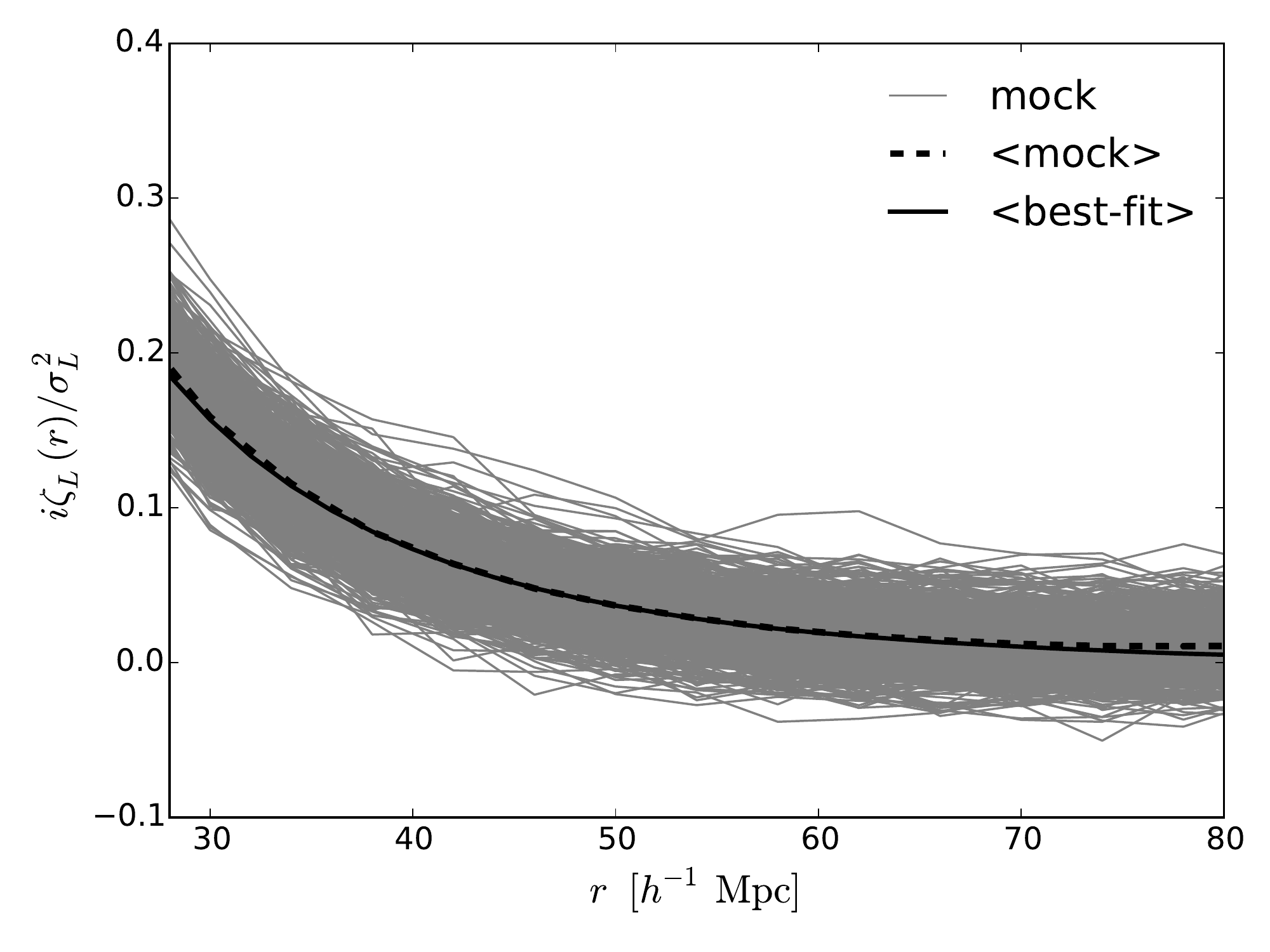}
\caption[Measurements of $\xi(r)$ and $\iz_L(r)/\sigma_L^2$ of PTHalos mock
catalogs in real space]
{(Top left) $\xi(r)$ of the mocks in real space. The gray lines show individual
realizations, while the dashed line shows the mean. The black solid line shows
the best-fitting model. (Top right) $\chi^2$-histogram of the 600 mocks jointly
fitting the models to $\xi(r)$ and $\iz_L(r)/\sigma_L^2$ in real space. The dashed
line shows the $\chi^2$-distribution with d.o.f.=36. (Bottom left) $\iz_L(r)/\sigma_L^2$
of the mocks in real space for $220\hMpc$ subvolumes. (Bottom right) Same as the
bottom left panel, but for $120\hMpc$ subvolumes.}
\label{fig:ch5_mock_r}
\end{figure}

We now fit models of $\xi(r)$ and $\iz_L(r)/\sigma_L^2$ to the measurements in
$30\hMpc\le r\le78\hMpc$. We choose this fitting range because there are less
galaxy pairs at larger separations due to the subvolume size, and the nonlinear
effect becomes too large for our SPT predictions to be applicable at smaller
separations. For the two-point function, we take the Fourier transform of
\cite{crocce/scoccimarro:2008}
\be
 P_g(k)=b_1^2[P_l(k)e^{-k^2\sigma_v^2}+A_{\rm MC}P_{\rm MC}(k)]~,
\ee
where $b_1$ is the linear bias, $P_l(k)$ is the linear power spectrum,
$A_{\rm MC}$ is the mode coupling constant, and
\be
 P_{\rm MC}(k)=2\int\frac{d^3q}{(2\pi)^3}~P_l(q)P_l(|\vk-\vq|)[F_2(\vq,\vk-\vq)]^2 ~.
\label{eq:ch5_xi_model}
\ee
Hence, $\xi_g(r)=b_1^2[\xi_{l,\sigma_v}(r)+A_{\rm MC}\xi_{\rm MC}(r)]$ with
\be
 \xi_{l,\sigma_v}(r)=\int\frac{d^3k}{(2\pi)^3}~P_l(k)e^{-k^2\sigma_v^2}e^{i\vk\cdot\vr}\,, ~~~~~
 \xi_{\rm MC}(r)=\int\frac{d^3k}{(2\pi)^3}~P_{\rm MC}(k)e^{i\vk\cdot\vr} \ .
\label{eq:ch5_xi_model_2}
\ee
We use a fixed value of $\sigma_v^2=20.644$. Varying it has only small
effect on the other fitted parameters. For the integrated three-point
function, we use the SPT calculation with the boundary effect correction,
which is given by
\be
 \frac{\iz_{L,g}(r)}{\sigma_L^2}=\frac{b_1\iz_{L,{\rm SPT}(r)}+b_2\iz_{L,b_2}(r)}{\sigma_{L,l}^2}
 +\frac1{f_{L,\rm bndry}(r)}~,
\label{eq:ch5_iz_model}
\ee
where $\iz_{L,{\rm SPT}}(r)$ and $\iz_{L,b_2}(r)$ are computed from \refeq{ch5_iz}
with \refeq{ch2_zeta_spt} and \refeq{ch2_zeta_b2}, respectively, and
$\sigma_{L,l}^2$ is the variance of the linear power spectrum computed from
\refeq{ch2_sigmalL2}, using the subvolume sizes of $L=220$ and $120\hMpc$
and the redshift of $z=0.57$. Note that the size of the subvolumes affects
the values of $\sigma_{L,l}^2$. We determine $L$ by first measuring $b_1^2$
using the real-space two-point function of the entire survey, and then find
$L$ such that $b_1^2\sigma_{L,l}^2=\sigma_L^2$ assuming the cubic top-hat
window function\footnote{In principle, the shape of the window function also
affects $\sigma_{L,l}^2$, but we ignore this small effect.}. We find that
these values ($L=220$ and $120\hMpc$) agree well with the cubic root of the
total survey volume divided by the number of subvolumes, to within a few percent.

We fit the models to $\xi(r)$ and $\iz_L(r)/\sigma_L^2$ of both subvolumes
simultaneously by minimizing
\be
 \chi^2=\sum_{ij}C^{-1}_{ij}(D_i-M_i)(D_j-M_j)\,,
\label{eq:ch5_chi2}
\ee
where $C^{-1}$ is the inverse covariance matrix computed from the 600 mocks,
$D_i$ and $M_i$ are the data and the model in the $i^{\rm th}$ bin, respectively.
The models contain three fitting parameters $b_1$, $b_2$, and $A_{\rm MC}$.

The models computed with the mean of the best-fitting parameters of 600 mocks
are shown as the black solid lines in \reffig{ch5_mock_r}. The best-fitting
parameters are $b_1=1.971\pm0.076$, $b_2=0.58\pm0.31$, and $A_{\rm MC}=1.44\pm0.93$,
where the error bars are 1-$\sigma$ standard deviations. The agreement between
the models and the mocks is good, with a difference much smaller than the scatter
among 600 mocks. Upon scrutinizing, the difference in $\xi(r)$ is larger for larger
separations because the fit is dominated by the small separations with smaller error
bars. On the other hand, for $\iz(r)/\sigma_L^2$ the agreement is good for two sizes
of subvolumes at all scales of interest. This indicates that the SPT calculation is
sufficient to capture the three-point function of the mocks in real space.

The data points in \reffig{ch5_mock_r} are highly correlated. To quantify the
quality of the fit, we compute the $\chi^2$-histogram from 600 mocks, and compare
it with the $\chi^2$-distribution with the corresponding degrees of freedom (d.o.f.).
There are 13 fitting points for each measurement ($\xi(r)$ and two sizes of subvolumes
for $\iz_L(r)/\sigma_L^2$) and three fitting parameters, so d.o.f.=36. The top right
panel of \reffig{ch5_mock_r} shows the $\chi^2$-histogram. The dashed line shows the
$\chi^2$-distribution with d.o.f.=36. The agreement is good, and we conclude that
our models well describe both $\xi(r)$ and $\iz_L(r)/\sigma_L^2$ of the mocks in
real space.

Our $b_1$ is in good agreement with the results presented in figure 16 of
\cite{gilmarin/etal:2014b}, whereas our $b_2$ is smaller than theirs, which
is $\simeq 0.95$, by 1.2$\sigma$. This may be due to the difference in the
bispectrum models. While we restrict to the local bias model and the tree-level
bispectrum, \cite{gilmarin/etal:2014b} includes a non-local tidal bias
\cite{mcdonald/roy:2009,baldauf/etal:2012,sheth/chan/scoccimarro:2013}
and uses more sophisticated bispectrum modeling using the effective $F_2$
kernel \cite{gilmarin/etal:2012,gilmarin/etal:2014a}. In \refapp{ft_b2},
we show that using the effective $F_2$ kernel and the non-local tidal
bias in the model increases the value of $b_2$, but the changes are well
within the 1-$\sigma$ uncertainties. Also, the differences of the goodness
of fit for various models are negligible.

The fitting range as well as the shapes of the bispectrum may also affect
the results: the integrated correlation function is sensitive only to the
squeezed configurations, whereas \cite{gilmarin/etal:2014b} includes more
equilateral and collapsed triangle configurations. Understanding this
difference merits further investigations.

\subsection{Measurements in redshift space}
\label{sec:ch5_mock_z}
\refFig{ch5_mock_z} shows the measurements of $\xi(r)$ (top left) and $\iz_L(r)/\sigma_L^2$
($220\hMpc$ in the bottom-left and $120\hMpc$ in the bottom-right panels) of
the mocks in redshift space. The gray lines show individual realizations, while
the dashed lines show the mean. Similar to the analysis in \refsec{ch5_mock_r},
we fit the models in redshift space to the measurements in $30\hMpc\le r\le78\hMpc$.
In this section, we use General Relativity to compute the growth rate,
$f(z)\approx\Omega_m(z)^{0.55}$, which yields $f(z=0.57)=0.751$. We shall allow
$f$ to vary when interpreting the measurements in the actual data.

\begin{figure}[t]
\centering
\includegraphics[width=0.495\textwidth]{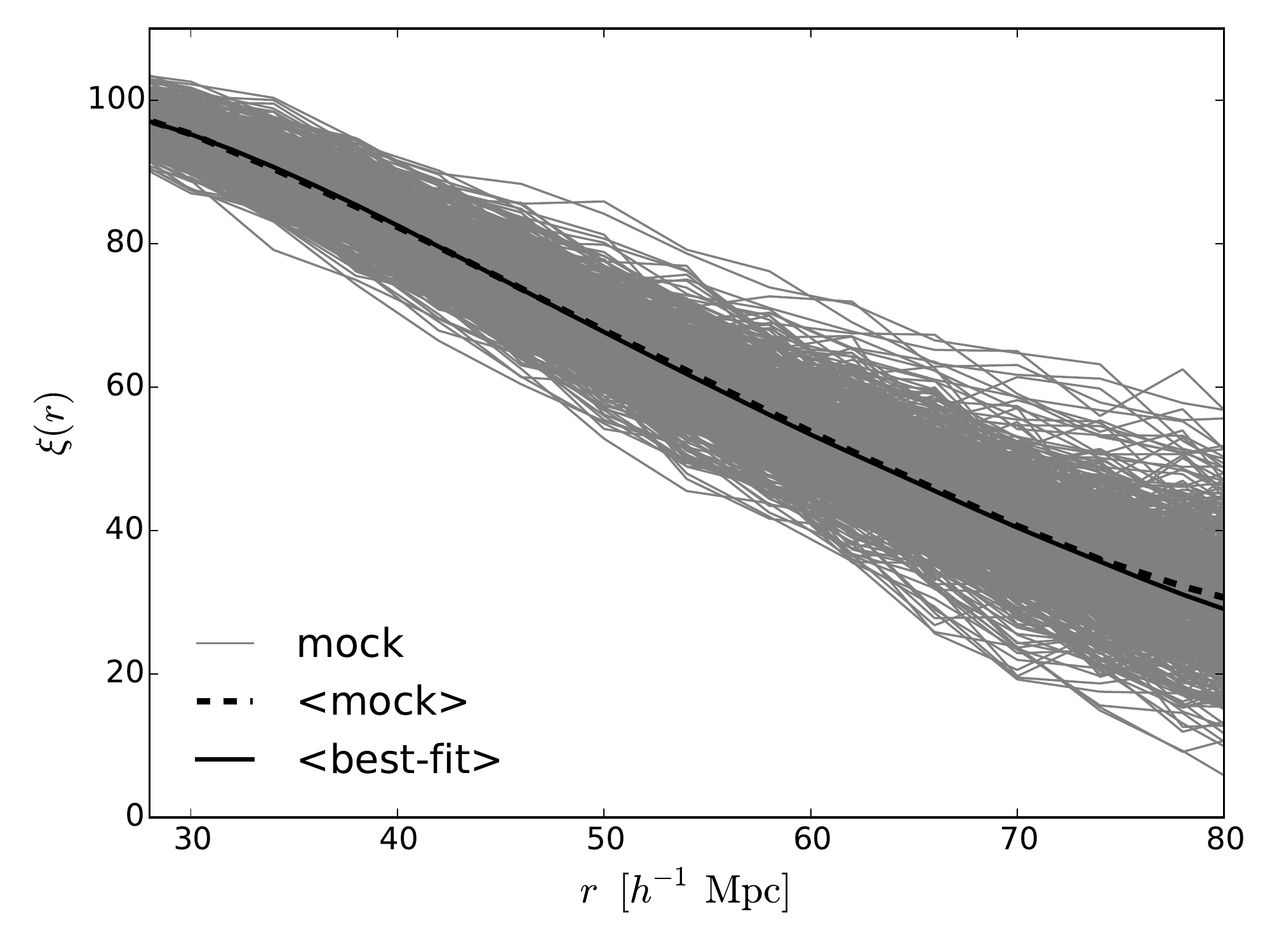}
\includegraphics[width=0.495\textwidth]{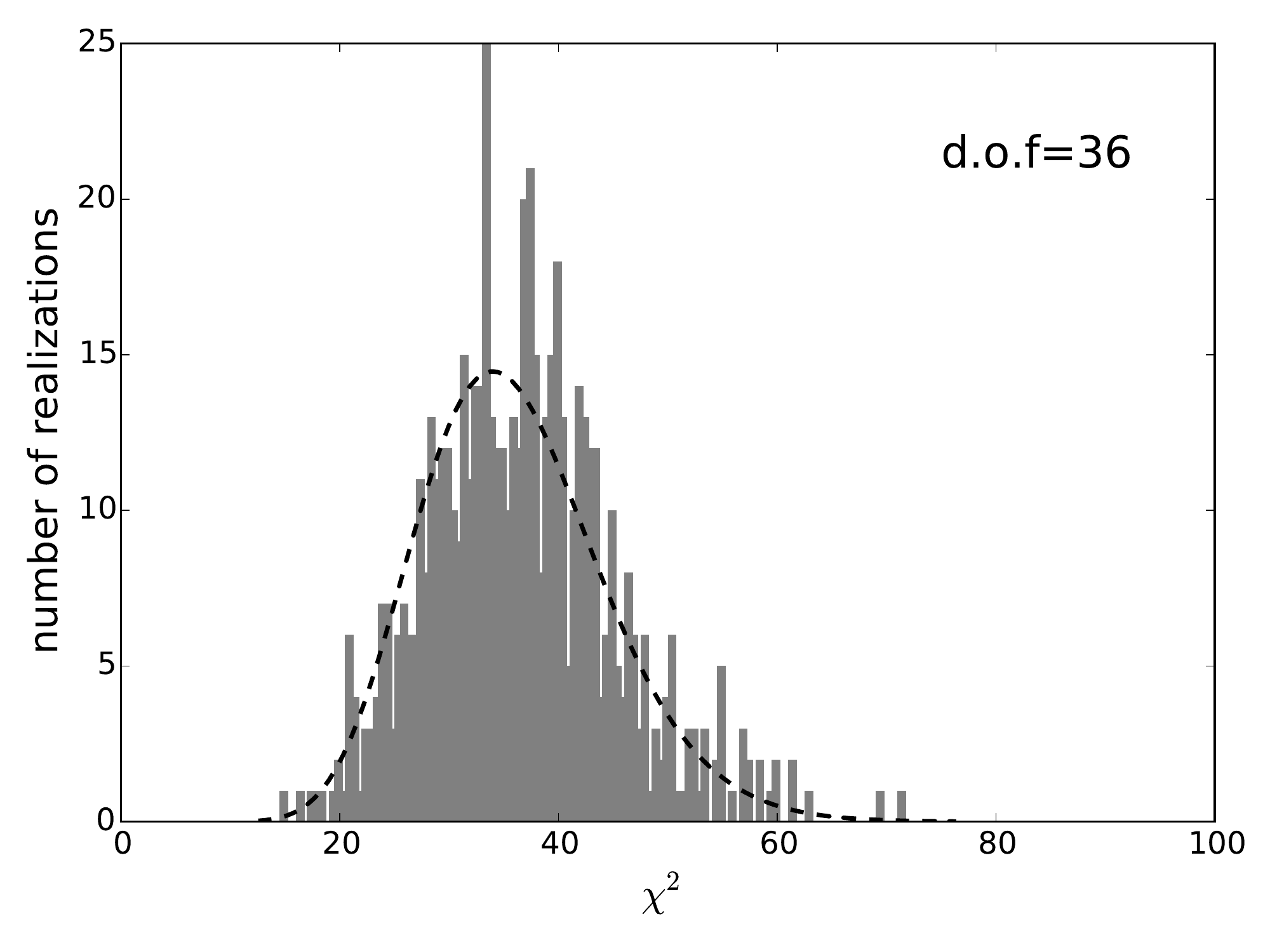}
\includegraphics[width=0.495\textwidth]{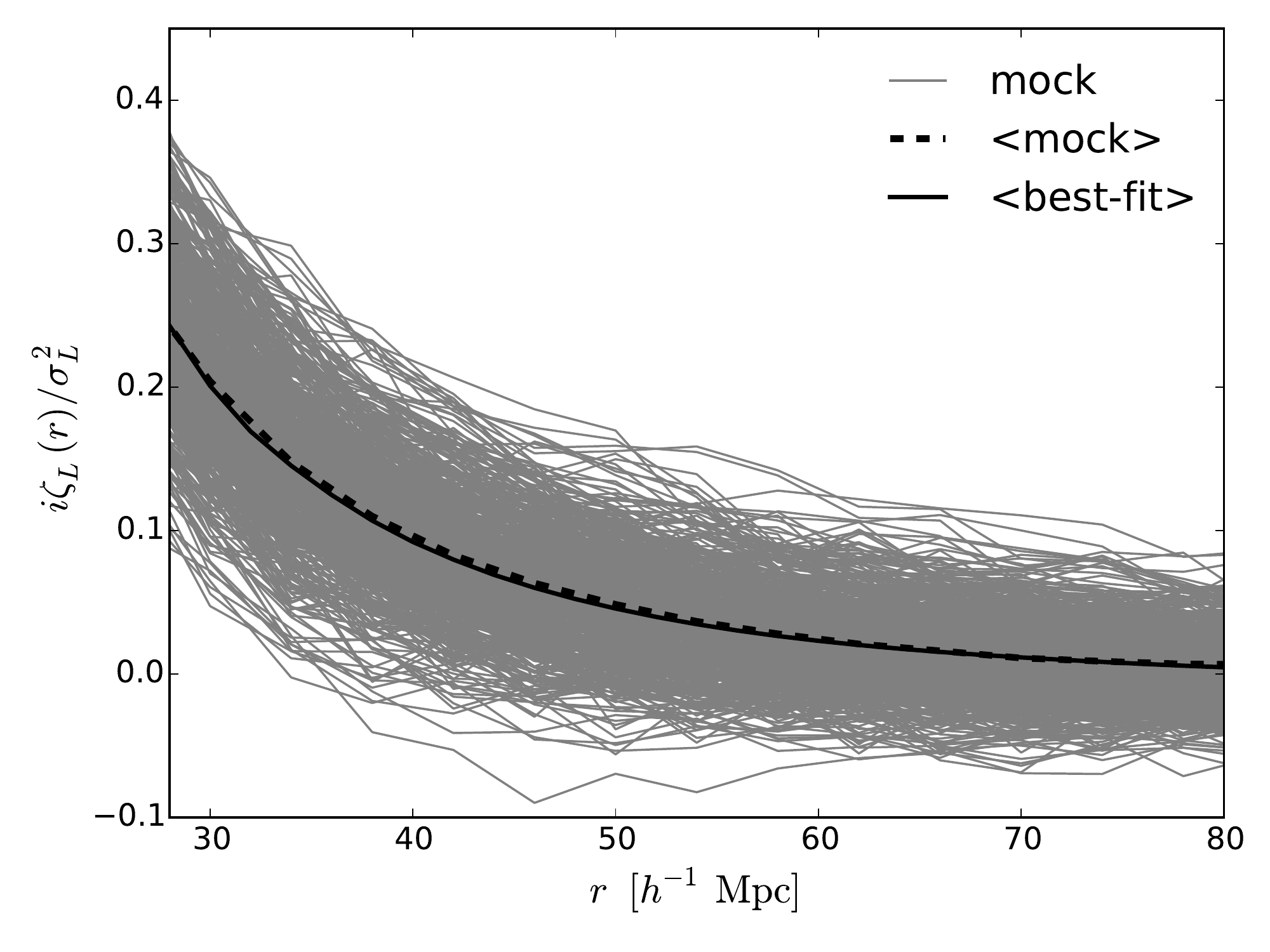}
\includegraphics[width=0.495\textwidth]{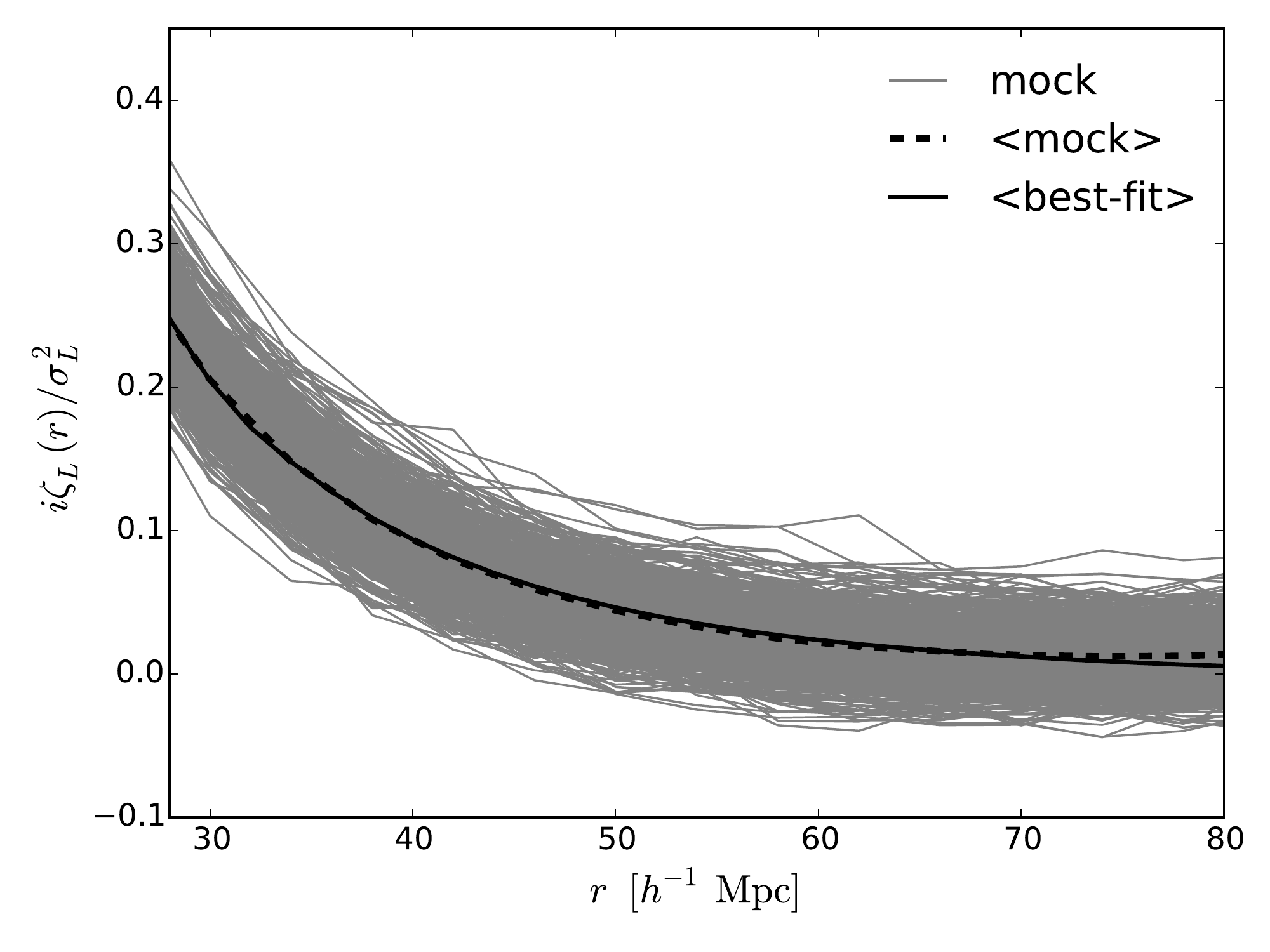}
\caption[Measurements of $\xi(r)$ and $\iz_L(r)/\sigma_L^2$ of PTHalos mock
catalogs in redshift space]
{Same as figure~\ref{fig:ch5_mock_r} but in redshift space.}
\label{fig:ch5_mock_z}
\end{figure}

Since there is no baryonic acoustic oscillation feature on the scales we are
interested in, we model the redshift-space two-point correlation function as
\be
 \xi_{g,z}(r)=b_1^2\left[\xi_{l,\sigma_v}(r)+A_{\rm MC}\xi_{\rm MC}(r)\right]K ~,
\label{eq:ch5_xiz_model}
\ee
where $\xi_{l,\sigma_v}(r)$ and $\xi_{\rm MC}(r)$ are given in \refeq{ch5_xi_model_2} and
\be
 K\equiv 1+\frac23\beta+\frac15\beta^2\,,
\label{eq:kaiser}
\ee
is the Kaiser factor with $\beta\equiv f/b_1$ \cite{kaiser:1987}.
As we do not include the subdominant term proportional to $b_2$ in the two-point
function, it only gives constraint on $b_1$, which we can then use to break the
degeneracy with $b_2$ in the integrated three-point function. We find that this
simple modeling yields unbiased $b_1$ and fulfills the demand.
We calculate the redshift-space integrated three-point function by first evaluating
the integrated bispectrum using SPT at the tree level (the explicit expression of
the redshift-space bispectrum is given in \refapp{trz_kernel}), compute its one-dimensional
Fourier transform, as \refeq{ch5_iz_ib_ang_avg}, and then correct for the boundary
effect. The $\sigma_L^2$ of the mocks in redshift space agrees with $b_1^2 K\,\sigma_{L,l}^2$
to percent level. The redshift-space models thus contain, as before in real space,
the three fitting parameters, $b_1$, $b_2$, and $A_{\rm MC}$. We then simultaneously
fit $\xi(r)$ and $\iz_L(r)/\sigma_L^2$ of both subvolumes by minimizing \refeq{ch5_chi2}.
\refFig{ch5_corr} shows the correlation matrix ($C_{ij}$ in $\chi^2$, normalized by
$\sqrt{C_{ii}C_{jj}}$) estimated from the 600 mocks in redshift space. Because we
normalize the integrated three-point function by $\sigma_L^2$, the covariance between
$\iz_L(r)/\sigma_L^2$ and $\sigma_L^2$ is negligible. On the other hand, the covariances
between $\iz_L(r)/\sigma_L^2$ and $\xi(r)$, between $\xi(r)$, and between $\iz_L(r)/\sigma_L^2$
for two sizes of subvolumes are significant.

\begin{figure}[t!]
\centering
\includegraphics[width=0.5\textwidth]{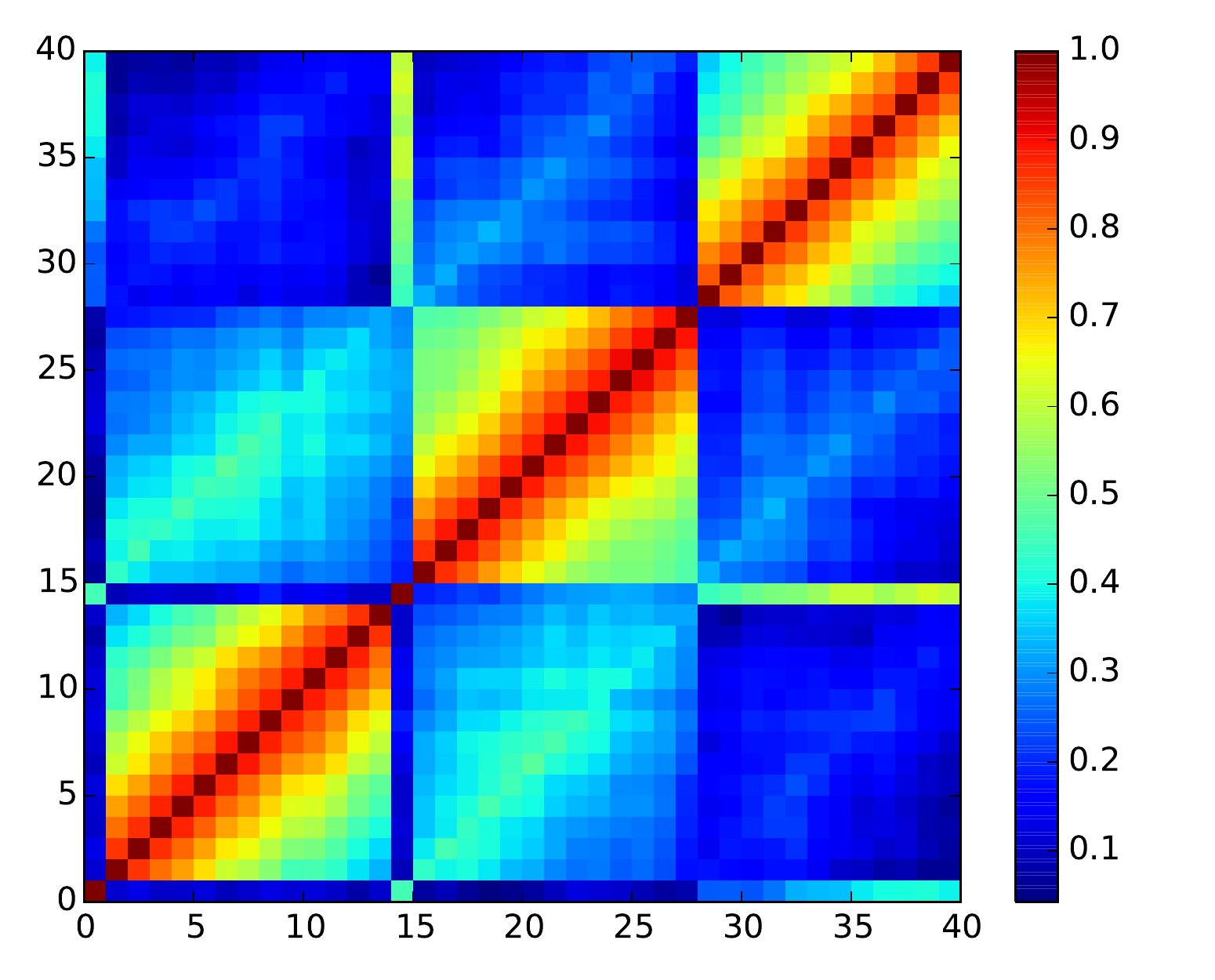}
\caption[Correlation matrix estimated from 600 mocks in redshift space]
{Correlation matrix estimated from 600 mocks in redshift space. The figure shows
$\sigma_L^2$ and $\iz_L(r)/\sigma_L^2$ of $220\hMpc$ subvolumes from bin 0 to 13,
$\sigma_L^2$ and $\iz_L(r)/\sigma_L^2$ of $120\hMpc$ subvolumes from bin 14 to 27,
and $\xi(r)$ from bin 28 to 40.}
\label{fig:ch5_corr}
\end{figure}

The models computed with the mean of the best-fitting parameters of 600 mocks are shown
as the thick solid lines in \reffig{ch5_mock_z}.
The best-fitting parameters are $b_1=1.931\pm0.077$, $b_2=0.54\pm0.35$,
and $A_{\rm MC}=1.37\pm0.82$. The agreement between the models and the
measurements in redshift space is as good as in real space.

Again, our $b_1$ is in good agreement with the results presented in figure 16
of \cite{gilmarin/etal:2014b}, whereas our $b_2$ is smaller than theirs, which
is $\simeq 0.75$, but still well within the 1-$\sigma$ uncertainty. As noted
in \refsec{ch5_mock_r}, the adopted models of the bispectrum are different.
In \refapp{ft_b2}, we show that using the effective $F_2$ and $G_2$ kernels
and the non-local tidal bias in the model increases the value of $b_2$. However,
the changes are within the uncertainties, and the goodness of the fit is similar
for different models. Thus, in this paper we shall primarily use the SPT at the
tree level with local bias for simpler interpretation of the three-point function,
but also report the results for the extended models.

\section{Measurement of the BOSS DR10 CMASS sample}
\label{sec:ch5_data}
We now present measurements of the position-dependent correlation function from
the BOSS DR10 CMASS sample\footnote{Catalogs of galaxies and the random samples
can be found in \url{http://www.sdss3.org}.} in NGC. The detailed description
of the observations can be found in \cite{ahn/etal:2014,anderson/etal:2014}.
Briefly, the sample contains 392,372 galaxies over 4,892 deg$^2$ in the redshift
range of $0.43<z<0.7$, which corresponds to the comoving volume of approximately
$2~h^{-3}~{\rm Gpc}^3$. We also weight the galaxies by $w_{\rm BOSS}$ to correct
for the observational systematics. We follow \refsec{ch5_division} to divide the
observations into subvolumes. However, the observations have their own set of
random samples, which are different from the ones of the mocks (the random samples
of the mocks have slightly higher $\bar n$ and different $\bar n(z)$), so we adjust
the redshift cuts to be $z_{\rm cut}=0.5108$, 0.5717 and $z_{\rm cut}=0.48710$,
0.52235, 0.55825, 0.60435 for the two resolutions, respectively. The resulting
properties of subvolumes of the observations and mocks are similar.

The mocks are constructed to match the two-point function of the observed galaxies,
but not for the three-point function. Hence there is no guarantee that the three-point
function of mocks agrees with the observations. We can test this using our measurements.

\begin{figure}[t!]
\centering
\includegraphics[width=0.495\textwidth]{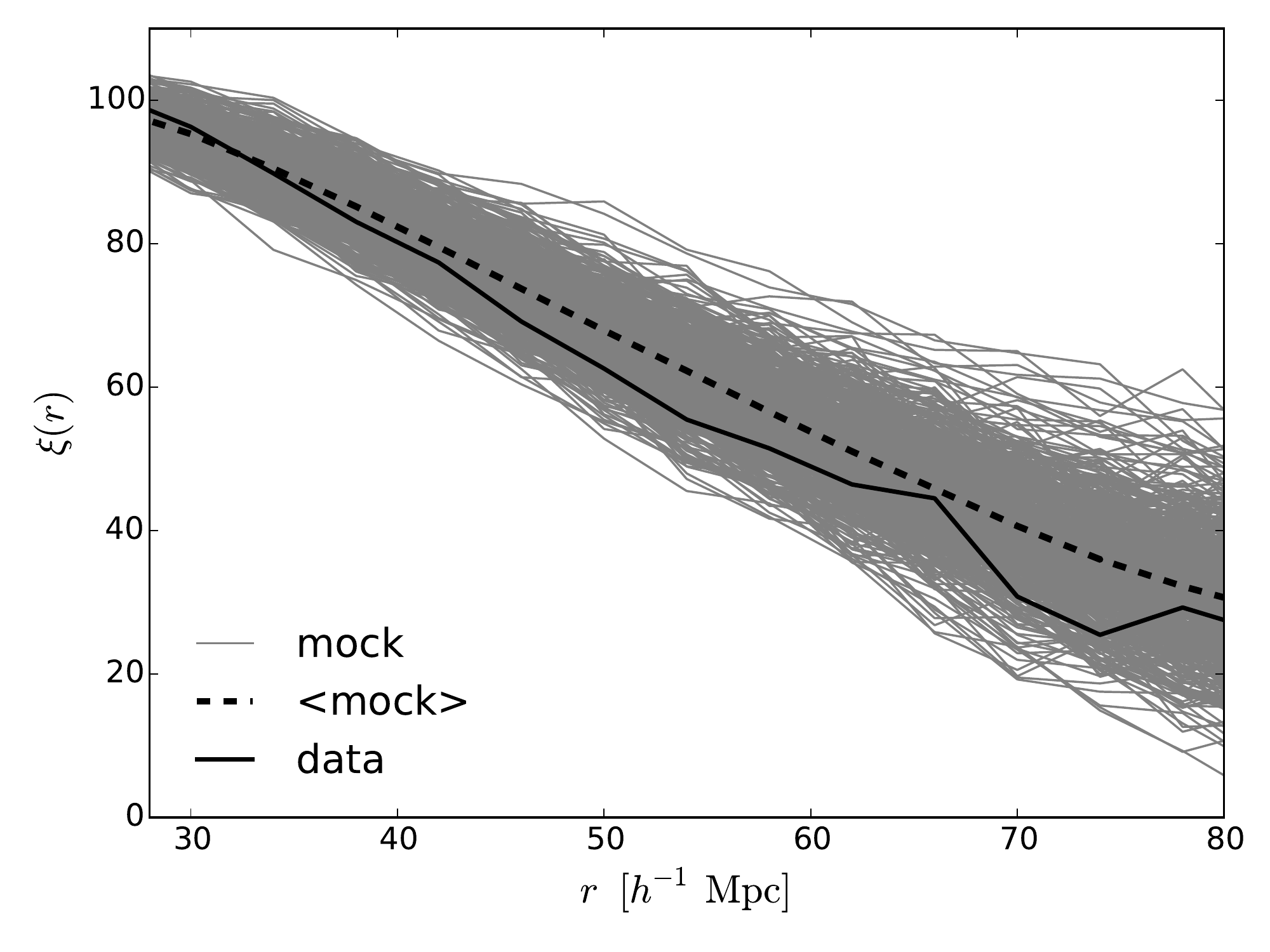}
\includegraphics[width=0.495\textwidth]{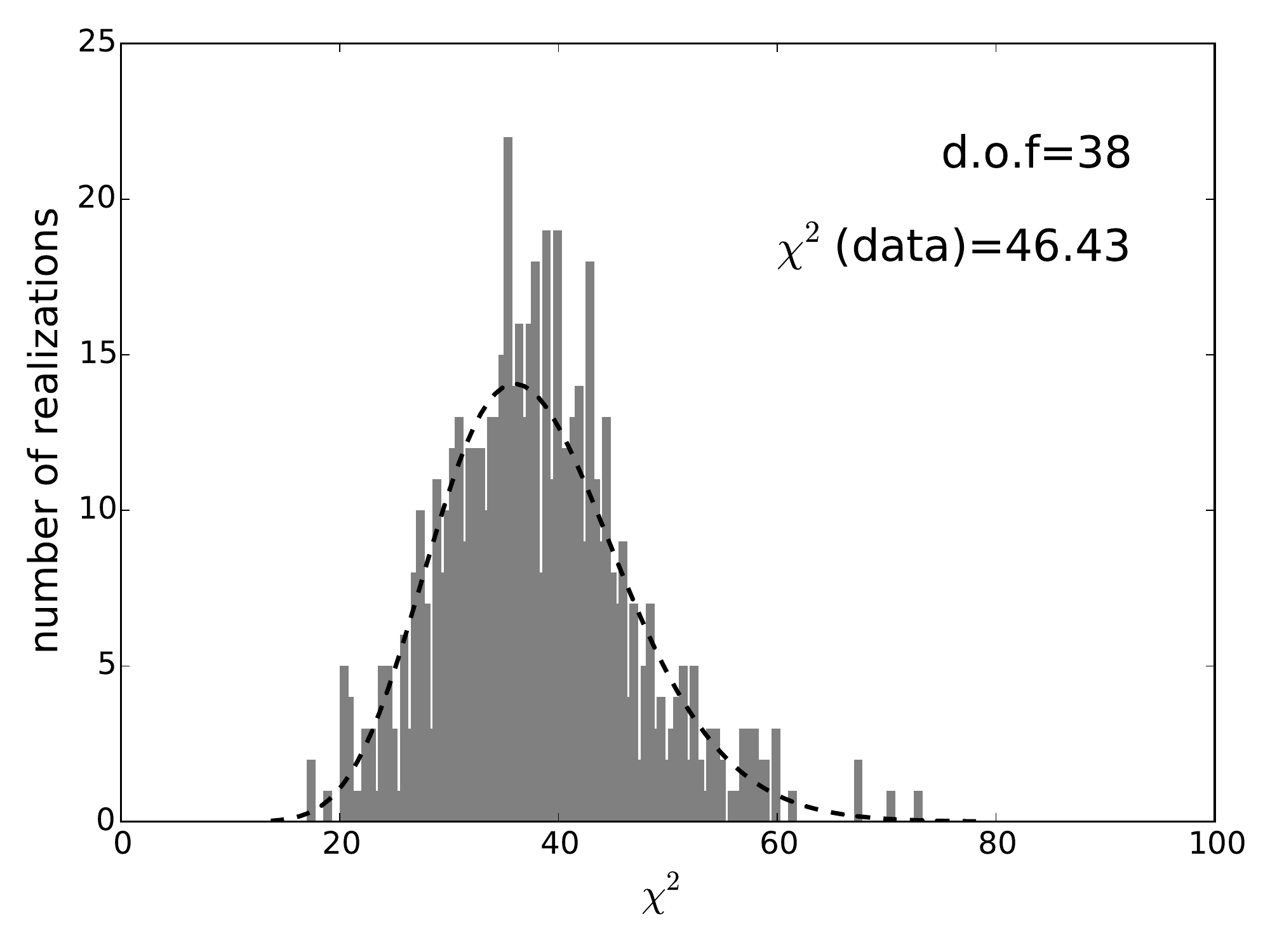}
\includegraphics[width=0.495\textwidth]{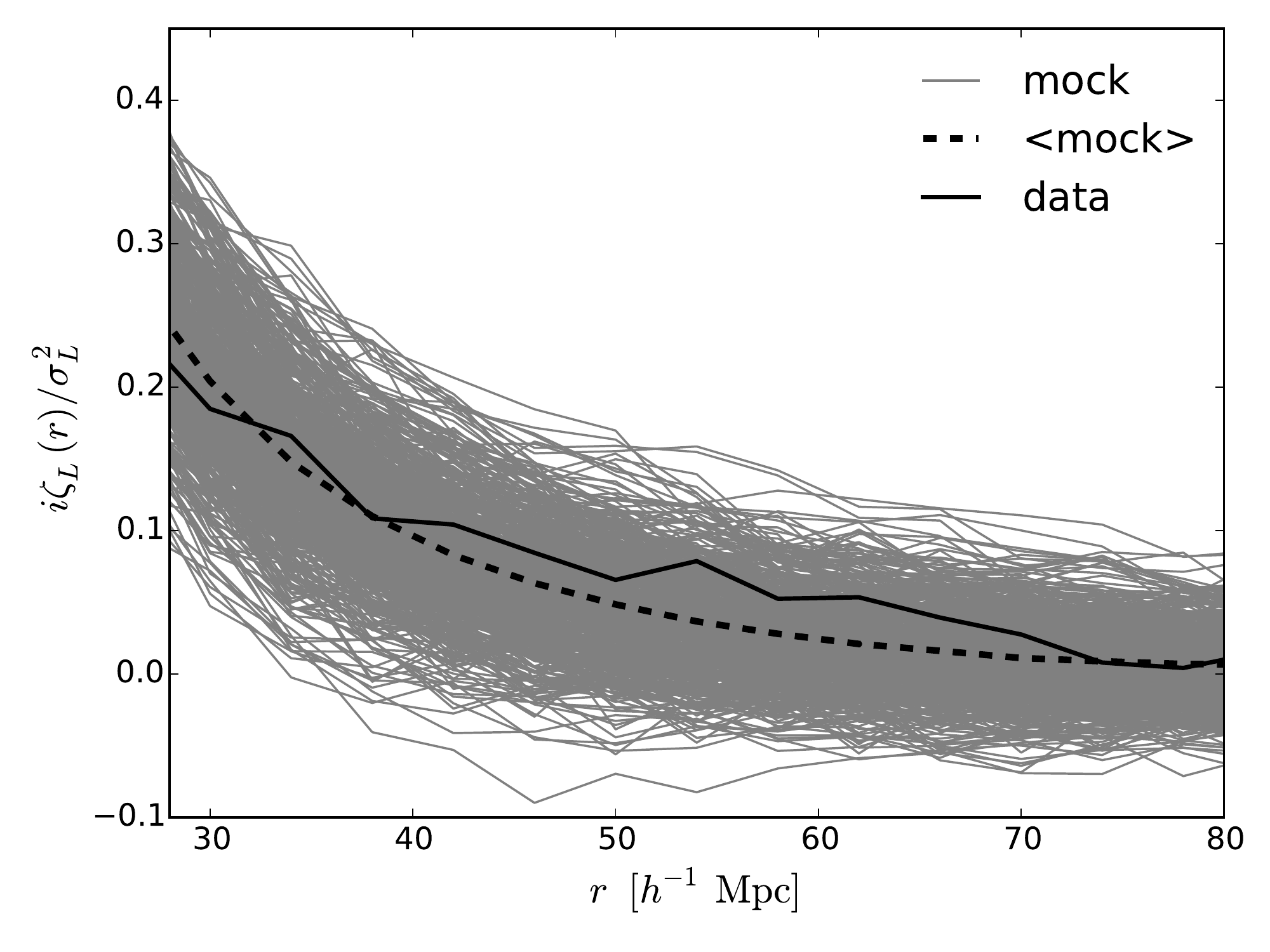}
\includegraphics[width=0.495\textwidth]{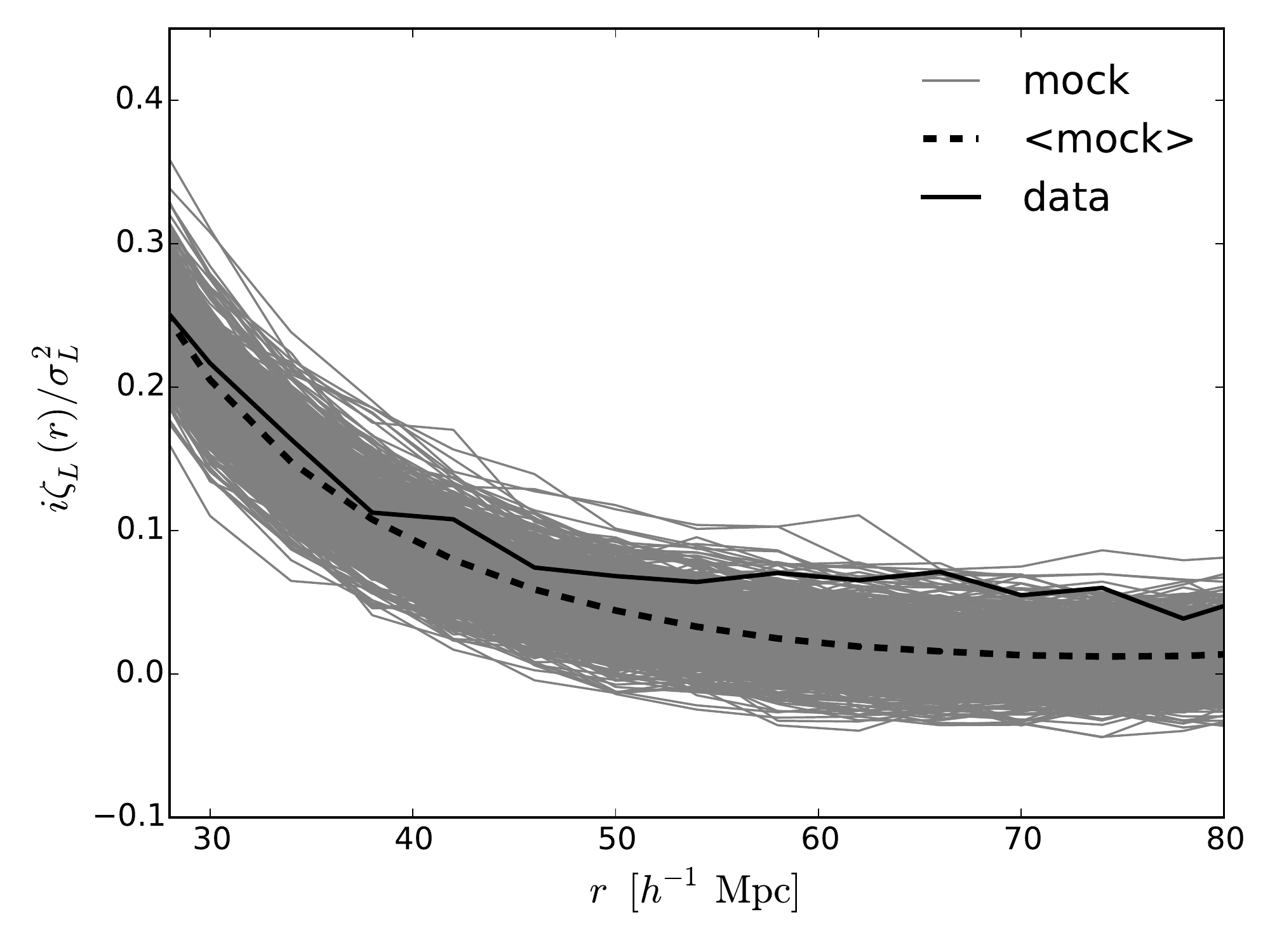}
\caption[Measurements of $\xi(r)$ and $\iz_L(r)/\sigma_L^2$ of BOSS DR10 CMASS sample]
{Measurements of the BOSS DR10 CMASS sample (black solid lines). The gray lines
show individual mocks in  redshift space and the dashed line shows the mean of
mocks. (Top left) $\xi(r)$, (Bottom left) $\iz_L(r)/\sigma_L^2$ for $220\hMpc$
subvolumes, and (Bottom right) $\iz_L(r)/\sigma_L^2$ for $120\hMpc$ subvolumes.
(Top right) $\chi^2$-histogram of the 600 mocks jointly fitting the three amplitudes
to $\xi(r)$ and $\iz_L(r)/\sigma_L^2$ in redshift space. The dashed line shows
the $\chi^2$-distribution with d.o.f.=38. The $\chi^2$ value measured from the
BOSS DR10 CMASS sample is 49.3.}
\label{fig:ch5_data}
\end{figure}

\begin{table}[t]
\centering
\begin{tabular}{ | c | c c c | }
\hline
& avg$[\sigma_{L,\rm mock}^2]$ & var$[\sigma_{L,\rm mock}^2]$ & $\sigma_{L,\rm data}^2$ \\
\hline
 $220\hMpc$ & $4.6\times10^{-3}$ & $5.6\times10^{-4}$ & $4.9\times10^{-3}$ \\
 $120\hMpc$ & $2.4\times10^{-2}$ & $1.3\times10^{-3}$ & $2.5\times10^{-2}$ \\
\hline
\end{tabular}
\caption[Measurements of $\sigma_L^2$ of the mock catalogs and the BOSS DR10 CMASS sample]
{Measurements of $\sigma_L^2$ of the mock catalogs and the BOSS DR10 CMASS sample.}
\label{tab:ch5_sigma2}
\end{table}

The measurements of $\xi(r)$ and $\iz_L(r)/\sigma_L^2$ from the observations are
shown as the solid lines in \reffig{ch5_data}; the measurements of $\sigma_L^2$
is summarized in \reftab{ch5_sigma2}. The measurements are consistent visually
with the mocks within the scatter of the mocks\footnote{These measurements of
$\iz_L(r)/\sigma_L^2$ are done for one effective redshift. We compare $\iz_L(r)/\sigma_L^2$
of the observations and mocks in different redshift bins in \refapp{zbin_iz},
finding that the observations and mocks are consistent at all redshift bins to
within the scatter of the mocks.}, and we shall quantify the goodness of fit
using $\chi^2$ statistics later.

To quantify statistical significance of the detection of $\iz_L(r)/\sigma_L^2$
and the goodness of fit, we use the mean of the mocks as the model (instead
of the model based on perturbation theory used in \refsec{ch5_mock_z}), and
fit only the amplitudes of $\iz_L(r)/\sigma_L^2$, $\xi(r)$, and $\sigma_L^2$
to the observations and the 600 mocks by minimizing \refeq{ch5_chi2}. Specifically,
we use $O_i(r) = A_i\,O_i^{\rm mock}(r)$ as the model, where $O_1(r)=\iz_L(r)/\sigma_L^2$,
$O_2(r)=\xi(r)$, and $O_3=\sigma_L^2$, with the amplitudes $A_1,\,A_2,\,A_3$.

\begin{table}
\centering
\begin{tabular}{ | c | c c c | }
\hline
& $A_1$ & $A_2$ & $A_3$ \\
\hline
1-$\sigma$ error & 0.12 & 0.03 & 0.04 \\
best-fit (DR10) & 0.89 & 1.02 & 1.08 \\
\hline
\end{tabular}
~~~~~~~
\begin{tabular}{ | c | c c c | }
\hline
& $(A_1,A_2)$ & $(A_1,A_3)$ & $(A_2,A_3)$ \\
\hline
corr & 0.34 & 0.09 & 0.36 \\
\hline
\end{tabular}
\caption[Fitted amplitudes of $\iz_L(r)/\sigma_L^2$, $\xi(r)$, and $\sigma_L^2$
of BOSS DR10 CMASS sample]
{Results of fitting the amplitudes: $A_1$ is $\iz_L(r)/\sigma_L^2$, $A_2$ is
$\xi(r)$, and $A_3$ is $\sigma_L^2$. (Left) The 1-$\sigma$ uncertainties of
the amplitudes estimated from the mocks, and the best-fitting amplitudes of
BOSS DR10 CMASS sample with respect to the mean of the mocks. (Right) The
correlation coefficients of the amplitudes.}
\label{tab:ch5_fit_amp}
\end{table}

\refTab{ch5_fit_amp} summarizes the fitted amplitudes. The 1-$\sigma$ uncertainties
and the correlations are estimated from the 600 mocks. Since we normalize $\iz_L(r)$
by $\sigma_L^2$, the correlation between $A_1$ and $A_3$ is small. On the other hand,
$A_2$ and $A_3$ are correlated significantly because $\sigma_L^2$ is an integral of
the two-point function [\refeq{ch2_sigmalL2}].

Comparing the BOSS DR10 CMASS sample to the mean of the mocks, we find that
$\iz_L(r)/\sigma_L^2$ is 1-$\sigma$ lower, $\xi(r)$ is unbiased (by construction
of the mocks), and $\sigma_L^2$ is 2-$\sigma$ higher. The result of $A_1$ for the
data is driven by the correlation between different separations of $\iz_L(r)/\sigma_L^2$.
On the other hand, the result of $A_3$ is driven by the positive correlation between
$\xi(r)$ and $\sigma_L^2$. While $\sigma_L^2$ of the data for two subvolumes are
larger than that of the mocks but still at the boundary of the variances (see
\reftab{ch5_sigma2}), it requires an even higher $A_3$ to minimize $\chi^2$ when
we jointly fit the three amplitudes. The fact that $A_3$ is larger than $A_2$ is
also possibly due the contributions to $\sigma_L^2$ from small separations (including
stochasticity at zero separations), where the mocks were not optimized. We find
$A_1=0.89\pm 0.12$, i.e., a 7.4$\sigma$ detection of the integrated three-point
function of the BOSS DR10 CMASS sample.

In order to assess the goodness of fit, we use the distribution of $\chi^2$,
a histogram of which is shown in the top right panel of \reffig{ch5_data}.
In total there are 41 fitting points (13 fitting points for $\xi(r)$ and two
sizes of subvolumes for $\iz_L(r)/\sigma_L^2$, and two fitting points for
$\sigma_L^2$) with three fitting parameters, so d.o.f.=38. The $\chi^2$ value
of the observations is 49.3, and the probability to exceed this $\chi^2$ value
is more than 10\%. Given the fact that the mocks are constructed to match only
the two-point function of the observations, this level of agreement for both
the two-point and integrated three-point correlation functions is satisfactory.

\section{Interpretation for the measurement of the integrated three-point function}
\label{sec:ch5_interpretation}
What can we learn from the measured $\iz_L(r)/\sigma_L^2$? In \refsec{ch5_mock_z},
we show that the prediction for $\iz_L(r)/\sigma_L^2$ based on SPT at the tree-level
in redshift space provides an adequate fit to the mocks to within the scatter of
the mocks; thus, we can use this prediction to infer cosmology from $\iz_L(r)/\sigma_L^2$.
Note that any unmodeled effects in the integrated three-point function such as
nonlinearities of the matter density, nonlocal bias parameters, and redshift-space
distortions beyond the Kaiser factor, will tend to bias our measurement of cosmological
parameters based on $\iz_L(r)$. We will discuss caveats at the end of this section.

Since the linear two-point and the tree-level three-point functions are proportional
to $\se^2$ and $\se^4$, respectively, and $\sigma_L^2$ is proportional to $\se^2$,
the scaling of the redshift-space correlation functions is
\ba
 \xi_{g,z}(r)\:&=b_1^2K\left[\xi^{\rm fid}_{l,\sigma_v}(r)\left(\frac{\se}{\sef}\right)^2
 +A_{\rm MC}\xi^{\rm fid}_{\rm MC}(r)\left(\frac{\se}{\sef}\right)^4\right] ~, \vs
 \frac{\iz_{L,g,z}(r)}{\sigma_L^2}\:&=
\frac{\iz^{\rm fid}_{L,g,z}(r)}{b_1^2\sigma_{L,l}^2K}\left(\frac{\se}{\sef}\right)^2\frac1{f_{L,\rm bndry}(r)} ~,
\label{eq:ch5_obs_zspace}
\ea
where ``fid'' denotes the quantities computed with the fiducial value of $\se$.
Note that $\xi_{\rm MC}(r)$ is proportional to $\se^4$ because it is an integral
of two linear power spectra (see \refeq{ch5_xi_model}). Since $\xi_{l,\sigma_v}(r)$
dominates the signal, the parameter combinations $b_1\se$ and $K=1+2\beta/3+\beta^2/5$
are degenerate in the two-point  function. That is, the amplitude of the two-point
function measures only $(b_1\se)^2+\frac{2}{3}(b_1\se)(f\se)+\frac{1}{5}(f\se)^2$.
This degeneracy can be lifted by including the quadrupole of the two-point function
in redshift space. See \cite{samushia/etal:2014,tojeiro/etal:2014,sanchez/etal:2014,beutler/etal:2014}
for the latest measurements using the BOSS DR11 sample.

As for the three-point function, \reffig{ch2_iz_norm} shows that the $b_1^3$ and
$b_1^2 b_2$ terms are comparable for $b_1\approx b_2$. This means that, at the
three-point function level, the nonlinear bias appears in the leading order,
so the amplitude of the three-point function measures a linear combination of
$b_1$ and $b_2$. This provides a wonderful opportunity to determine $b_2$. The
challenge is to break the degeneracy between $b_2$, $b_1$, $f$, and $\se$. For
this purpose, we combine our results with the two-point function in redshift
space and the weak lensing measurements of BOSS galaxies. We take the constraints
on $b_1\se(z=0.57)=1.29\pm0.03$ and $f(z=0.57)\se(z=0.57)=0.441\pm0.043$ from
table~2 of \cite{samushia/etal:2014}. To further break the degeneracy between $b_1$,
$f$, and $\se$, we take the constraint on  $\se=0.785\pm0.044$ from \cite{miyatake/etal:2013,more/etal:2014},
where they jointly analyze the clustering and the galaxy-galaxy lensing using the
BOSS DR11 CMASS sample and the shape catalog from Canada France Hawaii Telescope
Legacy Survey.

We assume Gaussian priors on $b_1\se$, $f\se$, and $\se$ with the known covariance
between $b_1\se$ and $f\se$. The cross-correlation coefficient between $b_1\se$
and $f\se$ is $-0.59$, as shown in figure~6 of \cite{samushia/etal:2014}.
We then run the Markov Chain Monte Carlo with the Metropolis-Hastings algorithm
to fit he model \refeq{ch5_obs_zspace} to the observed $\iz(r)/\sigma_L^2$. We
find $b_2=0.41\pm0.41$, and the results for the extended models are summarized
in \reftab{ch5_b2_data}.

\begin{table}
\centering
\begin{tabular}{ | c | c c c c | }
\hline
 & baseline & eff kernel & tidal bias & both \\
\hline
 $b_2$ & $0.41\pm0.41$ & $0.51\pm0.41$ & $0.48\pm0.41$ & $0.60\pm0.41$ \\
\hline
\end{tabular}
\caption[Best-fitting $b_2$ and their uncertainties for BOSS DR10 CMASS
sample for the extended models]
{Best-fitting $b_2$ and their uncertainties for BOSS DR10 CMASS sample for
the extended models. The detailed description of the extended models is in
\refapp{ft_b2}.}
\label{tab:ch5_b2_data}
\end{table}

The value of $b_2$ we find is lower than the mean of the mocks, $b_2^{\rm mock}=0.54\pm0.35$.
The difference is mainly due to two reasons. First, the amplitude of the integrated
three-point function of the observations is lower than that of the mocks by 10\%
($A_1=0.89\pm0.12$). Second, the priors from the correlation function and lensing
constraint $b_1$ to be close to 2.18, which is larger than that of the mocks,
$b_1^{\rm mock}=1.93$. Thus, it requires a smaller $b_2$ to fit the three-point
function. The argument is similar for the extended models. Note, however, that
the nonlinear bias of the data is still statistically consistent with the mocks.

Let us conclude this section by listing three caveats regarding our cosmological
interpretation of the measured integrated three-point function.
\begin{enumerate}
\item The models we use, \refeq{ch5_obs_zspace}, are based on tree-level perturbation
		theory, the lowest order redshift-space distortion treatment, as well as
		on the local bias parametrization. While this simple model describes the
		mocks well, as shown in \refsec{ch5_mock_r} and \ref{sec:ch5_mock_z}, we
		discuss in \refapp{ft_b2} that using the effective $F_2$ and $G_2$ and the
		non-local tidal bias brings $b_2$ closer to that of \cite{gilmarin/etal:2014b}.
		We, however, find similar goodness of fit for various models, and thus
		we cannot distinguish between these models.

\item Covariances between the integrated three-point function, monopole and quadrupole
		two-point function, and weak lensing signals are ignored. This can and should
		be improved by performing a joint fit to all the observables.

\item The cosmology is fixed throughout the analysis, except for $f$ and $\se$.
		In principle, marginalizing over the cosmological parameters is necessary
		to obtain self-consistent results, although the normalized integrated three-point
		function is not sensitive to cosmological parameters such as $\Omega_m$ as
		shown in \reffig{ch4_cosmo_dep}.
\end{enumerate}
These caveats need to be addressed in the future work.

\section{Discussion and conclusion}
\label{sec:ch5_conclusion}
In this chapter, we have reported on the first measurement of the three-point function
with the position-dependent correlation function from the SDSS-III BOSS DR10 CMASS
sample. The correlation between the position-dependent correlation function measured
within subvolumes and the mean overdensities of those subvolumes is robustly detected
at 7.4$\sigma$.

Both the position-dependent correlation function and the mean overdensity are easier
to measure than the three-point function. The computational expense for the two-point
function is much cheaper than the three-point function estimator using the triplet-counting
method. In addition, for a fixed size of the subvolume, the integrated three-point
function depends only on one variable (i.e., separation), unlike the full three-point
function which depends on three separations. This property allows for a useful compression
of information in the three-point function in the squeezed configurations, and makes
physical sense because the integrated three-point function measures how the small-scale
two-point function, which depends only on the separation, responds to a long-wavelength
fluctuation, as introduced in \refchp{ch4_posdeppk}. As there are only a small number of
measurement bins, the covariance matrix of the integrated three-point function is easier
to estimate than that of the full three-point function from a realistic number of mocks.
We have demonstrated this advantage in this chapter.

Of course, since this technique measures the three-point function with one long-wavelength
mode (mean overdensity in the subvolumes) and two relatively small-wavelength modes
(position-dependent correlation function), it is not very sensitive to the three-point
function of other configurations, which were explored in \cite{gilmarin/etal:2014b}.

We have used the mock galaxy catalogs, which are constructed to match the two-point
function of the SDSS-III BOSS DR10 CMASS sample in redshift space, to validate our
method and theoretical model. We show that in both real and redshift space, the
integrated three-point function of the mocks can be well described by the tree-level
SPT model. However, the nonlinear bias which we obtain from the mocks is higher than
that reported in \cite{gilmarin/etal:2014b}. This is possibly due to the differences
in the scales and configurations of the three-point function used for the analyses.
As discussed in \refsec{ch5_interpretation}, any unmodeled nonlinear effects in the
redshift-space integrated three-point function of CMASS galaxies will tend to bias
$b_2$, and will bias this parameter differently than the measurement of \cite{gilmarin/etal:2014b}.

Taking the mean of the mocks as the model, and treating the amplitudes of two-
and three-point functions as free parameters, we find  the best-fit amplitudes
of $\iz(r)/\sigma_L^2$, $\xi(r)$, and $\sigma_L^2$ of the CMASS sample. With
respect to the mean of the mocks, the observations show a somewhat smaller
$\iz(r)/\sigma_L^2$ ($A_1=0.89\pm 0.12$) and larger $\sigma_L^2$, while the
ensemble two-point function $\xi(r)$ matches the mocks. Given that the mocks
are generated to match specifically the two-point function of the BOSS DR10
CMASS sample within a certain range of separations, the level of agreement
between the observations and mocks is satisfactory.

Finally, by combining the integrated three-point function and the constraints
from the anisotropic clustering  ($b_1\se$ and $f\se$ in \cite{samushia/etal:2014})
and from the weak lensing measurements ($\se$ in \cite{more/etal:2014}), we break
the degeneracy between $b_1$, $b_2$, $f$, and $\sigma_8$. We find $b_2=0.41\pm0.41$
for the BOSS DR10 CMASS sample. The caveat of this result is that our model,
\refeq{ch5_obs_zspace}, relies on a rather simple model in redshift space as
well as on the local bias parametrization. We leave the extension of the model
to improved bias and redshift-space distortion modeling (especially in light of
the comparison with the results in \cite{gilmarin/etal:2014b}) for future work.

In summary, we have demonstrated that the integrated three-point function is
a new observable which can be measured straightforwardly from galaxy surveys
using basically the existing and routinely applied machinery to compute the
two-point function, and has the potential to yield a useful constraint on the
quadratic nonlinear bias parameter. Moreover, since the integrated three-point
function is most sensitive to the bispectrum in the squeezed configurations,
it is sensitive to primordial non-Gaussianity of the local type (see \refsec{ch2_ib_models}
and \refsec{ch4_fisher} for more details), thereby offering a probe of the
physics of inflation. We plan to extend this work to search for the signature
of primordial non-Gaussianity in the full BOSS galaxy sample.

}

%% file: kap_06.tex
{

\renewcommand{\v}[1]{\mathbf{#1}}
\newcommand{\vr}{\v{r}}
\newcommand{\vx}{\v{x}}
\newcommand{\vk}{\v{k}}
\newcommand{\vq}{\v{q}}

\renewcommand{\d}{\delta}
\newcommand{\dL}{\delta_L}
\newcommand{\drho}{\delta_{\rho}}
\newcommand{\da}{\delta_a}
\newcommand{\dH}{\delta_H}
\newcommand{\ta}{\tilde a}
\newcommand{\tP}{\tilde P}
\newcommand{\tD}{\tilde D}
\newcommand{\tH}{\tilde H}
\newcommand{\tO}{\tilde\Omega}
\newcommand{\tK}{\tilde K}
\newcommand{\cS}{\mathcal{S}}
\newcommand{\cO}{\mathcal{O}}
\newcommand{\hD}{\hat D}
\newcommand{\hk}{\hat k}
\newcommand{\rhob}{\bar{\rho}}
\newcommand{\hMpc}{~h~{\rm Mpc}^{-1}}
\newcommand{\ihMpc}{~h^{-1}~{\rm Mpc}}
\newcommand{\tou}{t_{\rm out}}
\newcommand{\Om}{\Omega_m}
\renewcommand{\[}{\left[}
\renewcommand{\]}{\right]}
\renewcommand{\(}{\left(}
\renewcommand{\)}{\right)}
\newcommand{\<}{\left\langle}
\renewcommand{\>}{\right\rangle}
\renewcommand{\{}{\left\lbrace}
\renewcommand{\}}{\right\rbrace}

\chapter{The angle-averaged squeezed limit of nonlinear matter $N$-point functions and separate universe simulations}
\label{chp:ch6_npt_sq_sep}
In the previous chapters, we have introduced the position-dependent two-point
statistics and shown the measurements from $N$-body simulations and observations.
In this chapter, we shall generalize the study to the angle-averaged squeezed
limit of nonlinear matter $N$-point functions, and demonstrate how to use the
``separate universe simulations'' to address this issue. Specifically, we consider
the case of the squeezed limit such that there is a hierarchy between two large
wavenumbers $\vk$ and $\vk'$, and $N-2$ small wavenumbers $\vk_1,\cdots,\vk_n$.
The configuration is sketched in \reffig{ch6_squ_conf}.

The squeezed limit of matter $N$-point functions has been the subject of a large
body of work in the context of the so-called ``consistency relations'' for the
large-scale structure \cite{kehagias/riotto:2013,peloso/pietroni:2013,creminelli/etal:2013,
kehagias/etal:2014,valageas:2014a,valageas:2014,kehagias/perrier/riotto:2014,creminelli/etal:2014a,
creminelli/etal:2014b,nishimichi/valageas:2014,ben-dayan/etal:2015,horn/hui/xiao:2015,
nishimichi/valageas:2015}. The contributions to $N$-point functions in the squeezed
limit are ordered by the ratio of wavenumbers $k_i/k$, which is assumed to be much
less than one. The lowest order contributions, up $\propto (k_i/k)^{-1}$ when the
$N$-point function is written in terms of the overdensity $\d$, are fixed by the
requirement that a uniform potential perturbation as well as a uniform velocity
(boost) do not lead to any locally observable effect on the density field, as demanded
by the equivalence principle \cite{kehagias/riotto:2013,peloso/pietroni:2013,valageas:2014a,dai/pajer/schmidt:2015b}.
They are also referred to as ``kinematic contributions''.

The next order contribution, $\propto (k_i/k)^0$, is the lowest order at which a
\emph{physical} coupling of long- and short-wavelength modes happens. More precisely,
the contributions at this order correspond to the impact of a uniform long-wavelength
density or tidal perturbation. When considering equal-time $N$-point functions and
subhorizon perturbations $k_i \gg aH$, the kinematic contributions disappear, and
the physical $(k_i/k)^0$ contributions are the leading contribution to the $N$-point
function in the squeezed limit.

\begin{figure}[t!]
\centering
\includegraphics[width=0.8\textwidth,trim=0cm 19cm 0cm 0cm]{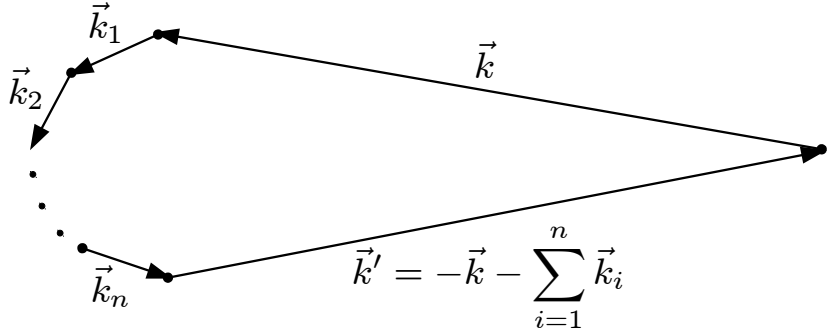}
\caption[Sketch of the squeezed-limit configuration of $N$-point functions]
{Sketch of the squeezed-limit configuration of $N$-point functions considered
in this chapter. $\vk_1,\cdots,\vk_n$ denote the long-wavelength modes which
are spherically averaged in \refeq{ch6_Sdef}, while $\vk$ and $\vk'$ denote
the small-scale modes which are allowed to be fully nonlinear.}
\label{fig:ch6_squ_conf}
\end{figure}

In this chapter, we disregard tidal fields, which leads us to first angle-average over
the $N-2$ small momenta (wavenumbers) in the $N$-point function. Specifically, we consider
$\cS_{N-2}$ defined through
\ba
 \cS_{N-2}(k, k'; k_1,\cdots,k_{N-2})\:&\equiv\int\frac{d^2\hk_1}{4\pi}
 \cdots\int\frac{d^2\hk_{N-2}}{4\pi}~
 \<\d(\vk)\d(\vk')\d(\vk_1)\cdots\d(\vk_{N-2})\>'_c ~
\label{eq:ch6_Sdef}
\ea
where $\hk_i$ are unit vectors and $\<\d(\vk_1)\cdots\d(\vk_{N})\>'_c$ denotes the
nonlinear connected matter $N$-point function with the momentum constraint
$(2\pi)^3\d_D(\vk_1+\cdots+\vk_N)$ dropped. Note that the momentum constraint fixes
$\vk'$ in terms of $\vk$ and $\vk_1,\dots,\vk_{N-2}$. We now let $k_1,\dots, k_{N-2}$
go to zero, and normalize the result by the nonlinear power spectrum $P(k)$ and the
linear power spectra $P_l(k_1)\cdots P_l(k_{N-2})$ to obtain a dimensionless quantity:
\be
 R_{N-2}(k)=\lim_{k_i\to 0}~
 \frac{\cS_{N-2}(k,k';k_1,\cdots,k_{N-2})}{P(k) P_l(k_1)\cdots P_l(k_{N-2})} ~.
\label{eq:ch6_Rnpoint}
\ee
Note that in this limit, spatial homogeneity enforces $\vk'=-\vk+\cO(k_i/k)$, so that
(for statistically isotropic initial conditions) the right-hand-sight of \refeq{ch6_Rnpoint}
depends only on $k$.

In \refapp{sq_npoint} (see also \cite{valageas:2014}), we show that $R_n(k)$ \emph{correspond
exactly to the power spectrum response functions}, which quantify the change in the nonlinear
matter power spectrum to an infinite-wavelength density perturbation. These response functions
are defined as the coefficients of the expansion of the power spectrum in the \emph{linearly
extrapolated initial overdensity} $\d_{L0}$:
\be
 P(k,t|\d_{L0})=\sum_{n=0}^{\infty}\frac1{n!}R_n(k,t)\[\d_{L0}\hD(t)\]^n P(k, t) ~,
\label{eq:ch6_Pkexpansion}
\ee
where $P(k,t|\d_{L0})$ is the nonlinear matter power spectrum at time $t$ in the presence
of a homogeneous (infinite-wavelength) density perturbation, and $\hD(t)$ is the linear
growth factor normalized to unity today. We have set $R_0(k,t)=1$ by definition. Thus, by
measuring $R_n$, we measure the angle-averaged squeezed limit (\refeq{ch6_Rnpoint}) of the
nonlinear matter $(n+2)$-point function. For $n=1$, the response $R_1$ describes the angle-averaged
squeezed-limit bispectrum, as discussed in \refchp{ch2_posdep_pk_xi} and \refchp{ch4_posdeppk}.

Independently of the derivation of \refeqs{ch6_Rnpoint}{ch6_Pkexpansion}, we also present
accurate measurements of $R_n$ for $n=1, 2, 3$ using $N$-body simulations which do not rely
on approximations in \refsec{ch6_sepuni_sim}. Specifically, we resort to $N$-body simulations
with an external homogeneous overdensity imposed via the \emph{separate universe} approach
described in \refchp{ch3_sepuni}. A flat FLRW universe with a homogeneous overdensity is
exactly equivalent to a different, curved FLRW universe, so that $N$-body simulations in
this modified cosmology provide, in principle, the exact result for the response functions
$R_n(k)$. This in turn corresponds to the exact (in the limit of infinite volume and resolution)
measurement of the squeezed-limit $N$-point function (\refeq{ch6_Rnpoint}). The measurements
of $R_1(k)$ are presented in \cite{li/hu/takada:2014}. We shall extend to $n=2$ and 3.

Many semi-analytical approaches to nonlinear large-scale structure assume that nonlinear
matter statistics can be described as a unique function of the linear matter power spectrum,
i.e. the power spectrum of initial fluctuations linearly extrapolated to a given time. In
the context of consistency relations, this approximation has been studied in e.g. \cite{valageas:2014,
kehagias/perrier/riotto:2014}. This ansatz is motivated by the fact that in Einstein-de
Sitter (flat matter-dominated universe), and to a very good approximation in $\Lambda$CDM,
the perturbation theory predictions factorize into powers of the linear growth factor and
convolutions of products of the initial matter power spectra and time-independent functions.
Another way to phrase this ansatz is that nonlinear large-scale structure only depends on
the normalization of the fluctuations at a given time, and not on the growth \emph{history}.

In the context of squeezed-limit $N$-point functions, this ansatz can be tested quantitatively
by comparing the outputs of separate universe simulations at a given time with simulations in
which the initial amplitude of fluctuations is rescaled to match the \emph{linear} power spectrum
at the same time. The difference between these ``rescaled initial amplitude'' simulations and
the separate universe simulations corresponds to the error made in the ansatz of assuming that
the linear power spectrum at a given time uniquely describes nonlinear large-scale structure
at the same time. For $n=1$, it is studied in \cite{li/hu/takada:2014b} and found that the two
simulations differ in the nonlinear regime. A closely related test using the matter bispectrum 
is shown in \cite{nishimichi/valageas:2014}. We shall study this comparison in more detail and
for $n=1, 2$ and 3 in \refsec{ch6_rescaled_sim}.

This chapter is organized as follows. In \refsec{ch6_pk_resp}, we develop semi-analytic
predictions for the power spectrum response. In \refsec{ch6_sepuni_sim}, we describe the
methodology of performing the separate universe simulations. In \refsec{ch6_results},
we compare the measurements from $N$-body simulations to the semi-analytic predictions.
In \refsec{ch6_rescaled_sim}, we compare the rescaled simulations to the separate universe
simulations. We conclude in \refsec{ch6_conclusion}. 

\section{Power spectrum response}
\label{sec:ch6_pk_resp}
We define the $n$-th order response function $R_n(k)$ of the power spectrum as the
$n$-th derivative of the power spectrum with respect to the linearly extrapolated
(or Lagrangian) overdensity $\dL$, normalized by the power spectrum. The definition,
consistent with \refeq{ch6_Pkexpansion}, is
\be
 R_n(k,t)=\frac{1}{P(k)}\frac{d^n P(k,t|\dL)}{d\[\dL(t)\]^n}\Bigg|_{\dL=0} ~,
\label{eq:ch6_def_response}
\ee
where $\dL(t)\equiv\delta_{L0}\hD(t)$. In the following, we will frequently suppress
the time argument for clarity. Analogously, one can define the power spectrum response
functions with respect to the fully evolved (or Eulerian) nonlinear overdensity $\drho$.
Since we can expand the nonlinear overdensity in powers of $\dL$ with known coefficients
via the spherical collapse (see \refchp{ch3_sepuni}), the $n$-th order Eulerian response
function is given by a sum of $R_m$ with $m\le n$. Motivated by the relation \refeq{ch6_Rnpoint},
we mainly consider the Lagrangian response functions. In the remainder of this section,
we develop semi-analytic models for the response functions based on the separate universe picture.

\subsection{Separate universe picture}
\label{sec:ch6_sepuni}
The idea of absorbing an infinitely-wavelength overdensity perturbation $\drho$ into the
background modified (curved) cosmology is extensively discussed in \refchp{ch3_sepuni},
and the response of the power spectrum to the overdensity is also discussed in
\refchp{ch4_posdeppk} (but restricted to the linear order). In this section, we shall
first summarize results, and generalize the study to higher order response, i.e. $n>1$.

Because of the overdensity, the expansion slows down, and the overdense region behaves
as a universe with positive curvature. The scale factor in the modified cosmology can
be written as
\be
 \ta(t)=a(t)\[1+\da(t)\] ~,
\label{eq:ch6_ta}
\ee
where the quantities in the modified cosmology are denoted with a tilde. Since the
\emph{physical} distance is the same in both cosmologies, \refeq{ch6_ta} implies the
change of the \emph{comving} distance to be
\be
 \tilde\vx=\frac{a(t)}{\ta(t)}\vx=\[1+\da(t)\]^{-1} ~.
\label{eq:ch6_tx}
\ee
Furthermore, due to mass conservation, the fractional difference in the scale factor
is related to the overdensity $\drho$ by
\be
 1+\drho(t)=\[1+\da(t)\]^{-3} ~.
\label{eq:ch6_drho}
\ee

Using the separate universe picture, we consider the matter power spectrum in this
patch just as that of a region with no homogeneous overdensity but properly modified
cosmology. The modification of the cosmology is such that the shape of the linear
power spectrum is unchanged, since the ratio of photon, baryon, and cold dark matter
densities is unmodified; moreover, the transfer function parameters are unchanged:
$\tO_m\tilde h^2=\Om h^2$ and $\tO_b\tilde h^2=\Omega_b h^2$.
Thus, only the growth of structure is affected.

The power spectrum that enters in the response given by \refeq{ch6_def_response} is
defined with respect to the background density and comoving coordinates of the fiducial
cosmology. Hence, the power spectrum calculated for the modified cosmology has to be
mapped to that with respect to the background density and comoving coordinates of the
fiducial cosmology. As discussed in \refchp{ch4_posdeppk}, this mapping yields the
``reference density'' and ``dilation'' contributions to the response. These can be
calculated exactly at any scale $k$ to any given order given the nonlinear matter
power spectrum in the \emph{fiducial} cosmology. That is, we do not need to run
$N$-body simulations to calculate these effects. They are thus merely ``projection
effects'', unlike the effect of the modified cosmology on the growth of structure,
which requires simulations in order to provide an accurate estimate.

Let us denote the power spectrum for the modified cosmology as $\tP(\tilde k)$.
Then, the reference density effect simply rescales the power spectrum as,
\be
 P(k)\stackrel{\rm ref.~density}{=}\(1+\drho\)^2 \tP(k) ~,
\ee
where the argument of $\tP(k)$ is not modified. The dilation effect due to the change
in the coordinates given by \refeq{ch6_tx} implies $k\to\tilde k=(1+\da)k$ and changes
the power spectrum by (see \refchp{ch4_posdeppk} for detailed derivation)
\be
 P(k)\stackrel{\rm dilation}{=}\(1+\da\)^3\tP(\[1+\da\]k) ~.
\ee
Putting the two together and using \refeq{ch6_drho} yields
\be
 P(k)=\(1+\drho\)\tP\([1+\da]k\) ~,
\label{eq:ch6_refdensity_dilation}
\ee
where all quantities are evaluated at some fixed time $t$. Note that one prefactor of
$(1+\drho)$ cancels, since the effect of the increased density is partially canceled
by the corresponding decrease in physical volume. As derived in \refchp{ch3_sepuni},
for an Einstein de-Sitter fiducial universe (and to high accuracy in $\Lambda$CDM)
$\da(t)$ and $\drho(t)$ have series solutions of the form
\be
 \da(t)=\sum_{n=1}^\infty e_n\[\d_{L0}\hD(t)\]^n ~,~~~
 \drho(t)=\sum_{n=1}^\infty f_n\[\d_{L0}\hD(t)\]^n ~,
\label{eq:ch6_seriessol}
\ee
where $\hD(t)=D(t)/D(t_0)$ is the fiducial growth factor normalized to one at the
epoch $t_0$ to which we extrapolate $\d_{L0}=\dL(t_0)$. Note that the values of
$e_n$ and $f_n$ are given in \refeq{ch3_en} and \refeq{ch3_fn}, respectively.

The third contribution to $R_n$ comes from the effect of the modified cosmology
on the growth of structure, which as mentioned above is the physical contribution
which requires $N$-body simulations for an accurate measurement. We thus define
a set of \emph{growth-only response functions} $G_n(k)$ which isolate the nontrivial
effect of the long-wavelength perturbation on the growth of small-scale structure,
\be
 G_n(k) \equiv\frac{1}{P(k)}\frac{d^n\tP(k)}{d\dL^n}\bigg|_{\d_L=0} ~.
\label{eq:ch6_def_Gn}
\ee
That is, $G_n$ are defined as $R_n$ without the contributions from the reference
density and dilation given by \refeq{ch6_refdensity_dilation}. This definition is
an extension of the similar decomposition for $n=1$ shown in \cite{sherwin/zaldarriaga:2012,
kehagias/perrier/riotto:2014,ben-dayan/etal:2015,li/hu/takada:2014b}. Thus, the
formula for the power spectrum (with respect to global coordinates) in the presence
of a long-wavelength overdensity is given by
\be
 P(k|\dL)=\(1+\drho\)\[\(1+\sum_{n=1}^\infty\frac1{n!}G_n(\tilde k)\dL^n\)P(\tilde k)\]_{\tilde k=(1+\da)k} ~.
\label{eq:ch6_PkdL}
\ee
Clearly, by the Leibniz rule, at any given order $n$ the total or ``full'' response
$R_n(k)$, \refeq{ch6_def_response}, is composed of the functions $G_m(k)$ and the
numbers $e_m,\,f_m$ with $1 \le m \le n$, where the $e_m$ multiply derivatives of
$G_l(k)$ and $P(k)$ with respect to $k$ (up to the $n^{\rm th}$ derivative).
Specifically, the first three full response functions are given by
\ba
\label{eq:ch6_R1}
 R_1(k) =\:&f_1+e_1\frac{k P'(k)}{P(k)}+G_1(k) ~, \\
\label{eq:ch6_R2}
 \frac{R_2(k)}{2}=\:&f_2+e_2\frac{k P'(k)}{P(k)}+e_1^2\frac{k^2 P''(k)}{2P(k)}
 +\frac{G_2(k)}{2}+f_1e_1\frac{k P'(k)}{P(k)} \vs
 &+f_1G_1(k)+e_1\frac{kP'(k)}{P(k)}G_1(k)+e_1kG_1'(k) ~, \\
\label{eq:ch6_R3}
 \frac{R_3(k)}{6}=\:&f_1G_1(k)e_1\frac{kP'(k)}{P(k)}+f_3+\frac{G_3(k)}{6}
 +e_3\frac{k P'(k)}{P(k)}+f_1\frac{G_2(k)}{2}+f_1e_2\frac{kP'(k)}{P(k)} \vs
 &+f_1e_1^2\frac{k^2 P''(k)}{2P(k)}+f_2G_1(k)+f_2e_1\frac{k P'(k)}{P(k)}
 +(f_1e_1+e_2)kG_1'(k)+e_1^2\frac{k^2G_1''(k)}{2} \vs
 &+e_1k\frac{G_2'(k)}{2}+e_1^2\frac{kP'(k)}{P(k)}kG_1'(k)
 +e_1^3\frac{k^3P'''(k)}{6 P(k)}+2e_1e_2\frac{k^2P''(k)}{2P(k)} \vs
 &+e_1\frac{kP'(k)}{P(k)}\frac{G_2(k)}{2}+G_1(k)\(e_2\frac{kP'(k)}{P(k)}
 +e_1^2\frac{k^2P''(k)}{2P(k)}\) ~,
\ea
where the primes denote derivatives with respect to $k$.

\subsection{Linear power spectrum predictions}
\label{sec:ch6_resp_ln}
We now evaluate \refeq{ch6_PkdL} for the simplest case, i.e., the response of
the linear matter power spectrum. In linear theory, the growth is scale-independent
and given by the linear growth factor. Thus, the growth-only response functions
are scale-independent and just described by the linear growth factor in the modified
cosmology $\tD(t)$,
\be
 G_n^{\rm linear} = \frac{1}{D^2}\frac{d^n(\tD^2)}{d\dL^n}\bigg|_{\dL=0} ~.
\label{eq:ch6_Gn_linear}
\ee
As for $\da(t)$ and $\drho(t)$, $\tD$ also has the perturbative expansion in powers
of $\dL$ as
\be
 \tD(t) = D(t)\{1+\sum_{n=1}^\infty g_n\[\d_{L0}\hD(t)\]^n\} ~.
\label{eq:ch6_Dtilde}
\ee
Thus, for an Einstein-de Sitter fiducial universe (and to high accuracy in $\Lambda$CDM),
the linear response functions are simply constants. Inserting the values of $g_n$ from
\refeq{ch3_gn}, we obtain
\be
 \{G_n^{\rm linear}\}_{n=1,\cdots,4}=
 \{\frac{26}{21}~,~\frac{3002}{1323}~,~\frac{240272}{43659}~,~
 \frac{197919160}{11918907}\} ~.
\label{eq:ch6_derivsGO}
\ee
\refeq{ch6_PkdL} evaluated for the linear matter power spectrum $P_l(k,t)$ then becomes
\be
 P_l(k,t|\dL)=[1+\drho(t)]\(\frac{\tD(t)}{D(t)}\)^2P_{l,\rm fid}([1+\da(t)]k,t) ~.
\ee
Inserting the series expansions, we obtain
\ba
 P_l(k,t|\d_{L0})=\:&\(1+\sum_{n=1}^\infty f_n[\d_{L0}\hD(t)]^n\)
 \(1+\sum_{n=1}^\infty g_n\[\d_{L0}\hD(t)\]^n\)^2 \vs
 &\times P_{l,\rm fid}\(\[1+\sum_{n=1}^\infty e_n[\d_{L0}\hD(t)]^n\]k,t\) ~.
\label{eq:ch6_Plinresponse}
\ea
\refEq{ch6_Plinresponse} allows for a consistent expansion in $\d_{L0}$. Specifically,
$d^nP_l(k)/d\d_{L0}^n$ is given by the $n$-th order coefficient in this expansion,
multiplied by $n!$.

\subsection{Nonlinear power spectrum predictions}
\label{sec:ch6_resp_nl}
Beyond the linear matter power spectrum, the growth coefficients $G_n$ will become
scale-dependent functions $G_n(k)$. Consider now what standard perturbation theory
(SPT) predicts. The power spectrum prediction is given by a series
\be
 P^{\rm SPT}(k)=P_l(k)+P^{1-\rm loop}(k)+P^{2-\rm loop}(k)+\cdots ~,
\ee
where $P^{n-\rm loop}$ scales as $[P_l]^n$. In an Einstein-de Sitter universe, one
can show (e.g., \cite{bernardeau/etal:2002}) that the time- and scale-dependence
of each order in perturbation theory factorizes, so that one can write
\be
 P^{\rm SPT}(k,t)=\hD^2(t)P_l(k,t_0)+\hD^4(t)P^{1-\rm loop}(k,t_0)
 +\hD^6P^{2-\rm loop}(k,t_0)+\cdots ~,
\label{eq:ch6_PSPT_EdS}
\ee
where $P^{n-\rm loop}(k,t_0)$ is a convolution of $n$ factors of $P_l(k,t_0)$ with
\emph{time-independent} coefficients. While \refeq{ch6_PSPT_EdS} is only strictly
correct in Einstein-de Sitter, it is used very commonly for $\Lambda$CDM as well,
since departures from the exact result are typically of order 1\% or less, and since
it simplifies the calculation significantly. Various variants of SPT, such as the
renormalized perturbation theory (RPT) \cite{crocce/scoccimarro:2006}, share the
same property.

\refEq{ch6_PSPT_EdS} allows for a very simple evaluation of the growth-only response:
as discussed above, the shape of the linear power spectrum in the modified cosmology
is unchanged, and hence $\tP^{\rm SPT}(\tilde k)$ can be simply evaluated by replacing
the fiducial $\hD(t)$ in \refeq{ch6_PSPT_EdS} with the modified one, \refeq{ch6_Dtilde}.
This is equivalent to assuming that the entire late-time cosmology dependence of the
nonlinear matter power spectrum enters through the linear growth factor
\cite{valageas:2014, kehagias/perrier/riotto:2014,ben-dayan/etal:2015}.
In \refsec{ch6_rescaled_sim}, we shall test this prescription to all orders in SPT
calculations by performing simulations with a rescaled initial power spectrum.

Apart from the SPT calculation, we can also apply this approximation to any
prescription that maps a given linear power spectrum to a nonlinear one.
In particular, we will show results for halofit \cite{smith/etal:2003}.
In this case, where the dependence on the linear growth factor is not explicit,
we instead compute the derivative with respect to the normalization of the
linear power spectrum,
\be
 \frac{d}{d\tD}\rightarrow\frac{d\tilde\sigma_8}{d\tD}\frac{d}{d\tilde\sigma_8} ~,
 \label{eq:ch6_growth_approx}
\ee
which at the redshift considered yields the equivalent change of the linear
matter power spectrum. This leads to
\be
 D^n\frac{d^nP(k)}{d\tD^n}\rightarrow\sigma_8^n\frac{d^nP(k)}{d\tilde\sigma_8^n} ~.
\ee
We use a five-point stencil with a step size of $0.75\%$ in $\sigma_8$ to compute
numerically the derivatives with respect to $\sigma_8$. In conjunction with the
change of the linear growth factor \refeq{ch6_Dtilde}, this allows us to compute
the growth-only response $G_n(k)$ for perturbation theory as well as fitting
formulae of the nonlinear matter power spectrum.

\subsection{Halo model predictions}
\label{sec:ch6_resp_hm}
In \refsec{ch4_halomodel_res}, we have derived the linear response of the power
spectrum under the halo model framework, in which all matter is assumed to be
contained within halos with a certain distribution of mass (given by the mass
function) and density profile. In this section, we shall focus on generalizing
the derivation to higher order responses for the halo model approach.

Adopting the notation in \refsec{ch4_halomodel_res}, the halo model powe
 spectrum, $P^{\rm HM}$, is given by
\ba
 P^{\rm HM}(k) =P^{\rm 2h}(k)+P^{\rm 1h}(k) ~,~~
 P^{\rm 2h}(k) =\[I^1_1(k)\]^2 P_l(k) ~,~~
 P^{\rm 1h}(k) =I^0_2(k,k) ~,
\label{eq:ch6_PkHM}
\ea
where
\be
 I^n_m(k_1,\cdots,k_m)\equiv\int d\ln M~n(\ln M)\(\frac{M}{\rhob}\)^mb_n(M)u(M|k_1)\cdots u(M|k_m) ~,
\label{eq:ch6_Inmdef}
\ee
and $n(\ln M)$ is the mass function (comoving number density per interval in log
mass), $M$ is the halo mass, $b_n(M)$ is the $n$-th order local bias parameter,
and $u(M|k)$ is the dimensionless Fourier transform of the halo density profile,
for which we use the NFW profile. One can find more details in \refsec{ch4_halomodel_res}.

To derive how the power spectrum given in \refeq{ch6_PkHM} responds to a homogeneous
(infinitely long-wavelength) density perturbation $\d_L$, we consider the one-halo
($P^{\rm 1h}$) and two-halo ($P^{\rm 2h}$) terms separately. The key physical assumption
we make is that halo profiles in \emph{physical} coordinates are unchanged by $\dL$.
That is, halos at a given mass $M$ in the presence of $\dL$ have the same scale radius
$r_s$ and scale density $\rho(r_s)$ as in the fiducial cosmology. We will discuss this
assumption later in this section. Given this assumption, the density perturbation $\dL$
then mainly affects the linear power spectrum, which determines the halo-halo clustering
(two-halo term), and the abundance of halos at a given mass.

For the two-halo term, as the response of the linear power spectrum is derived in 
\refeq{ch6_Plinresponse}, the remaining task is to consider the effect on $I^1_1$.
By the assumption that the density profile does not change in the presence of $\dL$,
$I^1_1$ only changes through the bias $b_1(M)$ and the mass function $n(\ln M)$.
The bias $b_N(M)$ quantifies the $N$-th order response of the mass function $n(\ln M)$
to $\dL$:
\be
 b_N(M)=\frac{1}{n(\ln M)}\frac{\partial^N n(\ln M)}{\partial\dL^N}\Big|_0 ~,~~{\rm or}~~
 \frac{\partial^Nn(\ln M)}{\partial\dL^N}\Big|_0=b_N(M) n(\ln M) ~.
\label{eq:ch6_bNdef}
\ee
Thus,
\ba
 \frac{\partial^N}{\partial\dL^N}I^1_1(k)\Big|_{\dL=0}=\:& \int d\ln M \: \(\frac{M}{\rhob}\)
 \frac{\partial^N}{\partial\dL^N}\[b_1(M)n(\ln M)\]\Big|_{\dL=0} u(M|k)=I^{N+1}_1(k) ~.
\label{eq:ch6_dI11}
\ea
Note that in the large-scale limit, $k\to0$, this vanishes for $N\ge1$ by way of
the halo model consistency relation
\be
 I_1^N(k)=\int d\ln M\: n(\ln M)\(\frac{M}{\rhob}\) b_N(M) =
 \bigg\lbrace
 \begin{array}{ll}
 1, & N = 1 ~, \\
 0, & N > 1 ~.
 \end{array}
\ee
For finite $k$ \refeq{ch6_dI11} does not vanish, we thus have
\be
 I^1_1(k,t|\d_{L0})=\sum_{n=0}^\infty\frac1{n!}I_1^{n+1}(k,t)[\hD(t)\d_{L0}]^n ~.
\ee
Thus, the two-halo term in the presence of $\d_{L0}$ becomes
\ba
\label{eq:ch6_P2hresponse}
 P^{\rm 2h}(k,t|\d_{L0}) =\:&\(1 +\sum_{n=1}^\infty f_n[\d_{L0}\hD(t)]^n\)
 \(1+\sum_{n=1}^\infty g_n\[\d_{L0}\hD(t)\]^n\)^2 \\
 &\times\(\sum_{n=0}^\infty\frac1{n!}I_1^{n+1}(k,t)[\hD(t)\d_{L0}]^n\)^2
 P_{l,\rm fid}\(\[1+\sum_{n=1}^\infty e_n[\d_{L0}\hD(t)]^n\]k,t\) ~. \nonumber
\ea
Note that we recover the tree-level result given in \refeq{ch6_Plinresponse} in the
large-scale limit, where only $n=0$ of the third term in \refeq{ch6_P2hresponse}
survives. Note also that in \refeq{ch6_P2hresponse} the dilation effect only enters
in the \emph{linear}, not 2-halo, power spectrum. This is a consequence of our
assumption that halo profiles do not change due to the long-wavelength density
perturbation.

For the one-halo term, due to our assumption about density profiles, the only effect
is the change in the mass function, which through \refeq{ch6_bNdef} becomes
\be
 \frac{\partial^N}{\partial\d_L^N}I^0_2(k,k) = I^N_2(k,k) ~,
\ee
and thus
\ba
 P^{\rm 1h}(k,t|\d_{L0})=\sum_{n=0}^\infty\frac1{n!}I_2^{n}(k,k,t)[\hD(t)\d_{L0}]^n ~.
\label{eq:ch6_P1hresponse}
\ea

Putting one-halo and two-halo terms together, we obtain
\ba
 P^{\rm HM}(k,t |\d_{L0})=\:&\(1+\sum_{n=1}^\infty f_n[\d_{L0}\hD(t)]^n\)
 \(1+\sum_{n=1}^\infty g_n\[\d_{L0}\hD(t)\]^n\)^2 \vs
 &\times\(\sum_{n=0}^\infty\frac1{n!}I_1^{n+1}(k,t)[\hD(t)\d_{L0}]^n\)^2
 P_{l,\rm fid}\(\[1+\sum_{n=1}^\infty e_n[\d_{L0}\hD(t)]^n\]k,t\) \vs
 &+\sum_{n=0}^\infty\frac1{n!}I_2^{n}(k,k,t)[\hD(t)\d_{L0}]^n ~.
\label{eq:ch6_PHMresponse}
\ea
The contribution $\propto I_1^{n+1}$ (for $n>0$) is numerically much smaller than
the other terms (see also the discussion in \refsec{ch4_halomodel_res}). Since it
is much smaller than the overall accuracy of the halo model description, we will
neglect it in the following. This yields
\ba
 P^{\rm HM}(k,t|\d_{L0})=\:&\(1+\sum_{n=1}^\infty f_n[\d_{L0}\hD(t)]^n\)
 \(1+\sum_{n=1}^\infty g_n\[\d_{L0}\hD(t)\]^n\)^2 \vs
 &\times\(I_1^1(k,t)\)^2P_{l,\rm fid}\(\[1 +\sum_{n=1}^\infty e_n[\d_{L0}\hD(t)]^n\]k,t\) \vs
 &+\sum_{n=0}^\infty\frac1{n!}I_2^{n}(k,k,t)[\hD(t)\d_{L0}]^n ~.
\label{eq:ch6_PHMsimplified}
\ea
Explicitly, the first and second order full response functions are given by
\ba
 R_1^{\rm HM}(k)=\:&\[f_1+2g_1+e_1\frac{d\ln P_l(k,t)}{d\ln k}\]P^{\rm 2h}(k,t)+I_2^1(k,k,t) \vs
 R_2^{\rm HM}(k)=\:&\bigg[2f_2+2f_1g_1+(f_1+2g_1)e_1\frac{d\ln P_l(k,t)}{d\ln k}+2g_1^2+4g_2 \vs
 &+2e_2\frac{d\ln P_l(k,t)}{d\ln k}+ e_1^2\frac1{P}\frac{d^2P_l(k,t)}{d(\ln k)^2 }\bigg]
 P^{\rm 2h}(k,t)+I_2^2(k,k,t) ~.
\ea

We also derive the growth-only response functions in the halo model approach. Since
the halo profiles are assumed fixed in physical coordinates, this means that we need
to rescale the halo model terms, $I^n_m$, accordingly. Following our discussion in
\refsec{ch6_sepuni}, we have $\tilde{k}=(1+\da)k$, where $\tilde k$ is the comoving
wavenumber with respect to the modified cosmology. We then obtain
\be
 I^n_m\Big|_{\rm growth~only}(\tilde k_1,\cdots,\tilde k_m)
 =I^n_m\Big|_{\rm physical}\(\frac{k_1}{1+\d_a(t)},\cdots,\frac{k_m}{1+\da(t)}\) ~.
\label{eq:ch6_InmGO}
\ee
Inserting this into \refeq{ch6_PHMresponse} and performing a series expansion of $\da$
in $\dL$ then allows us to derive the growth-only response functions $G^{\rm HM}_n(k)$.
Note that the NFW profile we assume is uniquely determined by the scale radius $r_s(M)$
for a halo of mass $M$, which enters the coefficients defined in \refeq{ch6_Inmdef} in
the combination $k r_s(M)$. Thus, it is easily possible to include a dependence of the
scale radius $r_s(M)$, or equivalently the halo concentration, on the long-wavelength
density in a similar way. We will leave this for future work.

Quantitatively, the main contribution of the rescaling \refeq{ch6_InmGO} is from the
one-halo term $\propto I_2^n(k,k)$, i.e. the term in the last line of \refeq{ch6_PHMsimplified}.
The rescaling of the other instances of $I^n_m$ only changes the response at the sub-percent
level and we will neglect them in the following. We then obtain for the growth-only contribution
to the halo model power spectrum
\ba
 P^{\rm HM}(k,t|\d_{L0})\stackrel{\rm growth~only}{=}\:&
 \(1+\sum_{n=1}^\infty g_n\[\d_{L0}\hD(t)\]^n\)^2\[I_1^1(k,t)\]^2P_{l,\rm fid}(k,t) \vs
 &+\sum_{n=0}^\infty\frac1{n!}I_2^{n}\[A(\d_{L0},t)\,k,A(\d_{L0},t)k,t\][\hD(t)\d_{L0}]^n ~,
\label{eq:ch6_PHMgo}
\ea
where
\be
 A(\d_{L0},t)=\(1+\sum_{n=1}^\infty e_n [\d_{L0}\hD(t)]^n\)^{-1} ~.
\ee

\section{Separate universe simulations}
\label{sec:ch6_sepuni_sim}
To test our semi-analytical models of the power spectrum response to the homogeneous
overdensity, particularly for the growth of structure due to the change of the cosmology,
we run separate universe simulations and measure the power spectrum response functions
directly. In this section, we describe the details for performing the separate universe
simulations. We shall first introduce the straightforward modifications, cosmological
parameters and the initial conditions, and then the non-trivial choices, comoving or
physical coordinates.

For usual $N$-body codes, one needs to specify the cosmological parameters at present
time and the output time $\tou$ normally specified by the scale factor. As discussed
in \refchp{ch3_sepuni}, in the presence of the overdensity $\d_{L0}$, the cosmological
parameters at $\ta(\tilde t_0)=1$ are modified as
\ba
 \tH_0=\:&H_0(1+\dH) ~,~~ \tO_m=\Om(1+\dH)^{-2} ~,~~ \tO_{\Lambda}=\Omega_{\Lambda}(1+\dH)^{-2} \vs
 \tO_K=\:&1-(1+\dH)^2 ~,~~ \dH=\(1-\frac{\tK}{H_0^2}\)^{1/2}-1 ~,~~
 \frac{\tK}{H_0^2}=\frac53\frac{\Om}{D(t_0)}\d_{L0} ~.
\label{eq:ch6_params}
\ea
For the output time, because we want to compare the simulations at the same \emph{physical}
time, we need to determine the corresponding scale factor in the modified cosmology
such that $\ta(\tou)=\[1+\da(\tou)\]$. We can numerically solve the ordinary differential
equation of $\da$ in the modified cosmologies \refeq{ch3_da_diff}; alternatively we can
numerically evaluate $\ta(\tou)$ by
\be
 \tou=\int_0^{a(\tou)}~\frac{da}{aH(a)}
 =\int_0^{\ta(\tou)}~\frac{d\ta}{\ta\tH(\ta)}
 =\int_0^{a(\tou)[1+\da(\tou)]}~\frac{d\ta}{\ta\tH(\ta)} ~.
\label{eq:ch6_da_int}
\ee

In order to generate the initial conditions for $N$-body simulations of the modified
cosmologies, we need the linear power spectrum at the initial redshift. The initial
power spectrum has to be generated for the cosmology $[\tO_m,\tO_{\Lambda},\tH_0]$
with the same amplitude of the primordial scalar curvature perturbation $\mathcal{A}_s$
as for the fiducial cosmology. Since the transfer function only involves the physical
matter and radiation densities quantified by $\tO_m\tH_0^2$ and so on, it is identical
in the modified and fiducial cosmologies. Therefore, the linear power spectra differ
only through the difference in the linear growth factor. We use CAMB \cite{lewis/challinor/lasenby:2000,lewis/bridle:2002}
to compute the power spectrum of the fiducial cosmology at $z=0$, and rescale it by
\be
 \[\tD(\ta_i)\frac{\tD(\ta=1)}{D(a=1)}\]^2 ~,
\label{eq:ch6_rescaled_zi}
\ee
where $\ta_i$ is the scale factor for which the initial conditions are generated.\footnote{In
order to recover the correct linear power spectrum at low redshifts, we compute
the growth functions ($D$ and $\tD$) without taking radiation into account. This
is because $N$-body codes do not include the effect of cosmological radiation. In
our procedure, we also neglect the effect of curvature on the transfer function
at very low wavenumbers $k\sim\sqrt{|K|}$, since terms of similar order are neglected
in the Poisson equation used in $N$-body codes. For sub-horizon box sizes these
effects are completely negligible.} Next, we generate a Gaussian realization of
the density field following the initial power spectrum. The positions and velocities
of the particles are computed by the second-order Lagrangian perturbation theory
\cite{crocce/pueblas/scoccimarro:2006}.

Given a fixed box size for the fiducial cosmology, there are two reasonable choices
for the box sizes of the modified cosmologies. Either we match the respective comoving
box sizes, i.e. the box size is $500~\tilde h/h$ in units of $\tilde h^{-1}$ Mpc comoving,
or we choose the box sizes such that their physical sizes coincide with that of the
fiducial simulation at one specific output time $\tou$, i.e. $500~\tilde ha(\tou)/[h\ta(\tou)]$
in units of $\tilde h^{-1}$ Mpc comoving. The former choice is adequate if we are interested
in the power spectrum response functions at the same comoving wavenumber, i.e. without
the ``dilation'' effect. By using the mean density of the separate universe cosmology
as the reference density when computing the power spectrum, we are further removing the
``reference density'' effect and are left with the growth-only response. The results of
these simulations are presented in \refsec{ch6_grow_resp}.

In order to measure the full response functions, we run simulations for which we match
the physical box size. We focus on two different output times $\tou$ corresponding to
$z=0$ and $z=2$ in the fiducial cosmology. As the physical size can only be matched at
one specific time, we have to run a new set of simulations for each output time. The
results of these simulations are presented in \refsec{ch6_full_resp}.

Finally, we summarize the common features for the separate universe simulations. All
simulations are gravity-only simulations and are carried out using the Tree-PM code
Gadget-2 \cite{springel:2005}. The starting redshift is $z=49$ ($a_i=0.02$), and the
particle load for each simulation is $512^3$. For the fiducial cosmology ($\d_{L0}=0$),
we choose a flat $\Lambda$CDM cosmology with cosmological parameters consistent with
the current observational constraints: $\Om=0.27$, $h=0.7$, $\Omega_b h^2=0.023$,
$n_s=0.95$, $\sigma_8=0.8$, and a comoving box size of $500\hMpc$.

We simulate separate universes corresponding to the linearly-evolved present-day overdensities
of $\d_{L0}=0$, $\pm0.01$, $\pm0.02$, $\pm0.05$, $\pm0.07$, $\pm0.1$, $\pm0.2$, $\pm0.5$,
$\pm0.7$, and $\pm1$. Then, for the separate universes, the Hubble constant and the curvature
fraction vary between $\tilde h$: 0.447 to 0.883 and  $\tO_K$: $-2.45$ to 0.372, respectively.
The physical densities $\tO_m\tilde h^2$, $\tO_\Lambda\tilde h^2$, and $\tO_b\tilde h^2$ as
well as $n_s$ and the amplitude of the primordial curvature power spectrum remain the same.

\section{Results of separate universe simulations}
\label{sec:ch6_results}
For the power spectrum computation, we first estimate the density contrast $\d({\vx})$
on a $1024^3$ grid using the cloud-in-cell mass assignment scheme, then apply a Fast
Fourier transform, and angular average the squared amplitude $|\d_{\vk}|^2$. The density
contrast $\d(\vx)=\rho(\vx)/\rhob-1$ describes the overdensity with respect to the
reference density $\rhob$. When we are interested in the growth-only response function,
$\rhob$ is equal to the mean density of the separate universe. When we compute the full
response function, $\rhob$ is equal to the mean density of the fiducial cosmology.
Similarly, for the growth-only response, distances are measured using the comoving
coordinates of the respective cosmology\footnote{Note, however, that the unit of length
is always $h^{-1}$Mpc, where $h$ corresponds to the fiducial cosmology.}, whereas, for
the full response, the power spectrum is always measured in comoving coordinates of the
fiducial cosmology.

We only report results up to a maximum wavenumber of $2\ihMpc$. A convergence study
with simulations with 8 times lower mass resolution shows differences in $G_1$, $G_2$
and $G_3$ of only 1 (3) to 5 (10) percent at $z=0$ ($z=2$) up to that wavenumber,
where the deviations increase from the linear response function to the higher-order
response functions. The results for the full response functions $R_1$, $R_2$ and $R_3$
are converged to an even better degree. We therefore expect that the simulation results
presented here converge to a sub-percent to a few percent level.

In order to compute the first three response functions, we fit a polynomial in $\dL$
to the fractional difference in the measured power spectrum $\Delta_k(\dL)\equiv P(k|\dL)/P(k|\dL=0)-1$
for each $k$-bin. For the fit, we only include results from separate universe simulations
with $|\dL(t_{\rm out})|\le0.5$ and use a polynomial with degree 6 to be unbiased from
higher-order response functions. As the random realization of the initial density field
is the same across different $\dL$ values, the corresponding power spectra are strongly
correlated. By considering the ratio, or the relative difference, of two power spectra
a large fraction of the noise cancels. However, for the same realization, the measured
fractional differences $\Delta_k(\dL)$ are still correlated over $\dL$. As the number
of realizations (16) is not large enough to reliably estimate the covariance between
different $\dL$ values, we cannot include this correlation in the polynomial fitting.
Instead, we construct quasi-decorrelated samples of $\Delta_k(\dL)$ by randomly choosing
a realization for each $\dL$ value. Fitting many of those subsamples allows for a robust
error estimation of the derived response functions.

\subsection{Growth-only response functions}
\label{sec:ch6_grow_resp}
\begin{figure}[t!]
\centering
\includegraphics[width=1\textwidth]{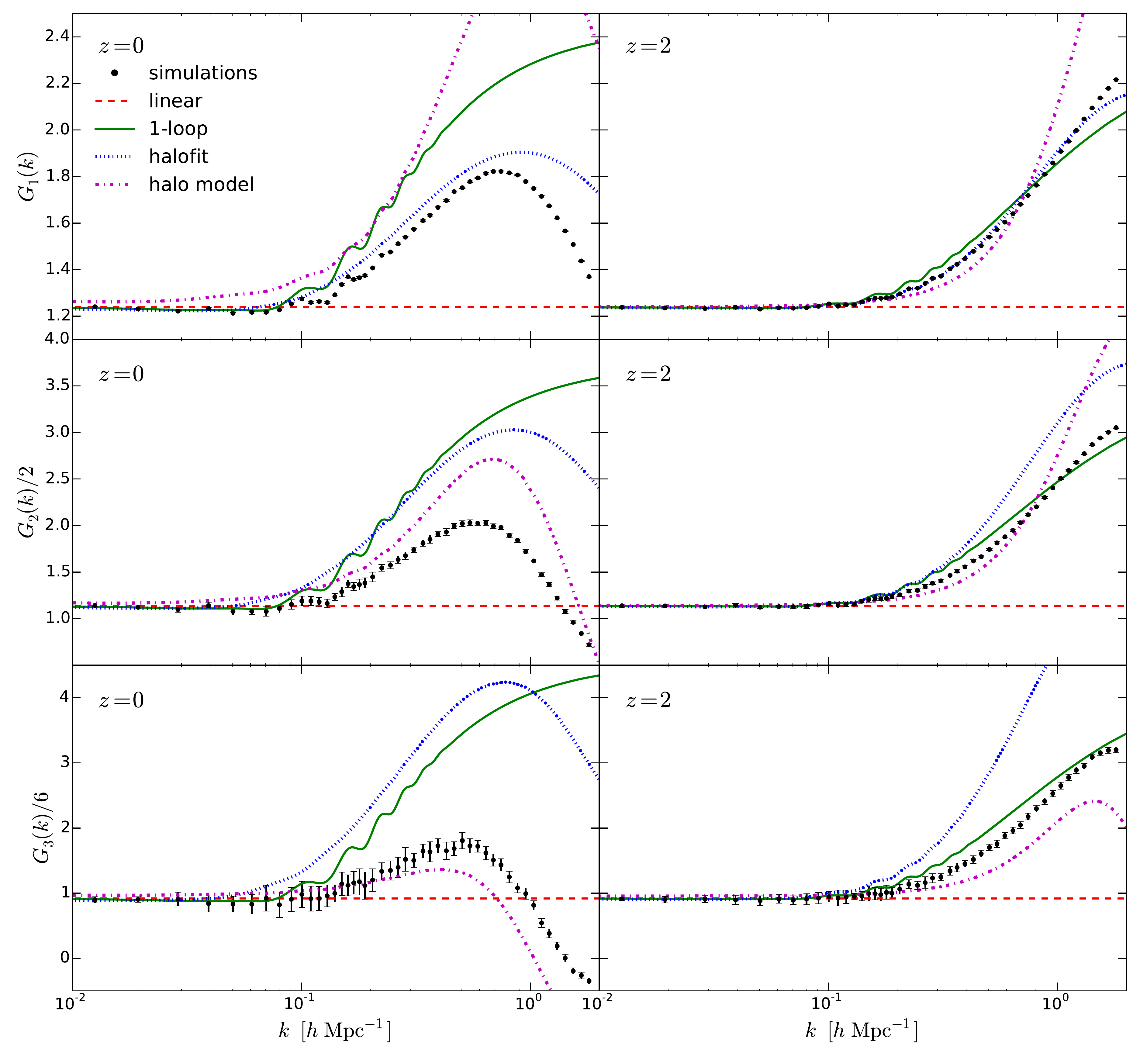}
\caption[The first three growth-only response functions]
{The first three growth-only response functions of the power spectrum measured
from the separate universe simulations at $z=0$ (left) and $z=2$ (right). The
error bars show the statistical error derived by random resampling of the data
(see text). For data points apparently without error bars, the statistical error
is smaller than the size of a dot.}
\label{fig:ch6_growth_only}
\end{figure}

\refFig{ch6_growth_only} shows the first three growth-only response functions measured
from the simulations at $z=0$ (left column) and $z=2$ (right column). These correspond
to the fully nonlinear squeezed limit bispectrum (three-point function), trispectrum
(four-point function) and five-point function. The small wiggles in the growth-only
response functions result from the damping of the baryon acoustic oscillations (BAO),
which depends on the amplitude of density fluctuations and thus on $\dL$.

Let us compare the simulation results to the theoretical predictions discussed in
\refsec{ch6_pk_resp}. On sufficiently large scales, the perturbation theory predictions
are the most accurate, as expected. At high redshift, the 1-loop predictions best
describe the results overall. The 1-loop predictions also show a BAO damping effect.
At $z=0$, the growth-only response is captured best by the halofit prescription (in
case of $G_1$) or the halo model (in case of $G_2,\,G_3$). We see that the halofit
prescription describes the simulation results of the linear response well at both
redshifts, but performs significantly worse for the higher-order response functions.
The BAO damping effect is essentially absent in both halofit and halo model predictions.
Overall, none of the models is able to accurately describe the simulation data in the
nonlinear regime, with discrepancies at $z=0$ ranging from 20\% in the best case to
a factor of several. These discrepancies are not surprising given that we are looking
at scales beyond the validity of perturbation theory and at higher $N$-point functions
for which the semi-analytical approaches were not tuned.

The halo model prediction does not asymptote exactly to the linear result in the $k\to0$
limit. This is because the one-halo term asymptotes to a white noise contribution in this
limit, and since the one-halo term contributes to $G_n$ due to the dependence of the halo
mass function on $\dL$ (\refsec{ch6_resp_hm}), this induces a correction to the linear
prediction which contributes on large scales.  Physically, this occurs because the halo
model does not enforce momentum conservation of the matter density field. This issue can
be fixed by introducing a ``mass compensation scale'' \cite{mohammed/seljak:2014}.

The halo model predictions can be tuned to better match the simulation results by allowing
for a dependence of the halo profiles on the long-wavelength density, which is expected
on physical grounds (see also \cite{li/hu/takada:2014b}). Specifically, if the scale
radius of halos at fixed mass increases in the presence of a long-wavelength density
perturbation, this lowers the peak in the response and thus could lead to better agreement
with the simulations results.

\subsection{Full response functions}
\label{sec:ch6_full_resp}
\begin{figure}[t!]
\centering
\includegraphics[width=1\textwidth]{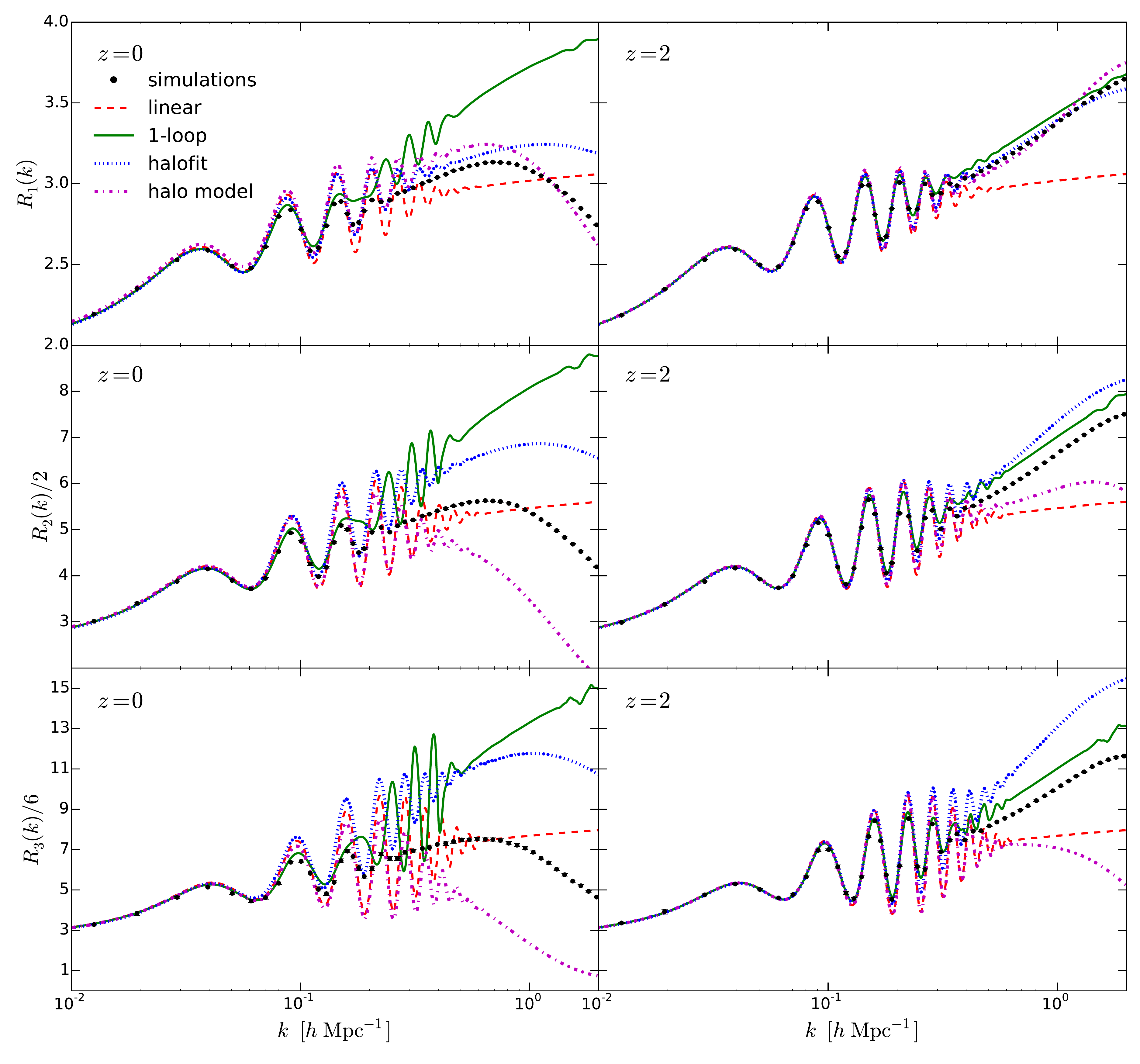}
\caption[The first three full response functions]
{The first three full response functions of the power spectrum measured from the
separate universe simulations at $z=0$ (left) and $z=2$ (right).}
\label{fig:ch6_full}
\end{figure}

We now turn to the results for the full response functions, i.e. including the ``dilation''
and ``reference density'' effects. The results of the simulations and the model predictions
are shown in \reffig{ch6_full}. The oscillations in the response functions can be traced back
to the BAOs in the power spectrum.  The BAOs propagate to the response functions primarily
by the ``dilation'' effect, which yields derivatives of the power spectrum with respect to
$k$ (see \refeqs{ch6_R1}{ch6_R3}). The 1-loop perturbation theory predictions describe the
simulation results accurately up to $k\le0.15\ihMpc$ and $k\le0.3\ihMpc$ at $z=0$ and $z=2$,
respectively. As the other theoretical models do not include the damping of the BAOs in the
nonlinear power spectrum, they predict oscillations in the response functions which are too
large. To improve the accuracy of those models around the BAO scale, one would need to put
in the BAO damping by hand. In the nonlinear regime, none of the models is able to reproduce
the simulation data. In principle, one could build a hybrid model for the full response by
combining an accurate prediction of the nonlinear power spectrum of the fiducial cosmology
and the growth-only response functions $G_n(k)$ discussed in the previous section. However,
we do not pursue this approach.

\subsection{Eulerian response functions}
\label{sec:ch6_eulr_resp}
\begin{figure}[t!]
\centering
\includegraphics[width=1\textwidth]{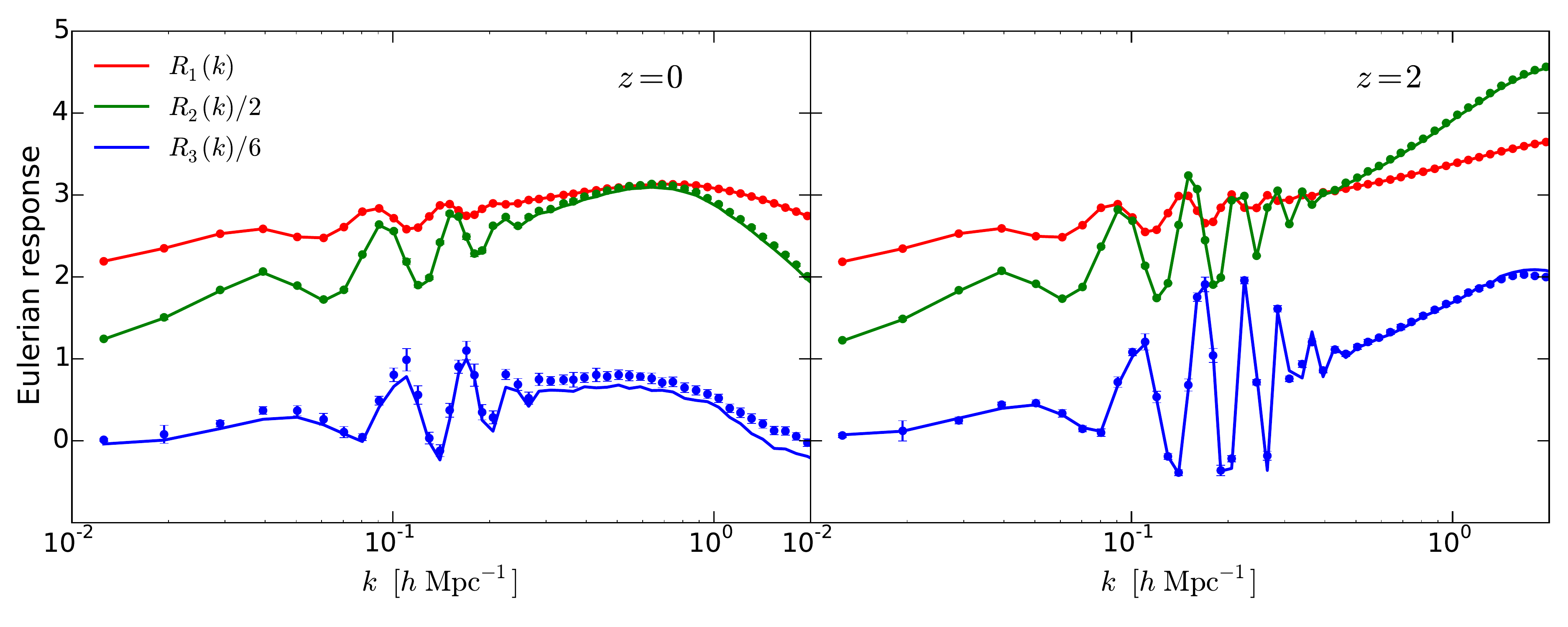}
\caption[The first three Eulerian response functions]
{The first three Eulerian response functions of the power spectrum measured
from the separate universe simulations (data points) at $z=0$ (left) and $z=2$
(right). The lines show the corresponding linear combinations of the Lagrangian
response functions using the $f_n$ coefficients derived for the Einstein-de
Sitter universe (see \refeq{ch6_Reuler} and \refeq{ch3_fn}).}
\label{fig:ch6_eulerian}
\end{figure}

So far, we have always considered the response to the linearly-extrapolated initial
(Lagrangian) overdensity $\dL$.   We now consider the corresponding response to the
\emph{evolved nonlinear} (Eulerian) overdensity $\drho$. Using the expansion derived
for the Einstein-de Sitter universe, \refeq{ch3_fn}, we find
\ba
 R_1^{\rm Eulerian}(k)&=R_1(k) ~, \vs
 R_2^{\rm Eulerian}(k)&=R_2(k)-2f_2R_1(k) ~, \vs
 R_3^{\rm Eulerian}(k)&=R_3(k)-6f_2 R_2(k)+6\(2f_2^2-f_3\)R_1(k) ~.
\label{eq:ch6_Reuler}
\ea
In \reffig{ch6_eulerian}, we compare the directly measured Eulerian response functions
with the appropriate linear combinations of the measured Lagrangian response functions.
The agreement is excellent as expected, especially at high redshift at which the $\Lambda$CDM
universe is very well approximated by the Einstein-de Sitter universe.

Interestingly, the higher-order Eulerian response functions are much smaller than in
the Lagrangian case. That is, the response of the nonlinear matter power spectrum to
a uniform nonlinear final-time density $\drho$ is close to linear. This is most likely
due to the fact that the growth-only response functions are subdominant compared to the
rescaling and reference density contributions, especially at higher order. In this case,
\refeq{ch6_refdensity_dilation} implies a close to linear scaling with $\drho$.

\section{Simulations with rescaled initial amplitudes}
\label{sec:ch6_rescaled_sim}
All models for the growth-only response functions that we have presented in
\refsec{ch6_pk_resp} and \refsec{ch6_grow_resp} are based on the approximation
that we can trade the effect of $\dL$ for an appropriate change to the linear
growth factor (or equivalently, the linear power spectrum). But how well does
this approximation work?

To investigate how well the effect of a homogeneous overdensity on the growth of structure
can be modeled by a change in the amplitude of the linear power spectrum, we additionally
run a set of simulations for which we always assume the fiducial cosmology but vary the
amplitude of the initial power spectrum. Specifically, for each $\delta_{L0}$ value for
which we simulate a separate universe, we also simulate the fiducial cosmology with the
initial power spectrum amplitude multiplied by $\tD(t_0)^2/D(t_0)^2$, where $\tD(t_0)$
is the linear growth factor in the corresponding separate universe cosmology.

Using these ``rescaled-amplitude simulations'', we can explicitly test the
approximation that $\dL$ effects the growth of structure only through the
change in the linear growth factor on all scales including the nonlinear regime.

\subsection{Comparison to separate universe simulations}
\label{sec:ch6_rescal_res}
In \reffig{ch6_rescale}, we show the growth-only response functions measured from
two different sets of simulations. In case of $G_1$, this comparison was also shown in
figure~6 of \cite{li/hu/takada:2014b}, and our results agree with theirs.\footnote{Note
that the in \cite{nishimichi/valageas:2014} a different comparison is performed using
the time derivative of the nonlinear power spectrum in simulations of the fiducial
cosmology.} The rescaled-amplitude simulations all assume the fiducial cosmology but
vary the amplitude of the linear power spectrum used to initialize the simulations so as
to match the linear power spectrum in the modified cosmology at the given output times
[\refeq{ch6_Dtilde} and \refeq{ch6_growth_approx}]. On linear scales, these simulations
thus agree with the separate universe simulations by construction. As the simulations
share the same random realization of the initial density, the sample variance (noise
in the upper panels) gets vastly reduced when considering the difference of the measured
response functions, $\Delta G_n=G_n^{\rm rescaled}-G_n^{\rm separate}$. This is shown
in the lower subpanels of \reffig{ch6_rescale}, where we have divided $\Delta G_n$ by
the corresponding linear growth-only response, i.e. the prediction in the $k\to 0$ limit.

\begin{figure}[t!]
\centering
\includegraphics[width=1\textwidth]{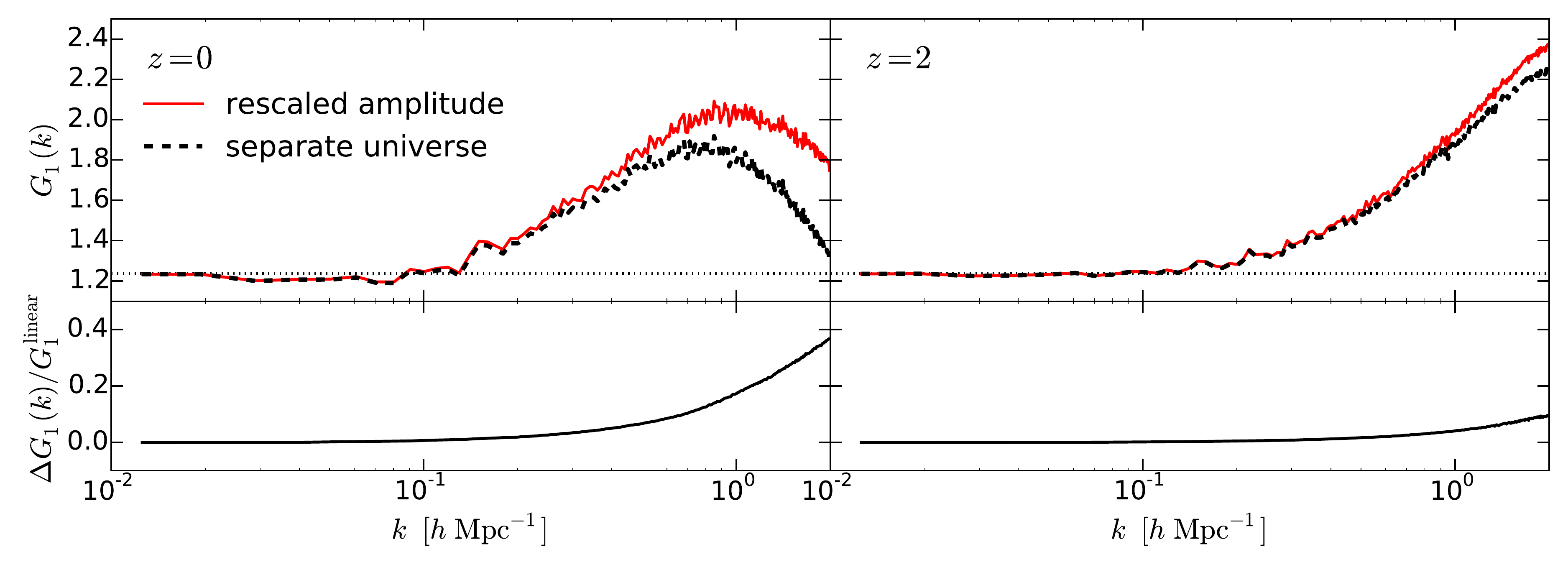}
\includegraphics[width=1\textwidth]{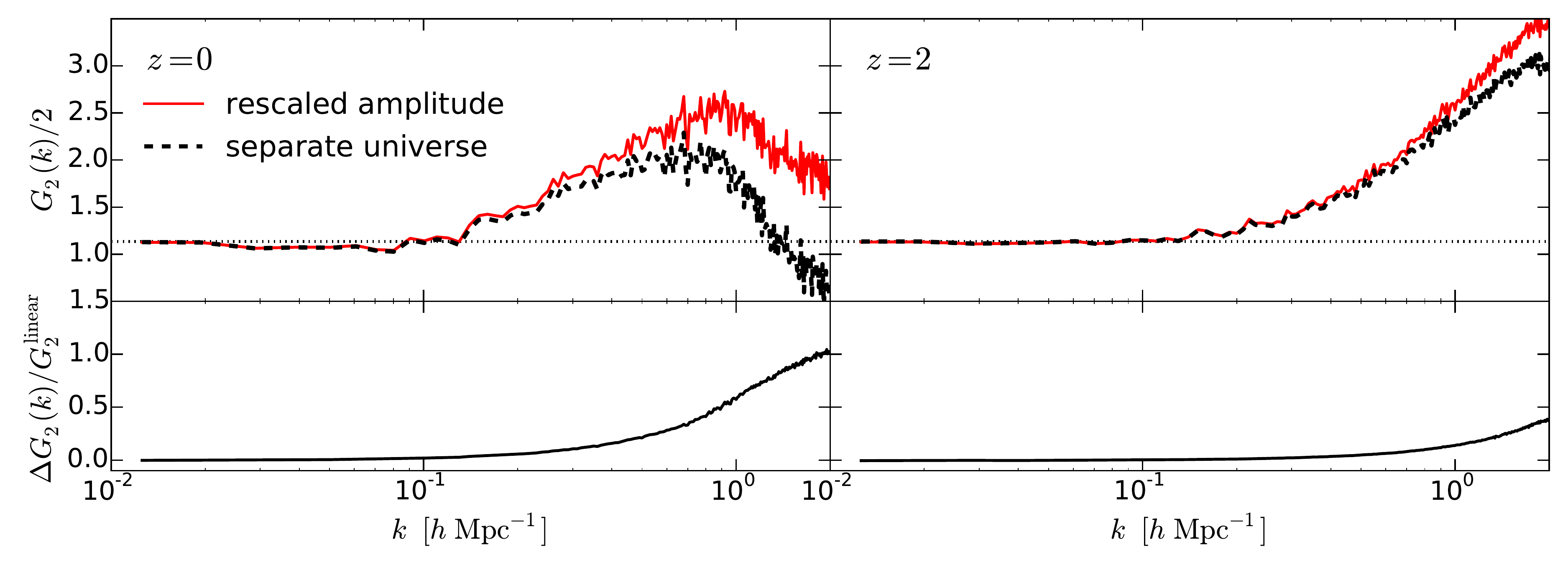}
\includegraphics[width=1\textwidth]{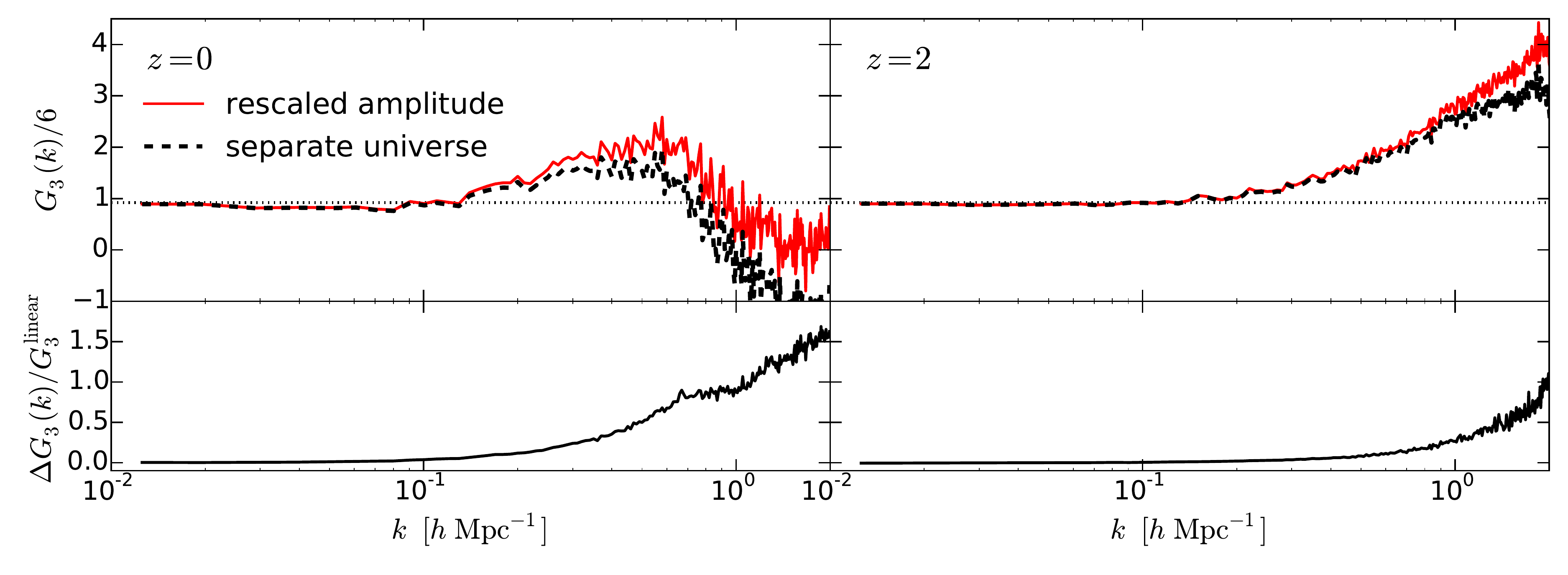}
\caption[Growth-only response functions of separate universe simulations and rescaled-amplitude simulations]
{Comparison of the growth-only response functions $G_1,\,G_2,\,G_3$ (top to bottom)
measured at $z=0$ (left column) and $z=2$ (right column) from one realization of
the separate universe simulations and from the same realization simulated by varying
the initial amplitude. The bottom subpanels show the difference,
$\Delta G_n=G_n^{\rm rescaled}-G_n^{\rm separate}$, divided by the response of the
linear matter power spectrum, $G_n^{\rm linear}$.}
\label{fig:ch6_rescale}
\end{figure}

The differences seen in \reffig{ch6_rescale} are caused by the different growth history,
which is not captured by the rescaling of the initial amplitude. Following the discussion
in \refsec{ch6_resp_nl}, the commonly used SPT approach factorizing the growth factor and
scale dependence assumes at all orders that a long wavelength density perturbation enters
exclusively through the modified linear growth. Thus, \emph{even when calculated to all
orders}, the best that this SPT calculation could do is to reproduce the rescaled amplitude
result in \reffig{ch6_rescale}, which deviates from the actual response at $z=0$ by 10\%
at $k\simeq0.5\ihMpc$ and 20\% at $k\simeq1\ihMpc$ for $G_1$, and significantly worse for
the higher-order response functions. At $z=2$ on the other hand, the rescaled-amplitude
$G_1$ matches the separate universe response to better than 10\% even beyond $k=1\ihMpc$,
and for $G_2,\,G_3$ performs significantly better as well.

There are two possible explanations for these discrepancies in the SPT context. First,
using the SPT kernels derived for an Einstein-de Sitter universe (which have time-independent
coefficients), with the $\Lambda$CDM linear growth factor replacing the Einstein-de Sitter
$a(t)$, could become highly inaccurate for $\Lambda$CDM at higher orders. Note that the
same issue exists for a fiducial flat Einstein-de Sitter universe, since for $\dL\neq 0$
the quantity $\Om/f^2$ is no longer 1 ($d(\Om/f^2)/d\dL=-5/21$ \cite{ben-dayan/etal:2015};
see also the discussion in \cite{nishimichi/valageas:2014}). There is no indication of
such a strong effect at low orders in perturbation theory, where this approximation
typically performs to better than a percent \cite{bernardeau/etal:2002}. Furthermore,
it is found in \cite{ben-dayan/etal:2015} that a cancellation in the curvature contribution
to the growth integral suppresses this effect. Finally, it is shown in \cite{li/hu/takada:2014b}
that the growth-only response of the power spectrum to a change in the Hubble constant
while keeping $\Om h^2$ fixed follows the separate universe response very closely (figure~6
there). If the much larger discrepancies between separate universe response and rescaled
amplitude response were due to the cosmology dependence of the SPT kernels, one would not
expect this to be the case. Nevertheless, we do not claim to be able to rigorously exclude
this possibility.

The other possibility, more likely in our opinion, is that the discrepancy between rescaled
amplitude and full separate universe simulations is due to effective non-perfect fluid terms,
such as pressure and anisotropic stress, in the dark matter fluid \cite{baumann/etal:2012}. 
The effective fluid properties depend on highly nonlinear small scales which are not described
by the Euler-Poisson system. Their value can depend on the growth history (as well as the
power spectrum shape) thus leading to a discrepancy between rescaled amplitude and separate
universe simulations. Assuming this interpretation is correct, \reffig{ch6_rescale} explicitly
shows the breakdown of SPT on nonlinear scales as effective pressure, anisotropic stress and
sound speed need to be included. Separate universe simulations can be used to measure the
response of these effective terms to a long-wavelength overdensity, which is crucial when
modeling $(N>2)$-point functions. The results shown in \reffig{ch6_rescale} are analogous
to what has been found for the mass function of halos which is a key ingredient in the halo
model description of the nonlinear matter density field. The mass function shows departures
from being a simple function of the linear matter power spectrum at the 5--10\% level
\cite{tinker/etal:2008,bhattacharya/etal:2011}.

In an Einstein-de Sitter cosmology with scale-invariant initial power spectrum $P_l(k)\propto k^n$,
there is only one characteristic spatial scale at any given time, which corresponds to the
scale at which the density field becomes order 1 \cite{pajer/zaldarriaga:2013}. Let us denote
this wavenumber as $k_{\rm NL}(t)$. Then, the response functions have to follow a universal
function of $k/k_{\rm NL}(t)$, i.e. $G_m(k,t)=G_m(k/k_{\rm NL})$ (keeping the index of the
initial power spectrum fixed). Thus, in this specific case, separate universe simulations
and rescaled-amplitude simulations will give exactly the same result when compared at fixed
$k/k_{\rm NL}$. The departures shown in \reffig{ch6_rescale} can thus be seen as a consequence
of the $\Lambda$CDM background and the departure from scale-invariance of the initial power
spectrum. It would be interesting to disentangle the two effects, e.g. by performing separate
universe simulations in $\Lambda$CDM with scale-invariant initial conditions. We leave this
for future work, but point out that when plotting the differences shown in the lower panel
of \reffig{ch6_rescale} as a function of $k/k_{\rm NL}$, we still find a factor of several
difference in the $z=0$ and $z=2$ results.


\section{Discussion and conclusion}
\label{sec:ch6_conclusion}
In this chapter, we described in detail the procedure for performing $N$-body
simulations with the separate universe technique. Using the separate universe
simulations, we compute the response of the  nonlinear matter power spectrum
to a homogeneous overdensity superimposed on a flat FLRW universe. The response
functions we computed give the squeezed limits of the 3-, 4-, and 5-point functions,
in which all but two wavenumbers are taken to be small and are angle-averaged.
By virtue of the separate universe technique, we reach an unprecedented accuracy
of these nonlinear matter $N$-point functions.

The response function consists of three parts: changing the reference density
with respect to which the power spectrum is defined; rescaling of comoving
coordinates; and the effect on the growth of structure. The former two effects
can be calculated trivially, whereas the third one requires separate universe
simulations. We have compared the simulation results with analytical and semi-analytical
results, in particular standard perturbation theory (SPT), the empirical fitting
function halofit, and the halo model, finding that SPT typically yields the best
results at high redshifts. The fitting function and halo model, while qualitatively
describe the trends seen in the response functions, give a poor quantitative
description on nonlinear scales.

A fundamental assumption of all of the analytical and semi-analytical methods
used in this chapter, including standard perturbation theory at any order, is
that nonlinear matter statistics at a given time are given solely by the linear
power spectrum at the same time, and do not depend on the growth history otherwise.
As was done in \cite{li/hu/takada:2014b} for the response function for $n=1$,
we were able to test this assumption for $n=2$ and 3 quantitatively by comparing
the separate universe simulations with simulations with a rescaled initial power
spectrum amplitude. We find that this assumption fails at the level of 10\% at
$k\simeq 0.2-0.5\ihMpc$ for $5$- to $3$-point functions at $z=0$. The failure
occurs at higher wavenumbers at $z=2$. In the context of SPT, this may signal
a breakdown of the perfect fluid description of the dark matter density field
at and beyond these wavenumbers. In other words, even if computed to all orders,
SPT (and its variants such as RPT \cite{crocce/scoccimarro:2006}) fails to describe
the nonlinear structure formation beyond these wavenumbers. Therefore, our results
yields a quantitative estimate for the scales at which effective fluid corrections
become important in the bispectrum and higher $N$-point functions, and at which
one should stop trusting pure SPT calculations.

Finally, we point out that the approach presented here can be augmented to
measure more general squeezed-limit $N$-point functions, by including the
response to long-wavelength tidal fields and by considering the response
of small-scale $n$-point functions in addition to the small-scale power
spectrum considered here.

}

%% file: kap_07.tex
\chapter{Summary and outlook}
\label{chp:ch7_summary}
In this dissertation, we have proposed and developed in detail a new observable, position-dependent
power spectrum, to extract the squeezed-limit bispectrum of the
large-scale structure by measuring the correlation between the position-dependent
two-point statistics and the mean overdensity in the subvolumes of a survey
volume. Since this new technique requires essentially measurements of the
two-point statistics and mean overdensity, it sidesteps the complexity of
the traditional three-point function estimation.

The correlation between the position-dependent two-point statistics and the
mean overdensity can be regarded as how the small-scale structure formation
responds to a long-wavelength mode. In \refchp{ch3_sepuni}, we have shown that
the long-wavelength overdensity compared to the scale of interest can be absorbed
into the background cosmology, and the small-scale structure formation evolves
in the corresponding modified cosmology. This \emph{separate universe approach}
thus provides an intuitive way to model the squeezed-limit bispectrum, i.e. the
response of the small-scale power spectrum to a long-wavelength mode.

In \refchp{ch4_posdeppk}, we have measured the position-dependent power spectrum
from cosmological $N$-body simulations, and compared the measurements to different
theoretical modeling. In particular, we have shown that it is not only straightforward
to combine the separate universe approach with various power spectrum models, but
the separate universe approach also describes nonlinearity in the squeezed-limit
bispectrum due to gravitational evolution better than the traditional approach
based on the perturbation theory. This would enable us to measure the primordial
non-Gaussianity in the large-scale structure because we must distinguish the
primordial signal from the contamination due to the late-time contribution that
we have computed precisely in this dissertation.

In \refchp{ch5_posdepxi}, we have reported on the first measurement of the three-point
function with the position-dependent correlation from the SDSS-III BOSS DR10 CMASS
sample. Since the integrated three-point function of a given subvolume size depends
on only one separation (unlike the full three-point function which depends on three
separations), the covariance matrix, which is necessary for the statistical interpretation
of the cosmological information, can be well estimated with 600 PTHalos mock catalogs.
This allows the detection of the amplitude of the three-point function of the BOSS
CMASS galaxies at $7.4\sigma$, which can be turned into the constraint on the nonlinear
bias $b_2=0.41\pm0.41$ when combining with the anisotropic clustering and the weak
lensing signal.

We have generalized the study of the response of the small-scale power spectrum
to $m$ long-wavelength overdensities in \refchp{ch6_npt_sq_sep}. This response
can be linked to the angle-average squeezed-limit $(m+2)$-point function, where
two modes have wavenumbers much larger than the other ones. We have used separate
universe simulations, where $N$-body simulations are performed in the presence
of a long-wavelength overdensity by modifying the cosmological parameters, to
test the separate universe approach on fully nonlinear scales to the unprecedented
accuracy. We have also tested the standard perturbation theory hypothesis that
the nonlinear $n$-point function is completely predicted by the linear power
spectrum at the equal time. We have found the discrepancies of 10\% at
$k\simeq0.2-0.5~h~{\rm Mpc}^{-1}$ for five- to three-point functions at $z=0$,
suggesting that the standard perturbation theory fails to describe the correct
dynamics of collisionless particles beyond these wavenumbers, even if it is
calculated to all orders in perturbations.

While the topic of \refchp{ch6_npt_sq_sep} seems somewhat academic because
even measuring the three-point function from galaxy surveys is already
challenging at the moment, the idea of separate universe simulations can
largely alleviate the computational resources for studying nonlinearities
in the squeezed-limit $n$-point functions. That is, we do not need to
perform $N$-body simulations with a huge volume to simulate the mode
coupling between long and short wavelength modes. As nonlinearity due
to gravitational evolution is the dominant contamination for extracting
the primordial non-Gaussianity from the large-scale structure bispectrum,
being able to accurately model the gravity induced bispectrum is currently
the most important challenge in this field. We can better construct and
test the models for the squeezed-limit bispectrum with separate universe
simulations.

The quantum origin of all the structures we observe today is one of the
most amazing ideas in history of physics. Such a bold claim requires careful
investigations and validations. Upcoming galaxy surveys contain data with
unprecedented amount and quality, which allow critical tests of this
paradigm. The soon-to-be-public BOSS DR12 CMASS sample contains approximately
50\% more observed galaxies and volume than the DR10 sample. We have
used the Fisher matrix to show that the BOSS DR12 CMASS sample can
potentially constrain the local-type primordial non-Gaussianity to be
$\sigma(f_{\rm NL})\sim17$ (95\% C.L.). Thus, we plan to apply in the
near future the same technique to the BOSS DR12 CMASS sample, and obtain
better constraints on the nonlinear bias, as well as on the logarithmic
growth rate and the primordial non-Gaussianity.

%% file: anh_01.tex
{

\renewcommand{\v}[1]{\mathbf{#1}}
\newcommand{\vx}{\v{x}}
\newcommand{\vk}{\v{k}}
\newcommand{\vq}{\v{q}}
\newcommand{\vu}{\v{u}}
\newcommand{\vv}{\v{v}}

\newcommand{\los}{\parallel}
\newcommand{\hx}{\hat x}
\newcommand{\hs}{\hat s}
\newcommand{\aH}{\mathcal H}
\newcommand{\tpc}{(2\pi)^3}
\newcommand{\tu}{\tilde{\theta}}
\newcommand{\tv}{\theta}
\newcommand{\dP}{\frac{d\ln P_l(k)}{d\ln k}}
\newcommand{\ds}{\delta_s}
\renewcommand{\[}{\left[}
\renewcommand{\]}{\right]}
\renewcommand{\(}{\left(}
\renewcommand{\)}{\right)}
\newcommand{\<}{\left\langle}
\renewcommand{\>}{\right\rangle}
\renewcommand{\{}{\left\lbrace}
\renewcommand{\}}{\right\rbrace}
\newcommand{\hMpc}{~h^{-1}~{\rm Mpc}}
\newcommand{\ihMpc}{~h~{\rm Mpc}^{-1}}

\chapter{Tree-level redshift-space bispectrum}
\label{app:trz_bi}
In this appendix, we summarize the tree-level redshift-space bispectrum
following \cite{scoccimarro/couchman/frieman:1999,bernardeau/etal:2002}.

\section{Mapping between real and redshift space}
\label{app:trz_mapping}
In redshift space, the observed radial position of an object (galaxy) is the combination
of the Hubble flow and its peculiar velocity, which is known as redshift-space distortion.
The mapping between the real-space position $\vx$ and the redshift-space position $\v{s}$ is
\be
 a\v{s}=a\vx+\frac{\tilde{v}_{\los}(\vx)}{H}\hx_{\los} ~~~{\rm or}~~~
 \v{s}=\vx-fu_{\los}(\vx)\hx_{\los} ~,
\label{eq:trz_s_to_r}
\ee
where $\vx$ and $\v{s}$ are in the comoving coordinates, $\hx_{\los}$ is the line-of-sight
direction, $\tilde{v}_{\los}=\tilde{\vv}\cdot\hx_{\los}$ is the line-of-sight component
of the \emph{physical} peculiar velocity field, $\vu=-\tilde{\vv}/(\aH f)$ is the \emph{rescaled}
peculiar velocity field, and $\aH=aH=a'/a$ is the conformal Hubble parameter with prime
being the derivative with respect to conformal time
\be
 \tau=\int_0^t~\frac{dt'}{a(t')} ~.
\label{eq:trz_tau}
\ee

The density fluctuation in redshift space, $\ds(\v{s})$, is related to the real-space
one, $\delta(\vx)$, by mass conservation, i.e.
\be
 [1+\ds(\v{s})]d^3s=[1+\delta(\vx)]d^3r ~.
\label{eq:trz_ms_cons}
\ee
Since the Jacobian, $J(\vx)=d^3s/d^3r$, is known exactly through \refeq{trz_s_to_r},
the redshift-space density fluctuation can be written as
\be
 \ds(\v{s})=\frac{1+\delta(\vx)}{J(\vx)}-1
 =\frac{\delta(\vx)+f\nabla_{\los}u_{\los}(\vx)}{J(\vx)} ~,
\label{eq:trz_deltas_r}
\ee
where $J(\vx)=1-f\nabla_{\los}u_{\los}(\vx)$ and $\nabla_{\los}\equiv d/dr_{\los}$.
In Fourier space, the redshift-space density fluctuation is
\ba
 \ds(\vk)\:&=\int d^3s~\ds(\v{s})e^{-i\vk\cdot\v{s}}
 =\int d^3s~\frac{\delta(\vx)+f\nabla_{\los}u_{\los}(\vx)}{J(\vx)}
 e^{-i\vk\cdot\[\vx-fu_{\los}(\vx)\hx_{\los}\]} \vs
 \:&=\int d^3r~\[\delta(\vx)+f\nabla_{\los}u_{\los}(\vx)\]
 e^{ifk_{\los}u_{\los}(\vx)}e^{-i\vk\cdot\vx} ~,
\label{eq:trz_deltas_k}
\ea
where $k_{\los}=\vk\cdot\hx_{\los}$. Note that in \refeq{trz_deltas_k} the only
approximation is the plane-parallel approximation, and so it describes the fully
nonlinear density fluctuation in redshift space. The term in the square brackets
describes the so-called ``squashing effect'', i.e. the increase of the clustering
amplitude due to infall into the gravitational potential \cite{kaiser:1987};
the term in the exponent encodes the ``Finger-of-God effect'' which erases power
due to the velocity dispersion along the line-of-sight \cite{jackson:1972}.

To proceed, we define the divergence of the rescaled peculiar velocity field as
$\tu(\vx)\equiv\nabla\cdot\vu(\vx)$, and so
\ba
 \int d^3r~u_{\los}(\vx)e^{-i\vk\cdot\vx}\:&=\frac{-i\vk\cdot\hx_{\los}}{k^2}\tu(\vk)
 =\frac{-i\mu_k}{k}\tu(\vk) \vs
 \int d^3r~\nabla_{\los}u_{\los}(\vx)e^{-i\vk\cdot\vx}\:&=
 \(\frac{\vk\cdot\hx_{\los}}{k}\)^2\tu(\vk)=\mu_k^2\tu(\vk) ~,
\label{eq:trz_tu}
\ea
where $\mu_k\equiv\vk\cdot\hx_{\los}/k=k_{\los}/k$ is the cosine of the angle between
$\vk$ and the line-of-sight. We then perturbatively expand
$e^{ifk_{\los}u_{\los}(\vx)}$ and use \refeq{trz_tu} to get
\ba
 \ds(\vk)\:&=\int d^3r~e^{-i\vk\cdot\vx}\[\delta(\vx)+f\nabla_{\los}u_{\los}(\vx)\]
 \{\sum_{n=0}^{\infty}\frac{\[ifk_{\los}u_{\los}(\vx)\]^n}{n!}\} \vs
 \:&=\int d^3r~e^{-i\vk\cdot\vx}\int\frac{d^3q}{\tpc}~
 \[\delta(\vq)+f\mu_q^2\tu(\vq)\]e^{i\vq\cdot\vx} \vs
 &~~~~\times\Bigg[1+\sum_{n=1}^{\infty}\frac{(if\mu_kk)^n}{(n)!}
 \int\frac{d^3q_1}{\tpc}\cdots\int\frac{d^3q_n}{\tpc} \vs
 &\hspace{5cm}\(-i\frac{\mu_{q_1}}{q_1}\)\tu(\vq_1)\cdots
 \(-i\frac{\mu_{q_n}}{q_n}\)\tu(\vq_n)e^{i(\vq_1+\cdots+\vq_n)\cdot\vx}\Bigg] \vs
 \:&=\[\delta(\vk)+f\mu_k^2\tu(\vk)\] \vs
 &~~~~+\sum_{n=2}^{\infty}\int\frac{d^3q_1}{\tpc}\cdots\int\frac{d^3q_n}{\tpc}~
 \[\delta(\vq_1)+f\mu_{q_1}^2\tu(\vq_1)\]\[\delta_D\]_n \vs
 &\hspace{5.5cm}\times\frac{(f\mu_kk)^{n-1}}{(n-1)!}
 \frac{\mu_{q_2}}{q_2}\tu(\vq_2)\cdots\frac{\mu_{q_n}}{q_n}\tu(\vq_n) ~,
\ea
where $\[\delta_D\]_n\equiv\tpc\delta_D(\vk-\vq_1-\cdots-\vq_n)$.

\section{Redshift-space kernel}
\label{app:trz_kernel}
To obtain $\ds(\vk)$, we need to solve the velocity divergence. Since we are
interested in scales smaller than the Jeans length, we shall treat dark matter
and baryons as pressureless fluid. Moreover, as the peculiar velocity is much
smaller than the speed of light and the scales of density fluctuations are much
smaller than the horizon size, the system can be treated by Newtonian dynamics.
The equations of the system are (see, e.g. \cite{jeong/komatsu:2006})
\ba
 \:&\delta'+\nabla\cdot[(1+\delta)\vv]=0 ~, \vs
 \:&\vv'+\(\vv\cdot\nabla\)\vv=-\aH-\nabla\phi ~, \vs
 \:&\nabla^2\phi=4\pi Ga^2\bar\rho\delta ~,
\label{eq:trz_newtonian}
\ea
where $\vv=d\vx/d\tau=\tilde{\vv}$ is the peculiar velocity field in the \emph{comoving
coordinate with conformal time} and is equivalent to the physical peculiar velocity field,
$\phi$ is the peculiar gravitational potential from density fluctuations and $\bar\rho$
is the mean density. Combining \refeq{trz_newtonian} and the Friedmann equation (i.e.
$4\pi G\bar\rho(\tau)=\frac32H^2\Omega_m(\tau)$), the continuity equation (the first
line in \refeq{trz_newtonian}) and the Euler equation (the second line in \refeq{trz_newtonian})
in Fourier space can be written as
\ba
 \:&\delta'(\vk,\tau)+\tv(\vk,\tau)=-\int\frac{d^3k_1}{\tpc}\int\frac{d^3k_1}{\tpc}~
 [\delta_D]_2\frac{\vk\cdot\vk_1}{k_1^2}\delta(\vk_2,\tau)\tv(\vk_1,\tau) ~, \vs
 \:&\tv'(\vk,\tau)+\aH\tv(\vk,\tau)+\frac32\aH^2\Omega_m(\tau)\delta(\vk,\tau)
 =-\int\frac{d^3k_1}{\tpc}\int\frac{d^3k_1}{\tpc}~[\delta_D]_2
 \frac{k^2(\vk_1\cdot\vk_2)}{2k_1^2k_2^2}\tv(\vk_1,\tau)\tv(\vk_2,\tau) ~,
\label{eq:trz_con_eul}
\ea
where $\tv\equiv\nabla\cdot\vv$ is the comoving velocity divergence. Note that
$\tv=-aHf\tu$.

To proceed further, we assume that the universe is Einstein de-Sitter, i.e.
$\Omega_m(\tau)=1$ and $a(\tau)\propto\tau^2$. Then, $\delta$ and $\tv$
can be solved perturbatively as \cite{fry:1984,goroff/etal:1986}
\ba
 \delta(\vk,\tau)\:&=\sum_{n=1}^{\infty}a^n(\tau)\int\frac{d^3q_1}{\tpc}\cdots
 \int\frac{d^3q_n}{\tpc}~[\delta_D]_nF_n^{\rm unsym}(\vq_1,\cdots,\vq_n)
 \delta_l(\vq_1)\cdots\delta_l(\vq_n) ~, \vs
 \tv(\vk,\tau)\:&=-\sum_{n=1}^{\infty}a'(\tau)a^{n-1}(\tau)\int\frac{d^3q_1}{\tpc}
 \cdots\int\frac{d^3q_n}{\tpc}~[\delta_D]_n
 G_n^{\rm unsym}(\vq_1,\cdots,\vq_n)\delta_l(\vq_1)\cdots\delta_l(\vq_n) ~,
\label{eq:trz_d_t_1}
\ea
where $\delta_l$ is the linear density field, $F_n^{\rm unsym}$ and $G_n^{\rm unsym}$
are the (unsymmetrized) kernels of \refeq{trz_con_eul}, and the recursion relation
can be found in \cite{jain/bertschinger:1994,bernardeau/etal:2002}. Strictly speaking,
\refeq{trz_d_t_1} is only valid in Einstein de-Sitter universe, but a good approximation
is to replace $a=D$ and $a'=D^2Hf$ (see appendix~B.3 in \cite{scoccimarro/etal:1998}),
where $D$ is the linear growth factor and $f=d\ln D/d\ln a$ is the logarithmic growth
rate, and \refeq{trz_d_t_1} can be written as
\ba
 \delta(\vk,\tau)\:&=\sum_{n=1}^{\infty}D^n(\tau)\int\frac{d^3q_1}{\tpc}\cdots
 \int\frac{d^3q_n}{\tpc}~[\delta_D]_nF_n^{\rm unsym}(\vq_1,\cdots,\vq_n)
 \delta_l(\vq_1)\cdots\delta_l(\vq_n) ~, \vs
 \tu(\vk,\tau)\:&=\sum_{n=1}^{\infty}D^n(\tau)\int\frac{d^3q_1}{\tpc}
 \cdots\int\frac{d^3q_n}{\tpc}~[\delta_D]_n
 G_n^{\rm unsym}(\vq_1,\cdots,\vq_n)\delta_l(\vq_1)\cdots\delta_l(\vq_n) ~.
\label{eq:trz_d_t_2}
\ea
One can then define
\ba
 \delta_n(\vk,\tau)\:&=D^n(\tau)\int\frac{d^3q_1}{\tpc}\cdots\int\frac{d^3q_n}{\tpc}~
 [\delta_D]_nF^{\rm unsym}_n(\vq_1,\cdots,\vq_n)\delta_l(\vq_1)\cdots\delta_l(\vq_n) ~, \vs
 \tu_n(\vk,\tau)\:&=D^n(\tau)\int\frac{d^3q_1}{\tpc}\cdots\int\frac{d^3q_n}{\tpc}~
 [\delta_D]_nG^{\rm unsym}_n(\vq_1,\cdots,\vq_n)\delta_l(\vq_1)\cdots\delta_l(\vq_n)
\label{eq:trz_d_t_3}
\ea
such that $\delta(\vk,\tau)=\sum_{n=1}^{\infty}\delta_n(\vk,\tau)$ and $\tu(\vk,\tau)
=\sum_{n=1}^{\infty}\tu_n(\vk,\tau)$, and thus $\delta_n$ and $\tu_n$ are the $n^{\rm th}$
order in linear density field.

For observation, however, one cannot probe dark matter directly, but only the biased
tracers (e.g. halos or galaxies). Let us assume that the halo (galaxy) density fluctuations
can be parametrized by the local bias model as \cite{fry/gaztanaga:1993}
\be
 \delta_h(\vx)=\sum_{n=0}\frac{b_n}{n!}\delta^n(\vx) ~,
\label{eq:trz_local_bias}
\ee
where $b_n$ are local bias parameters and $b_0$ assures $\langle\delta_h\rangle=0$.
On the other hand, because of the conservation of mass and momentum, we assume that
the halo peculiar velocity field is identical to the underlying matter peculiar velocity
field, i.e. $\tu_h=\tu$ (see e.g. \cite{chan/scoccimarro/sheth:2012}).\footnote{Note
however, that using $N$-body simulations \cite{baldauf/desjacques/seljak:2014} recently
shows the evidence for linear \emph{statistical} halo velocity bias which remains constant
with time, as predicted by the peak model \cite{desjacques/sheth:2010,desjacques/etal:2010}.
It is argued in \cite{baldauf/desjacques/seljak:2014,biagetti/etal:2014} that the Euler
equation has to be modified to correctly describe the coevolution between dark matter
and halos.} The redshift-space halo density fluctuations can thereby be written as
\ba
 \delta_{h,s}(\vk)=\:&\sum_{n=1}^{\infty}\int\frac{d^3q_1}{\tpc}\int\frac{d^3q_2}{\tpc}\cdots
 \int\frac{d^3q_n}{\tpc}~[\delta_D]_n[\delta_h(\vq_1)+f\mu_{q_1}^2\tu(\vq_1)] \vs
 &\hspace{5.5cm}\times\frac{(f\mu_kk)^{n-1}}{(n-1)!}\frac{\mu_{q_2}}{q_2}\tu(\vq_2)\cdots\frac{\mu_{q_n}}{q_n}\tu(\vq_n) ~.
\label{eq:trz_dhs_1}
\ea
Combining \refeqs{trz_d_t_2}{trz_dhs_1}, one can expand $\delta_{h,s}(\vk)$ as
\ba
 \delta_{h,s}(\vk)\:&=[b_1\delta_1(\vk)+f\mu_k^2\tu_1(\vk)] \vs
 \:&+\Bigg\lbrace\[b_1\delta_2(\vk)+\frac{b_2}{2}\int\frac{d^3q}{\tpc}~
 \delta_1(\vq)\delta_1(\vk-\vq)+f\mu_k^2\tu_2(\vk)\] \vs
 \:&+\int\frac{d^3q_1}{\tpc}\int\frac{d^3q_2}{\tpc}~[\delta_D]_2
 [b_1\delta_1(\vq_1)+f\mu_{q_1}^2\tu_1(\vq_1)](f\mu_kk)\frac{\mu_{q_2}}{q_2}\tu_1(\vq_2)\Bigg\rbrace \vs
 \:&+\cdots ~,
\label{eq:trz_dhs_2}
\ea
for which we expand up to the second order of $\delta_l$.

For simplicity, one may define the redshift-space kernel $Z_n(\vq_1,\cdots,\vq_n)$ such that
\be
 \delta_{h,s}(\vk,\tau)=\sum_{n=1}^{\infty}D^n(\tau)
 \int\frac{d^3q_1}{(2\pi)^3}\cdots\int\frac{d^3q_n}{(2\pi)^3}~[\delta_D]_n
 Z_n(\vq_1,\cdots,\vq_n)\delta_l(\vq_1)\cdots\delta_l(\vq_n)
\label{eq:trz_dhs_n}
\ee
with
\ba
 Z_1(\vq_i)\:&=b_1+f\mu_{q_i}^2 \vs
 Z_2(\vq_1,\vq_2)\:&=b_1F_2(\vq_1,\vq_2)+\frac{b_2}{2}+f\tilde\mu^2G_2(\vq_1,\vq_2)
 +\frac{f\tilde\mu\tilde q}{2}\[\frac{\mu_{q_1}}{q_1}(b_1+f\mu_{q_2}^2)
 +\frac{\mu_{q_2}}{q_2}(b_1+f\mu_{q_1}^2)\] ~,
\label{eq:trz_zn}
\ea
where $\tilde\mu=\tilde\vq\cdot\hx_{\los}/\tilde q$ with $\tilde\vq=\vq_1+\cdots+\vq_n$
at the $n^{\rm th}$ order, and $F_n$ and $G_n$ are the symmetrized kernels and obtained
by taking the mean of all possible permutations of the unsymmetrized kernels.
\refEqs{trz_dhs_n}{trz_zn} are the main results of this appendix.

The redshift-space halo bispectrum is defined as
\be
 \langle\delta_{h,s}(\vk_1)\delta_{h,s}(\vk_2)\delta_{h,s}(\vk_3)\rangle
 =(2\pi)^3\delta_D(\vk_1+\vk_2+\vk_3)B_{h,s}(\vk_1,\vk_2,\vk_3) ~.
\label{eq:trz_bhz_1}
\ee
Using the redshift-space kernels, one obtains the leading order terms as
\be
 B_{h,s}(\vk_1,\vk_2,\vk_3)=
 2[Z_1(\vk_1)Z_1(\vk_2)Z_2(\vk_1,\vk_2)P_l(k_1)P_l(k_2)+2~{\rm cyclic}] ~,
\label{eq:trz_bhz_2}
\ee
in which we use the fact that $Z_2(-\vk_1,-\vk_2)=Z_2(\vk_1,\vk_2)$ and $\delta_l$ is Gaussian
so that $\langle\delta_l(\vk_1) \delta_l(\vk_2)\rangle=(2\pi)^3\delta_D(\vk_1+\vk_2)P_l(k_1)$
and $\langle\delta_l^n \rangle=0$ for odd $n$.

It is useful to separate $B_{h,s}$ into four categories: the linear squashing
terms (SQ1) which are proportional to $F_2(\vk_1,\vk_2)$; the second-order
squashing terms (SQ2) which are proportional to $G_2(\vk_1,\vk_2)$; the nonlinear
bias terms (NLB) which are related to $b_2$; the damping terms (FOG) due to the
velocity dispersion which are not related to $F_2(\vk_1,\vk_2)$, $G_2(\vk_1,\vk_2)$,
and $b_2$. That is, $B_{h,s}=B_{\rm SQ1}+B_{\rm SQ2}+B_{\rm NLB}+B_{\rm FOG}$ such that
\ba
 B_{\rm SQ1}(\vk_1,\vk_2,\vk_3)\:&=b_1^3\sum_{i=1}^3\beta^{i-1}
 B_{{\rm SQ1},i}(\vk_1,\vk_2,\vk_3) ~, \vs
 B_{\rm SQ2}(\vk_1,\vk_2,\vk_3)\:&=b_1^3\beta\sum_{i=1}^3\beta^{i-1}
 B_{{\rm SQ2},i}(\vk_1,\vk_2,\vk_3) ~, \vs
 B_{\rm NLB}(\vk_1,\vk_2,\vk_3)\:&=b_1^2b_2\sum_{i=1}^3\beta^{i-1}
 B_{{\rm NLB},i}(\vk_1,\vk_2,\vk_3) ~, \vs
 B_{\rm FOG}(\vk_1,\vk_2,\vk_3)\:&=b_1^4\beta[B_{{\rm FOG},1}(\vk_1,\vk_2,\vk_3)
 +\beta B_{{\rm FOG},2}(\vk_1,\vk_2,\vk_3)+\beta B_{{\rm FOG},3}(\vk_1,\vk_2,\vk_3) \vs
 \:&\hspace{0.8cm}+\beta^2 B_{{\rm FOG},4}(\vk_1,\vk_2,\vk_3)
 +\beta^2 B_{{\rm FOG},5}(\vk_1,\vk_2,\vk_3)+\beta^3 B_{{\rm FOG},6}(\vk_1,\vk_2,\vk_3)] ~,
\label{eq:trz_brsd}
\ea
where $\beta=f/b_1$. The explicit expressions are: for SQ1,
\ba
 B_{{\rm SQ1},1}(\vk_1,\vk_2,\vk_3)\:&=2[F_2(\vk_1,\vk_2)P_l(k_1)P_l(k_2)+2~{\rm cyclic}] ~, \vs
 B_{{\rm SQ1},2}(\vk_1,\vk_2,\vk_3)\:&=2[(\mu_1^2+\mu_2^2)F_2(\vk_1,\vk_2)P_l(k_1)P_l(k_2)+2~{\rm cyclic}] ~, \vs
 B_{{\rm SQ1},3}(\vk_1,\vk_2,\vk_3)\:&=2[\mu_1^2\mu_2^2F_2(\vk_1,\vk_2)P_l(k_1)P_l(k_2)+2~{\rm cyclic}] ~;
\label{eq:trz_bsq1}
\ea
for SQ2,
\ba
 B_{{\rm SQ2},1}(\vk_1,\vk_2,\vk_3)\:&=2[\mu^2G_2(\vk_1,\vk_2)P_l(k_1)P_l(k_2)+2~{\rm cyclic}] ~, \vs
 B_{{\rm SQ2},2}(\vk_1,\vk_2,\vk_3)\:&=2[(\mu_1^2+\mu_2^2)\mu^2G_2(\vk_1,\vk_2)P_l(k_1)P_l(k_2)+2~{\rm cyclic}] ~, \vs
 B_{{\rm SQ2},3}(\vk_1,\vk_2,\vk_3)\:&=2[\mu_1^2\mu_2^2\mu^2G_2(\vk_1,\vk_2)P_l(k_1)P_l(k_2)+2~{\rm cyclic}] ~;
\label{eq:trz_bsq2}
\ea
for NLB,
\ba
 B_{{\rm NLB},1}(\vk_1,\vk_2,\vk_3)\:&=P_l(k_1)P_l(k_2)+2~{\rm cyclic} ~, \vs
 B_{{\rm NLB},2}(\vk_1,\vk_2,\vk_3)\:&=(\mu_1^2+\mu_2^2)P_l(k_1)P_l(k_2)+2~{\rm cyclic} ~, \vs
 B_{{\rm NLB},3}(\vk_1,\vk_2,\vk_3)\:&=\mu_1^2\mu_2^2P_l(k_1)P_l(k_2)+2~{\rm cyclic} ~;
\label{eq:trz_bnlb}
\ea
for FOG,
\ba
 B_{{\rm FOG},1}(\vk_1,\vk_2,\vk_3)\:&=k\mu\(\frac{\mu_1}{k_1}+\frac{\mu_2}{k_2}\)
 P_l(k_1)P_l(k_2)+2~{\rm cyclic} ~, \vs
 B_{{\rm FOG},2}(\vk_1,\vk_2,\vk_3)\:&=2\[k\mu\mu_1\mu_2\(\frac{\mu_2}{k_1}+\frac{\mu_1}{k_2}\)
 P_l(k_1)P_l(k_2)+2~{\rm cyclic}\] ~, \vs
 B_{{\rm FOG},3}(\vk_1,\vk_2,\vk_3)\:&=k\mu\(\frac{\mu_1^3}{k_1}+\frac{\mu_2^3}{k_2}\)
 P_l(k_1)P_l(k_2)+2~{\rm cyclic} ~, \vs
 B_{{\rm FOG},4}(\vk_1,\vk_2,\vk_3)\:&=2\[k\mu\mu_1^2\mu_2^2\(\frac{\mu_1}{k_1}+\frac{\mu_2}{k_2}\)
 P_l(k_1)P_l(k_2)+2~{\rm cyclic}\] ~, \vs
 B_{{\rm FOG},5}(\vk_1,\vk_2,\vk_3)\:&=k\mu\mu_1\mu_2\(\frac{\mu_2^3}{k_1}+\frac{\mu_1^3}{k_2}\)
 P_l(k_1)P_l(k_2)+2~{\rm cyclic} ~, \vs
 B_{{\rm FOG},6}(\vk_1,\vk_2,\vk_3)\:&=k\mu\mu_1^3\mu_2^3\(\frac{\mu_2}{k_1}+\frac{\mu_1}{k_2}\)
 P_l(k_1)P_l(k_2)+2~{\rm cyclic} ~.
\label{eq:trz_bfog}
\ea
Note that in \refeqs{trz_bsq1}{trz_bfog} we shall simplify the notations
$\mu_n=\mu_{k_n}$ and $\tilde\mu=\mu$ for clarity.

\section{Tree-level redshift-space integrated bispectrum in the squeezed-limit}
\label{app:trz_sq}
Let us define the integrated bispectrum of each component as
\ba
 iB_{{\rm X},i}(k)=\:&\frac{1}{V_L^2}\int\frac{d^2\hat{k}}{4\pi}
 \int\frac{d^3q_a}{(2\pi)^3}\int\frac{d^3q_b}{(2\pi)^3}~
 B_{{\rm X},i}(\vk-\vq_a,-\vk+\vq_a+\vq_b,-\vq_b) \vs
 &\hspace{5cm}\times W_L(\vq_a)W_L(-\vq_a-\vq_b)W_L(\vq_b) ~,
\label{eq:trz_ib_x}
\ea
where $X$ refers to SQ1, SQ2, NLB, or FOG. We numerically evaluate all the components
of the integrated bispectrum through \refeq{trz_ib_x} and \refeqs{trz_bsq1}{trz_bfog}.
\refFig{trz_ib_zspace} shows the ratios of the components at $z=0$ with $L=200\hMpc$.
The left and middle panels of \reffig{trz_ib_zspace} show $iB_{{\rm X},j}(k)/iB_{{\rm X},1}(k)$
for X=SQ1, SQ2, NLB, and FOG with $j=2$ and 3; the right panel shows $iB_{{\rm X},1}(k)/iB_{{\rm SQ1},1}(k)$
for X=SQ2, NLB, and FOG. We find that the ratios become quite scale-independent for
$k\gtrsim0.5\ihMpc$, indicating that all components have very similar scale-dependencies
when the squeezed limit is reached ($k\gg1/L=0.005\ihMpc$). Note that in principle the
ratios depend on the window function and the power spectrum. Fortunately, the anisotropy
of the window function can be neglected in the squeezed limit.

\begin{figure}[t!]
\centering
\includegraphics[width=1\textwidth]{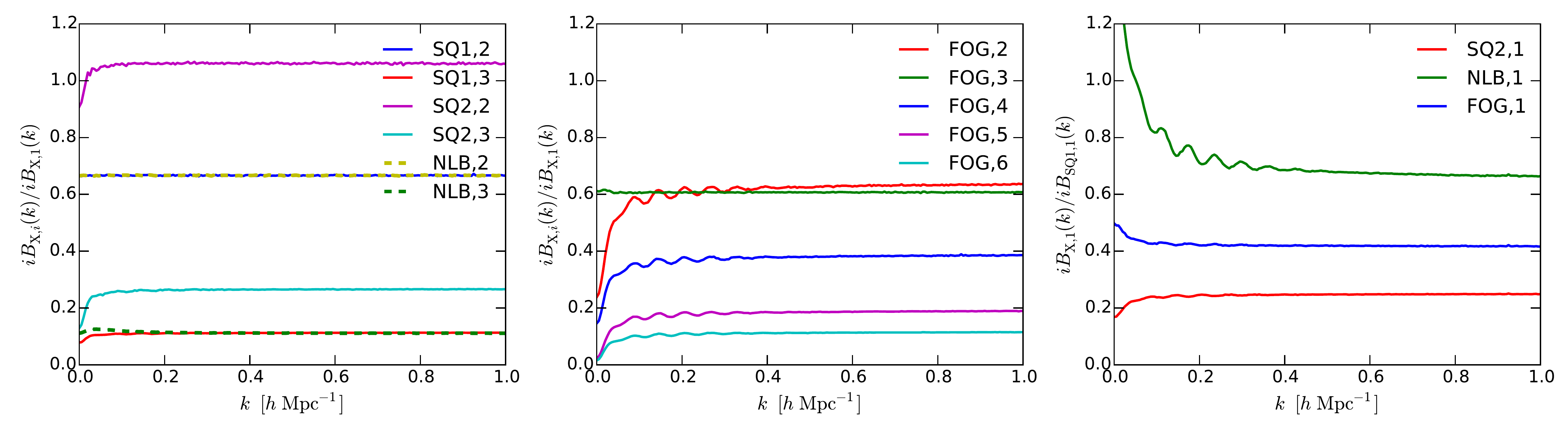}
\caption[Tree-level redshift-space integrated bispectrum]
{Ratios of the components of the tree-level redshift-space halo bispectrum at $z=0$
with $L=200\hMpc$, and the line-of-sight direction is $\hat z$. Left and middle
panels: $iB_{{\rm X},j}(k)/iB_{{\rm X},1}(k)$ for X=SQ1, SQ2, NLB, and FOG, and $j=2$
and 3. Right panel: $iB_{{\rm X},1}(k)/iB_{{\rm SQ1},1}(k)$ for X=SQ2, NLB, and FOG.}
\label{fig:trz_ib_zspace}
\end{figure}

To understand the similar scale-dependences for different terms of the tree-level
redshift-space integrated bispectrum in the squeezed limit, let us now consider
the three wavenumbers to be $\vk_1=\vk-\vq_a$, $\vk_2=-\vk+\vq_a+\vq_b$, and
$\vk_3=\vq_b$ for the $k$-configuration of the integrated bispectrum. In the
squeezed limit, where $k\gg q_a,q_b$, we can expand all quantities in series of
$(q_a/k)$ and $(q_b/k)$, and in the leading order (up to $(q_{a,b}/k)^0$) we get
\ba
 \:&k_1=k\(1-\mu_{ak}\frac{q_a}{k}\) ~,~~
 k_2=k\(1-\mu_{ak}\frac{q_a}{k}-\mu_{bk}\frac{q_b}{k}\) ~,~~
 k_3=q_b ~, \vs
 \:&P_l(k_1)=P_l(k)\[1-\frac{q_a\mu_{ak}}{k}\dP\] ~,~~
 P_l(k_2)=P_l(k)\[1-\frac{q_a\mu_{ak}+q_b\mu_{bk}}{k}\dP\] ~, \vs
 \:&P_l(k_3)=P_l(q_3) ~,~~  F_2(\vk_1,\vk_2)=0 ~,~~ G_2(\vk_1,\vk_2)=0 ~,\vs
 \:&F_2(\vk_1,\vk_3)=\frac{5}{7}-\frac{k\mu_{bk}-q_a\mu_{ab}}{2q_b}+\frac{2}{7}\mu_{bk}^2 ~,~~
 F_2(\vk_2,\vk_3)=\frac{3}{14}+\frac{k\mu_{bk}-q_a\mu_{ab}}{2q_b}+\frac{2}{7}\mu_{bk}^2 ~, \vs
 \:&G_2(\vk_1,\vk_3)=\frac{3}{7}-\frac{k\mu_{bk}-q_a\mu_{ab}}{2q_b}+\frac{4}{7}\mu_{bk}^2 ~,~~
 G_2(\vk_2,\vk_3)=\frac{-1}{14}+\frac{k\mu_{bk}-q_a\mu_{ab}}{2q_b}+\frac{4}{7}\mu_{bk}^2 ~,\vs
 \:&\mu_1=\mu_k+\frac{q_a\mu_{ak}\mu_k-q_a\mu_a}{k} ~,~~
 \mu_2=-\mu_k+\frac{q_a\mu_{ak}+q_b\mu_{bk}-q_a\mu_a\mu_k-q_b\mu_b\mu_k}{k} ~,~~ \mu_3=-\mu_b ~,
\label{eq:trz_series}
\ea
where $\mu_k=\hat{k}\cdot\hx_{\los}$, $\mu_a=\hat{q}_a\cdot\hx_{\los}$,
$\mu_b=\hat{q}_b\cdot\hx_{\los}$, $\mu_{bk}=\hat{q}_b\cdot\hat{k}$, and
$\mu_{ab}=\hat{q}_a\cdot\hat{q}_b$. The redshift-space bispectrum can
then be expanded in series of $(q_a/k)$ and $(q_b/k)$ as (at the leading
order up to $(q_{a,b}/k)^0$)
\ba
 B_{\rm SQ1,1,sq}\:&=P_l(k)P_l(q_b)\[\frac{13}{7}+\frac{8}{7}\mu_{bk}^2-\dP\mu_{bk}^2\] ~, \vs
 B_{\rm SQ1,2,sq}\:&=P_l(k)P_l(q_b)\bigg[\frac{13}{7}\mu_k^2+\frac{22}{7}\mu_{bk}^2\mu_k^2
 -2\mu_{bk}\mu_k\mu_b+\frac{13}{7}\mu_b^2+\frac{8}{7}\mu_{bk}^2\mu_b^2 \vs
 &\hspace{5cm}-\dP(\mu_{bk}^2\mu_k^2+\mu_{bk}^2\mu_b^2)\bigg] ~, \vs
 B_{\rm SQ1,3,sq}\:&=P_l(k)P_l(q_b)\[\frac{13}{7}\mu_k^2\mu_b^2+\frac{22}{7}\mu_{bk}^2\mu_k^2\mu_b^2
 -2\mu_{bk}\mu_k\mu_b^3-\dP\mu_{bk}^2\mu_k^2\mu_b^2\] ~, \vs
 B_{\rm SQ2,1,sq}\:&=P_l(k)P_l(q_b)\[\frac{5}{7}\mu_k^2+\frac{2}{7}\mu_{bk}^2\mu_k^2
 +2\mu_{bk}\mu_k\mu_b-\dP\mu_{bk}^2\mu_b^2\] ~, \vs
 B_{\rm SQ2,2,sq}\:&=P_l(k)P_l(q_b)\bigg[\frac{5}{7}\mu_k^4+\frac{16}{7}\mu_{bk}^2\mu_k^4
 +\frac{5}{7}\mu_k^2\mu_b^2+\frac{2}{7}\mu_{bk}^2\mu_k^2\mu_b^2+2\mu_{bk}\mu_k\mu_b^3 \vs
 &\hspace{5cm}-\dP(\mu_{bk}^2\mu_k^4+\mu_{bk}^2\mu_k^2\mu_b^2)\bigg] ~, \vs
 B_{\rm SQ2,3,sq}\:&=P_l(k)P_l(q_b)\[\frac{5}{7}\mu_k^4\mu_b^2+\frac{16}{7}\mu_{bk}^2\mu_k^4\mu_b^2-\dP\mu_{bk}^2\mu_k^4\mu_b^2\] ~, \vs
 B_{\rm NLB,1,sq}\:&=P_l(k)P_l(q_b)[2] ~, \vs
 B_{\rm NLB,2,sq}\:&=P_l(k)P_l(q_b)[2\mu_k^2+2\mu_b^2] ~, \vs
 B_{\rm NLB,3,sq}\:&=P_l(k)P_l(q_b)[2\mu_k^2\mu_b^2] ~, \vs
 B_{\rm FOG,1,sq}\:&=P_l(k)P_l(q_b)\[2\mu_k^2+\mu_b^2-\dP\mu_{bk}\mu_k\mu_b\] ~, \vs
 B_{\rm FOG,2,sq}\:&=P_l(k)P_l(q_b)\[4\mu_{bk}\mu_k^3\mu_b+2\mu_k^2\mu_b^2-2\dP\mu_{bk}\mu_k^3\mu_b\] ~, \vs
 B_{\rm FOG,3,sq}\:&=P_l(k)P_l(q_b)\[2\mu_k^4+\mu_b^4-\dP\mu_{bk}\mu_k\mu_b^3\] ~, \vs
 B_{\rm FOG,4,sq}\:&=P_l(k)P_l(q_b)\[4\mu_k^4\mu_b^2+4\mu_{bk}\mu_k^3\mu_b^3
 -2\mu_k^2\mu_b^4-2\dP\mu_{bk}\mu_k^3\mu_b^3\] ~, \vs
 B_{\rm FOG,5,sq}\:&=P_l(k)P_l(q_b)\[4\mu_{bk}\mu_k^5\mu_b-3\mu_k^4\mu_b^2
 +2\mu_k^2\mu_b^4-\dP\mu_{bk}\mu_k^5\mu_b\] ~, \vs
 B_{\rm FOG,6,sq}\:&=P_l(k)P_l(q_b)\[4\mu_{bk}\mu_k^5\mu_b^3-\mu_k^4\mu_b^4-\dP\mu_{bk}\mu_k^5\mu_b^3\] ~,
\label{eq:trz_bz_series}
\ea
where $\mu_k=\hat{k}\cdot\hx_{\los}$. Note that \refeq{trz_bz_series} is
independent of $\vq_a$, so we can simplify the integrated bispectrum as
\ba
 iB(k)=\:&\frac{1}{V_L^2}\int\frac{d^2\hat k}{4\pi}\int\frac{d^3q_a}{(2\pi)^3}
 \int\frac{d^3q_b}{(2\pi)^3}~B(\vk-\vq_a,-\vk+\vq_a+\vq_b,-\vq_b) \vs
 &\hspace{5cm}\times W_L(\vq_a)W_L(-\vq_a-\vq_b)W_L(\vq_b) \vs
 \stackrel{k \gg q_a,q_b}{=}\:&\frac{1}{V_L^2}\int\frac{d^2\hat k}{4\pi}
 \int\frac{d^3q_b}{(2\pi)^3}~B(\vk-\vq_a,-\vk+\vq_a+\vq_b,-\vq_b)W_L^2(\vq_b) ~,
\ea
where we use the fact that
\be
 \int\frac{d^3q_a}{(2\pi)^3}~W_L(\vq_a)W_L(-\vq_a-\vq_b)=W_L(\vq_b) ~.
\label{eq:trz_WL}
\ee

If the window function is isotropic, we can angular average both $\hat{k}$
and $\hat{q}_b$ over \refeq{trz_bz_series} and obtain
\ba
 B_{\rm SQ1,1,sq}\:&=P_l(k)P_l(q_b)\[\frac{47}{21}-\frac13\dP\] ~,~
 B_{\rm SQ1,2,sq}=P_l(k)P_l(q_b)\[\frac{94}{63}-\frac29\dP\] ~, \vs
 B_{\rm SQ1,3,sq}\:&=P_l(k)P_l(q_b)\[\frac{17}{75}-\frac{11}{225}\dP\] ~,~
 B_{\rm SQ2,1,sq}=P_l(k)P_l(q_b)\[\frac{31}{63}-\frac19\dP\] ~, \vs
 B_{\rm SQ2,2,sq}\:&=P_l(k)P_l(q_b)\[\frac{274}{525}-\frac{26}{225}\dP\] ~, \vs
 B_{\rm SQ2,3,sq}\:&=P_l(k)P_l(q_b)\[\frac{149}{1225}-\frac{17}{525}\dP\] ~,~
 B_{\rm NLB,1,sq}=P_l(k)P_l(q_b)\[2\] ~, \vs
 B_{\rm NLB,2,sq}\:&=P_l(k)P_l(q_b)\[\frac43\] ~,~
 B_{\rm NLB,3,sq}=P_l(k)P_l(q_b)\[\frac29\] ~, \vs
 B_{\rm FOG,1,sq}\:&=P_l(k)P_l(q_b)\[1-\frac19\dP\] ~,~
 B_{\rm FOG,2,sq}=P_l(k)P_l(q_b)\[\frac{22}{45}-\frac{2}{15}\dP\] ~, \vs
 B_{\rm FOG,3,sq}\:&=P_l(k)P_l(q_b)\[\frac{3}{5}-\frac{1}{15}\dP\] ~,~
 B_{\rm FOG,4,sq}=P_l(k)P_l(q_b)\[\frac{22}{75}-\frac{2}{25}\dP\] ~, \vs
 B_{\rm FOG,5,sq}\:&=P_l(k)P_l(q_b)\[\frac{13}{105}-\frac{1}{21}\dP\] ~,~
 B_{\rm FOG,6,sq}=P_l(k)P_l(q_b)\[\frac{13}{175}-\frac{1}{35}\dP\] ~.
\label{eq:trz_bz_sq_ang_avg}
\ea
Using \refeq{trz_bz_sq_ang_avg}, the integrated bispectrum can further be simplified as
\be
 iB(k)\stackrel{k \gg q_a,q_b}{=}P_l(k)\sigma_{l,L}^2H_{\rm X}(k) ~,
\label{eq:trz_ibz_sq_ang_avg}
\ee
where $\sigma_{l,L}^2=\frac{1}{V_L^2}\int\frac{dq_b}{2\pi^2}~q_b^2P_l(q_b)W_L^2(q_b)$
and $H_{\rm X}(k)$ corresponds to the terms of the four categories in the square brackets
in \refeq{trz_bz_sq_ang_avg}. For $n_s=0.95$, in the squeezed limit $\dP\sim n_s-4=-3.05$.
Plugging in the value of $\dP$, we find that the analytical (\refeq{trz_bz_sq_ang_avg})
and numerical (\reffig{trz_ib_zspace}) results agree well.

}

%% file: anh_02.tex
{

\renewcommand{\v}[1]{\mathbf{#1}}
\newcommand{\vx}{\v{x}}
\newcommand{\vr}{\v{r}}
\newcommand{\vk}{\v{k}}
\newcommand{\vq}{\v{q}}

\renewcommand{\d}{\delta}
\newcommand{\hMpc}{~h^{-1}~{\rm Mpc}}
\newcommand{\ihMpc}{~h~{\rm Mpc}^{-1}}
\newcommand{\hd}{\hat{\delta}}
\newcommand{\hdb}{\hat{\bar\delta}}
\newcommand{\hP}{\hat{P}}
\newcommand{\hib}{\hat{iB}}

\newcommand{\Dr}{\Delta r}
\newcommand{\Dk}{\Delta k}
\newcommand{\vl}{\v{l}}
\newcommand{\vm}{\v{m}}
\newcommand{\vn}{\v{n}}
\newcommand{\vj}{\v{j}}
\newcommand{\vL}{\v{L}}
\newcommand{\vJ}{\v{J}}
\newcommand{\vS}{\v{S}}
\newcommand{\vg}{\v{g}}
\newcommand{\tpc}{(2\pi)^3}

\chapter{Variance of the integrated bispectrum estimator}
\label{app:ib_var}
In this section we compute the variance of the integrated bispectrum estimator.
In a survey/simulation with volume $V_r$, we first estimate the mean overdensity
and the position-dependent power spectrum in the $j^{\rm th}$ subvolume with volume
$V_L$ by
\be
 \hdb_j=\frac{1}{N_{rL}}\sum_{\vl\Dr\in V_{Lj}}\delta_{r,\vl} ~,~~
 \hP_j(k)=\frac{1}{V_LN_{kL}}\sum_{|\vl k_{FL}|\in k\pm\Delta k/2}
 \delta_{kj,\vl}\delta_{kj,\vl}^* ~,
\label{eq:ibv_bd_dj_est1}
\ee
where $N_{rL}$ is the number of meshes in $V_{Lj}$, $\Dr=(V_L/N_{rL})^{1/3}$ is the
mesh size, $\delta_{r,\vl}$ is the discrete density fluctuation field at the integer
vector $\vl$, $N_{kL}$ is the number of Fourier modes in $(k-\Dk/2,k+\Dk/2)$ of $V_L$,
$k_{FL}=2\pi/L$ is the fundamental frequency of $V_L$, and $\delta_{kj}$ is the local
Fourier transformation of the density fluctuation field in the $j^{\rm th}$ sub-volume.
The estimated integrated bispectrum is then
\be
 \hib_L(k)=\frac1{N_{\rm cut}^3}\sum_{j=1}^{N_{\rm cut}^3}\hdb_j\hP_j(k) ~,
\label{eq:ibv_ib_est}
\ee
where $N_{\rm cut}^3=V_r/V_L$ is the number of subvolumes.

We can rewrite \refeq{ibv_bd_dj_est1} using the window function $W_{rj,\vl}$
($W_{rj,\vl}=1$ if $\vl\Dr\in V_{Lj}$ and 0 otherwise) as
\ba
 \hdb_j\:&=\frac{1}{N_{rL}}\sum_{\vl\Dr\in V_r}\delta_{r,\vl}W_{rj,\vl}
 =\left(\frac{k_F}{2\pi}\right)^3\frac{1}{V_L}\sum_{\vm k_F\in V_k}\delta_{k,m}^*W_{kj,\vm} ~, \vs
 \delta_{kj,\vl}\:&=(\Dr)^3\sum_{\vm\Dr\in V_{Lj}}\delta_{r,\vm}e^{-i(\vm-\vn_{sj})\cdot\vl M_L}
 =\left(\frac{k_F}{2\pi}\right)^3e^{i\vn_{sj}\cdot\vl M_L}
 \sum_{\vn k_F\in V_k}\delta_{k,\vL-\vn}W_{kj,\vn} ~,
\label{eq:ibv_bd_di_est2}
\ea
where $V_k$ is the Fourier volume of $V_r$ with the fundamental frequency $k_F=2\pi/V_r^{1/3}=2\pi/L_r$,
$M_r=\Dr k_F=2\pi/N_{rL,1}$ with $N_{rL,1}=N_{rL}^{1/3}$ being the mesh number of $L_r$,
$M_L=\Dr k_{FL}=M_rN_{\rm cut}$, $\vL=\vl N_{\rm cut}$, and exponential term of $\delta_{jk,\vl}$
reflects the phase of the local Fourier transform (which will then be canceled out when
computing the local power spectrum $\hP_j(k)$). We shall use the capital letters to
denote the integer vector multiplied by $N_{\rm cut}$. $W_{ki,\vl}$ is the window
function in Fourier space, which can be written as
\be
 W_{ki,\vl}=(\Dr)^3\sum_{\vm\Dr\in V_r}W_{ri,\vm}e^{-i\vm\cdot\vl M_r}
 =W_L(\vl k_F) e^{-i\vl k_F\vr_{Li}} ~,
\label{eq:ibv_Wk_disc}
\ee
where $W_L(\vk)=V_L\prod_{i=0}^2{\rm sinc}(k_iL/2)$ is the window function of $V_L$.
\refEq{ibv_Wk_disc} is true in the continuous limit, i.e. $k_F\rightarrow0$.
Namely, there is a slight difference between the discrete Fourier transform of
the window function and $W_L(\vk)$. However we shall ignore this difference and
use the continuous limit in the derivation for simplicity.

Combining the above equations, the integrated bispectrum can be estimated as
\ba
 \hib_L(k)\:&=\left(\frac{k_F}{2\pi}\right)^9\frac{1}{V_L^2N_{kL}}\sum_{|\vj|\in k\pm\Dk/2}
 \sum_{(\vl,\vm,\vn)\in V_k}\delta_{k,\vl}^*\delta_{k,\vJ-\vm}\delta_{k,\vJ+\vn}^*
 W_L(\vl)W_L(\vm)W_L(\vn) \vs
 \:&\hspace{6cm}\times\frac{1}{N_{\rm cut}^3}
 \sum_{i=1}^{N_{\rm cut}^3}e^{-i(\vl+\vm+\vn)k_F\cdot\vr_{Li}} ~.
\label{eq:ibv_ib_est1}
\ea
Here, to simplify the notation, we drop all the fundamental units ($\Dr$ and $k_F$)
of the integer vectors in the summation and the window function. We find that the
$p^{\rm th}$ axis of $\vr_{Li}$ ($\vr_L$ is the center of the $i^{\rm th}$ subvolume)
is $r_{Li_p,p}=(i_p+1/2)L$ with $i_p$ being the order of the $i^{\rm th}$ subvolume
in the $p^{\rm th}$ axis, hence the last term of \refeq{ibv_ib_est1} can be simplified as
\be
 \sum_{i=1}^{N_{\rm cut}^3}e^{-i(\vl+\vm+\vn)k_F\cdot\vr_{Li}}
 =\prod_{p=0}^2(-1)^{l_p+m_p+n_p}\frac{\sin[(l_p+m_p+n_p)\pi]}{\sin[(l_p+m_p+n_p)\pi/N_{\rm cut}]} ~.
\label{eq:ibv_sum_exp}
\ee
\refEq{ibv_sum_exp} is non-zero only if $l_p+m_p+n_p=N_{\rm cut}s_p$ with $s_p$ being an
integer, and the value is $(-1)^{s_0+s_1+s_2}N_{\rm cut}^3$ for even $N_{\rm cut}$ and
$N_{\rm cut}^3$ for odd $N_{\rm cut}$. We shall assume odd $N_{\rm cut}$ for simplifying
the notation, but the results (variance of the estimator) would be identical for both
cases, as we will show later. We can thus rewrite \refeq{ibv_sum_exp} as
\be
 N_{\rm cut}^3\delta^K_{\vl+\vm+\vn,\v{s}N_{\rm cut}}=N_{\rm cut}^3\delta^K_{\vl+\vm+\vn,\vS}~,
\label{eq:ibv_kroneker}
\ee
where $\delta^K_{a,b}$ is the Kronecker delta and $\vS\equiv\v{s}N_{\rm cut}$. Finally,
the estimator of the integrated bispectrum becomes
\ba
 \hib_L(k)=\:&\left(\frac{k_F}{2\pi}\right)^9\frac{1}{V_L^2N_{kL}}
 \sum_{|\vj|\in k\pm\Dk/2}\sum_{(\vm,\vn)\in V_k}\sum_{\v{s}=-\infty}^\infty
 \delta_{k,\vS-\vm-\vn}^*\delta_{k,\vJ-\vm}\delta_{k,\vJ+\vn}^* \vs
 &\hspace{6.5cm}\times W_L(\v{S}-\vm-\vn)W_L(\vm)W_L(\vn) ~.
\label{eq:ibv_ib_est2}
\ea
\refEq{ibv_ib_est2} is an unbiased estimator because
\ba
 \langle\hib_L(k)\rangle\:&=\left(\frac{k_F}{2\pi}\right)^9\frac{1}{V_L^2N_{kL}}
 \sum_{|\vj|\in k\pm\Dk/2}\sum_{(\vm,\vn)\in V_k}\sum_{\v{s}=-\infty}^\infty
 \langle\delta_{k,\vS-\vm-\vn}^*\delta_{k,\vJ-\vm}\delta_{k,\vJ+\vn}^*\rangle \vs
 &\hspace{7cm}\times W_L(\vS-\vm-\vn)W_L(\vm)W_L(\vn) \vs
 \:&=\frac{1}{V_L^2N_{kL}}\sum_{|\vj|\in k\pm\Dk/2}\int\frac{d^3q_1}{\tpc}\int\frac{d^3q_2}{\tpc}~
 B(+\vq_1+\vq_2,\vJ k_F-\vq_1,-\vJ k_F-\vq_2) \vs
 \:&\hspace{6.5cm}\times W_L(\vq_1)W_L(\vq_2)W_L(-\vq_1-\vq_2) \vs
 \:&=\frac1{N_{kL}}\sum_{|\vj|\in k\pm\Dk/2}iB(\vJ k_F)=iB(\vk) ~,
\ea
where we replace the discrete Fourier transform with the continuous one as
$[k_F^3/\tpc]\sum_{\vm\in V_k}\rightarrow\int d^3q/\tpc$. Note that only
$\v{s}=0$ contributes to $\langle\hib_L(k)\rangle$.

Replacing the discrete Fourier transform with the integral form, the variance
of the integrated bispectrum estimator can be computed by
\ba
 \:&\langle[\hib_L(k)-\langle\hib_L(k)\rangle]^2\rangle=
 \langle[\hib_L(k)]^2\rangle-[\langle\hib_L(k)\rangle]^2 \vs
 \:&\hspace{-0.2cm}=\left(\frac{k_F}{2\pi}\right)^{18}\frac{1}{V_L^4N_{kL}^2}
 \sum_{|\vj_1,\vj_2|\in k\pm\Dk/2}\sum_{\v{s}_1,\v{s}_2=-\infty}^\infty
 \int\frac{d^3q_1}{\tpc}\cdots\int\frac{d^3q_4}{\tpc}\vs
 \:&\hspace{1cm}\langle\delta(\vk_1-\vq_1)\delta(-\vk_1-\vq_2)\delta(\vq_1+\vq_2-\vg_1)
 \delta(\vk_2-\vq_3)\delta(-\vk_2-\vq_4)\delta(\vq_3+\vq_4-\vg_2)\rangle \vs
 \:&\hspace{1.3cm}\times
 W_L(\vq_1)W_L(\vq_2)W_L(\vg_1-\vq_1-\vq_2)W_L(\vq_3)W_L(\vq_4)W_L(\vg_2-\vq_3-\vq_4) ~,
\label{eq:ibv_ib_var_1}
\ea
where $\vk_n=\vJ_nk_F$ and $\vg_n=\vS_nk_F$. We shall assume that the dominant
component of the six-point function is the disconnected part, and thus the only
non-zero component has the wavenumber combinations as\footnote{Note that another
seemingly non-zero component has the wavenumber combinations as
\be
 \vk_1-\vq_1+\vk_2-\vq_3=0 ~,~ -\vk_1-\vq_2-\vk_2-\vq_4=0 ~,~
 \vq_1+\vq_2-\vg_1+\vq_3+\vq_4-\vg_2=0 ~.
\label{eq:ibv_ib_nzero_comb_false}
\ee
However, this requires $\vk_1=-\vk_2$ or $\vj_1=-\vj_2$. Since we count only the
independent Fourier modes, only half of the Fourier space is counted (or only
the hemisphere is considered), and so it is impossible to have $\vj_1=-\vj_2$.}
\be
 \vk_1-\vq_1-\vk_2-\vq_4=0 ~,~ -\vk_1-\vq_2+\vk_2-\vq_3=0 ~,~
 \vq_1+\vq_2-\vg_1+\vq_3+\vq_4-\vg_2=0 ~.
\label{eq:ibv_ib_nzero_comb}
\ee
This gives three delta functions as
\be
 (2\pi)^9\delta_D(\vk_1-\vk_2-\vq_1-\vq_4)\delta_D(-\vk_1+\vk_2-\vq_2-\vq_2)
 \delta_D(\vg_1+\vg_2) ~.
\label{eq:ibv_ib_nzero_delta}
\ee
The last delta function in \refeq{ibv_ib_nzero_delta} requires $\vg_1=-\vg_2$ or
$\v{s}_1=-\v{s}_2$. This means that even if even $N_{\rm cut}$ has a parity
term $(-1)^{s}$, it would be canceled because $\v{s}_1=-\v{s}_2$.

With the above results, the variance of the integrated bispectrum estimator is
given by ($\vg=\vS k_F=\v{s}N_{\rm cut}k_F$)
\ba
 \:&\left(\frac{k_F}{2\pi}\right)^3\frac{1}{V_L^4N_{kL}^2}
 \sum_{|\vj_1,\vj_2|\in k\pm\Dk/2}\sum_{\v{s}=-\infty}^\infty
 \int\frac{d^3q_1}{\tpc}\int\frac{d^3q_2}{\tpc}~P(\vk_1-\vq_1)P(-\vk_1-\vq_2)P(\vq_1+\vq_2-\vg) \vs
 \:&\times W_L(\vq_1)W_L(\vq_2)W_L(\vg-\vq_1-\vq_2)W_L(\vk_1-\vk_2-\vq_1)W_L(-\vk_1+\vk_2-\vq_2)W_L(-\vg+\vq_1+\vq_2) ~.
\label{eq:ibv_ib_var2}
\ea
Note that \refeq{ibv_ib_var2} is non-zero only if $\vk_1=\vk_2$ or $\vj_1=\vj_2$,
so the variance of the integrated bispectrum estimator can be simplified as
($\vk=\vJ k_F=\vj N_{\rm cut}k_F$)
\ba
 \:&\frac{1}{V_rV_L^4N_{kL}^2}\sum_{|\vj|\in k\pm\Dk/2}\sum_{\v{s}=-\infty}^\infty
 \int\frac{d^3q_1}{\tpc}\int\frac{d^3q_2}{\tpc}~P(\vk-\vq_1)P(-\vk-\vq_2)P(\vq_1+\vq_2-\vg) \vs
 \:&\hspace{6cm}\times |W_L(\vq_1)|^2|W_L(\vq_2)|^2|W_L(\vg-\vq_1-\vq_2)|^2 ~,
\label{eq:ibv_ib_var3}
\ea
where we replace $k_F^3/(2\pi)^3$ with $V_r$. To proceed further, let us consider
the sum over $\v{s}$. Since $\vg=\v{s}N_{\rm cut}k_F=\v{s}k_{F,L}$, we replace the
discrete sum with the integral as
\ba
 \sum_{\v{s}=-\infty}^\infty P(\vq-\vg)|W_L(\vg-\vq)|^2\rightarrow\:&
 \left(\frac{k_{F,L}}{2\pi}\right)^{-3}\int\frac{d^3g}{\tpc}~P(\vq-\vg)|W_L(\vg-\vq)|^2 \vs
 =\:&\left(\frac{k_{F,L}}{2\pi}\right)^{-3}V_L^2\sigma_L^2=V_L^3\sigma_L^2 ~,
\label{eq:sum_s}
\ea
where $\sigma_L^2=\frac{1}{V_L^2}\int\frac{d^3q}{\tpc}~P(q)|W_L(\vq)|^2$
is the variance of the fluctuation in the volume $V_L$.

Finally, the variance of the integrated bispectrum estimator is simply
\ba
 \:&\frac{1}{V_rV_L^4N_{kL}^2}\sum_{|\vj|\in k\pm\Dk/2}
 V_L^3\sigma_L^2\left[\int\frac{d^3q}{\tpc}~P(\vk-\vq)|W_L(\vq)|^2\right]^2 \vs
 \:&=\frac{V_L}{V_rN_{kL}^2}\sigma_L^2\sum_{|\vj|\in k\pm\Dk/2}[P_L(\vk)]^2
 =\frac{V_L}{V_rN_{kL}}\sigma_L^2[P_L(k)]^2 ~,
\label{eq:ibv_ib_var4}
\ea
where $P_L(\vk)=\frac{1}{V_L}\frac{d^3q}{\tpc}~P(\vk-\vq)|W_L(\vq)|^2$ is the
convolved power spectrum of the subvolume $V_L$. Note that the previous derivation
ignores the shot noise contribution. If the shot noise is Poisson like, i.e.
$P_{\rm shot}=\bar n^{-1}$ with $\bar n$ being the number density of the discrete
particles, then it is trivial to add back as
\be
 \frac{V_L}{V_rN_{kL}}\left[\sigma_L^2+\frac{P_{\rm shot}}{V_L}\right][P_L(k)+P_{\rm shot}]^2 ~.
\label{eq:ibv_ib_var5}
\ee
For the normalized integrated bispectrum, we assume that the variance is dominated
by the bispectrum term instead of the normalization which contains $P_L(k)$ and
$\sigma_L^2$, and so its variance is
\be
 \frac{V_L}{V_rN_{kL}}\frac{\left[\sigma_L^2+\frac{P_{\rm shot}}{V_L}\right]}{\sigma_L^4}
 \frac{[P_L(k)+P_{\rm shot}]^2}{P_L^2(k)} ~.
\label{eq:ibv_ib_norm_var}
\ee

\refFig{ibv_ib_corr} shows the correlation matrices measured from 160 dark matter
$N$-body simulations at $z=0$ and 2. The details of the simulations are given in
\refsec{ch4_nbody}. There are four subvolume sizes: 600 (bin 0 to 47 in the axis
labels), 400 (bin 48 to 79), 300 (bin 80 to 103), and $200\hMpc$ (bin 104 to 119).
The bottom (top) half of the correlation matrices is the unnormalized (normalized)
integrated bispectrum. We find that the cross-correlation between different subvolumes
is much smaller than that within the same subvolumes. This is expected because different
subvolumes have different long-wavelength modes, which are uncorrelated. At $z=2$
the correlation matrices are more diagonal than at $z=0$ because of the smaller
nonlinearity. We also find that the normalization largely diagonalizes the correlation
matrices, particularly at $z=2$ where the off-diagonal components nearly vanish. Note
that there are stripes, which are more obvious in the normalized integrated bispectrum
at $z=2$, across the diagonal elements of the cross covariances between different sizes
of subvolumes. This is because these components have the same short-wavelength modes,
i.e. the scales of the position-dependent power spectrum. Consequently, the correlation
is stronger.

\begin{figure}[t!]
\centering
\includegraphics[width=0.495\textwidth]{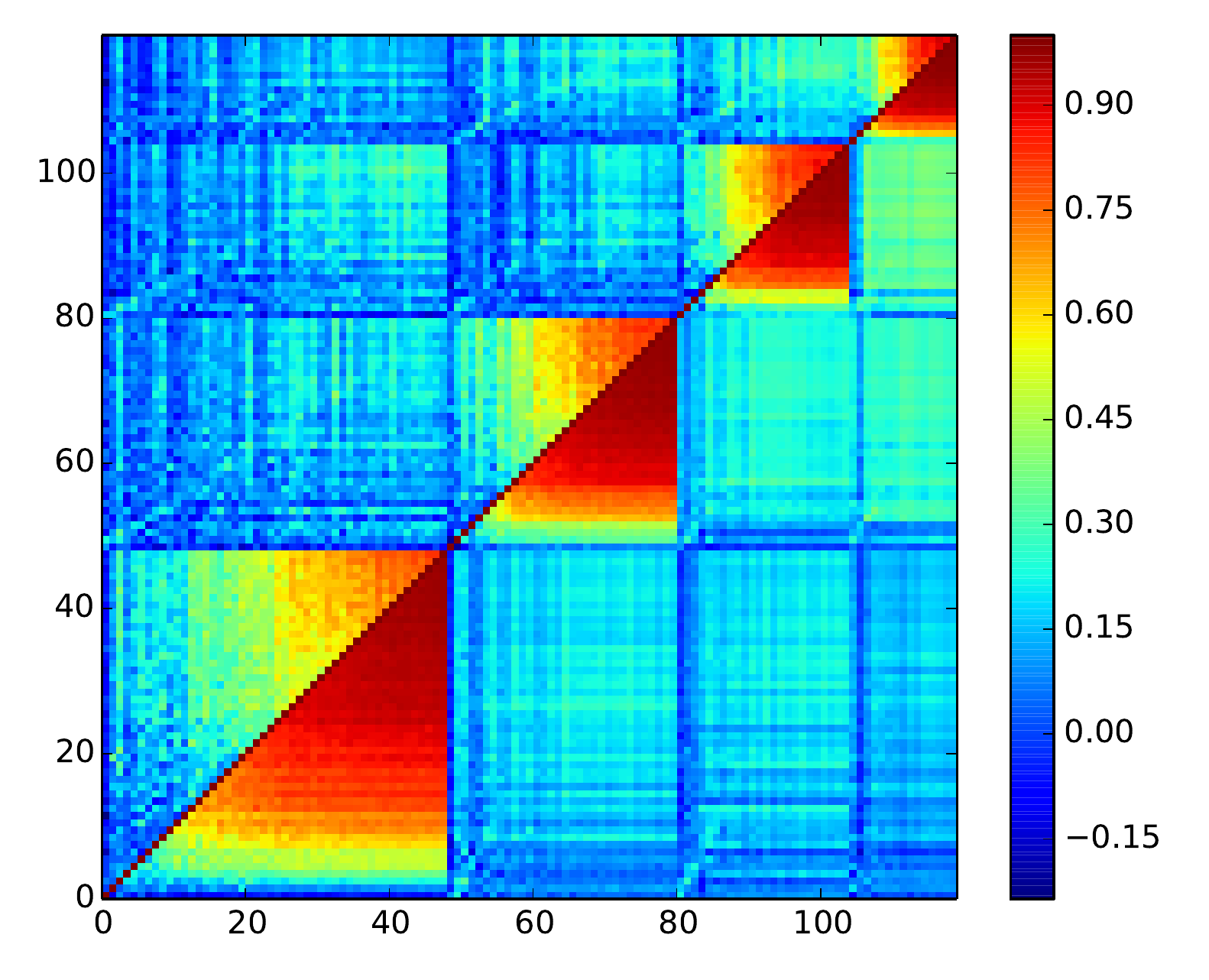}
\includegraphics[width=0.495\textwidth]{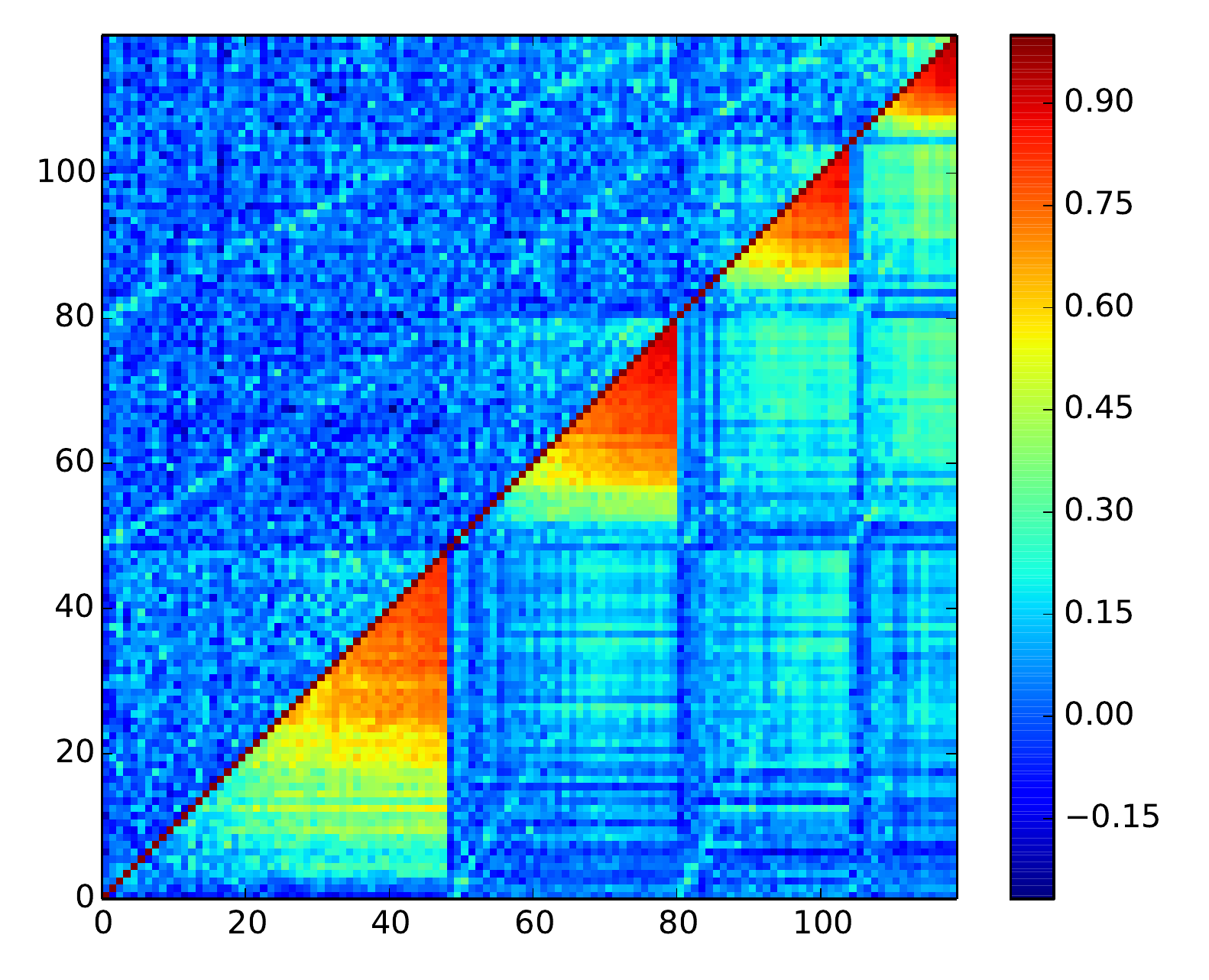}
\caption[Correlation matrices of the unnormalized and normalized
integrated bispectra measured from 160 dark matter $N$-body simulations]
{Correlation matrices measured from 160 dark matter $N$-body simulations at
$z=0$ (left) and 2 (right) for $N_{\rm cut}=4$ (bin 0 to 47 in the axis labels),
6 (bin 48 to 79), 8 (bin 80 to 103), 12 (bin 104 to 119), which correspond to
600, 400, 300, and $200\hMpc$, respectively. The bottom half and top half of
the correlation matrices show the unnormalized and normalized integrated
bispectrum.}
\label{fig:ibv_ib_corr}
\end{figure}

\refFig{ibv_ib_norm_var} shows the square root of the variances of the normalized
integrated bispectra at $z=0$ (left) and 2 (right). The data points are measured
from simulations, and the solid lines are computed using \refeq{ibv_ib_norm_var}
with the linear power spectrum and zero shot noise. We find that on large scales
as well as at high redshift, the agreement between simulations and analytical
calculation is good. The agreement between the data points and the solid lines
confirm our calculation of the variance of the integrated bispectrum estimator.

\begin{figure}[t!]
\centering
\includegraphics[width=0.495\textwidth]{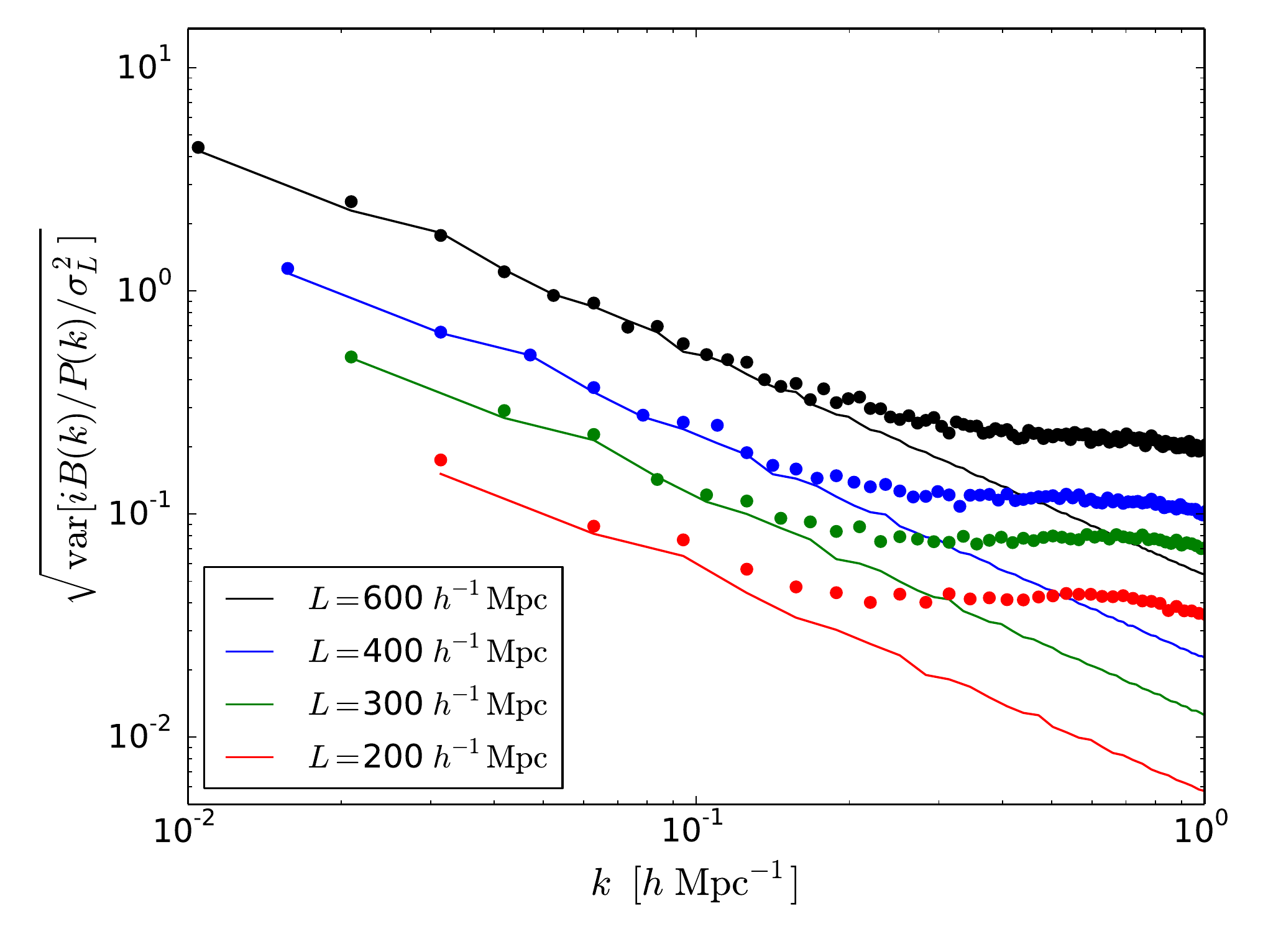}
\includegraphics[width=0.495\textwidth]{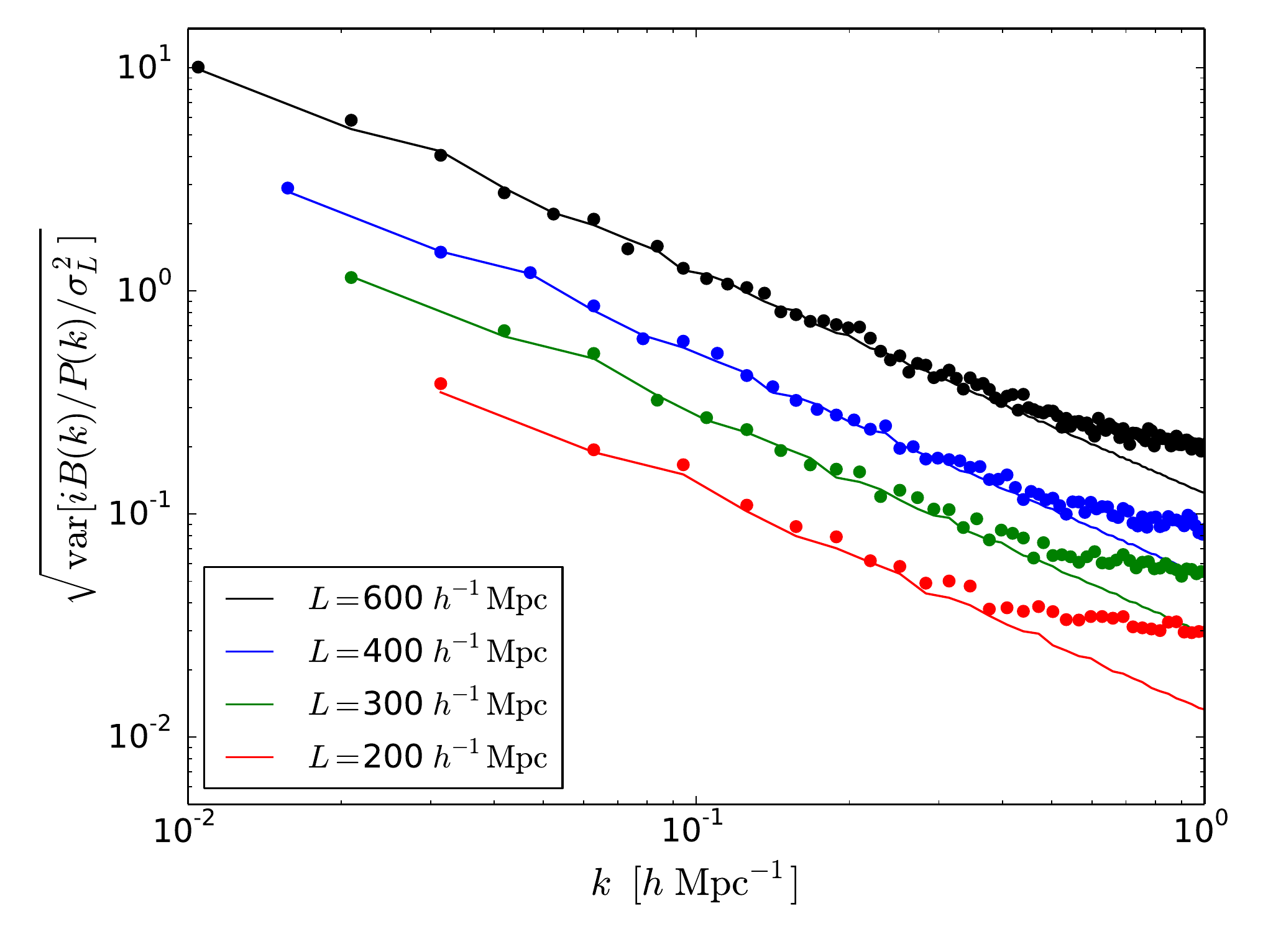}
\caption[Square root of the variances of the normalized integrated bispectrum]
{Square root of the variances of the normalized integrated bispectrum
at $z=0$ (left) and 2 (right). The data points show the measurements
from the simulations and the solid lines show \refeq{ibv_ib_norm_var}
with the linear power spectrum and zero shot noise.}
\label{fig:ibv_ib_norm_var}
\end{figure}

}

%% file: anh_03.tex
{

\newcommand{\iz}{i\zeta}
\newcommand{\sL}{\sigma_L^2}
\newcommand{\hMpc}{~h^{-1}~{\rm Mpc}}

\chapter{Testing the integrated three-point function estimator
with Gaussian realizations and the local bias model}
\label{app:mock_halo}
We shall demonstrate that our integrated three-point function estimator is unbiased.
To do this, we first generate the matter density field, $\delta_m({\bf r})$, by Gaussian
realizations\footnote{Since $\delta_m({\bf r})$ follows the Gaussian statistics,
it is possible that $\delta_m({\bf r})<-1$, which is unphysical. However, as we only
compute the power spectrum without Poisson sampling the density field, this effect
can be neglected.} with $P_l(k)$ at $z=0$ for the volume $V_r$ of $1200~h^{-3}~{\rm Mpc}^3$
and a mesh size of $4\hMpc$. We then compute a mock ``halo'' density field using the
local bias model via
\be
 \delta_h({\bf r})=b_1\delta_m({\bf r})+\frac{b_2}2\left[\delta_m^2({\bf r})
 -\frac{\sum_{{\bf r}\in V_r}\delta_m^2({\bf r})}{\sum_{{\bf r}\in V_r}1}\right] ~,
\ee
where we set $b_1=3$ and $b_2=1$, and $\sum_{{\bf r}\in V_r}$ denotes a sum over grid cells
in the entire volume. Note that $\sum_{{\bf r}\in V_r}\delta_h({\bf r})=0$. We then divide the
entire volume $V_r$ into $N_s=12^3=1728$ subvolumes $V_L$ of $100~h^{-3}~{\rm Mpc}^3$ and
$N_s=6^3=216$ subvolumes $V_L$ of $200~h^{-3}~{\rm Mpc}^3$. The two-point function in the
subvolumes and the integrated three-point function are estimated by
\be
 \xi_h(r,{\bf r}_L)=\frac{\sum_{{\bf x}+{\bf r},{\bf x}\in V_L}
 \delta_h({\bf x}+{\bf r})\delta_h({\bf x})}{\sum_{{\bf x}+{\bf r},{\bf x}\in V_L}1} ~,~~
 \iz_{L,h}(r)=\sum_{i=1}^{N_s}\xi_h(r,{\bf r}_L)\bar\delta_h({\bf r}_L) ~,
\label{eq:xih_izh}
\ee
where $\bar\delta_h({\bf r}_L)$ is the mean halo overdensity in the subvolume centered at
${\bf r}_L$. Note that the denominator in the estimator of $\xi_h(r,{\bf r}_L)$ takes the
boundary effect into account so $\langle\xi_h(r,{\bf r}_L)\rangle=\xi_h(r)$ without
$f_{L,\rm bndry}(r)$. This means that the theoretical model of the integrated three-point
function computed by \refeq{ch2_iz} has to be divided by $f_{L,\rm bndry}(r)$. Since
$\delta_m({\bf r})$ is Gaussian, the only contribution to the three-point function is
from the nonlinear bias term, and so the estimated integrated three-point function is
exactly given by $\frac{\iz_{L,b_2}(r)}{f_{L,\rm bndry}(r)}$.

\begin{figure}[t]
\centering
\includegraphics[width=0.495\textwidth]{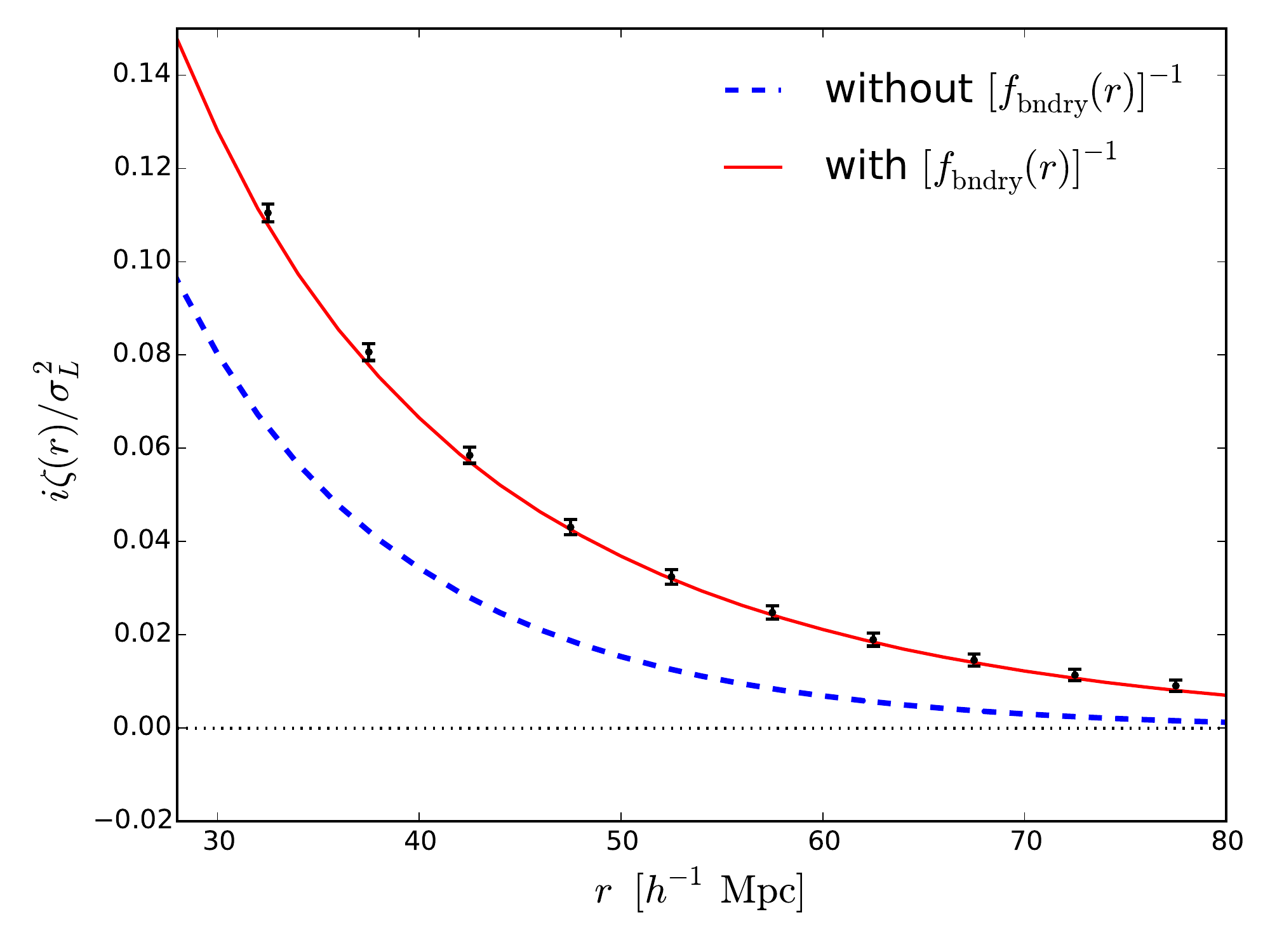}
\includegraphics[width=0.495\textwidth]{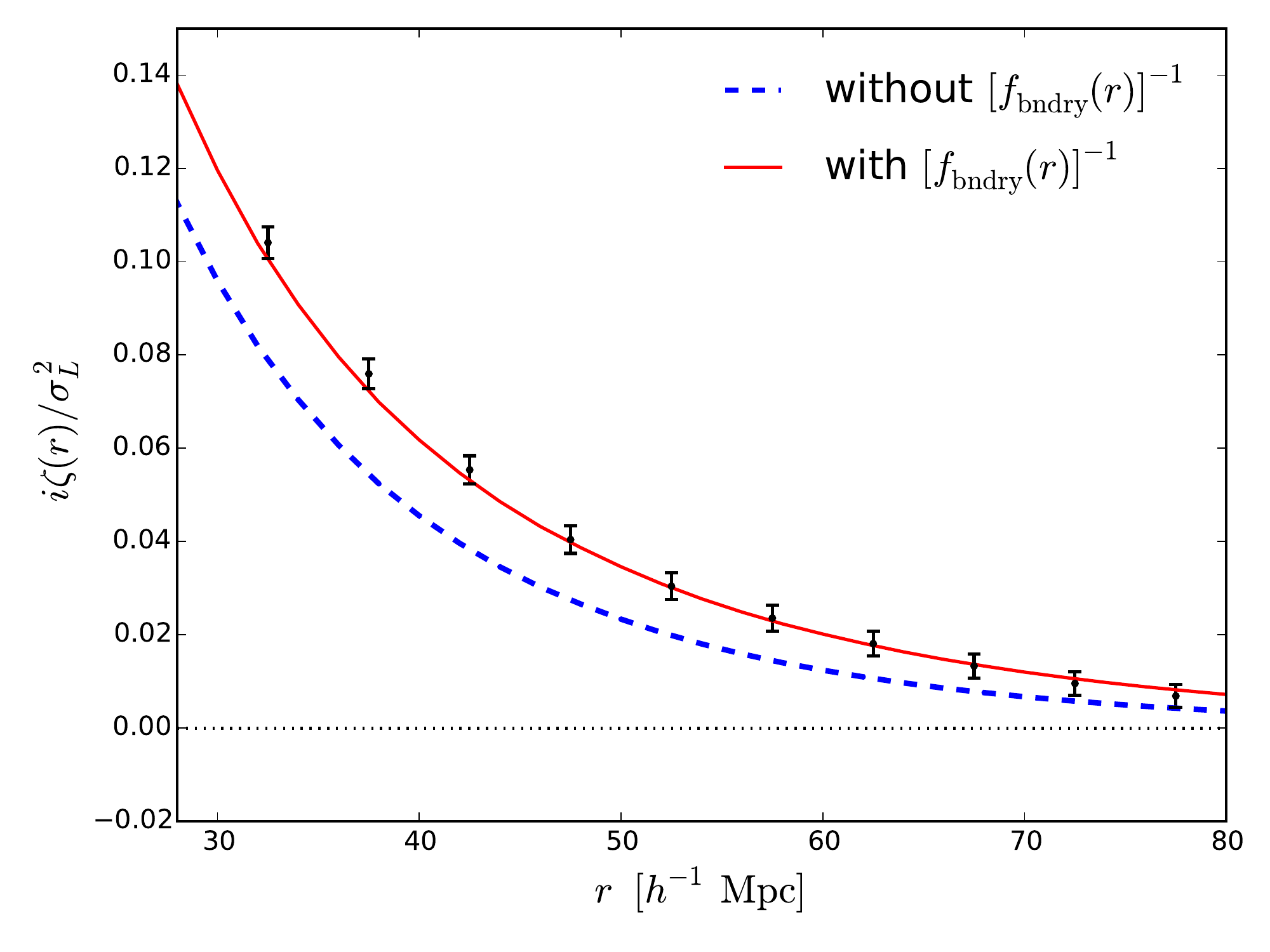}
\caption[Normalized integrated three-point functions of the mock halo density field]
{The normalized integrated three-point functions of the mock halo density field
with $b_1=3$ and $b_2=1$. The left and right panels are for $V_L=100~h^{-3}~{\rm Mpc}^3$
and $200~h^{-3}~{\rm Mpc}^3$, respectively. The data points show the mean of 300
Gaussian realizations, and the error bars are the variances of the mean (but note
that the data points are highly correlated). The blue dashed and red solid lines
are the theoretical models ($\iz_{L,b_2}$) without and with $[f_{L,\rm bndry}(r)]^{-1}$,
respectively.}
\label{fig:mock_halo}
\end{figure}

\refFig{mock_halo} shows the normalized integrated three-point functions of the
mock halo density field with $b_1=3$ and $b_2=1$ from 300 Gaussian realizations.
The measurements are in excellent agreement with $\frac{\iz_{L,b_2}(r)}{f_{L,\rm bndry}(r)}$.
This test gives us the confidence that our estimator is unbiased.

\chapter{Effects of effective $F_2$ and $G_2$ kernels and non-local tidal bias}
\label{app:ft_b2}
In this appendix, we show how the inferred value of $b_2$ changes
when extending our baseline model for the bispectrum based on SPT
at the tree level with local bias to the model used in the analysis
of \cite{gilmarin/etal:2014b}.

Their model replaces $F_2$ and $G_2$ in \refeq{trz_zn} with
``effective'' kernels, $F_2^{\rm eff}$ \cite{gilmarin/etal:2012}
and $G_2^{\rm eff}$ \cite{gilmarin/etal:2014a}, which are calibrated
to match the nonlinear matter bispectrum in of N-body simulations.
Their model also adds a non-local galaxy bias caused by tidal fields
\cite{mcdonald/roy:2009,baldauf/etal:2012,sheth/chan/scoccimarro:2013}
to $Z_2$, i.e.,
\be
 Z_2\to Z_2+\frac12b_{s^2}\left[(\hat{k}_1\cdot\hat{k}_2)^2-\frac13\right] ~,
\ee
where $b_{s^2}=-(4/7)(b_1-1)$. We use this model to compute the integrated
three-point function, and find $b_2$ of the mocks in real and redshift
space by performing a joint fit with the two-point function as described
in \refsec{ch5_mock_r} and \ref{sec:ch5_mock_z}.


\begin{table}[t]
\centering
\begin{tabular}{ | c | c c | }
\hline
r-space & $b_1$ & $b_2$ \\
\hline
baseline & $1.971\pm0.076$ & $0.58\pm0.31$ \\
eff kernel & $1.973\pm0.076$ &  $0.62\pm0.31$ \\
tidal bias & $1.971\pm0.076$ &  $0.64\pm0.31$ \\
both & $1.973\pm0.076$ &  $0.68\pm0.31$ \\
\hline
\end{tabular}
~~~~~
\begin{tabular}{ | c | c c | }
\hline
z-space & $b_1$ & $b_2$ \\
\hline
baseline & $1.931\pm0.077$ & $0.54\pm0.35$ \\
eff kernel & $1.933\pm0.077$ &  $0.65\pm0.35$ \\
tidal bias & $1.932\pm0.077$ &  $0.60\pm0.35$ \\
both & $1.933\pm0.077$ &  $0.71\pm0.35$ \\
\hline
\end{tabular}
\caption[Best-fitting values of $b_1$ and $b_2$ of the mock catalogs for various bispectrum models]
{Best-fitting values of $b_1$ and $b_2$ and their uncertainties for mock
catalogs, obtained using different models of the bispectrum in real space
(left) and redshift space (right).}
\label{tab:ft_b1b2}
\end{table}

\refTab{ft_b1b2} summarizes the results. The ``baseline model'' refers to
the model based on SPT and local bias. The ``eff kernel'' refers to the
model with $F^{\rm eff}_2$, $G^{\rm eff}_2$, and local bias. The ``tidal
bias'' refers to the model with $F_2$, $G_2$, local bias, and tidal bias.
Finally, ``both'' refers to the model with $F^{\rm eff}_2$, $G^{\rm eff}_2$,
local bias, and tidal bias.

Both the effective kernels and the non-local tidal bias result in a larger
nonlinear bias, which is in better agreement with \cite{gilmarin/etal:2014b}.
The changes of the best-fitting nonlinear bias, however, are still within the
$1-\sigma$ uncertainties, and all the results are consistent with \cite{gilmarin/etal:2014b}.
We also calculate the goodness of the fit for all the models in both real and
redshift space by comparing the mean of the mocks and the best-fitting models,
as well as the $\chi^2$-distribution. We find that all models perform equally
well; thus, we shall primarily use the simplest model, i.e. the SPT at the
tree level with local bias for modeling the three-point function, but also
report the results for the extended models.

\chapter{Comparison for $\iz(r)/\sL$ of BOSS DR10 CMASS sample and
PTHalo mock catalogs in different redshift bins}
\label{app:zbin_iz}
The BOSS DR10 CMASS sample and the mocks have different sets of random samples
with slightly different $\bar n(z)$, hence the properties of the observations
and the mocks may not agree well in all redshift bins. Moreover, as mentioned
in \cite{more/etal:2014}, the CMASS sample is flux-limited, and thus the observed
galaxies statistically have larger stellar masses at higher redshift (see figure~1
in \cite{more/etal:2014}). This may cause redshift evolution of the bias, and so
the correlation functions. We shall measure $\iz(r)/\sigma_L^2$ as a function of
redshift to test this.

The measurements in the subvolumes are mostly the same as introduced in \refsec{ch5_sub_quan},
except that we now measure $\alpha(z_j)$ as a function of redshift bin $z_j$,
and the average is done in the individual redshift bin. Namely,
\be
 \alpha(z_j)=\frac{\sum_{i\in z_j}w_{g,i}}{\sum_{i\in z_j}w_{r,i}}
 =\frac{w_{r,z_j}}{w_{g,z_j}}\,, ~~~~~
 \bar{g}(z_j)=\frac{1}{w_{r,z_j}}\sum_{i\in z_j}g_iw_{r,i} ~.
\ee
This assures that $\bar\delta(z_j)=0$ for all redshift bins.

\refFig{zbin_iz_norm_zi_1} and \reffig{zbin_iz_norm_zi_2} show $\iz(r)/\sigma_L^2$
at different redshift bins for 220 and $120\hMpc$ subvolumes, respectively. We find
no clear sign that $\iz(r)/\sigma_L^2$ of the observations has different redshift
evolution relative to the mocks. Thus, it is justified to study $\iz(r)/\sigma_L^2$
using one effective redshift for the BOSS DR10 CMASS sample. With the upcoming DR12
sample with a larger volume, the redshift evolution of $\iz(r)/\sigma_L^2$ can be
better studied.

\begin{figure}[t!]
\centering
\includegraphics[width=0.325\textwidth]{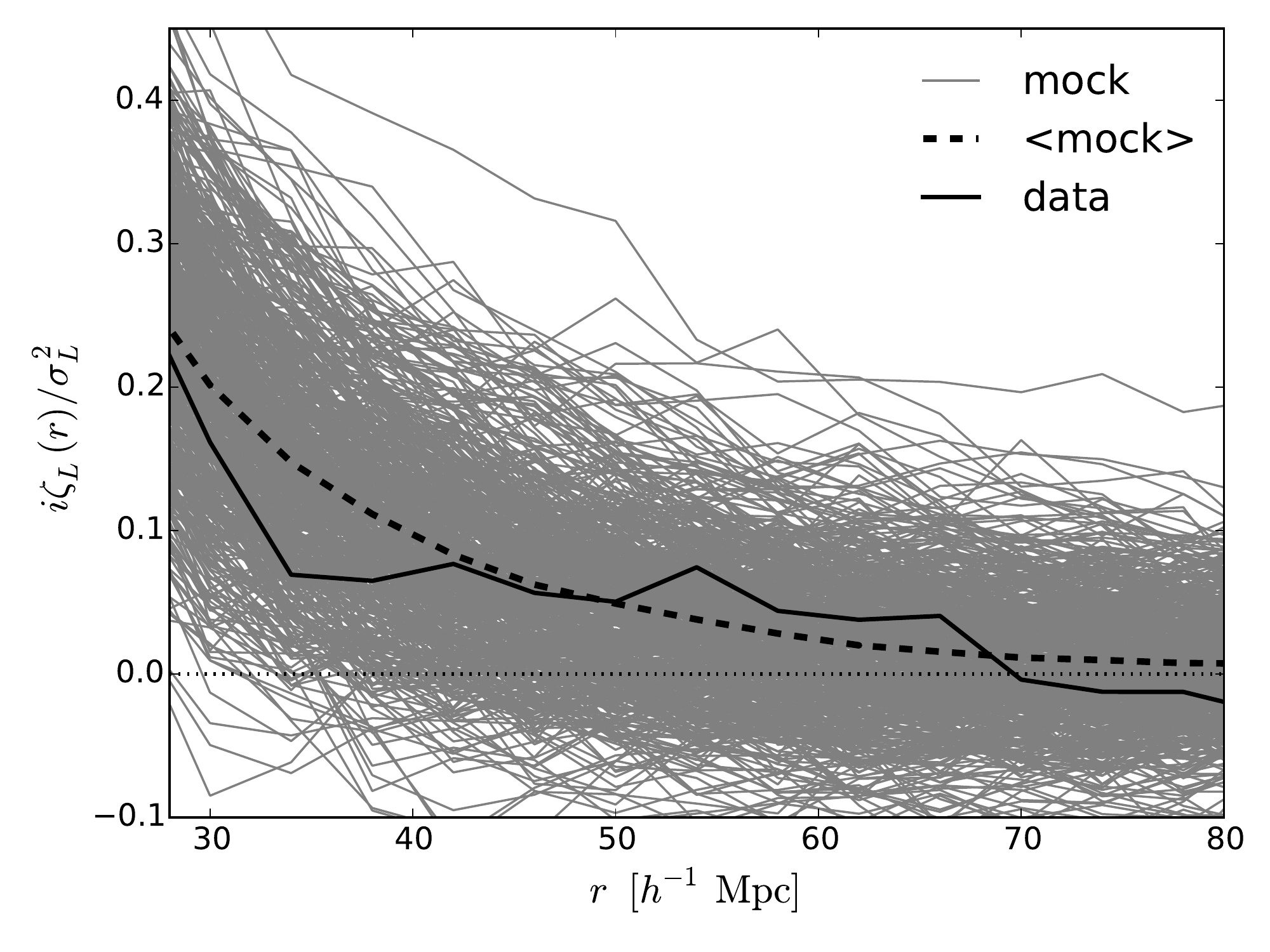}
\includegraphics[width=0.325\textwidth]{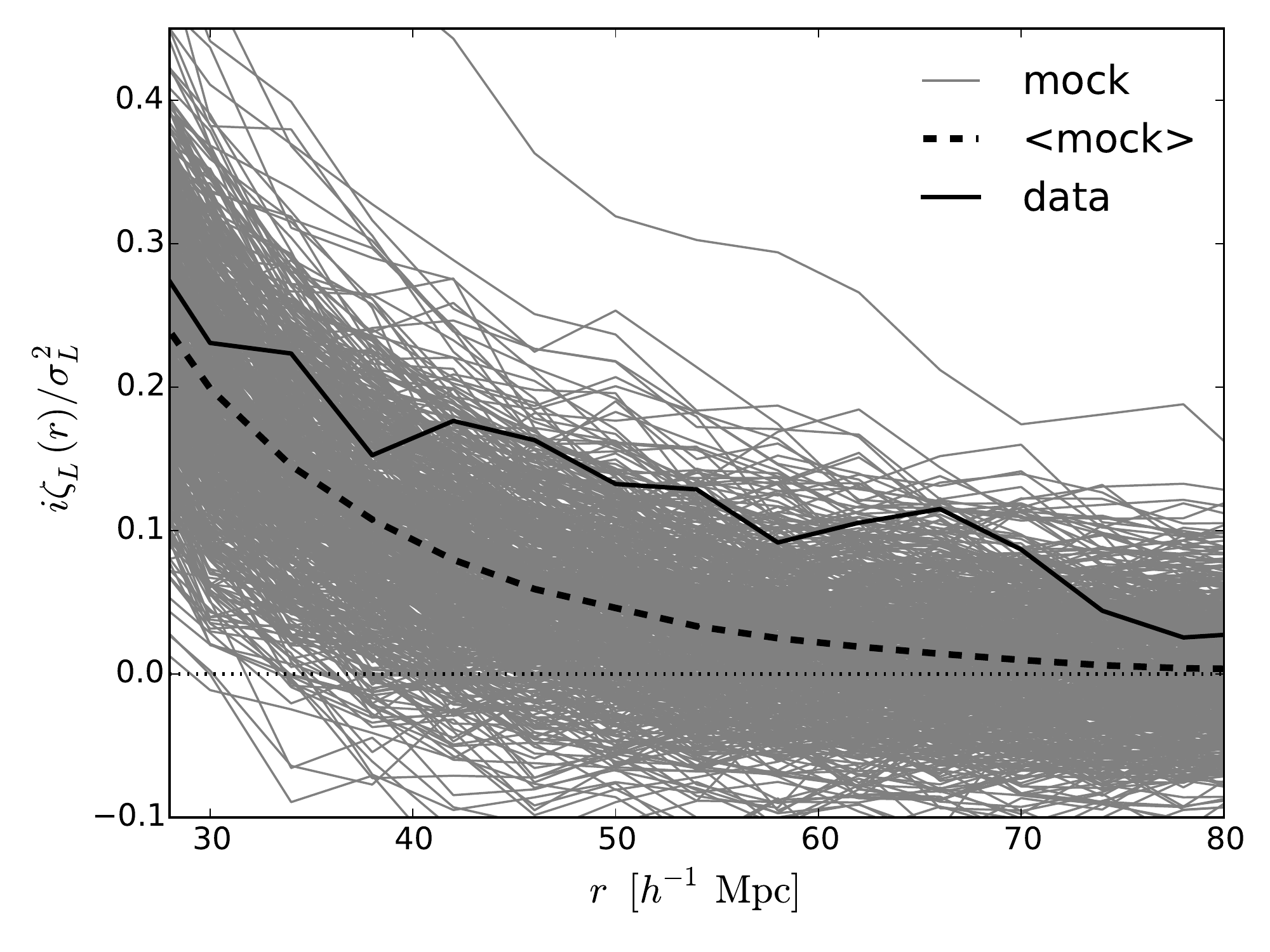}
\includegraphics[width=0.325\textwidth]{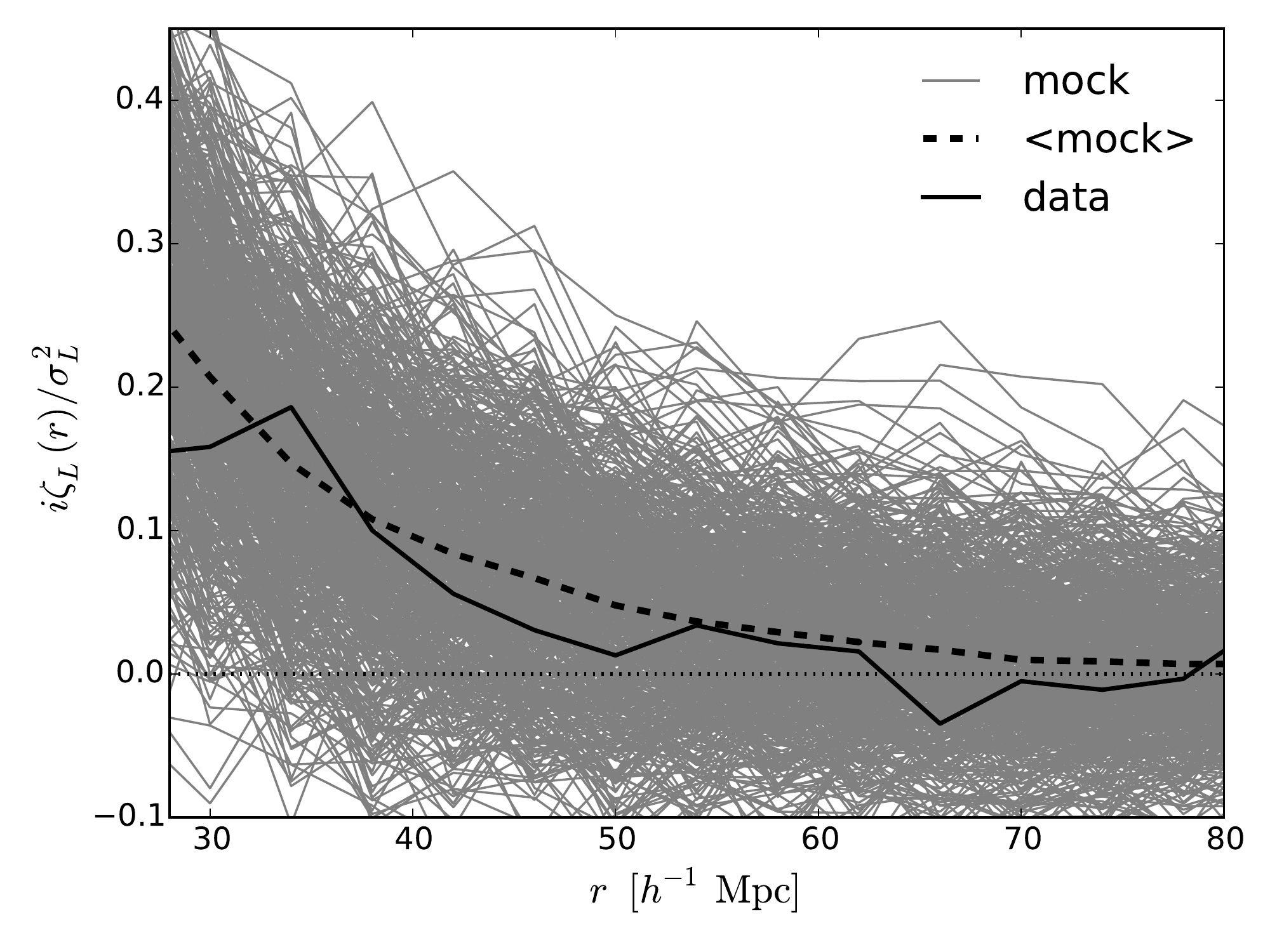}
\caption[Normalized integrated three-point function of $220\hMpc$ subvolumes
in different redshift bins]
{$\iz(r)/\sigma_L^2$ of $220\hMpc$ subvolumes in different redshift bins. The
redshift bins increase from left to right, with the redshift cuts quoted in
the beginning of \refsec{ch5_mock} and \refsec{ch5_data}.}
\label{fig:zbin_iz_norm_zi_1}
\end{figure}

\begin{figure}[t!]
\centering
\includegraphics[width=0.325\textwidth]{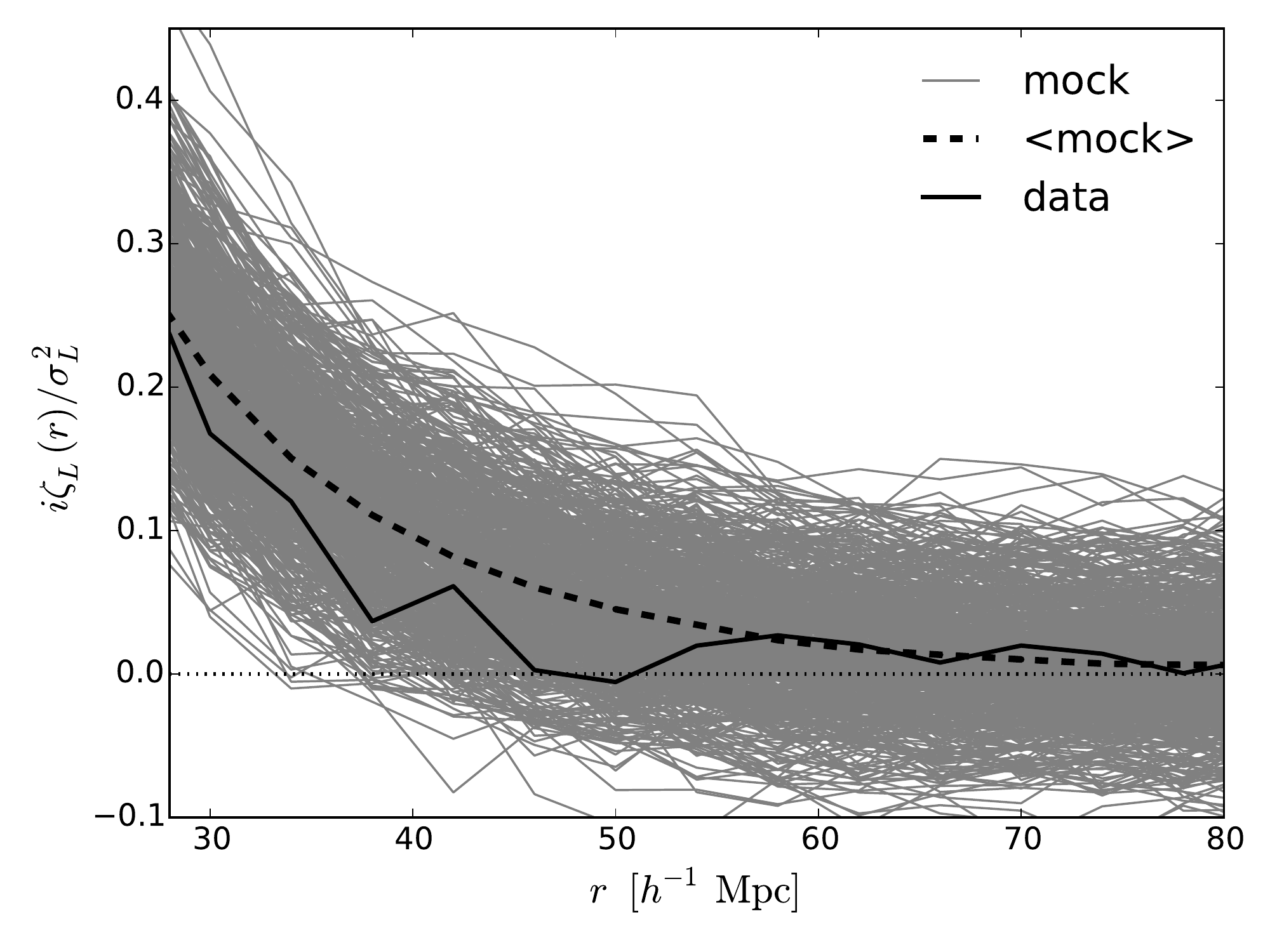}
\includegraphics[width=0.325\textwidth]{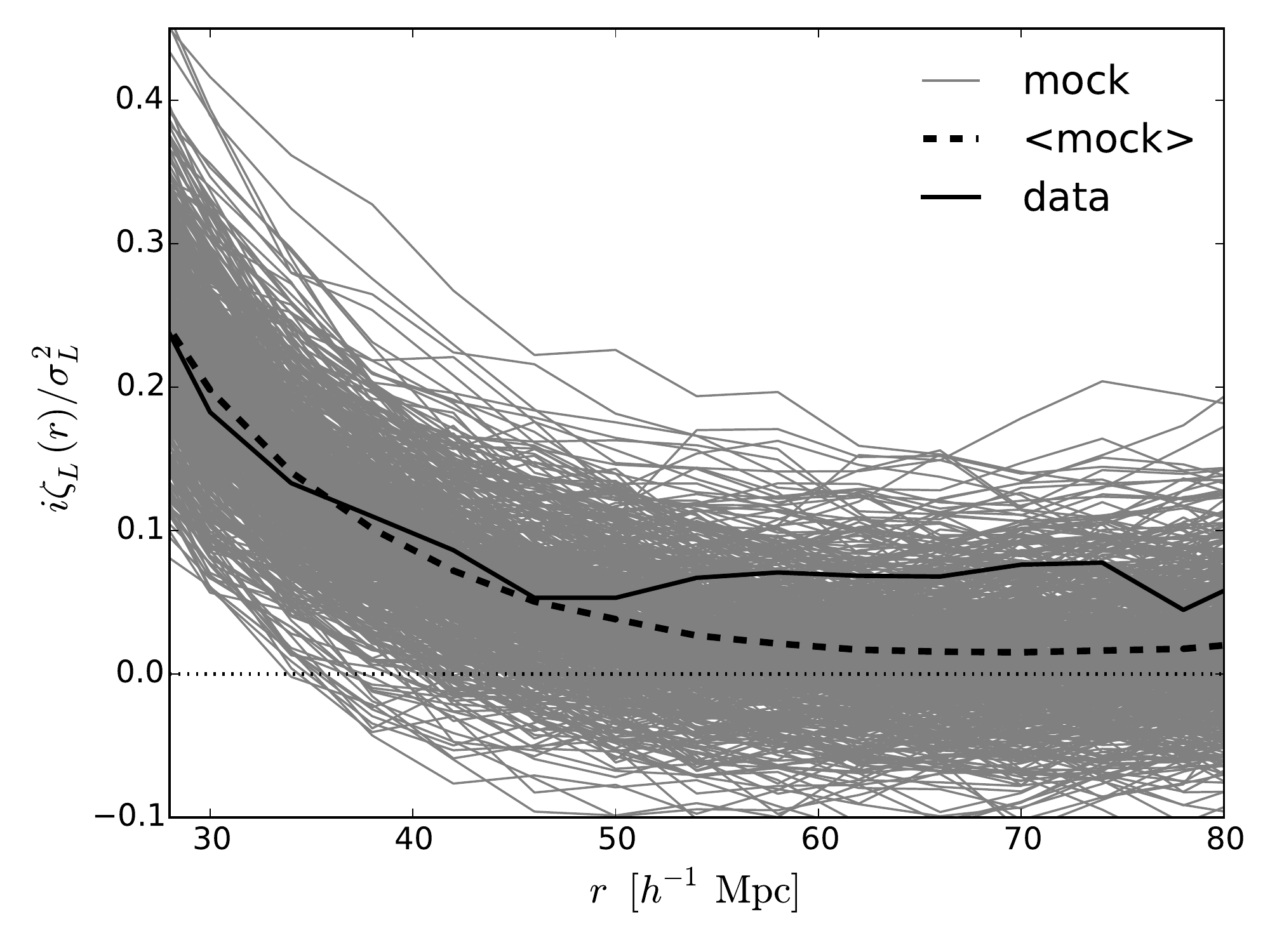}
\includegraphics[width=0.325\textwidth]{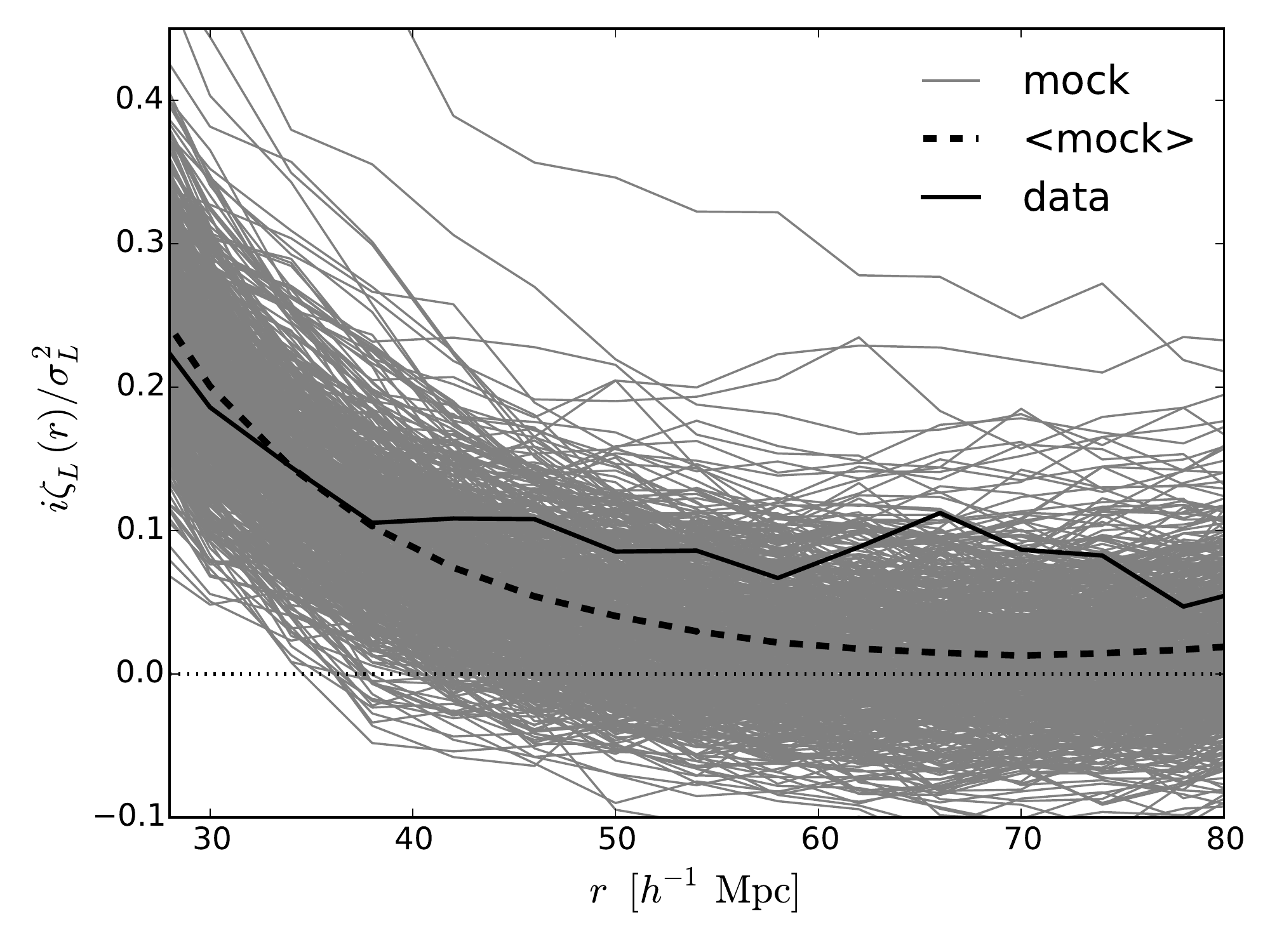}
\includegraphics[width=0.325\textwidth]{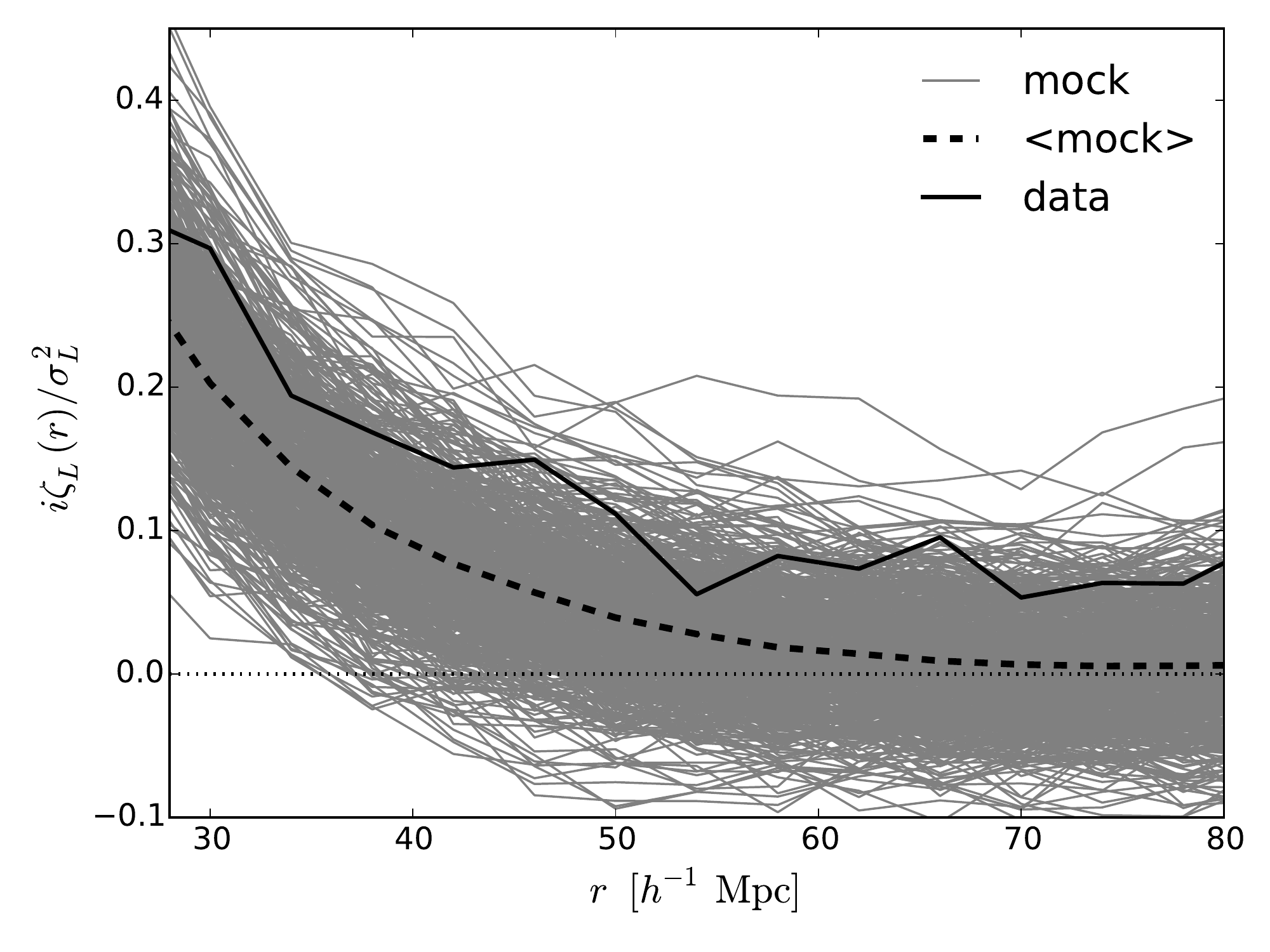}
\includegraphics[width=0.325\textwidth]{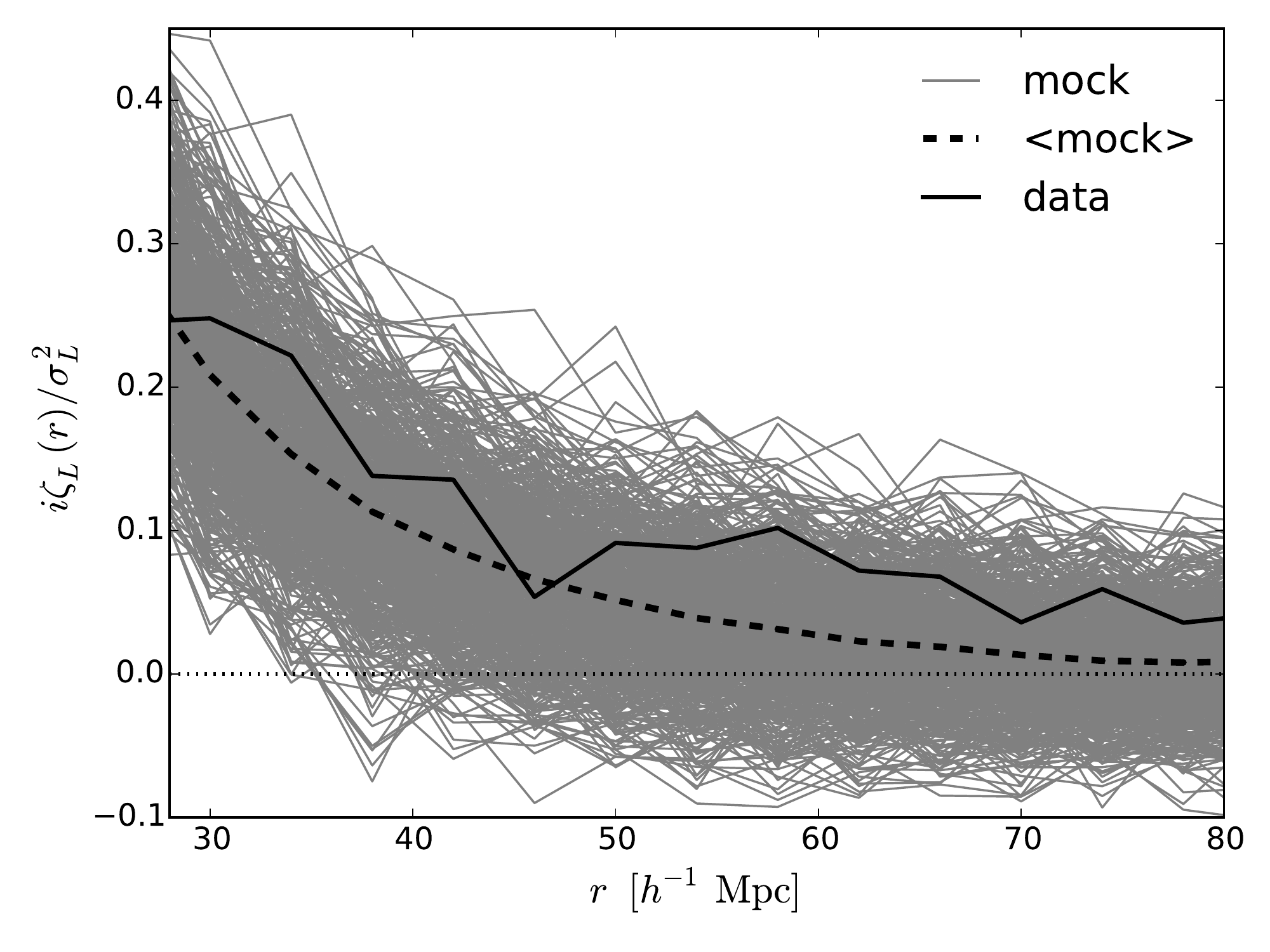}
\caption[Normalized integrated three-point function of $120\hMpc$ subvolumes
in different redshift bins]
{Same as \reffig{zbin_iz_norm_zi_1}, but for $120\hMpc$ subvolumes. The redshift bins
increase from top left to bottom right.}
\label{fig:zbin_iz_norm_zi_2}
\end{figure}

}

%% file: anh_04.tex
{

\renewcommand{\v}[1]{\mathbf{#1}}
\newcommand{\vr}{\v{r}}
\newcommand{\vx}{\v{x}}
\newcommand{\vk}{\v{k}}
\newcommand{\vq}{\v{q}}

\renewcommand{\d}{\delta}
\newcommand{\cS}{\mathcal S}
\newcommand{\cO}{\mathcal O}
\newcommand{\hk}{\hat k}
\newcommand{\hx}{\hat x}
\newcommand{\<}{\left\langle}
\renewcommand{\>}{\right\rangle}
\newcommand{\dP}{\frac{d\ln P_l(k)}{d\ln k}}

\chapter{Squeezed-limit $N$-point functions and power spectrum response}
\label{app:sq_npoint}
In this appendix, we prove the relation between the power spectrum response and
the squeezed limit $N$-point functions, as discussed in \refchp{ch6_npt_sq_sep}.
We only consider equal-time $N$-point functions, to which there are no boost-type
contributions from kinematic consistency relations. Further, we assume that the
long-wavelength modes are well inside the horizon, removing gauge-dependent
terms present for horizon-scale modes.

We first expand the power spectrum as a function of the {\it linearly
extrapolated initial overdensity} $\d_{L0}$ as
\be
 P(k,t|\d_{L0})=\sum_{n=0}^\infty\frac1{n!}R_n(k,t)\left[\d_{L0}\hat D(t)\right]^nP(k,t) ~,
\label{eq:sq_PkexpansionA}
\ee
where $R_n(k,t)$ are response functions with $R_0(k,t)=1$. At the same order
in derivatives, that is at the same order in $k_i/k$ of the squeezed-limit
$N$-point function, the power spectrum will also depend on the long-wavelength
tidal field which can be parametrized through
\be
 K_{ij}(\vk)\equiv\left(\frac{k_i k_j}{k^2}-\frac13\d_{ij}\right)\d(\vk) ~.
\label{eq:sq_tidal}
\ee
As here we consider only the angle-averaged long-wavelength modes, the long-wavelength
tidal field contributions drop out.

In the following, we will suppress the time argument for clarity. The definition
of $\cS_n$ is given by
\ba
 \cS_n(k, k'; k_1,\cdots,k_n)\:&\equiv\int\frac{d^2\hk_1}{4\pi}\cdots
 \int\frac{d^2\hk_n}{4\pi} ~ \< \d(\vk)\d(\vk')\d(\vk_1)\cdots\d(\vk_n) \>'_c \ ~,
\label{eq:sq_Sdef}
\ea
where the prime denotes that the factor $(2\pi)^3\d_D(\vk+\vk'+\vk_{1\cdots n})$
is dropped and $\vk_{1\cdots n}=\sum_{i=1}^n\vk_i$. We consider the limit
\be
 \lim_{k_i\to 0}~\frac{\cS_n(k,k';k_1,\cdots,k_n)}{P(k)P_l(k_1)\cdots P_l(k_n)} ~,
\label{eq:sq_lim_S_1}
\ee
which means that {\it all} $|\vk_i|$ are taken to zero. In this limit, spatial
homogeneity enforces $\vk'=-\vk+\cO(k_i/k)$, so that (for statistically isotropic
initial conditions) the right-hand-side of \refeq{sq_lim_S_1} only depends on $k$.

In order to prove $R_n(k)$ is equivalent to \refeq{sq_lim_S_1}, we first note that
since we are interested in the limit $k_i\to 0$, we can replace $\d(\vk_i)$ in
\refeq{sq_Sdef} with the linear density field $\d_L(\vk_i)$. We further transform
$\vk_i$ into configuration space, writing
\ba
 \cS_n(k;k_1,\cdots,k_n)\:&=\int\frac{d^2\hk_1}{4\pi}\cdots\int\frac{d^2\hk_n}{4\pi}
 \prod_{i=1}^n\int d^3x_i~e^{i\vx_i\cdot\vk_i}\< \d(\vk)\d(\vk')\d(\vx_1)\cdots\d(\vx_n)\>'_c \vs
 \:&=\prod_{i=1}^n\int d^3x_i~e^{i\vx_i\cdot\vk_i}\tilde{\cS}_n(k,x_1,\cdots,x_n) ~,
\label{eq:sq_S_rspace}
\ea
where
\be
 \tilde{\cS}_n(k;x_1,\cdots x_n)\equiv\int\frac{d^2\hx_1}{4\pi}
 \cdots\int\frac{d^2\hx_n}{4\pi}~\< \d(\vk)\d(\vk')\d(\vx_1)\cdots\d(\vx_n)\>'_c ~.
\label{eq:sq_Stdef}
\ee
Note that the angle average is a linear operation and thus commutes with the Fourier
transform; in other words, the $k$-space angle average of the Fourier transform of a
function is the Fourier transform of the $x$-space angle average of the same function.

Now consider the limit $k_i\to0$, which implies that $x_i\to\infty$ in the argument
of $\tilde{\cS}_n$. Then $\tilde{\cS}_n(k)$ describes the modulation of the small-scale
power spectrum $P(k,\v{0})$ measured around $\vx=0$ by $n$ spherically symmetric
large-scale modes (recall that $\vk'\approx-\vk$). This statement can be formalized
by introducing an intermediate scale $R_L$ such that $1/k\ll R_L\ll|\vx_i|\sim1/k_i$
and defining $\d(\vk)\to\d_{R_L}(\vk)$ to be the Fourier transform within a cubic
volume of size $R_L$ around $\vx=0$. Then, $\d_{R_L}(\vk)=\d(\vk)+\cO(1/(kR_L))$,
while the long-wavelength modes are constant over the same volume with corrections
suppressed by $k_iR_L$. The corrections we expect in the end are thus of order $k_i/k$.
To lowest order in these corrections, $\tilde{\cS}_n(k)$ can be written as
\be
 \lim_{k_i\to 0}:~
 \tilde{\cS}_n(k;x_1,\cdots,x_n)=\int\frac{d^2\hx_1}{4\pi}\cdots\int\frac{d^2\hx_n}{4\pi}~
 \< P(k,\v{0})\d_L(\vx_1)\cdots\d_L(\vx_n)\>'_c  ~.
\label{eq:sq_lim_S_2}
\ee
Inserting the expression for the local power spectrum from \refeq{sq_PkexpansionA},
we obtain
\be
 \lim_{k_i\to0}:~\tilde{\cS}_n(k;x_1,\cdots,x_n)=\sum_{m=0}^\infty
 \frac1{m!}R_m(k)P(k)\int\frac{d^2\hx_1}{4\pi}\cdots\int\frac{d^2\hx_n}{4\pi}~
 \< \d_L^m(\v{0})\d_L(\vx_1)\cdots\d_L(\vx_n) \>'_{i-0} ~.
\label{eq:sq_rspace2}
\ee
Here, the subscript $i-0$ indicates that only contractions between $\v{0}$ and $\vx_i$
are to be taken, since the left-hand-side of \refeq{sq_rspace2} is defined through the
connected correlation function (all other contractions would contribute to the disconnected
part of  $\< \d(\vk)\d(\vk')\d(\vx_1)\cdots\d(\vx_n) \>$). Since all density fields in
the correlator in \refeq{sq_rspace2} are linear, limiting to the contractions between
$\v{0}$ and $\vx_i$ then constrains $m=n$.
Therefore,
\ba
 \lim_{k_i\to 0}:~\tilde{\cS}_n(k;x_1,\cdots,x_n)\:&=\frac1{n!}R_n(k)P(k)n!
 \prod_{i=1}^n\int\frac{d^2\hx_i}{4\pi}~\xi_L(\vx_i) \vs
 \:&=R_n(k)P(k)\prod_{i=1}^n\int\frac{d^3k_i}{(2\pi)^3}e^{i\vx_i\cdot\vk_i}P_L(k_i) ~.
\label{eq:sq_lim_S_3}
\ea
Here, $\xi_L$ and $P_L$ denote the linear matter correlation function and power spectrum,
respectively. Going back to Fourier space then immediately yields that for $k_i\to 0$,
\be
 \cS_n(k; k_1,\cdots,k_n)=R_n(k)P(k)\prod_{i=1}^nP_L(k_i)
 +\cO\left(\frac{k_i}{k},\frac{k_i}{k_{\rm NL}}\right) ~,
\label{eq:sq_S_final}
\ee
where $k_{\rm NL}$ is the nonlinear scale. Finally, we can show that
\be
 R_n(k)=\lim_{k_i\to 0}\frac{\cS_n(k;k_1,\cdots,k_n)}{P(k)P_L(k_1)\cdots P_L(k_n)} ~.
\label{eq:sq_Rnpoint}
\ee

This provides the connection between the response functions $R_n(k)$ and the angle-averaged
matter $(n+2)$-point function \refeq{sq_Sdef} in a certain limit squeezed limit (since $k_i\ll k$).
Note that no assumption about the magnitude of $k$ has been made, i.e. this value can be fully
nonlinear. In the following, we illustrate \refeq{sq_Rnpoint} at tree level in perturbation
theory for the cases $n=1$ (three-point function) and $n=2$ (four-point function).

\section{Tree-level result: $n=1$}
\label{app:sq_tr1}
At tree-level for $n=1$ we obtain
\be
 \lim_{k_i\to0}\cS_1(k; k_1)\stackrel{\rm tree-level}{=}
 2 \lim_{k_1\to0}\:\int \frac{d^2\hk_1}{4\pi}
 \left[F_{2}(\vk, \vk_1)P_l(k)+F_{2}(-\vk-\vk_{1},\vk_1)P_l(|\vk+\vk_{1}|)\right]P_l(k_1) ~.
\ee
\refEq{sq_Rnpoint} then yields
\be
 R_1(k) \stackrel{\rm tree-level}{=}\frac{2}{P_l(k)}
 \lim_{k_1\to 0}\int\frac{d^2\hk_1}{4\pi}
 \left[F_{2}(\vk,\vk_1)P_l(k)+F_2(-\vk-\vk_1,\vk_1)P_l(|\vk+\vk_1|)\right]
\label{eq:sq_R1F2}
\ee
with
\be
 F_2(\vk_1,\vk_2)=\frac57+\frac12\mu\left(\frac{k_1}{k_2}+\frac{k_2}{k_1}\right)+\frac27\mu^2 ~,
\ee
where $\mu$ is the cosine of $\vk_1$ and $\vk_2$. The term $\mu/2 (k/k_1)$
is problematic as we are sending $k_1\to 0$. Using that
\ba
 |\vk+\vk_1| \:&= k\left[1 + q \mu + \cO(q^2)\right] ~,~~ q = \frac{k_1}k \vs
 P_l(|\vk+\vk_1|) \:&= P_l(k) \left[1 +\dP q\mu+\cO(q^2)\right] ~,
\ea
the sum of the two IR-divergent terms in \refeq{sq_R1F2} becomes
\ba
 \frac12\left\{\mu\frac{k}{k_1}P_l(k)+\frac{-(\vk+\vk_1)\cdot\vk_1}{|\vk+\vk_1| k_1}P_l(|\vk+\vk_1|)\right\} =
 \frac12 P_l(k)\left[-\mu^2\dP-1\right]+\cO(q) ~.
\ea
As expected, the divergent pieces have canceled.  We have dropped terms of order
$k_1/k$ which are irrelevant in the limit we are interested in. We finally obtain
\ba
 R_1(k)\:&\stackrel{\rm tree-level}{=}2\int_{-1}^1\frac{d\mu}2
 \left[\frac{10}7-\frac12\left(\mu^2\dP+1\right)+\frac47\mu^2\right] \vs
 \:&\hspace{1cm}=2\left[\frac{10}7-\frac16\dP-\frac12+\frac4{21}\right]=\frac{47}{21}-\frac13\dP ~.
\label{eq:sq_R1F2r}
\ea
\refEq{sq_R1F2r} agrees with the linear prediction for $R_1$ shown in \refchp{ch6_npt_sq_sep},
with $f_1$, $e_1$, and $g_1$ computed in \refchp{ch3_sepuni} assuming Einstein-de Sitter universe.

\section{Tree-level result: $n=2$}
\label{app:sq_tr2}
At $n=2$, we have
\ba
 \cS_n(k; k_1,k_2)=\:&\int\frac{d^2\hk_1}{4\pi}\int\frac{d^2\hk_2}{4\pi}~
 \< \d(\vk)\d(\vk')\d(\vk_1)\d(\vk_2)\>'_c \vs
 \stackrel{k_1,k_2\to 0}{=}\:&\int\frac{d^2\hk_1}{4\pi}\int\frac{d^2\hk_2}{4\pi}~
 \Bigg\{6\Big[F_3(\vk,\vk_1,\vk_2)P_l(k)+F_3(-\vk-\vk_{12},\vk_1,\vk_2)P_l(|\vk+\vk_{12}|)\Big] \vs
 &\hspace{3cm}+4 \Big[ F_2(-\vk_1, \vk+\vk_1) F_2(\vk_2, \vk+\vk_1) P_l(|\vk+\vk_1|)\vs
 &\hspace{3cm}+F_2(\vk_1, \vk+\vk_2) F_2(-\vk_2,\vk+\vk_2) P_l(|\vk+\vk_2|) \Big] \Bigg\}P_l(k_1) P_l(k_2) ~,
\ea
where $F_3$ is the symmetrized third-order perturbation theory kernel, and the unsymmetrized
form can be found in \cite{goroff/etal:1986}. Both $F_2^2$ and $F_3$ contain formally IR-divergent
terms up to $\cO[(q_{1,2})^2]$ where $q_{1,2}=k_{1,2}/k$ for $k_{1,2} \to 0$ which cancel in
the end, so we should expand the power spectrum to $\cO[(k_{1,2})^2]$ to obtain the consistent
result at order $\cO[(q_{1,2})^0]$. We have
\ba
 P_l(|\vk+\vk_1|)=\:&P_l(k)\left[1+\left(q_1\mu_1+\frac{q_1^2}{2}[1-\mu_1^2]\right)\frac{k}{P_l(k)}\frac{dP_l(k)}{dk}
 +\frac{q_1^2\mu_1^2}{2}\frac{k^2}{P_l(k)}\frac{d^2P_l(k)}{dk^2}+\cO(q_1^3)\right] \vs
 P_l(|\vk+\vk_1+\vk_2|)=\:&P_l(k)\Bigg[1+\Big([q_1\mu_1+q_2\mu_2]
 +\frac{1}{2}\big[q_1^2(1-\mu_1^2)+q_2^2 (1-\mu_2^2) \vs
 &\hspace{2cm}+2q_1q_2(\mu_{12}-\mu_1\mu_2)\big]\Big)\frac{k}{P_l(k)}\frac{dP_l(k)}{d\ln k} \vs
 &\hspace{1.2cm}+\frac{1}{2}\left(q_1^2\mu_1^2+q_2^2\mu_2^2+2q_1q_2\mu_1\mu_2\right)
 \frac{k^2}{P_l(k)}\frac{d^2P_l(k)}{dk^2}+\cO[(q_{1,2})^3]\Bigg] ~,
\ea
where $\mu_{1,2}$ is the cosine of $\vk$ and $\vk_{1,2}$, and $\mu_{12}$ is the
cosine of $\vk_1$ and $\vk_2$. The leading order terms are
\ba
 R_2(k) \stackrel{\rm tree-level}{=}\:&\int\frac{d^2\hk_1}{4\pi}
 \int\frac{d^2\hk_2}{4\pi}\frac{1}{147}\Bigg[ (628+324\mu_1^2+112\mu_{12}^2-280\mu_1\mu_2\mu_{12} \vs
 &\hspace{5.5cm}+380\mu_2^2 +656\mu_1^2\mu_2^2-56\mu_2^2\mu_{12}^2) \vs
 &\hspace{4cm}+(-273\mu_1^2+147\mu_1\mu_2\mu_{12}-336\mu_2^2 \vs
 &\hspace{5.5cm}-483\mu_1^2\mu_2^2+63\mu_2^2\mu_{12}^2) \frac{k}{P_l(k)} \frac{dP_l(k)}{dk} \vs
 &\hspace{4cm}+ 147\mu_1^2\mu_2^2 \frac{k^2}{P_l(k)} \frac{d^2P_l(k)}{dk^2} \Bigg] \vs
 =\:&\frac{8420}{1323}-\frac{100}{63}\frac{k}{P_l(k)} \frac{dP_l(k)}{dk}
 +\frac{1}{9} \frac{k^2}{P_l(k)} \frac{d^2P_l(k)}{dk^2} ~.
\label{eq:sq_R2F2F3r}
\ea
\refEq{sq_R2F2F3r} also agrees with the linear prediction in \refchp{ch6_npt_sq_sep},
with $e_i,\,f_i,\,g_i$ $(i=1,2)$, computed in \refchp{ch3_sepuni} assuming Eistein-de
Sitter universe, are inserted.

}

%% file: danksagung.tex
\chapter*{Acknowledgment}
First, I would like to thank my advisor, Eiichiro Komatsu, for everything
he has done to me through my graduate school life, in particular for not
giving me up after my qualifying exam and for taking me from Austin to
Munich. Without his help, I will not be able to finish this dissertation.
His broad knowledge from observational astronomy to high energy physics
always amazes me, and sets a gold standard for me in my future career.
Not only advising me academically, he also cares about me personally as
my family, and takes me away from work when necessary. I will never ever
forget the immense gratitude from him.

I would like to thank Donghui Jeong, who is essentially my second advisor
and led me into the field of large-scale structure. Although we only
overlapped in Austin for one year, he taught me so many things in cosmology,
especially during our Friday-textbook-reading session, which were extremely
useful for a beginning student. I also got tremendous amount of help from
him for my first published paper \cite{chiang/etal:2013}.

I would like to thank Christian Wagner, whom I worked the most with after
moving to MPA. He taught me everything about simulations, and I really
enjoyed discussing with him on all topics in the large-scale structure.
Thank him for his patience when I found ``another'' bug in my code. This
dissertation is based on the work I started working with him in late 2012,
and thankfully he became interested in so we could publish a series of
work \cite{chiang/etal:2014,wagner/etal:2015a,wagner/etal:2015b,chiang/etal:2015}.

Faiban Schmidt is one of the most intelligent people I know. His way
of tackling problems as well as extracting the physical meaning always
inspires me. Because of his physical insight, the project that I started
working with Christian grew bigger and became even more interesting.
I benefited a lot from working and discussing with him.

I would like to thank Ariel S\'anchez and Marat Gilfanov for serving
as my committee members. I got many useful from the annual committee
meeting, especially their encouragement when I was still waiting for
the results from my job applications. Especially, I would also like
to thank Ariel, who taught me all the things about BOSS data, so I
can finish the work in \refchp{ch5_posdepxi}.

Being able to participate in this cool project is one of the greatest
experiences in my career. I would like to thank the HETDEX members, in
particular Gary Hill and Karl Gebhardt, for their guide when I first
joined the project and their useful comments on my sparse sampling work
\cite{chiang/etal:2013}. I would also like to thank Niv Drory, who
gave me a ``serious'' lesson on not missing the deadline. He probably
does not remember what happened, but it was a shocking experience for
me so now I always be prepared beforehand.

Masatoshi Shoji and Jonathan Ganc are my best academic siblings.
Sharing office RLM 16.206 is one of the best memories during my
graduate school life. I miss the time when we discussed various
issues in cosmology, made fun of each other, and made the most noise
in the hallway. Our short-lived Monday drinking tradition and Masa's
lung-exercising (smoking) time were truly memorable.

I would like to thank all the astronomy folks in Austin, especially
my classmates Jacob Hummel, Taylor Chonis, Sam Harrold, Rodolfo Santana,
Myoungwon Jeon, Mimi Song, Nalin Vutisalchavakul, and Alan Sluder.
I still remember the time when we were working hard together for
the homework problems. I am also benefited a lot from the Texas
cosmology group: Hyunbae Park, Jun Zhang, Andreas Pawlik, Tanja
Rindler-Daller, Jun Koda, Yi Mao, and Paul Shapiro.

I would like to thank all my friends in Austin. Without them my graduate
student life would be boring. In particular, I would like to thank
Ping-Chun Li and Ke-Yi Lin. They were always the people I talked to
when I felt down about work or life. I still remember the times when
we gathered at late night drinking and complaining the bad things
in our lives, but in the next morning we still had to deal with them!

Moving to Germany is not an easy thing for me. I would like to thank
Philipp Wullstein for all the help whenever I needed it, and Sonja Gr\"undl
for helping me find an apartment. Without their help I would not be
able to survive in Germany. I also want to thank my landlord Herr
Mooseder, who is extremely kind to me even if I do not speak German
at all. (I promised to learn, but gave up because of the difficulty
and time...) He lent his bike``s'' to me and fixed them whenever
I broke them... I genuinely appreciate what they have done for me.

For the years in MPA, I would like to thank everyone in physical
cosmology group: Jaiseung Kim, Xun Shi, Inh Jee, Aniket Agrawal,
Kari Helgason, Sam Ip, and Titouan Lazeyras. Having lunch with
them and discussing various things are always fun. I hope they
do not hate me for pushing them to read papers for the large-scale
structure journal club. In particular, I would like to thank Tsz Yan Lam,
who gave me many useful suggestions for job applications, and encouraged
me when I was still waiting for the results. It was very thoughtful.

I would like to thank all my friends in Munich, especially Yi-Hao Chen (who
also helped me a lot when I moved to Germany), Chia-Yu Hu, Chien-Hsiu Lee,
Ming-Yi Lin, Li-Ting Hsu, and I-Non Chiu. Our Friday study group (which you
should keep even without me) was a truly enjoyable experience, and I learned
a lot about different topics in astronomy, especially observations.

Last but most importantly, I would like to thank my parents and sister
for understanding me of pursuing my own dream, and supporting me endlessly.
I would like to show my deepest appreciation to my beloved fianc\'ee, Hua-Shu,
for her constant encouragement and warm love, especially during the job
application season. She is always the source of happiness and bliss in my life.

%% file: diss.bbl
\begin{thebibliography}{100}

\bibitem{aareth:1963}
S.J{.~}{Aarseth}, \emph{\mnras} \textbf{126}  (1963), 223.

\bibitem{agarwal/ho/shandera:2014}
N{.~}{Agarwal}, S{.~}{Ho} and S{.~}{Shandera}, \emph{\jcap} \textbf{2}  (2014),
  38.

\bibitem{ahn/etal:2014}
C.P{.~}{Ahn}, R{.~}{Alexandroff}, C{.~}{Allende Prieto}, F{.~}{Anders},
  S.F{.~}{Anderson}, T{.~}{Anderton}, B.H{.~}{Andrews}, {\'E}{.~}{Aubourg},
  S{.~}{Bailey}, F.A{.~}{Bastien} and et~al., \emph{\apjs} \textbf{211}
  (2014), 17.

\bibitem{bossdr12:2015}
S{.~}{Alam}, F.D{.~}{Albareti}, C{.~}{Allende Prieto}, F{.~}{Anders},
  S.F{.~}{Anderson}, B.H{.~}{Andrews}, E{.~}{Armengaud}, {\'E}{.~}{Aubourg},
  S{.~}{Bailey}, J.E{.~}{Bautista} and et~al., \emph{ArXiv e-prints}   (2015).

\bibitem{albrecht/steinhardt:1982}
A{.~}{Albrecht} and P.J{.~}{Steinhardt}, \emph{Physical Review Letters}
  \textbf{48}  (1982), 1220.

\bibitem{anderson/etal:2014}
L{.~}{Anderson}, {\'E}{.~}{Aubourg}, S{.~}{Bailey}, F{.~}{Beutler},
  V{.~}{Bhardwaj}, M{.~}{Blanton}, A.S{.~}{Bolton}, J{.~}{Brinkmann},
  J.R{.~}{Brownstein}, A{.~}{Burden}, C.H{.~}{Chuang}, A.J{.~}{Cuesta},
  K.S{.~}{Dawson}, D.J{.~}{Eisenstein}, S{.~}{Escoffier}, J.E{.~}{Gunn},
  H{.~}{Guo}, S{.~}{Ho}, K{.~}{Honscheid}, C{.~}{Howlett}, D{.~}{Kirkby},
  R.H{.~}{Lupton}, M{.~}{Manera}, C{.~}{Maraston}, C.K{.~}{McBride},
  O{.~}{Mena}, F{.~}{Montesano}, R.C{.~}{Nichol}, S.E{.~}{Nuza},
  M.D{.~}{Olmstead}, N{.~}{Padmanabhan}, N{.~}{Palanque-Delabrouille},
  J{.~}{Parejko}, W.J{.~}{Percival}, P{.~}{Petitjean}, F{.~}{Prada},
  A.M{.~}{Price-Whelan}, B{.~}{Reid}, N.A{.~}{Roe}, A.J{.~}{Ross},
  N.P{.~}{Ross}, C.G{.~}{Sabiu}, S{.~}{Saito}, L{.~}{Samushia},
  A.G{.~}{S{\'a}nchez}, D.J{.~}{Schlegel}, D.P{.~}{Schneider},
  C.G{.~}{Scoccola}, H.J{.~}{Seo}, R.A{.~}{Skibba}, M.A{.~}{Strauss},
  M.E.C{.~}{Swanson}, D{.~}{Thomas}, J.L{.~}{Tinker}, R{.~}{Tojeiro},
  M.V{.~}{Maga{\~n}a}, L{.~}{Verde}, D.A{.~}{Wake}, B.A{.~}{Weaver},
  D.H{.~}{Weinberg}, M{.~}{White}, X{.~}{Xu}, C{.~}{Y{\`e}che}, I{.~}{Zehavi}
  and G.B{.~}{Zhao}, \emph{\mnras} \textbf{441}  (2014), 24.

\bibitem{anderson/etal:2014a}
L{.~}{Anderson}, E{.~}{Aubourg}, S{.~}{Bailey}, F{.~}{Beutler},
  A.S{.~}{Bolton}, J{.~}{Brinkmann}, J.R{.~}{Brownstein}, C.H{.~}{Chuang},
  A.J{.~}{Cuesta}, K.S{.~}{Dawson}, D.J{.~}{Eisenstein}, S{.~}{Ho},
  K{.~}{Honscheid}, E.A{.~}{Kazin}, D{.~}{Kirkby}, M{.~}{Manera},
  C.K{.~}{McBride}, O{.~}{Mena}, R.C{.~}{Nichol}, M.D{.~}{Olmstead},
  N{.~}{Padmanabhan}, N{.~}{Palanque-Delabrouille}, W.J{.~}{Percival},
  F{.~}{Prada}, A.J{.~}{Ross}, N.P{.~}{Ross}, A.G{.~}{S{\'a}nchez},
  L{.~}{Samushia}, D.J{.~}{Schlegel}, D.P{.~}{Schneider}, H.J{.~}{Seo},
  M.A{.~}{Strauss}, D{.~}{Thomas}, J.L{.~}{Tinker}, R{.~}{Tojeiro},
  L{.~}{Verde}, D{.~}{Wake}, D.H{.~}{Weinberg}, X{.~}{Xu} and C{.~}{Yeche},
  \emph{\mnras} \textbf{439}  (2014), 83.

\bibitem{baldauf/desjacques/seljak:2014}
T{.~}{Baldauf}, V{.~}{Desjacques} and U{.~}{Seljak}, \emph{ArXiv e-prints}
  (2014).

\bibitem{baldauf/etal:2012}
T{.~}{Baldauf}, U{.~}{Seljak}, V{.~}{Desjacques} and P{.~}{McDonald},
  \emph{\prd} \textbf{86}  (2012), 083540.

\bibitem{baldauf/seljak/senatore:2011}
T{.~}{Baldauf}, U{.~}{Seljak} and L{.~}{Senatore}, \emph{\jcap} \textbf{4}
  (2011), 6.

\bibitem{baldauf/etal:2011}
T{.~}{Baldauf}, U{.~}{Seljak}, L{.~}{Senatore} and M{.~}{Zaldarriaga},
  \emph{\jcap} \textbf{10}  (2011), 31.

\bibitem{barriga/gaztanaga:2002}
J{.~}{Barriga} and E{.~}{Gazta{\~n}aga}, \emph{\mnras} \textbf{333}  (2002),
  443.

\bibitem{barrow/saich:1993}
J.D{.~}{Barrow} and P{.~}{Saich}, \emph{\mnras} \textbf{262}  (1993), 717.

\bibitem{baumann/etal:2012}
D{.~}{Baumann}, A{.~}{Nicolis}, L{.~}{Senatore} and M{.~}{Zaldarriaga},
  \emph{\jcap} \textbf{7}  (2012), 51.

\bibitem{ben-dayan/etal:2015}
I{.~}{Ben-Dayan}, T{.~}{Konstandin}, R.A{.~}{Porto} and L{.~}{Sagunski},
  \emph{\jcap} \textbf{2}  (2015), 26.

\bibitem{berlind/etal:2003}
A.A{.~}{Berlind}, D.H{.~}{Weinberg}, A.J{.~}{Benson}, C.M{.~}{Baugh},
  S{.~}{Cole}, R{.~}{Dav{\'e}}, C.S{.~}{Frenk}, A{.~}{Jenkins}, N{.~}{Katz} and
  C.G{.~}{Lacey}, \emph{\apj} \textbf{593}  (2003), 1.

\bibitem{bernardeau/etal:2002}
F{.~}{Bernardeau}, S{.~}{Colombi}, E{.~}{Gazta{\~n}aga} and R{.~}{Scoccimarro},
  \emph{\physrep} \textbf{367}  (2002), 1.

\bibitem{bernardeau/crocce/scoccimarro:2012}
F{.~}{Bernardeau}, M{.~}{Crocce} and R{.~}{Scoccimarro}, \emph{\prd}
  \textbf{85}  (2012), 123519.

\bibitem{beutler/etal:2014}
F{.~}{Beutler}, S{.~}{Saito}, H.J{.~}{Seo}, J{.~}{Brinkmann}, K.S{.~}{Dawson},
  D.J{.~}{Eisenstein}, A{.~}{Font-Ribera}, S{.~}{Ho}, C.K{.~}{McBride},
  F{.~}{Montesano}, W.J{.~}{Percival}, A.J{.~}{Ross}, N.P{.~}{Ross},
  L{.~}{Samushia}, D.J{.~}{Schlegel}, A.G{.~}{S{\'a}nchez}, J.L{.~}{Tinker} and
  B.A{.~}{Weaver}, \emph{\mnras} \textbf{443}  (2014), 1065.

\bibitem{bhattacharya/etal:2011}
S{.~}{Bhattacharya}, K{.~}{Heitmann}, M{.~}{White}, Z{.~}{Luki{\'c}},
  C{.~}{Wagner} and S{.~}{Habib}, \emph{\apj} \textbf{732}  (2011), 122.

\bibitem{biagetti/etal:2014}
M{.~}{Biagetti}, V{.~}{Desjacques}, A{.~}{Kehagias} and A{.~}{Riotto},
  \emph{\prd} \textbf{90}  (2014), 103529.

\bibitem{blake/etal:2011a}
C{.~}{Blake}, T{.~}{Davis}, G.B{.~}{Poole}, D{.~}{Parkinson}, S{.~}{Brough},
  M{.~}{Colless}, C{.~}{Contreras}, W{.~}{Couch}, S{.~}{Croom},
  M.J{.~}{Drinkwater}, K{.~}{Forster}, D{.~}{Gilbank}, M{.~}{Gladders},
  K{.~}{Glazebrook}, B{.~}{Jelliffe}, R.J{.~}{Jurek}, I.H{.~}{Li},
  B{.~}{Madore}, D.C{.~}{Martin}, K{.~}{Pimbblet}, M{.~}{Pracy}, R{.~}{Sharp},
  E{.~}{Wisnioski}, D{.~}{Woods}, T.K{.~}{Wyder} and H.K.C{.~}{Yee},
  \emph{\mnras} \textbf{415}  (2011), 2892.

\bibitem{blake/etal:2011b}
C{.~}{Blake}, E.A{.~}{Kazin}, F{.~}{Beutler}, T.M{.~}{Davis}, D{.~}{Parkinson},
  S{.~}{Brough}, M{.~}{Colless}, C{.~}{Contreras}, W{.~}{Couch}, S{.~}{Croom},
  D{.~}{Croton}, M.J{.~}{Drinkwater}, K{.~}{Forster}, D{.~}{Gilbank},
  M{.~}{Gladders}, K{.~}{Glazebrook}, B{.~}{Jelliffe}, R.J{.~}{Jurek},
  I.H{.~}{Li}, B{.~}{Madore}, D.C{.~}{Martin}, K{.~}{Pimbblet}, G.B{.~}{Poole},
  M{.~}{Pracy}, R{.~}{Sharp}, E{.~}{Wisnioski}, D{.~}{Woods}, T.K{.~}{Wyder}
  and H.K.C{.~}{Yee}, \emph{\mnras} \textbf{418}  (2011), 1707.

\bibitem{bullock/etal:2001}
J.S{.~}{Bullock}, T.S{.~}{Kolatt}, Y{.~}{Sigad}, R.S{.~}{Somerville},
  A.V{.~}{Kravtsov}, A.A{.~}{Klypin}, J.R{.~}{Primack} and A{.~}{Dekel},
  \emph{\mnras} \textbf{321}  (2001), 559.

\bibitem{carrasco/hertzberg/senatore:2012}
J.J.M{.~}{Carrasco}, M.P{.~}{Hertzberg} and L{.~}{Senatore}, \emph{Journal of
  High Energy Physics} \textbf{9}  (2012), 82.

\bibitem{chan/scoccimarro/sheth:2012}
K.C{.~}{Chan}, R{.~}{Scoccimarro} and R.K{.~}{Sheth}, \emph{\prd} \textbf{85}
  (2012), 083509.

\bibitem{chen:2010}
X{.~}{Chen}, \emph{Advances in Astronomy} \textbf{2010}  (2010), 72.

\bibitem{chiang/etal:2015}
C.T{.~}{Chiang}, C{.~}{Wagner}, A.G{.~}{S{\'a}nchez}, F{.~}{Schmidt} and
  E{.~}{Komatsu}, \emph{ArXiv e-prints}   (2015).

\bibitem{chiang/etal:2014}
C.T{.~}{Chiang}, C{.~}{Wagner}, F{.~}{Schmidt} and E{.~}{Komatsu}, \emph{\jcap}
  \textbf{5}  (2014), 48.

\bibitem{chiang/etal:2013}
C.T{.~}{Chiang}, P{.~}{Wullstein}, D{.~}{Jeong}, E{.~}{Komatsu},
  G.A{.~}{Blanc}, R{.~}{Ciardullo}, N{.~}{Drory}, M{.~}{Fabricius},
  S{.~}{Finkelstein}, K{.~}{Gebhardt}, C{.~}{Gronwall}, A{.~}{Hagen},
  G.J{.~}{Hill}, I{.~}{Jee}, S{.~}{Jogee}, M{.~}{Landriau}, E{.~}{Mentuch
  Cooper}, D.P{.~}{Schneider} and S{.~}{Tuttle}, \emph{\jcap} \textbf{12}
  (2013), 30.

\bibitem{chuang/etal:2014}
C.H{.~}{Chuang}, C{.~}{Zhao}, F{.~}{Prada}, E{.~}{Munari}, S{.~}{Avila},
  A{.~}{Izard}, F.S{.~}{Kitaura}, M{.~}{Manera}, P{.~}{Monaco}, S{.~}{Murray},
  A{.~}{Knebe}, C.G{.~}{Scoccola}, G{.~}{Yepes}, J{.~}{Garcia-Bellido},
  F.A{.~}{Marin}, V{.~}{Muller}, R{.~}{Skibba}, M{.~}{Crocce}, P{.~}{Fosalba},
  S{.~}{Gottlober}, A.A{.~}{Klypin}, C{.~}{Power}, C{.~}{Tao} and
  V{.~}{Turchaninov}, \emph{ArXiv e-prints}   (2014).

\bibitem{cole/kaiser:1989}
S{.~}{Cole} and N{.~}{Kaiser}, \emph{\mnras} \textbf{237}  (1989), 1127.

\bibitem{cole/etal:2005}
S{.~}{Cole}, W.J{.~}{Percival}, J.A{.~}{Peacock}, P{.~}{Norberg},
  C.M{.~}{Baugh}, C.S{.~}{Frenk}, I{.~}{Baldry}, J{.~}{Bland-Hawthorn},
  T{.~}{Bridges}, R{.~}{Cannon}, M{.~}{Colless}, C{.~}{Collins}, W{.~}{Couch},
  N.J.G{.~}{Cross}, G{.~}{Dalton}, V.R{.~}{Eke}, R{.~}{De Propris},
  S.P{.~}{Driver}, G{.~}{Efstathiou}, R.S{.~}{Ellis}, K{.~}{Glazebrook},
  C{.~}{Jackson}, A{.~}{Jenkins}, O{.~}{Lahav}, I{.~}{Lewis}, S{.~}{Lumsden},
  S{.~}{Maddox}, D{.~}{Madgwick}, B.A{.~}{Peterson}, W{.~}{Sutherland} and
  K{.~}{Taylor}, \emph{\mnras} \textbf{362}  (2005), 505.

\bibitem{colless/etal:2001}
M{.~}{Colless}, G{.~}{Dalton}, S{.~}{Maddox}, W{.~}{Sutherland},
  P{.~}{Norberg}, S{.~}{Cole}, J{.~}{Bland-Hawthorn}, T{.~}{Bridges},
  R{.~}{Cannon}, C{.~}{Collins}, W{.~}{Couch}, N{.~}{Cross}, K{.~}{Deeley},
  R{.~}{De Propris}, S.P{.~}{Driver}, G{.~}{Efstathiou}, R.S{.~}{Ellis},
  C.S{.~}{Frenk}, K{.~}{Glazebrook}, C{.~}{Jackson}, O{.~}{Lahav},
  I{.~}{Lewis}, S{.~}{Lumsden}, D{.~}{Madgwick}, J.A{.~}{Peacock},
  B.A{.~}{Peterson}, I{.~}{Price}, M{.~}{Seaborne} and K{.~}{Taylor},
  \emph{\mnras} \textbf{328}  (2001), 1039.

\bibitem{cooray/sheth:2002}
A{.~}{Cooray} and R{.~}{Sheth}, \emph{\physrep} \textbf{372}  (2002), 1.

\bibitem{creminelli/etal:2010}
P{.~}{Creminelli}, G{.~}{D'Amico}, J{.~}{Nore{\~n}a}, L{.~}{Senatore} and
  F{.~}{Vernizzi}, \emph{\jcap} \textbf{3}  (2010), 27.

\bibitem{creminelli/etal:2014b}
P{.~}{Creminelli}, J{.~}{Gleyzes}, L{.~}{Hui}, M{.~}{Simonovi{\'c}} and
  F{.~}{Vernizzi}, \emph{\jcap} \textbf{6}  (2014), 9.

\bibitem{creminelli/etal:2014a}
P{.~}{Creminelli}, J{.~}{Gleyzes}, M{.~}{Simonovi{\'c}} and F{.~}{Vernizzi},
  \emph{\jcap} \textbf{2}  (2014), 51.

\bibitem{creminelli/etal:2013}
P{.~}{Creminelli}, J{.~}{Nore{\~n}a}, M{.~}{Simonovi{\'c}} and F{.~}{Vernizzi},
  \emph{\jcap} \textbf{12}  (2013), 25.

\bibitem{crocce/pueblas/scoccimarro:2006}
M{.~}{Crocce}, S{.~}{Pueblas} and R{.~}{Scoccimarro}, \emph{\mnras}
  \textbf{373}  (2006), 369.

\bibitem{crocce/scoccimarro:2006}
M{.~}{Crocce} and R{.~}{Scoccimarro}, \emph{\prd} \textbf{73}  (2006), 063519.

\bibitem{crocce/scoccimarro:2008}
M{.~}{Crocce} and R{.~}{Scoccimarro}, \emph{\prd} \textbf{77}  (2008), 023533.

\bibitem{dai/pajer/schmidt:2015a}
L{.~}{Dai}, E{.~}{Pajer} and F{.~}{Schmidt}, \emph{ArXiv e-prints}   (2015).

\bibitem{dai/pajer/schmidt:2015b}
L{.~}{Dai}, E{.~}{Pajer} and F{.~}{Schmidt}, \emph{ArXiv e-prints}   (2015).

\bibitem{dalal/etal:2008}
N{.~}{Dalal}, O{.~}{Dor{\'e}}, D{.~}{Huterer} and A{.~}{Shirokov}, \emph{\prd}
  \textbf{77}  (2008), 123514.

\bibitem{davis/peebles:1983}
M{.~}{Davis} and P.J.E{.~}{Peebles}, \emph{\apj} \textbf{267}  (1983), 465.

\bibitem{deputter/etal:2012}
R{.~}{de Putter}, C{.~}{Wagner}, O{.~}{Mena}, L{.~}{Verde} and
  W.J{.~}{Percival}, \emph{\jcap} \textbf{4}  (2012), 19.

\bibitem{dehnen/read:2011}
W{.~}{Dehnen} and J.I{.~}{Read}, \emph{European Physical Journal Plus}
  \textbf{126}  (2011), 55.

\bibitem{desjacques/etal:2010}
V{.~}{Desjacques}, M{.~}{Crocce}, R{.~}{Scoccimarro} and R.K{.~}{Sheth},
  \emph{\prd} \textbf{82}  (2010), 103529.

\bibitem{desjacques/sheth:2010}
V{.~}{Desjacques} and R.K{.~}{Sheth}, \emph{\prd} \textbf{81}  (2010), 023526.

\bibitem{dodelson/schneider:2013}
S{.~}{Dodelson} and M.D{.~}{Schneider}, \emph{\prd} \textbf{88}  (2013),
  063537.

\bibitem{efstathiou/etal:1990}
G{.~}{Efstathiou}, N{.~}{Kaiser}, W{.~}{Saunders}, A{.~}{Lawrence},
  M{.~}{Rowan-Robinson}, R.S{.~}{Ellis} and C.S{.~}{Frenk}, \emph{\mnras}
  \textbf{247}  (1990), 10P.

\bibitem{eisenstein/seo/white:2007}
D.J{.~}{Eisenstein}, H.J{.~}{Seo} and M{.~}{White}, \emph{\apj} \textbf{664}
  (2007), 660.

\bibitem{eisenstein/etal:2005}
D.J{.~}{Eisenstein}, I{.~}{Zehavi}, D.W{.~}{Hogg}, R{.~}{Scoccimarro},
  M.R{.~}{Blanton}, R.C{.~}{Nichol}, R{.~}{Scranton}, H.J{.~}{Seo},
  M{.~}{Tegmark}, Z{.~}{Zheng}, S.F{.~}{Anderson}, J{.~}{Annis},
  N{.~}{Bahcall}, J{.~}{Brinkmann}, S{.~}{Burles}, F.J{.~}{Castander},
  A{.~}{Connolly}, I{.~}{Csabai}, M{.~}{Doi}, M{.~}{Fukugita},
  J.A{.~}{Frieman}, K{.~}{Glazebrook}, J.E{.~}{Gunn}, J.S{.~}{Hendry},
  G{.~}{Hennessy}, Z{.~}{Ivezi{\'c}}, S{.~}{Kent}, G.R{.~}{Knapp}, H{.~}{Lin},
  Y.S{.~}{Loh}, R.H{.~}{Lupton}, B{.~}{Margon}, T.A{.~}{McKay}, A{.~}{Meiksin},
  J.A{.~}{Munn}, A{.~}{Pope}, M.W{.~}{Richmond}, D{.~}{Schlegel},
  D.P{.~}{Schneider}, K{.~}{Shimasaku}, C{.~}{Stoughton}, M.A{.~}{Strauss},
  M{.~}{SubbaRao}, A.S{.~}{Szalay}, I{.~}{Szapudi}, D.L{.~}{Tucker},
  B{.~}{Yanny} and D.G{.~}{York}, \emph{\apj} \textbf{633}  (2005), 560.

\bibitem{feldman/etal:2001}
H.A{.~}{Feldman}, J.A{.~}{Frieman}, J.N{.~}{Fry} and R{.~}{Scoccimarro},
  \emph{Physical Review Letters} \textbf{86}  (2001), 1434.

\bibitem{feldman/kaiser/peacock:1994}
H.A{.~}{Feldman}, N{.~}{Kaiser} and J.A{.~}{Peacock}, \emph{\apj} \textbf{426}
  (1994), 23.

\bibitem{frieman/turner/huterer:2008}
J.A{.~}{Frieman}, M.S{.~}{Turner} and D{.~}{Huterer}, \emph{\araa} \textbf{46}
  (2008), 385.

\bibitem{fry:1984}
J.N{.~}{Fry}, \emph{\apj} \textbf{279}  (1984), 499.

\bibitem{fry/gaztanaga:1993}
J.N{.~}{Fry} and E{.~}{Gaztanaga}, \emph{\apj} \textbf{413}  (1993), 447.

\bibitem{giannantonio/percival:2014}
T{.~}{Giannantonio} and W.J{.~}{Percival}, \emph{\mnras} \textbf{441}  (2014),
  L16.

\bibitem{giannantonio/etal:2014}
T{.~}{Giannantonio}, A.J{.~}{Ross}, W.J{.~}{Percival}, R{.~}{Crittenden},
  D{.~}{Bacher}, M{.~}{Kilbinger}, R{.~}{Nichol} and J{.~}{Weller}, \emph{\prd}
  \textbf{89}  (2014), 023511.

\bibitem{gilmarin/etal:2014b}
H{.~}{Gil-Mar{\'{\i}}n}, J{.~}{Nore{\~n}a}, L{.~}{Verde}, W.J{.~}{Percival},
  C{.~}{Wagner}, M{.~}{Manera} and D.P{.~}{Schneider}, \emph{ArXiv e-prints}
  (2014).

\bibitem{gilmarin/etal:2014a}
H{.~}{Gil-Mar{\'{\i}}n}, C{.~}{Wagner}, J{.~}{Nore{\~n}a}, L{.~}{Verde} and
  W{.~}{Percival}, \emph{\jcap} \textbf{12}  (2014), 29.

\bibitem{gilmarin/etal:2012}
H{.~}{Gil-Mar{\'{\i}}n}, C{.~}{Wagner}, L{.~}{Verde}, C{.~}{Porciani} and
  R{.~}{Jimenez}, \emph{\jcap} \textbf{11}  (2012), 29.

\bibitem{goroff/etal:1986}
M.H{.~}{Goroff}, B{.~}{Grinstein}, S.J{.~}{Rey} and M.B{.~}{Wise}, \emph{\apj}
  \textbf{311}  (1986), 6.

\bibitem{gunn/gott:1972}
J.E{.~}{Gunn} and J.R{.~}{Gott}, III, \emph{\apj} \textbf{176}  (1972), 1.

\bibitem{guo/etal:2015}
H{.~}{Guo}, Z{.~}{Zheng}, Y.P{.~}{Jing}, I{.~}{Zehavi}, C{.~}{Li},
  D.H{.~}{Weinberg}, R.A{.~}{Skibba}, R.C{.~}{Nichol}, G{.~}{Rossi},
  C.G{.~}{Sabiu}, D.P{.~}{Schneider} and C.K{.~}{McBride}, \emph{\mnras}
  \textbf{449}  (2015), L95.

\bibitem{guo/etal:2013}
Q{.~}{Guo}, S{.~}{White}, R.E{.~}{Angulo}, B{.~}{Henriques}, G{.~}{Lemson},
  M{.~}{Boylan-Kolchin}, P{.~}{Thomas} and C{.~}{Short}, \emph{\mnras}
  \textbf{428}  (2013), 1351.

\bibitem{guth:1981}
A.H{.~}{Guth}, \emph{\prd} \textbf{23}  (1981), 347.

\bibitem{heitmann/etal:2014}
K{.~}{Heitmann}, E{.~}{Lawrence}, J{.~}{Kwan}, S{.~}{Habib} and D{.~}{Higdon},
  \emph{\apj} \textbf{780}  (2014), 111.

\bibitem{hill/etal:2008}
G.J{.~}{Hill}, K{.~}{Gebhardt}, E{.~}{Komatsu}, N{.~}{Drory},
  P.J{.~}{MacQueen}, J{.~}{Adams}, G.A{.~}{Blanc}, R{.~}{Koehler},
  M{.~}{Rafal}, M.M{.~}{Roth}, A{.~}{Kelz}, C{.~}{Gronwall}, R{.~}{Ciardullo}
  and D.P{.~}{Schneider}: \emph{{The Hobby-Eberly Telescope Dark Energy
  Experiment (HETDEX): Description and Early Pilot Survey Results}}.
\newblock \emph{{The Hobby-Eberly Telescope Dark Energy Experiment (HETDEX):
  Description and Early Pilot Survey Results}}, In \emph{Panoramic Views of
  Galaxy Formation and Evolution}, herausgegeben von T.~{Kodama}, T.~{Yamada}
  and K.~{Aoki}, Band 399 von \emph{Astronomical Society of the Pacific
  Conference Series}. (October 2008) Seite 115.

\bibitem{ho/etal:2013}
S{.~}{Ho}, N{.~}{Agarwal}, A.D{.~}{Myers}, R{.~}{Lyons}, A{.~}{Disbrow},
  H.J{.~}{Seo}, A{.~}{Ross}, C{.~}{Hirata}, N{.~}{Padmanabhan},
  R{.~}{O'Connell}, E{.~}{Huff}, D{.~}{Schlegel}, A{.~}{Slosar},
  D{.~}{Weinberg}, M{.~}{Strauss}, N.P{.~}{Ross}, D.P{.~}{Schneider},
  N{.~}{Bahcall}, J{.~}{Brinkmann}, N{.~}{Palanque-Delabrouille} and
  C{.~}{Y{\`e}che}, \emph{ArXiv e-prints}   (2013).

\bibitem{horn/hui/xiao:2015}
B{.~}{Horn}, L{.~}{Hui} and X{.~}{Xiao}, \emph{ArXiv e-prints}   (2015).

\bibitem{huchra/etal:1983}
J{.~}{Huchra}, M{.~}{Davis}, D{.~}{Latham} and J{.~}{Tonry}, \emph{\apjs}
  \textbf{52}  (1983), 89.

\bibitem{jackson:1972}
J.C{.~}{Jackson}, \emph{\mnras} \textbf{156}  (1972), 1P.

\bibitem{jain/bertschinger:1994}
B{.~}{Jain} and E{.~}{Bertschinger}, \emph{\apj} \textbf{431}  (1994), 495.

\bibitem{jeong/komatsu:2006}
D{.~}{Jeong} and E{.~}{Komatsu}, \emph{\apj} \textbf{651}  (2006), 619.

\bibitem{jing:2005}
Y.P{.~}{Jing}, \emph{\apj} \textbf{620}  (2005), 559.

\bibitem{jing/boerner:1997}
Y.P{.~}{Jing} and G{.~}{Boerner}, \emph{\aap} \textbf{318}  (1997), 667.

\bibitem{kaiser:1987}
N{.~}{Kaiser}, \emph{\mnras} \textbf{227}  (1987), 1.

\bibitem{kauffmann/white/guiderdoni:1993}
G{.~}{Kauffmann}, S.D.M{.~}{White} and B{.~}{Guiderdoni}, \emph{\mnras}
  \textbf{264}  (1993), 201.

\bibitem{kayo/etal:2004}
I{.~}{Kayo}, Y{.~}{Suto}, R.C{.~}{Nichol}, J{.~}{Pan}, I{.~}{Szapudi},
  A.J{.~}{Connolly}, J{.~}{Gardner}, B{.~}{Jain}, G{.~}{Kulkarni},
  T{.~}{Matsubara}, R{.~}{Sheth}, A.S{.~}{Szalay} and J{.~}{Brinkmann},
  \emph{\pasj} \textbf{56}  (2004), 415.

\bibitem{kehagias/etal:2014}
A{.~}{Kehagias}, J{.~}{Nore{\~n}a}, H{.~}{Perrier} and A{.~}{Riotto},
  \emph{Nuclear Physics B} \textbf{883}  (2014), 83.

\bibitem{kehagias/perrier/riotto:2014}
A{.~}{Kehagias}, H{.~}{Perrier} and A{.~}{Riotto}, \emph{Modern Physics Letters
  A} \textbf{29}  (2014), 50152.

\bibitem{kehagias/riotto:2013}
A{.~}{Kehagias} and A{.~}{Riotto}, \emph{Nuclear Physics B} \textbf{873}
  (2013), 514.

\bibitem{klypin/etal:2014}
A{.~}{Klypin}, G{.~}{Yepes}, S{.~}{Gottlober}, F{.~}{Prada} and S{.~}{Hess},
  \emph{ArXiv e-prints}   (2014).

\bibitem{komatsu:2010}
E{.~}{Komatsu}, \emph{Classical and Quantum Gravity} \textbf{27}  (2010),
  124010.

\bibitem{komatsu/etal:2011}
E{.~}{Komatsu}, K.M{.~}{Smith}, J{.~}{Dunkley}, C.L{.~}{Bennett}, B{.~}{Gold},
  G{.~}{Hinshaw}, N{.~}{Jarosik}, D{.~}{Larson}, M.R{.~}{Nolta}, L{.~}{Page},
  D.N{.~}{Spergel}, M{.~}{Halpern}, R.S{.~}{Hill}, A{.~}{Kogut}, M{.~}{Limon},
  S.S{.~}{Meyer}, N{.~}{Odegard}, G.S{.~}{Tucker}, J.L{.~}{Weiland},
  E{.~}{Wollack} and E.L{.~}{Wright}, \emph{\apjs} \textbf{192}  (2011), 18.

\bibitem{komatsu/spergel:2001}
E{.~}{Komatsu} and D.N{.~}{Spergel}, \emph{\prd} \textbf{63}  (2001), 063002.

\bibitem{kravtsov/etal:2004}
A.V{.~}{Kravtsov}, A.A{.~}{Berlind}, R.H{.~}{Wechsler}, A.A{.~}{Klypin},
  S{.~}{Gottl{\"o}ber}, B{.~}{Allgood} and J.R{.~}{Primack}, \emph{\apj}
  \textbf{609}  (2004), 35.

\bibitem{landy/szalay:1993}
S.D{.~}{Landy} and A.S{.~}{Szalay}, \emph{\apj} \textbf{412}  (1993), 64.

\bibitem{leistedt/peiris/roth:2014}
B{.~}{Leistedt}, H.V{.~}{Peiris} and N{.~}{Roth}, \emph{Physical Review
  Letters} \textbf{113}  (2014), 221301.

\bibitem{lemaitre:1933}
G{.~}{Lema{\^i}tre}, \emph{Annales de la Soci{\'e}t{\'e} Scientifique de
  Bruxelles} \textbf{53}  (1933), 51.

\bibitem{lesgourgues:2011}
J{.~}{Lesgourgues}, \emph{ArXiv e-prints}   (2011).

\bibitem{lewis/bridle:2002}
A{.~}{Lewis} and S{.~}{Bridle}, \emph{\prd} \textbf{66}  (2002), 103511.

\bibitem{lewis/challinor/lasenby:2000}
A{.~}{Lewis}, A{.~}{Challinor} and A{.~}{Lasenby}, \emph{\apj} \textbf{538}
  (2000), 473.

\bibitem{li/hu/takada:2014}
Y{.~}{Li}, W{.~}{Hu} and M{.~}{Takada}, \emph{\prd} \textbf{89}  (2014),
  083519.

\bibitem{li/hu/takada:2014b}
Y{.~}{Li}, W{.~}{Hu} and M{.~}{Takada}, \emph{\prd} \textbf{90}  (2014),
  103530.

\bibitem{linde:1982}
A.D{.~}{Linde}, \emph{Physics Letters B} \textbf{108}  (1982), 389.

\bibitem{lsst:2012}
{LSST Dark Energy Science Collaboration}, \emph{ArXiv e-prints}   (2012).

\bibitem{maddox/etal:1990}
S.J{.~}{Maddox}, G{.~}{Efstathiou}, W.J{.~}{Sutherland} and J{.~}{Loveday},
  \emph{\mnras} \textbf{242}  (1990), 43P.

\bibitem{mana/etal:2013}
A{.~}{Mana}, T{.~}{Giannantonio}, J{.~}{Weller}, B{.~}{Hoyle}, G{.~}{H{\"u}tsi}
  and B{.~}{Sartoris}, \emph{\mnras} \textbf{434}  (2013), 684.

\bibitem{manera/etal:2015}
M{.~}{Manera}, L{.~}{Samushia}, R{.~}{Tojeiro}, C{.~}{Howlett}, A.J{.~}{Ross},
  W.J{.~}{Percival}, H{.~}{Gil-Mar{\'{\i}}n}, J.R{.~}{Brownstein},
  A{.~}{Burden} and F{.~}{Montesano}, \emph{\mnras} \textbf{447}  (2015), 437.

\bibitem{manera/etal:2013}
M{.~}{Manera}, R{.~}{Scoccimarro}, W.J{.~}{Percival}, L{.~}{Samushia},
  C.K{.~}{McBride}, A.J{.~}{Ross}, R.K{.~}{Sheth}, M{.~}{White}, B.A{.~}{Reid},
  A.G{.~}{S{\'a}nchez}, R{.~}{de Putter}, X{.~}{Xu}, A.A{.~}{Berlind},
  J{.~}{Brinkmann}, C{.~}{Maraston}, B{.~}{Nichol}, F{.~}{Montesano},
  N{.~}{Padmanabhan}, R.A{.~}{Skibba}, R{.~}{Tojeiro} and B.A{.~}{Weaver},
  \emph{\mnras} \textbf{428}  (2013), 1036.

\bibitem{marin/etal:2013}
F.A{.~}{Mar{\'{\i}}n}, C{.~}{Blake}, G.B{.~}{Poole}, C.K{.~}{McBride},
  S{.~}{Brough}, M{.~}{Colless}, C{.~}{Contreras}, W{.~}{Couch},
  D.J{.~}{Croton}, S{.~}{Croom}, T{.~}{Davis}, M.J{.~}{Drinkwater},
  K{.~}{Forster}, D{.~}{Gilbank}, M{.~}{Gladders}, K{.~}{Glazebrook},
  B{.~}{Jelliffe}, R.J{.~}{Jurek}, I.h{.~}{Li}, B{.~}{Madore}, D.C{.~}{Martin},
  K{.~}{Pimbblet}, M{.~}{Pracy}, R{.~}{Sharp}, E{.~}{Wisnioski}, D{.~}{Woods},
  T.K{.~}{Wyder} and H.K.C{.~}{Yee}, \emph{\mnras} \textbf{432}  (2013), 2654.

\bibitem{matarrese/verde:2008}
S{.~}{Matarrese} and L{.~}{Verde}, \emph{\apjl} \textbf{677}  (2008), L77.

\bibitem{matsubara:2008b}
T{.~}{Matsubara}, \emph{\prd} \textbf{78}  (2008), 083519.

\bibitem{matsubara:2008a}
T{.~}{Matsubara}, \emph{\prd} \textbf{77}  (2008), 063530.

\bibitem{mcbride/etal:2011a}
C.K{.~}{McBride}, A.J{.~}{Connolly}, J.P{.~}{Gardner}, R{.~}{Scranton},
  J.A{.~}{Newman}, R{.~}{Scoccimarro}, I{.~}{Zehavi} and D.P{.~}{Schneider},
  \emph{\apj} \textbf{726}  (2011), 13.

\bibitem{mcbride/etal:2011b}
C.K{.~}{McBride}, A.J{.~}{Connolly}, J.P{.~}{Gardner}, R{.~}{Scranton},
  R{.~}{Scoccimarro}, A.A{.~}{Berlind}, F{.~}{Mar{\'{\i}}n} and
  D.P{.~}{Schneider}, \emph{\apj} \textbf{739}  (2011), 85.

\bibitem{mccullagh/jeong}
N{.~}McCullagh and D{.~}Jeong, \emph{in prep.}   (2015).

\bibitem{mcdonald:2003}
P{.~}{McDonald}, \emph{\apj} \textbf{585}  (2003), 34.

\bibitem{mcdonald/roy:2009}
P{.~}{McDonald} and A{.~}{Roy}, \emph{\jcap} \textbf{8}  (2009), 20.

\bibitem{miyatake/etal:2013}
H{.~}{Miyatake}, S{.~}{More}, R{.~}{Mandelbaum}, M{.~}{Takada},
  D.N{.~}{Spergel}, J.P{.~}{Kneib}, D.P{.~}{Schneider}, J{.~}{Brinkmann} and
  J.R{.~}{Brownstein}, \emph{ArXiv e-prints}   (2013).

\bibitem{mo/white:1996}
H.J{.~}{Mo} and S.D.M{.~}{White}, \emph{\mnras} \textbf{282}  (1996), 347.

\bibitem{mohammed/seljak:2014}
I{.~}{Mohammed} and U{.~}{Seljak}, \emph{\mnras} \textbf{445}  (2014), 3382.

\bibitem{more/etal:2014}
S{.~}{More}, H{.~}{Miyatake}, R{.~}{Mandelbaum}, M{.~}{Takada}, D{.~}{Spergel},
  J{.~}{Brownstein} and D.P{.~}{Schneider}, \emph{ArXiv e-prints}   (2014).

\bibitem{navarro/frenk/white:1997}
J.F{.~}{Navarro}, C.S{.~}{Frenk} and S.D.M{.~}{White}, \emph{\apj} \textbf{490}
   (1997), 493.

\bibitem{nishimichi/etal:2007}
T{.~}{Nishimichi}, I{.~}{Kayo}, C{.~}{Hikage}, K{.~}{Yahata}, A{.~}{Taruya},
  Y.P{.~}{Jing}, R.K{.~}{Sheth} and Y{.~}{Suto}, \emph{\pasj} \textbf{59}
  (2007), 93.

\bibitem{nishimichi/valageas:2014}
T{.~}{Nishimichi} and P{.~}{Valageas}, \emph{\prd} \textbf{90}  (2014), 023546.

\bibitem{nishimichi/valageas:2015}
T{.~}{Nishimichi} and P{.~}{Valageas}, \emph{ArXiv e-prints}   (2015).

\bibitem{padmanabhan:1993/struform}
T{.~}{Padmanabhan}: \emph{{Structure Formation in the Universe}}, May 1993.

\bibitem{pajer/schmidt/zaldarriaga:2013}
E{.~}{Pajer}, F{.~}{Schmidt} and M{.~}{Zaldarriaga}, \emph{\prd} \textbf{88}
  (2013), 083502.

\bibitem{pajer/zaldarriaga:2013}
E{.~}{Pajer} and M{.~}{Zaldarriaga}, \emph{\jcap} \textbf{8}  (2013), 37.

\bibitem{peebles:1973}
P.J.E{.~}{Peebles}, \emph{\apj} \textbf{185}  (1973), 413.

\bibitem{peebles:1974}
P.J.E{.~}{Peebles}, \emph{\aap} \textbf{32}  (1974), 391.

\bibitem{peloso/pietroni:2013}
M{.~}{Peloso} and M{.~}{Pietroni}, \emph{\jcap} \textbf{5}  (2013), 31.

\bibitem{planck/nonG:2015}
{Planck Collaboration}, P.A.R{.~}{Ade}, N{.~}{Aghanim}, M{.~}{Arnaud},
  F{.~}{Arroja}, M{.~}{Ashdown}, J{.~}{Aumont}, C{.~}{Baccigalupi},
  M{.~}{Ballardini}, A.J{.~}{Banday} and et~al., \emph{ArXiv e-prints}
  (2015).

\bibitem{roukema/etal:2015}
B.F{.~}{Roukema}, T{.~}{Buchert}, J.J{.~}{Ostrowski} and M.J{.~}{France},
  \emph{\mnras} \textbf{448}  (2015), 1660.

\bibitem{samushia/etal:2014}
L{.~}{Samushia}, B.A{.~}{Reid}, M{.~}{White}, W.J{.~}{Percival},
  A.J{.~}{Cuesta}, G.B{.~}{Zhao}, A.J{.~}{Ross}, M{.~}{Manera},
  {\'E}{.~}{Aubourg}, F{.~}{Beutler}, J{.~}{Brinkmann}, J.R{.~}{Brownstein},
  K.S{.~}{Dawson}, D.J{.~}{Eisenstein}, S{.~}{Ho}, K{.~}{Honscheid},
  C{.~}{Maraston}, F{.~}{Montesano}, R.C{.~}{Nichol}, N.A{.~}{Roe},
  N.P{.~}{Ross}, A.G{.~}{S{\'a}nchez}, D.J{.~}{Schlegel}, D.P{.~}{Schneider},
  A{.~}{Streblyanska}, D{.~}{Thomas}, J.L{.~}{Tinker}, D.A{.~}{Wake},
  B.A{.~}{Weaver} and I{.~}{Zehavi}, \emph{\mnras} \textbf{439}  (2014), 3504.

\bibitem{sanchez/etal:2014}
A.G{.~}{S{\'a}nchez}, F{.~}{Montesano}, E.A{.~}{Kazin}, E{.~}{Aubourg},
  F{.~}{Beutler}, J{.~}{Brinkmann}, J.R{.~}{Brownstein}, A.J{.~}{Cuesta},
  K.S{.~}{Dawson}, D.J{.~}{Eisenstein}, S{.~}{Ho}, K{.~}{Honscheid},
  M{.~}{Manera}, C{.~}{Maraston}, C.K{.~}{McBride}, W.J{.~}{Percival},
  A.J{.~}{Ross}, L{.~}{Samushia}, D.J{.~}{Schlegel}, D.P{.~}{Schneider},
  R{.~}{Skibba}, D{.~}{Thomas}, J.L{.~}{Tinker}, R{.~}{Tojeiro}, D.A{.~}{Wake},
  B.A{.~}{Weaver}, M{.~}{White} and I{.~}{Zehavi}, \emph{\mnras} \textbf{440}
  (2014), 2692.

\bibitem{sato:1981}
K{.~}{Sato}, \emph{Physics Letters B} \textbf{99}  (1981), 66.

\bibitem{sato/etal:2013}
T{.~}{Sato}, G{.~}{H{\"u}tsi}, G{.~}{Nakamura} and K{.~}{Yamamoto},
  \emph{International Journal of Astronomy and Astrophysics} \textbf{3}
  (2013), 243.

\bibitem{sato/huetsi/yamamoto:2011}
T{.~}{Sato}, G{.~}{H{\"u}tsi} and K{.~}{Yamamoto}, \emph{Progress of
  Theoretical Physics} \textbf{125}  (2011), 187.

\bibitem{saunders/etal:1991}
W{.~}{Saunders}, C{.~}{Frenk}, M{.~}{Rowan-Robinson}, A{.~}{Lawrence} and
  G{.~}{Efstathiou}, \emph{\nat} \textbf{349}  (1991), 32.

\bibitem{schaye/etal:2015}
J{.~}{Schaye}, R.A{.~}{Crain}, R.G{.~}{Bower}, M{.~}{Furlong}, M{.~}{Schaller},
  T{.~}{Theuns}, C{.~}{Dalla Vecchia}, C.S{.~}{Frenk}, I.G{.~}{McCarthy},
  J.C{.~}{Helly}, A{.~}{Jenkins}, Y.M{.~}{Rosas-Guevara}, S.D.M{.~}{White},
  M{.~}{Baes}, C.M{.~}{Booth}, P{.~}{Camps}, J.F{.~}{Navarro}, Y{.~}{Qu},
  A{.~}{Rahmati}, T{.~}{Sawala}, P.A{.~}{Thomas} and J{.~}{Trayford},
  \emph{\mnras} \textbf{446}  (2015), 521.

\bibitem{schmidt/jeong/desjacques:2013}
F{.~}{Schmidt}, D{.~}{Jeong} and V{.~}{Desjacques}, \emph{\prd} \textbf{88}
  (2013), 023515.

\bibitem{schmidt/etal:2009}
F{.~}{Schmidt}, M{.~}{Lima}, H{.~}{Oyaizu} and W{.~}{Hu}, \emph{\prd}
  \textbf{79}  (2009), 083518.

\bibitem{scoccimarro/etal:1998}
R{.~}{Scoccimarro}, S{.~}{Colombi}, J.N{.~}{Fry}, J.A{.~}{Frieman},
  E{.~}{Hivon} and A{.~}{Melott}, \emph{\apj} \textbf{496}  (1998), 586.

\bibitem{scoccimarro/couchman:2001}
R{.~}{Scoccimarro} and H.M.P{.~}{Couchman}, \emph{\mnras} \textbf{325}  (2001),
  1312.

\bibitem{scoccimarro/couchman/frieman:1999}
R{.~}{Scoccimarro}, H.M.P{.~}{Couchman} and J.A{.~}{Frieman}, \emph{\apj}
  \textbf{517}  (1999), 531.

\bibitem{scoccimarro/etal:2001}
R{.~}{Scoccimarro}, H.A{.~}{Feldman}, J.N{.~}{Fry} and J.A{.~}{Frieman},
  \emph{\apj} \textbf{546}  (2001), 652.

\bibitem{scoccimarro/sheth:2002}
R{.~}{Scoccimarro} and R.K{.~}{Sheth}, \emph{\mnras} \textbf{329}  (2002), 629.

\bibitem{sefusatti/komatsu:2007}
E{.~}{Sefusatti} and E{.~}{Komatsu}, \emph{\prd} \textbf{76}  (2007), 083004.

\bibitem{seljak/vlah:2015}
U{.~}{Seljak} and Z{.~}{Vlah}, \emph{ArXiv e-prints}   (2015).

\bibitem{shane/wirtanen:1967}
C.D{.~}{Shane} and C.A{.~}{Wirtanen}: \emph{{The distribution of galaxies}},
  1967.

\bibitem{sherwin/zaldarriaga:2012}
B.D{.~}{Sherwin} and M{.~}{Zaldarriaga}, \emph{\prd} \textbf{85}  (2012),
  103523.

\bibitem{sheth/chan/scoccimarro:2013}
R.K{.~}{Sheth}, K.C{.~}{Chan} and R{.~}{Scoccimarro}, \emph{\prd} \textbf{87}
  (2013), 083002.

\bibitem{sheth/tormen:1999}
R.K{.~}{Sheth} and G{.~}{Tormen}, \emph{\mnras} \textbf{308}  (1999), 119.

\bibitem{sirko:2005}
E{.~}{Sirko}, \emph{\apj} \textbf{634}  (2005), 728.

\bibitem{slosar/etal:2008}
A{.~}{Slosar}, C{.~}{Hirata}, U{.~}{Seljak}, S{.~}{Ho} and N{.~}{Padmanabhan},
  \emph{\jcap} \textbf{8}  (2008), 31.

\bibitem{smith/etal:2003}
R.E{.~}{Smith}, J.A{.~}{Peacock}, A{.~}{Jenkins}, S.D.M{.~}{White},
  C.S{.~}{Frenk}, F.R{.~}{Pearce}, P.A{.~}{Thomas}, G{.~}{Efstathiou} and
  H.M.P{.~}{Couchman}, \emph{\mnras} \textbf{341}  (2003), 1311.

\bibitem{springel:2005}
V{.~}{Springel}, \emph{\mnras} \textbf{364}  (2005), 1105.

\bibitem{springel/etal:2005}
V{.~}{Springel}, S.D.M{.~}{White}, A{.~}{Jenkins}, C.S{.~}{Frenk},
  N{.~}{Yoshida}, L{.~}{Gao}, J{.~}{Navarro}, R{.~}{Thacker}, D{.~}{Croton},
  J{.~}{Helly}, J.A{.~}{Peacock}, S{.~}{Cole}, P{.~}{Thomas}, H{.~}{Couchman},
  A{.~}{Evrard}, J{.~}{Colberg} and F{.~}{Pearce}, \emph{\nat} \textbf{435}
  (2005), 629.

\bibitem{stadel/etal:2009}
J{.~}{Stadel}, D{.~}{Potter}, B{.~}{Moore}, J{.~}{Diemand}, P{.~}{Madau},
  M{.~}{Zemp}, M{.~}{Kuhlen} and V{.~}{Quilis}, \emph{\mnras} \textbf{398}
  (2009), L21.

\bibitem{takada/etal:2014}
M{.~}{Takada}, R.S{.~}{Ellis}, M{.~}{Chiba}, J.E{.~}{Greene}, H{.~}{Aihara},
  N{.~}{Arimoto}, K{.~}{Bundy}, J{.~}{Cohen}, O{.~}{Dor{\'e}}, G{.~}{Graves},
  J.E{.~}{Gunn}, T{.~}{Heckman}, C.M{.~}{Hirata}, P{.~}{Ho}, J.P{.~}{Kneib},
  O.L{.~}{F{\`e}vre}, L{.~}{Lin}, S{.~}{More}, H{.~}{Murayama}, T{.~}{Nagao},
  M{.~}{Ouchi}, M{.~}{Seiffert}, J.D{.~}{Silverman}, L{.~}{Sodr{\'e}},
  D.N{.~}{Spergel}, M.A{.~}{Strauss}, H{.~}{Sugai}, Y{.~}{Suto}, H{.~}{Takami}
  and R{.~}{Wyse}, \emph{\pasj} \textbf{66}  (2014), 1.

\bibitem{takada/hu:2013}
M{.~}{Takada} and W{.~}{Hu}, \emph{\prd} \textbf{87}  (2013), 123504.

\bibitem{tasinato/etal:2014}
G{.~}{Tasinato}, M{.~}{Tellarini}, A.J{.~}{Ross} and D{.~}{Wands}, \emph{\jcap}
  \textbf{3}  (2014), 32.

\bibitem{tassev/zaldarriaga/eisenstein:2013}
S{.~}{Tassev}, M{.~}{Zaldarriaga} and D.J{.~}{Eisenstein}, \emph{\jcap}
  \textbf{6}  (2013), 36.

\bibitem{tegmark/taylor/heavens:1997}
M{.~}{Tegmark}, A.N{.~}{Taylor} and A.F{.~}{Heavens}, \emph{\apj} \textbf{480}
  (1997), 22.

\bibitem{teyssier/etal:2009}
R{.~}{Teyssier}, S{.~}{Pires}, S{.~}{Prunet}, D{.~}{Aubert}, C{.~}{Pichon},
  A{.~}{Amara}, K{.~}{Benabed}, S{.~}{Colombi}, A{.~}{Refregier} and
  J.L{.~}{Starck}, \emph{\aap} \textbf{497}  (2009), 335.

\bibitem{des:2005}
{The Dark Energy Survey Collaboration}, \emph{ArXiv Astrophysics e-prints}
  (2005).

\bibitem{tinker/etal:2008}
J{.~}{Tinker}, A.V{.~}{Kravtsov}, A{.~}{Klypin}, K{.~}{Abazajian},
  M{.~}{Warren}, G{.~}{Yepes}, S{.~}{Gottl{\"o}ber} and D.E{.~}{Holz},
  \emph{\apj} \textbf{688}  (2008), 709.

\bibitem{tojeiro/etal:2014}
R{.~}{Tojeiro}, A.J{.~}{Ross}, A{.~}{Burden}, L{.~}{Samushia}, M{.~}{Manera},
  W.J{.~}{Percival}, F{.~}{Beutler}, J{.~}{Brinkmann}, J.R{.~}{Brownstein},
  A.J{.~}{Cuesta}, K{.~}{Dawson}, D.J{.~}{Eisenstein}, S{.~}{Ho},
  C{.~}{Howlett}, C.K{.~}{McBride}, F{.~}{Montesano}, M.D{.~}{Olmstead},
  J.K{.~}{Parejko}, B{.~}{Reid}, A.G{.~}{S{\'a}nchez}, D.J{.~}{Schlegel},
  D.P{.~}{Schneider}, J.L{.~}{Tinker}, M.V{.~}{Maga{\~n}a} and M{.~}{White},
  \emph{\mnras} \textbf{440}  (2014), 2222.

\bibitem{valageas:2014}
P{.~}{Valageas}, \emph{\prd} \textbf{89}  (2014), 123522.

\bibitem{valageas:2014a}
P{.~}{Valageas}, \emph{\prd} \textbf{89}  (2014), 083534.

\bibitem{verde/etal:2002}
L{.~}{Verde}, A.F{.~}{Heavens}, W.J{.~}{Percival}, S{.~}{Matarrese},
  C.M{.~}{Baugh}, J{.~}{Bland-Hawthorn}, T{.~}{Bridges}, R{.~}{Cannon},
  S{.~}{Cole}, M{.~}{Colless}, C{.~}{Collins}, W{.~}{Couch}, G{.~}{Dalton},
  R{.~}{De Propris}, S.P{.~}{Driver}, G{.~}{Efstathiou}, R.S{.~}{Ellis},
  C.S{.~}{Frenk}, K{.~}{Glazebrook}, C{.~}{Jackson}, O{.~}{Lahav},
  I{.~}{Lewis}, S{.~}{Lumsden}, S{.~}{Maddox}, D{.~}{Madgwick}, P{.~}{Norberg},
  J.A{.~}{Peacock}, B.A{.~}{Peterson}, W{.~}{Sutherland} and K{.~}{Taylor},
  \emph{\mnras} \textbf{335}  (2002), 432.

\bibitem{vogeley/etal:1992}
M.S{.~}{Vogeley}, C{.~}{Park}, M.J{.~}{Geller} and J.P{.~}{Huchra},
  \emph{\apjl} \textbf{391}  (1992), L5.

\bibitem{vogelsberger/etal:2014}
M{.~}{Vogelsberger}, S{.~}{Genel}, V{.~}{Springel}, P{.~}{Torrey},
  D{.~}{Sijacki}, D{.~}{Xu}, G{.~}{Snyder}, D{.~}{Nelson} and L{.~}{Hernquist},
  \emph{\mnras} \textbf{444}  (2014), 1518.

\bibitem{wagner/etal:2015a}
C{.~}{Wagner}, F{.~}{Schmidt}, C.T{.~}{Chiang} and E{.~}{Komatsu},
  \emph{\mnras} \textbf{448}  (2015), L11.

\bibitem{wagner/etal:2015b}
C{.~}{Wagner}, F{.~}{Schmidt}, C.T{.~}{Chiang} and E{.~}{Komatsu}, \emph{ArXiv
  e-prints}   (2015).

\bibitem{york/etal:2000}
D.G{.~}{York}, J{.~}{Adelman}, J.E{.~}{Anderson}, Jr., S.F{.~}{Anderson},
  J{.~}{Annis}, N.A{.~}{Bahcall}, J.A{.~}{Bakken}, R{.~}{Barkhouser},
  S{.~}{Bastian}, E{.~}{Berman}, W.N{.~}{Boroski}, S{.~}{Bracker},
  C{.~}{Briegel}, J.W{.~}{Briggs}, J{.~}{Brinkmann}, R{.~}{Brunner},
  S{.~}{Burles}, L{.~}{Carey}, M.A{.~}{Carr}, F.J{.~}{Castander}, B{.~}{Chen},
  P.L{.~}{Colestock}, A.J{.~}{Connolly}, J.H{.~}{Crocker}, I{.~}{Csabai},
  P.C{.~}{Czarapata}, J.E{.~}{Davis}, M{.~}{Doi}, T{.~}{Dombeck},
  D{.~}{Eisenstein}, N{.~}{Ellman}, B.R{.~}{Elms}, M.L{.~}{Evans}, X{.~}{Fan},
  G.R{.~}{Federwitz}, L{.~}{Fiscelli}, S{.~}{Friedman}, J.A{.~}{Frieman},
  M{.~}{Fukugita}, B{.~}{Gillespie}, J.E{.~}{Gunn}, V.K{.~}{Gurbani}, E{.~}{de
  Haas}, M{.~}{Haldeman}, F.H{.~}{Harris}, J{.~}{Hayes}, T.M{.~}{Heckman},
  G.S{.~}{Hennessy}, R.B{.~}{Hindsley}, S{.~}{Holm}, D.J{.~}{Holmgren},
  C.h{.~}{Huang}, C{.~}{Hull}, D{.~}{Husby}, S.I{.~}{Ichikawa},
  T{.~}{Ichikawa}, {\v Z}{.~}{Ivezi{\'c}}, S{.~}{Kent}, R.S.J{.~}{Kim},
  E{.~}{Kinney}, M{.~}{Klaene}, A.N{.~}{Kleinman}, S{.~}{Kleinman},
  G.R{.~}{Knapp}, J{.~}{Korienek}, R.G{.~}{Kron}, P.Z{.~}{Kunszt},
  D.Q{.~}{Lamb}, B{.~}{Lee}, R.F{.~}{Leger}, S{.~}{Limmongkol},
  C{.~}{Lindenmeyer}, D.C{.~}{Long}, C{.~}{Loomis}, J{.~}{Loveday},
  R{.~}{Lucinio}, R.H{.~}{Lupton}, B{.~}{MacKinnon}, E.J{.~}{Mannery},
  P.M{.~}{Mantsch}, B{.~}{Margon}, P{.~}{McGehee}, T.A{.~}{McKay},
  A{.~}{Meiksin}, A{.~}{Merelli}, D.G{.~}{Monet}, J.A{.~}{Munn},
  V.K{.~}{Narayanan}, T{.~}{Nash}, E{.~}{Neilsen}, R{.~}{Neswold},
  H.J{.~}{Newberg}, R.C{.~}{Nichol}, T{.~}{Nicinski}, M{.~}{Nonino},
  N{.~}{Okada}, S{.~}{Okamura}, J.P{.~}{Ostriker}, R{.~}{Owen}, A.G{.~}{Pauls},
  J{.~}{Peoples}, R.L{.~}{Peterson}, D{.~}{Petravick}, J.R{.~}{Pier},
  A{.~}{Pope}, R{.~}{Pordes}, A{.~}{Prosapio}, R{.~}{Rechenmacher},
  T.R{.~}{Quinn}, G.T{.~}{Richards}, M.W{.~}{Richmond}, C.H{.~}{Rivetta},
  C.M{.~}{Rockosi}, K{.~}{Ruthmansdorfer}, D{.~}{Sandford}, D.J{.~}{Schlegel},
  D.P{.~}{Schneider}, M{.~}{Sekiguchi}, G{.~}{Sergey}, K{.~}{Shimasaku},
  W.A{.~}{Siegmund}, S{.~}{Smee}, J.A{.~}{Smith}, S{.~}{Snedden}, R{.~}{Stone},
  C{.~}{Stoughton}, M.A{.~}{Strauss}, C{.~}{Stubbs}, M{.~}{SubbaRao},
  A.S{.~}{Szalay}, I{.~}{Szapudi}, G.P{.~}{Szokoly}, A.R{.~}{Thakar},
  C{.~}{Tremonti}, D.L{.~}{Tucker}, A{.~}{Uomoto}, D{.~}{Vanden Berk},
  M.S{.~}{Vogeley}, P{.~}{Waddell}, S.i{.~}{Wang}, M{.~}{Watanabe},
  D.H{.~}{Weinberg}, B{.~}{Yanny}, N{.~}{Yasuda} and {SDSS Collaboration},
  \emph{\aj} \textbf{120}  (2000), 1579.

\bibitem{zeldovich:1970}
Y.B{.~}{Zel'dovich}, \emph{\aap} \textbf{5}  (1970), 84.

\bibitem{zwicky/etal:1961}
F{.~}{Zwicky}, E{.~}{Herzog}, P{.~}{Wild}, M{.~}{Karpowicz} and C.T{.~}{Kowal}:
  \emph{{Catalogue of galaxies and of clusters of galaxies, Vol. I}}, 1961.

\end{thebibliography}
